\documentclass[preprint,authoryear,11pt]{elsarticle}

\usepackage{epsfig, graphicx, lscape, amsmath, amsbsy, amstext, natbib, multirow}
\usepackage{amsfonts}
\usepackage{srcltx}
\usepackage{amsmath}
\usepackage{titlesec}
\usepackage{natbib}
\usepackage{enumerate,color}
\usepackage{overpic}
\usepackage[left=2.3cm, right=2.3cm, top=1.5cm, bottom=1.95cm]{geometry}
\usepackage{here}
\usepackage{multirow}
\usepackage{colortbl}
\usepackage{xcolor}
\usepackage{xfrac}
\usepackage{caption}
\usepackage{subcaption}
\usepackage{booktabs}
\usepackage{hyperref}
\newcommand{\ve}[1]{{\mbox{\boldmath ${#1}$}}}
\newcommand\bes{\begin{eqnarray}}
\newcommand\ees{\end{eqnarray}}
\newcommand{\ra}[1]{\renewcommand{\arraystretch}{#1}}

\usepackage{soul}

\journal{arXiv}

\begin{document}

\begin{frontmatter}

\title{PPA: Principal Parcellation Analysis for Brain Connectomes and Multiple Traits
 }

\author[RL]{Rongjie Liu}
 \author[ML]{Meng Li}
 \author[DD]{David B. Dunson}
\address[RL]{Department of Statistics, 
 Florida State University, Tallahassee, FL, USA}
 \address[ML]{Department of Statistics, 
 Rice University, Houston, TX, USA}
  \address[DD]{Department of Statistical Science, Duke University, Durham, NC, USA}
\begin{abstract}
Our understanding of the structure of the brain and its relationships with human traits is largely determined by how we represent the structural connectome.  Standard practice divides the brain into regions of interest (ROIs) and represents the connectome as an adjacency matrix having cells measuring connectivity between pairs of ROIs.  Statistical analyses are  then heavily driven by the (largely arbitrary) choice of ROIs.  In this article, we propose a novel tractography-based representation of brain connectomes, which clusters fiber endpoints to define a data adaptive parcellation targeted to explain variation among individuals and predict human traits. This representation leads to Principal Parcellation Analysis (PPA), representing individual brain connectomes by compositional vectors building on a basis system of fiber bundles that captures the connectivity at the population level. PPA reduces subjectivity and facilitates statistical analyses.
We illustrate the proposed approach through applications to data from the Human Connectome Project (HCP) and show that PPA connectomes improve power in predicting human traits over state-of-the-art methods based on classical connectomes, while dramatically improving parsimony and maintaining interpretability. Our PPA package is publicly available on GitHub, and can be implemented routinely for diffusion image data. 
\end{abstract}

\end{frontmatter}

\section{Introduction}
\label{sec:intro}

Image-based brain parcellation is a fundamental tool for understanding brain organization and function, in which the brain is divided into multiple non-overlapping and interacting regions \citep{eickhoff2018imaging}. Modern neuroimaging techniques enable the collection of whole-brain magnetic resonance (MR) scans in large samples of individuals. Several large studies, such as the Human Connectome Project \citep{van2013wu}, have collected such data along with human behavioral and cognitive information, leading to  
 a surge of interest in 
 relating 
  structural brain networks with various human traits \citep{glasser2016human,roine2019reproducibility,lin2020mapping,zhang2019tensor,apkarian2020neural}. A key issue in the understanding of brain connectivity, as noted by~\cite{Park2013}, is the way brain connectivity is measured, represented, and modeled. 
In structural connectomics, the brain connectome is typically defined as corresponding to the collection of white matter fiber tracts. Classically, the intricate spatial 
locations of all the fibers in the brain 
are summarized via an adjacency matrix, with the cells containing summaries of connections between pairs of regions of interest (ROIs) \citep{o2013fiber}. However, as we argue and demonstrate in this paper, there are other ways to summarize the brain connectome that have distinct advantages \citep{sporns2005thehuaman,toga2012mapping}. 
A method does not need to focus on an adjacency matrix representation for it to be an approach for analysis of brain connectomes.
This article contributes to the growing literature of structural connectomics by proposing a new representation of brain connectivity. 

Existing literature on brain parcellation and structural brain networks typically represents structural brain connectivity via anatomical connectivity that is estimated by tractography on diffusion-weighted images \citep{behrens2003non}. The associated anatomical parcellation analysis (APA)---one of the most popular approaches in connectivity-based parcellation analysis \citep{yao2015review}---obtains the connectivity map by calculating connectivities between all pairs of regions in a predetermined anatomical parcellation scheme. Based on the selected atlas, we may build the connectivity matrix by counting the number of fibers passing between each pair of ROIs after fiber tracking. This connectivity matrix can then be used as a matrix-valued predictor in statistical analyses studying relationships with human traits ~\citep{zhang2018mapping,wang2019symmetric,lin2020mapping,de2013parcellation}. 

However, such APA analyses require a particular anatomical ROI definition \citep{eickhoff2015connectivity}, and many different schemes are available involving different numbers and locations of ROIs, such as the automated anatomical labeling (AAL), automatic nonlinear imaging matching and anatomical labeling (ANIMAL) atlases, and many others \citep{tzourio2002automated,he2007small,he2008structural,yao2015review,wang2019symmetric}. 
Choosing which scheme to use in practice is challenging. Several studies have reported impacts of different atlases on brain networks \citep{zalesky2010whole,messe2020parcellation}, and evidence suggests that not only the connectivity maps but also the inferences relating connectomes to human traits 
are strongly sensitive to the parcellation strategy.

An additional major issue is that APA leads to connectome representations corresponding to high-dimensional adjacency matrices.  While there is a growing literature focused on statistical analysis of such replicated graph or network data \citep{schiffler2017cortex,bansal2018personalized}, such methods are under-developed and poorly understood relative to the rich literature on methods for high-dimensional vector-valued predictors, and computationally efficient methods that scale well in practice are lacking. For these reasons, it is common to simply vectorize the upper-triangular part of the adjacency matrix prior to statistical analysis.  However, this fails to exploit the network structure in performing dimension reduction \citep{wang2017bayesian,o2008analysis,hochberg2007whom}, and can suffer from substantial loss of efficiency and accuracy 
~\citep{wang2019symmetric}.

In this article, we propose a novel tractography-based representation of the connectome, which clusters fiber endpoints to define a data adaptive parcellation targeted to explain variation among individuals and predict human traits. This leads to Principal Parcellation Analysis (PPA), representing individual brain connectomes by compositional vectors building on a basis system of fiber bundles that captures the connectivity at the population level. Unlike APA connectomes, PPA connectomes do not rely on any anatomical atlas or choosing ROIs a priori, thus reducing subjectivity and leading to a substantially different representation of the connectome. 
This new representation of structural brain connectomes facilitates statistical analyses using well-established statistical methods designed for vector data.The PPA representation provides an alternative to the current standard ROI-based adjacency matrix representation in analyses studying how structural connectomes vary across individuals, both randomly and in relation to individual traits, and can accomplish these same inference goals at a fraction of the cost for implementing APA-based counterparts.
We illustrate the proposed approach through applications to data from the Human Connectome Project (HCP) and show that PPA connectomes, when combined with standard high-dimensional regression methods, improve power in predicting human traits over state-of-the-art methods based on classical connectomes, while dramatically improving parsimony and maintaining interpretability. 
Our proposed method is general for any data that consist of a collection of fibers. 
Our PPA package is publicly available on GitHub, and can be implemented routinely for diffusion image based parcellation analysis.

The rest of the paper is structured as follows. Section \ref{s:method} introduces
the proposed parcellation approach and PPA formulation.  
In Section \ref{s:realdata}, we compare PPA to state-of-the-art APA-based methods using HCP data, and focus on prediction, visualization, and interpretability. Section 
 \ref{s:discuss} contains a discussion.


\section{Methods}\label{s:method}
\subsection{The PPA framework} \label{sec:PPA} 

\begin{figure}[ht!]
	\centering
	\includegraphics[width=0.9\linewidth]{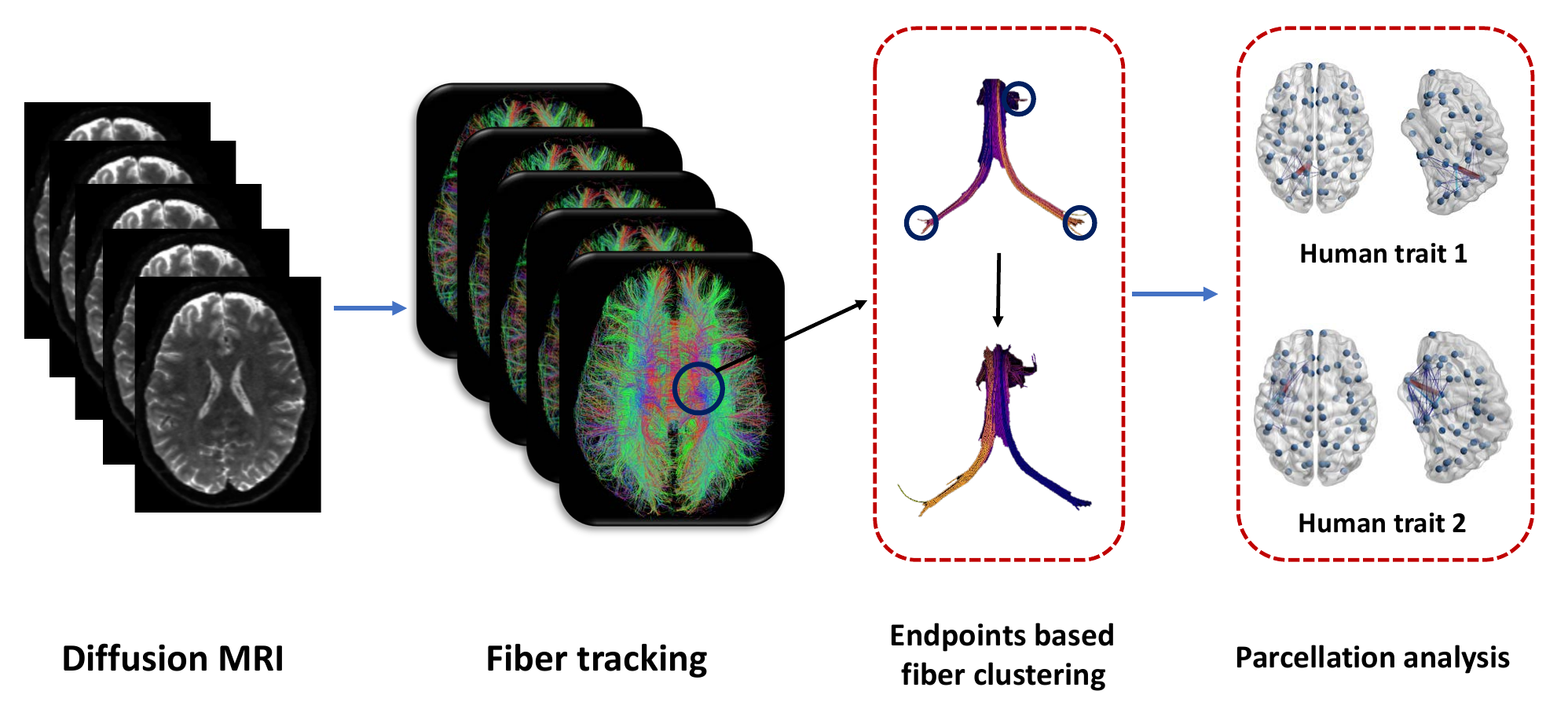}\\
	\caption{Pipeline of tractography-based Principal Parcellation Analysis.}\label{tppa_pipeline}
\end{figure}

Suppose that we observe structural MRI, diffusion MRI, and human traits from $n$ individuals. The PPA pipeline, as illustrated in Figure  \ref{tppa_pipeline}, consists of three modules: (i) reconstruction of fibers; (ii) representation of fibers and unsupervised clustering; and (iii) high-dimensional supervised learning adaptive to human traits. Each module of PPA encompasses a variety of choices, equipping PPA with easy extensibility.    
We first describe each module of PPA using initial default settings, 
 followed by a discussion on extensions. 

Let $\mathcal{F}=\{f_{ik}, k=1,\ldots,m_i, i=1,\ldots,n\}$, where $f_{ik}$ is the $k$-th fiber in the $i$-th individual's brain, and $m_i$ is the total number of fibers in the $i$-th subject. 
In addition, let $y_i(s)$ denote the $s$th `trait' of the $i$th individual with $\ve{y}(s) = (y_1(s),\ldots,y_n(s))^T$ for $s=1,\ldots,S$; traits can range from demographic characteristics, alcohol and drug exposures, to scores on cognitive, psychological and behavioral assessments.

In Module (i), we reconstruct fibers 
using the recently proposed 
TractoFlow method \citep{theaud2019tractoflow}; alternative fiber tracking algorithms can be used instead without changing the subsequent steps in the PPA pipeline.
TractoFlow takes raw diffusion weighted images as the input, consists of 14 steps for the diffusion weighted image (DWI) processing and 8 steps for the T1 weighted image processing, and outputs classical diffusion imaging measures.
The outlier fiber tracts are detected and removed using the method proposed in \cite{garyfallidis2012quickbundles}. 
Module (i) is also a key step in estimating APA connectomes.

In Module (ii), we formulate connectomes at the population level through basis networks in the form of fiber bundles that represent groups/clusters of streamlines, and represent individual connectomes via compositional vectors.  Let $\{\ve{a}_{ik}\}_{k=1}^{m_i}$ and $\{\ve{b}_{ik}\}_{k=1}^{m_i}$ be the 3D coordinates of two endpoints for each fiber from the $i$-th subject. Let $Z$ be a $6\times m$ matrix stacking all $(\ve{a}_{ik}^T,\ve{b}_{ik}^T)^T$ as columns for $k=1,\ldots,m_i$ and $i = 1, \ldots, n$, where $m=\sum_i m_i$ is the total number of fibers from all subjects. We perform a cluster analysis at the fiber level using the matrix $Z$, outputting a collection of partitions of $\mathcal{F}$, denoted by $\mathcal{A}_K=\{{A}_K^{(1)},\ldots,{A}_K^{(K)}\}$, where $K$ is the number of clusters and each set ${A}_K^{(k)}$ can be interpreted as a fiber bundle. The enormous number of fibers presents substantial computational challenges in clustering; for example, traditional $K$-Means does not scale well and requires large memory, leading to prohibitive computational cost. As such, we adopt mini-batch $K$-Means \citep{sculley2010web}, which reduces the memory use and converges to the optimal value orders of magnitude faster than the full-batch $K$-Means. For a fiber and cluster center, we define their distance as the minimum of their Euclidean distances considering two orderings of fiber endpoints, accounting for fiber bi-directionality similarly to fiber flipping~\citep{garyfallidis2012quickbundles}.
The number of clusters $K$ is an important hyperparameter. Our experiments have provided guidance on good values of $K$ in practice.  To automate the choice of $K$, we can rely on cross validation as illustrated in Figure \ref{mse1}.  Alternatively, we can decrease sensitivity to $K$ by choosing multiple values in a multi-scale representation of the brain connectome.

For a given $K$, an individual's connectome can be represented by the proportions of the individual's fibers belonging to each of the inferred population-level fiber bundles.  In particular, the $i$th individual's connectome is represented via the $K$-dimensional compositional vector 
$\ve{\omega}_i=\{\omega_{i1},\ldots,\omega_{iK}\}$, with 
 $\omega_{ik}$ the proportion of fibers belonging to the $k$th fiber bundle $A_K^{(k)}$, for $k=1,\ldots,K$.  The connectome data for all $n$ subjects is then contained in the matrix 
 $\ve{\omega}=(\ve{\omega}_1^T,\cdots,\ve{\omega}_n^T)^T$.  This provides a much simpler representation than the adjacency matrix-based APA approach. Note that the framework does not require all clusters $K$ to be present in all subjects since for certain individuals the number of fibers belonging to a particular cluster can be zero.


In Module (iii), we relate the connectome $\ve{\omega}_i$ to traits $y_i(s)$.  For simplicity in interpretation, we initially focus on trait-specific linear regression models: 
\begin{equation}
{y}_i(s)=\beta_0(s)+\sum_{k=1}^{K-1}\omega_{ik}\beta_k(s)+\epsilon_i(s),\label{tpaa_lm}
\end{equation}
where $\beta_0(s)$ is a trait specific intercept, which can be expanded to include non-connectome covariates, and $\beta_k(s)$ ($k=1,\ldots,K$) is a regression coefficient characterizing the relationship between the density of connections in the $k$th fiber bundle and the $s$th trait.  For a sufficiently flexible specification, one may choose $K$ to be large in which case many of the $\beta_k(s)$ coefficients are expected to be zero or close to zero.  Standard sparse learning methods can be used to estimate the coefficients while learning the sparsity pattern. 
This yields a set of estimated non-zero coefficients $\{\hat{\beta}_k: k \in \mathcal{K}(s)\}$, where $\mathcal{K}(s) = \{k_1, \ldots, k_{m(s)}\} \subset \{1, \ldots, K - 1\}$ collects the indices of 
the $m(s)$ fiber bundles having non-zero coefficients.  We refer to these fiber bundles as ``active'' for the $s$th trait.  Active bundles impact the response $y_i(s)$ via equation (\ref{tpaa_lm}), while inactive bundles have no impact on the response.
 


In our numerical experiments, we use LASSO~\citep{tibshirani1996regression}, one of the most popular high-dimensional regression methods, as a representative example. LASSO has been very widely studied and relatively efficient  algorithms are readily available. However, applying LASSO to the vectorized upper triangle portion of APA connectome adjacency matrices produces less reliable estimation and has worse predictive performance than an identical analysis using PPA instead of APA connectomes. This is consistent with previous results motivating complex statistical methods that take into account the graph structure of the APA connectomes
\citep{wang2019symmetric}.


\subsection{Extensions of PPA} \label{sec:extension}


In Module (i), other fiber tracking algorithms alternatives to TractoFlow can be considered, such as Euler Delta Crossings (EuDX) in \cite{garyfallidis2012quickbundles} and Sparse Fascicle Model (SFM) in \cite{rokem2015evaluating}. We will compare various fiber tracking algorithms in our analyses of HCP data.  In Module (ii), other clustering or factorization methods, including spectral clustering and non-negative matrix factorization (NMF), can also be adopted. In addition to the endpoints, the length and shape of the fibers may contain useful information~\citep{zhang2018mapping}, which can be incorporated in clustering analyses.  
Module (iii) can be modified building on the rich literature on high-dimensional supervised learning methods. Instead of LASSO, other sparse shrinkage methods, such as elastic net \citep{zou2005regularization} and smoothly clipped absolute deviation (SCAD) penalty \citep{fan2001variable}, can be used without complication. 



\section{Human Connectome Project Data Analyses}
\label{s:realdata}

In this section we use the HCP dataset to compare PPA-based methods with state-of-the-art APA-based approaches using various human traits, demonstrate how to choose $K$ in a data-driven manner, assess the robustness of PPA with respect to the fiber tracking algorithms and regularization strategies, and illustrate the interpretability of PPA. 

\subsection{HCP Data Description}

Data collection and sharing for this project was provided by the MGH-USC Human Connectome Project (HCP; Principal Investigators: Bruce Rosen, M.D., Ph.D., Arthur W. Toga, Ph.D., Van J. Weeden, MD). HCP funding was provided by the National Institute of Dental and Craniofacial Research (NIDCR), the National Institute of Mental Health (NIMH), and the National Institute of Neurological Disorders and Stroke (NINDS). HCP data are disseminated by the Laboratory of Neuro Imaging at the University of Southern California. 

We use the same set of 1065 HCP subjects as in \cite{wang2019symmetric}, including dMRI data along with human traits, downloaded from {\it HCP 1200 Subjects Data Release}\footnote{https://www.humanconnectome.org/study/hcp-young-adult/document/1200-subjects-data-release}. Details about the dMRI data acquisition and preprocessing can be found in \cite{van2012human,sotiropoulos2013advances}. For the human traits data, seven different scores were selected: receptive vocabulary, oral reading, list sorting, flanker, picture sequence memory, card sort, and processing speed. These scores can be used to study human cognition. Note that although we use task-based scores in this section, the proposed methods are broadly applicable for any measurement reflecting human traits. All the scores are age-adjusted, and their details can be found on the HCP website\footnote{https://wiki.humanconnectome.org/display/PublicData/}. A brief description of each trait is also included in the Appendix for easy reference. 

\subsection{Analysis using PPA and APA}


PPA and APA provide distinct representations of human brain connectomes. Performance in studying relationships between connectomes and traits depends on the downstream analysis methods after the connectomes are obtained. As such, we chose state-of-the-art methods developed under APA connectomes, and adopted one of the most standard analysis methods, LASSO, under PPA connectomes. Such comparisons give APA an advantage. We also implemented LASSO for the vectorized APA connectomes.

\begin{figure}[ht!]
	\centering
	\includegraphics[width=0.9\linewidth]{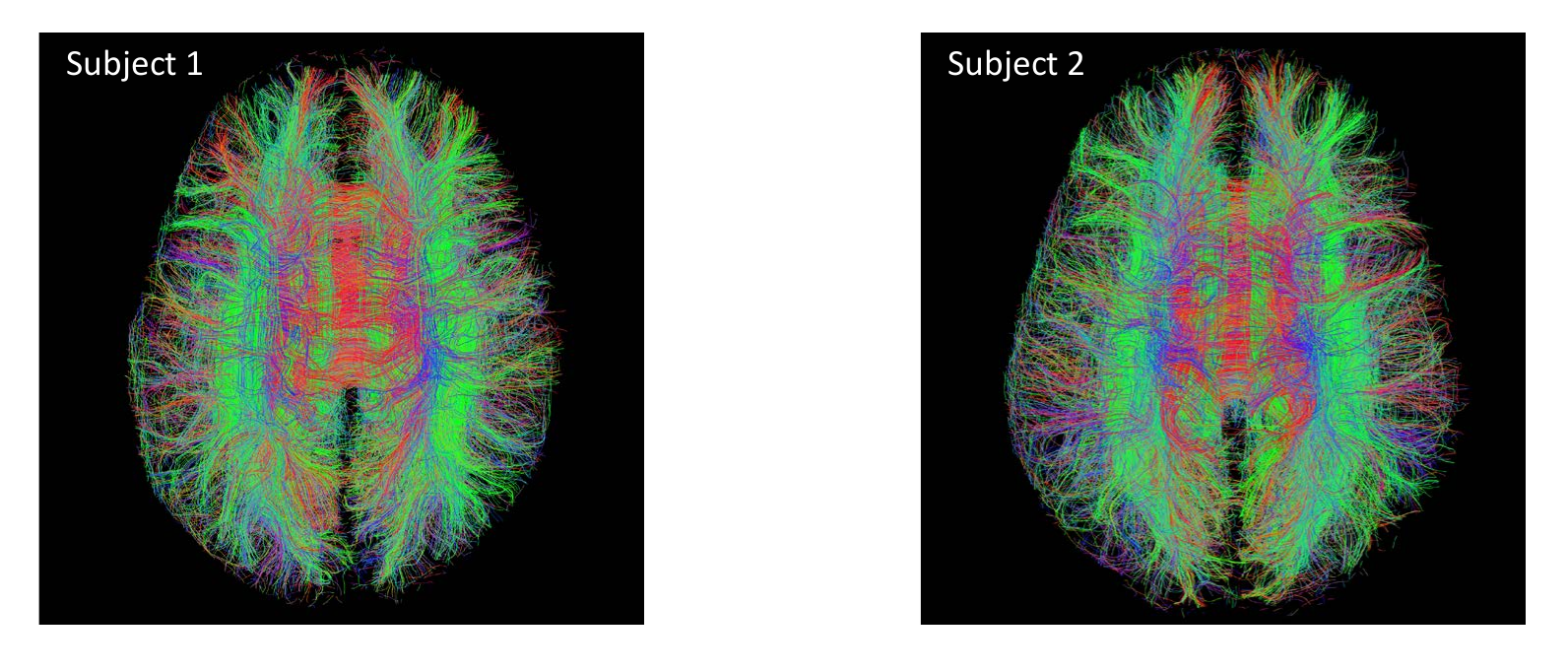}\\
	\caption{Examples of fiber tracking results for two randomly selected HCP subjects.}\label{fiber_tracking}
\end{figure}

We implemented PPA using the default choices in Section~\ref{s:method}. In particular, we used TractoFlow for fiber tracking, which depends on two main technologies: Nextflow and Singularity \citep{kurtzer2017singularity, di2017nextflow, garyfallidis2014dipy, tournier2019mrtrix3, avants2009advanced, jenkinson2012fsl}. 
We obtained around 2.8 billion fibers for the HCP subjects. The fiber tracking results for two randomly selected subjects are displayed in Figure \ref{fiber_tracking}. We clustered the fibers using mini-batch $K$-Means \citep{sculley2010web}. We set the batch size to 1000 and varied the number of clusters $K$ from 10 to 500 
 across $\{10,50,100,200,300,400,500\}$.  For each choice of $K$, we conducted analyses relating the PPA connectome to the trait of interest.  We can then assess, based on cross validation, which choice of $K$ is best for that particular trait.
For each $K$, we obtained $K$ fiber bundles $A_K^{(k)}$ for $k = 1, \ldots, K$, leading to PPA connectome $\ve{\omega}_i$ for each individual $i = 1, \ldots, n.$

\begin{figure}[h!]
\centering 
	\begin{tabular}{ccc}
		 \includegraphics[width=0.3\linewidth, height=5cm]{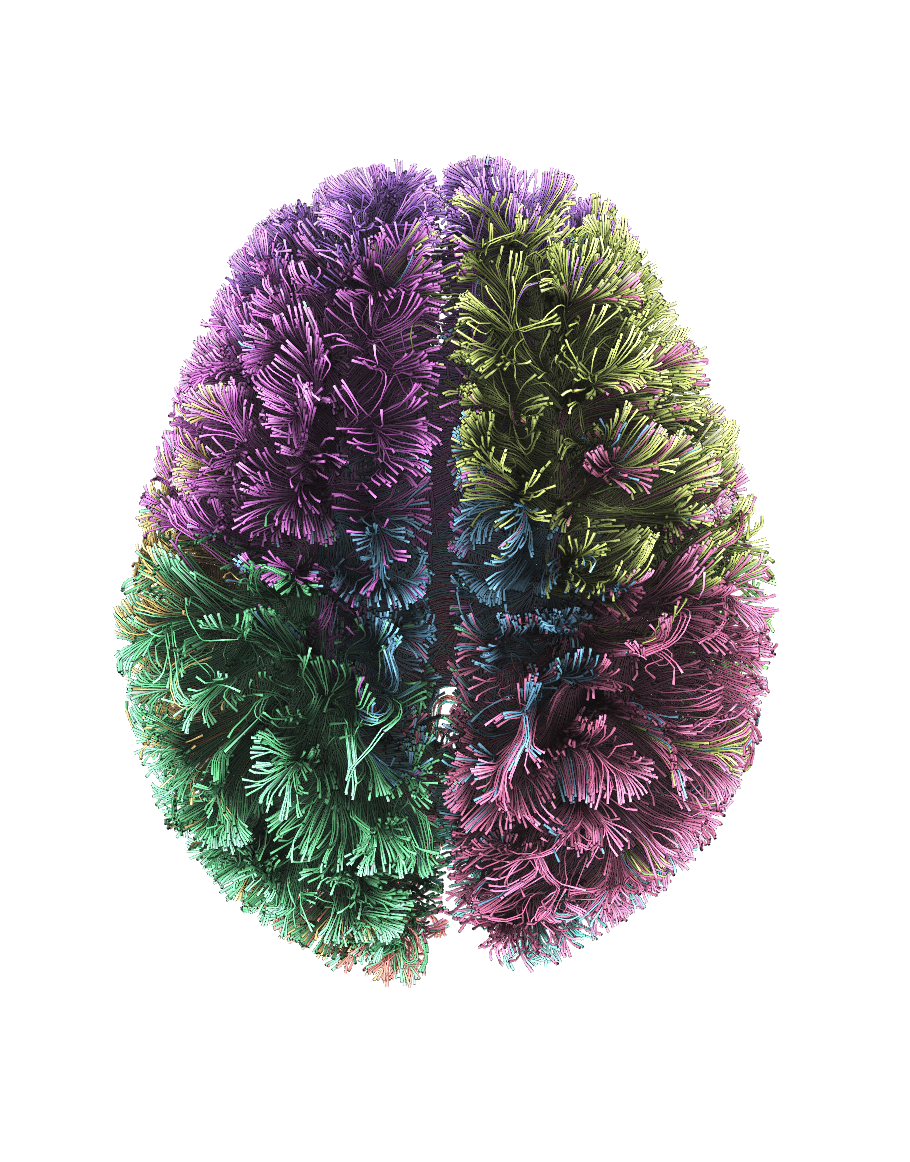} & 
		  \includegraphics[width=0.3\linewidth,height=5cm]{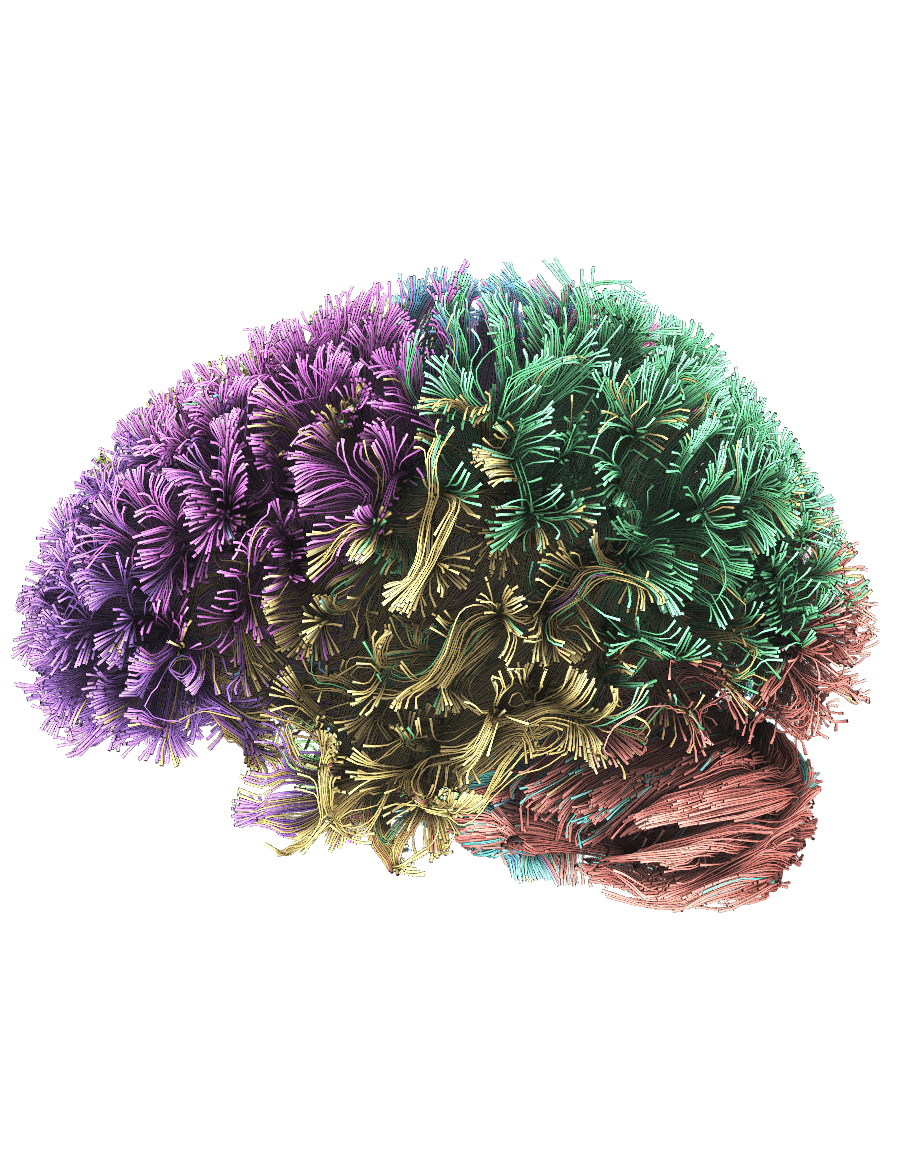}&
		  \includegraphics[width=0.3\linewidth,height=5cm]{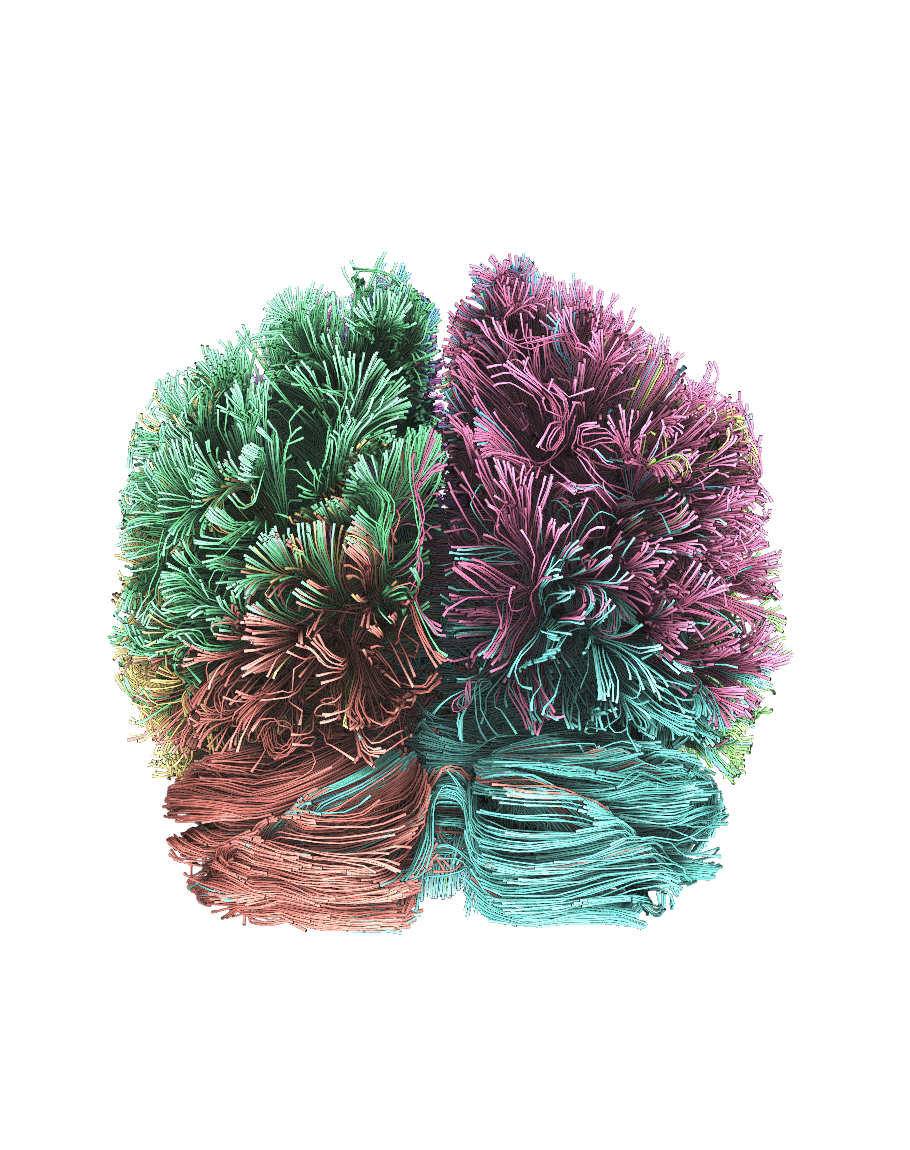}\\
		  \includegraphics[width=0.3\linewidth, height=5cm]{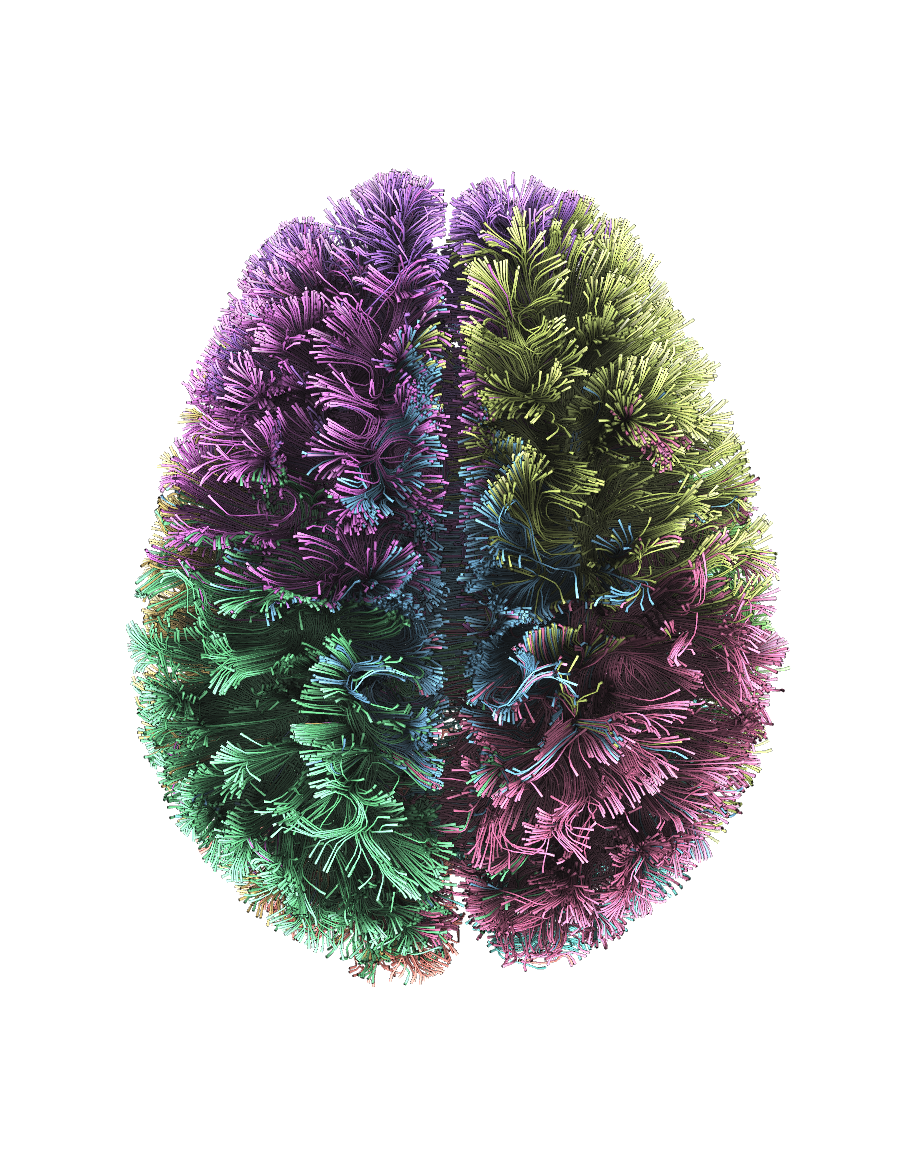} & 
		  \includegraphics[width=0.3\linewidth, height=5cm]{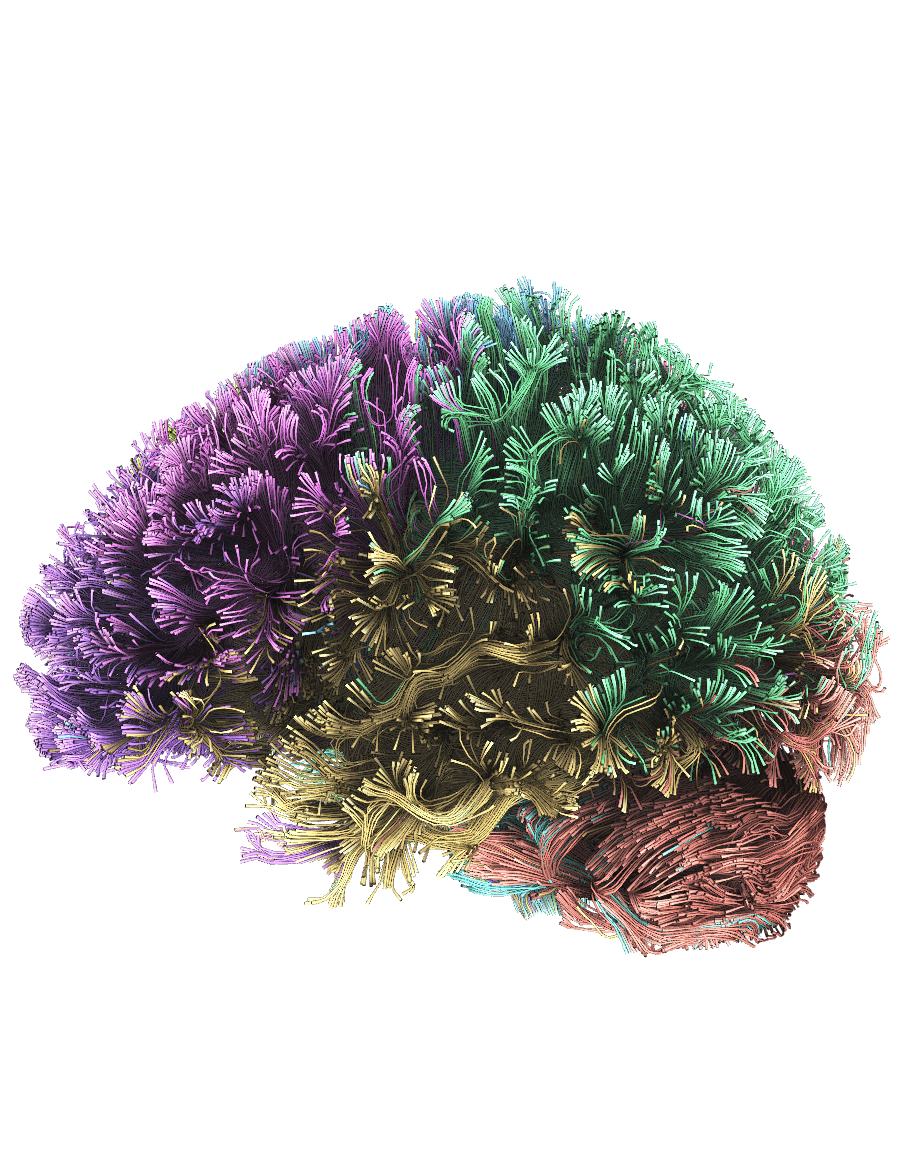}&
		  \includegraphics[width=0.3\linewidth, height=5cm]{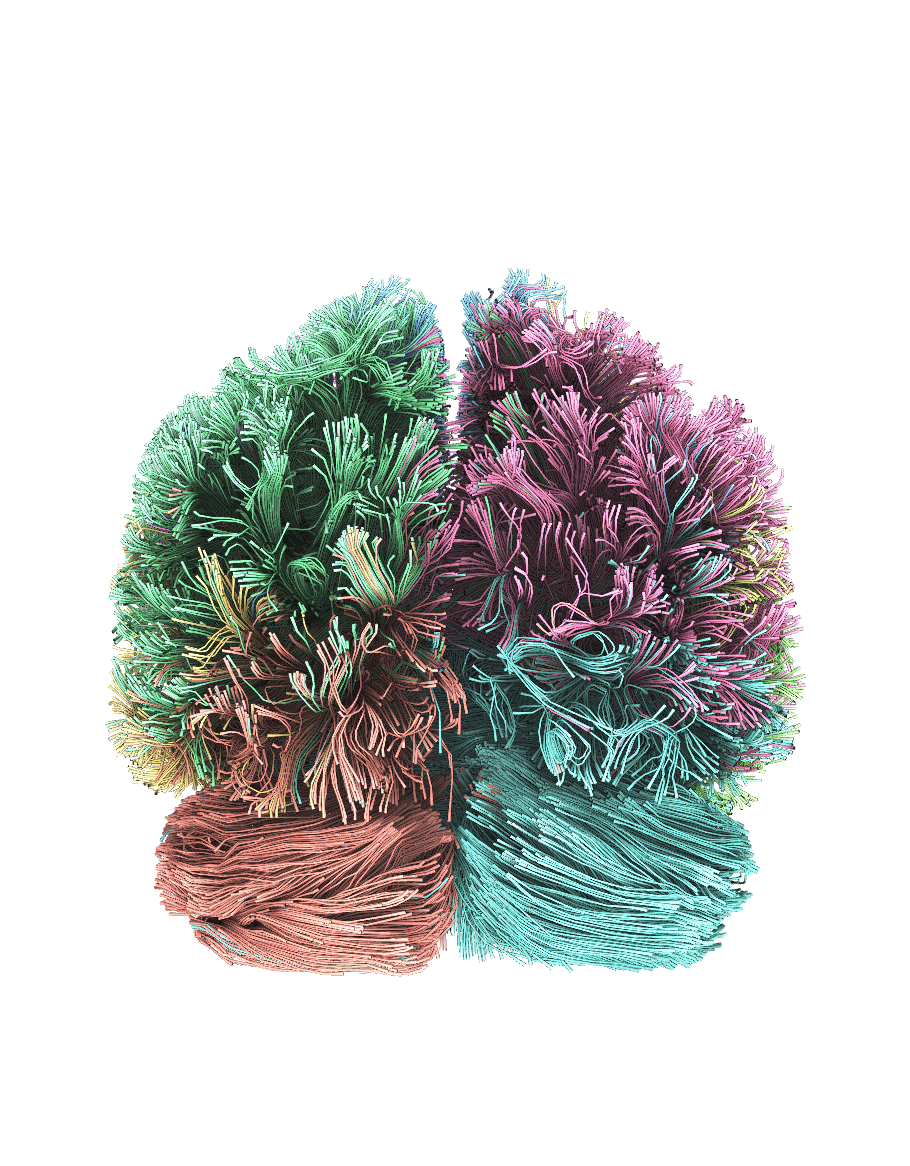}\\
		   {\small (a)  Axial} & {\small (b) Sagittal} & {\small (c) Coronal}\\
		  \end{tabular}
\caption{Fiber clusters ($K=10$) of two subjects (each row represent one subject) at three planes: (a) Axial, (b) Sagittal, and (c) Coronal. In each plane, each color denotes one cluster.}
\label{clusters} 
\end{figure}

\begin{figure}[h!]
\centering 
	\begin{tabular}{cccccccccc}
		 \includegraphics[width=0.09\linewidth, height=10cm]{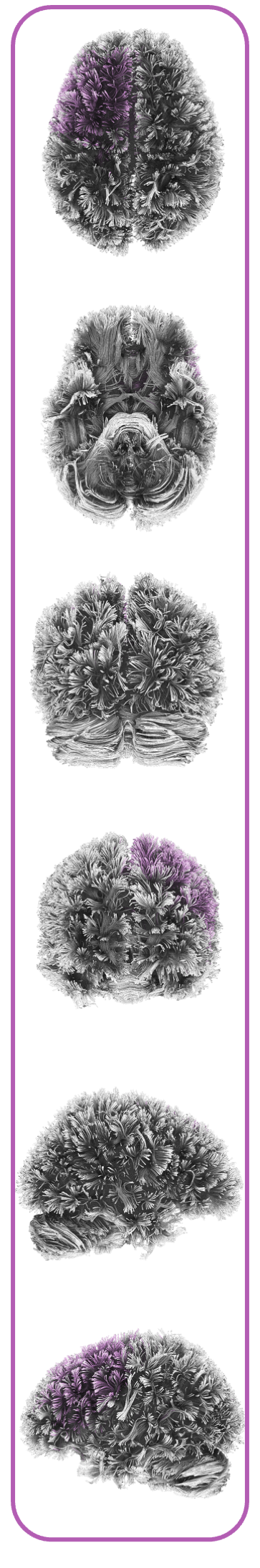}\hspace*{-1em}&
		  \includegraphics[width=0.09\linewidth,height=10cm]{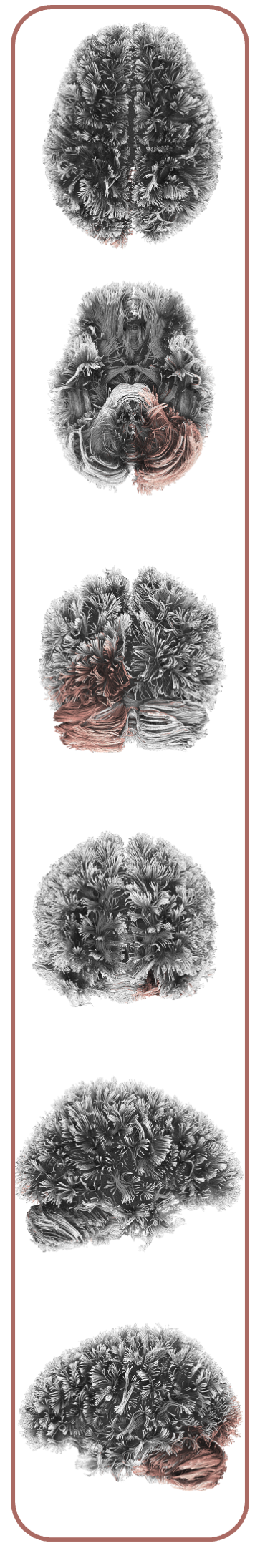}\hspace*{-1em}&
		   \includegraphics[width=0.09\linewidth, height=10cm]{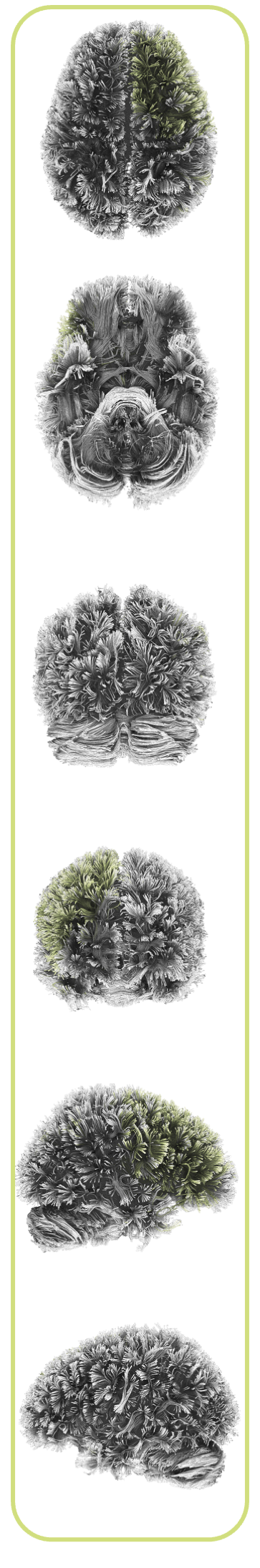}\hspace*{-1em}& 
		  \includegraphics[width=0.09\linewidth,height=10cm]{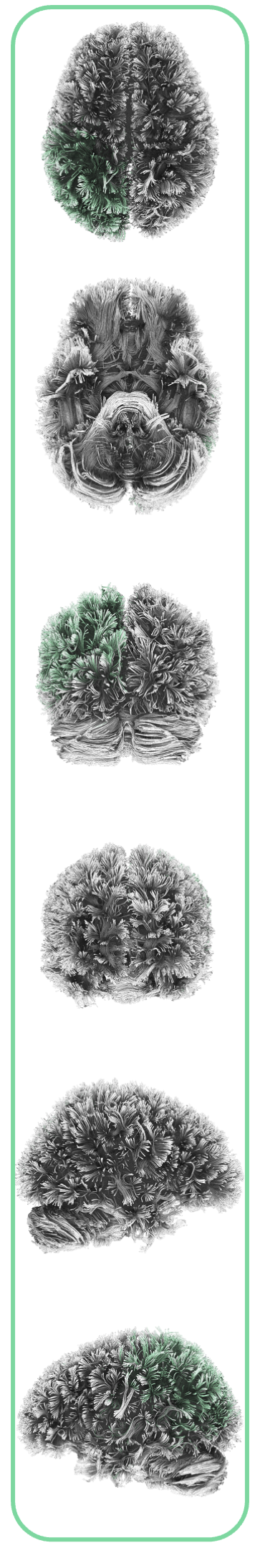}\hspace*{-1em}& 
		  \includegraphics[width=0.09\linewidth,height=10cm]{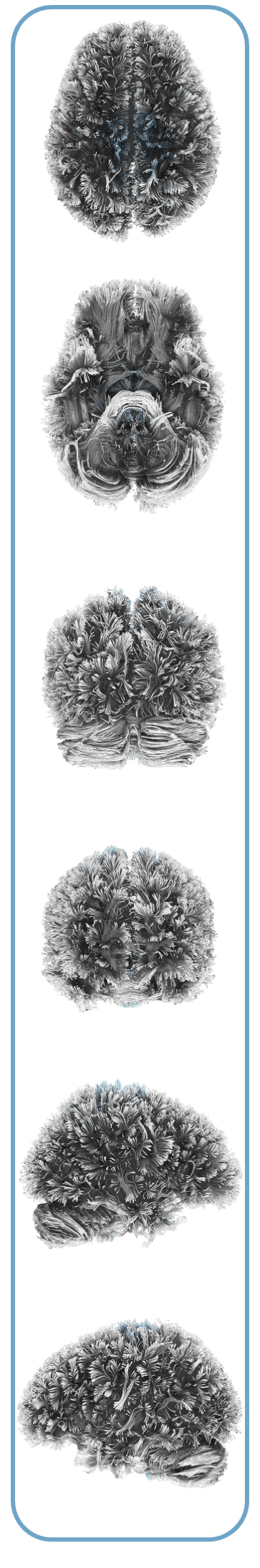}\hspace*{-1em}&
		  \includegraphics[width=0.09\linewidth, height=10cm]{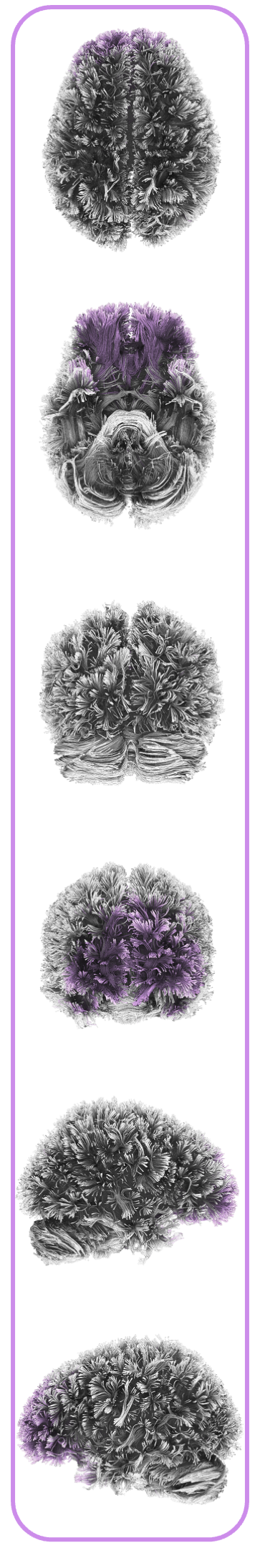}\hspace*{-1em}& 
		  \includegraphics[width=0.09\linewidth,height=10cm]{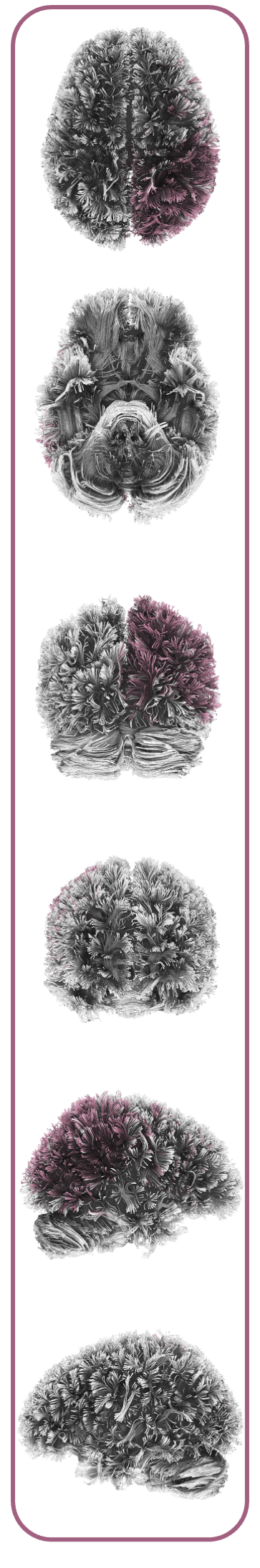}\hspace*{-1em}&
		   \includegraphics[width=0.09\linewidth, height=10cm]{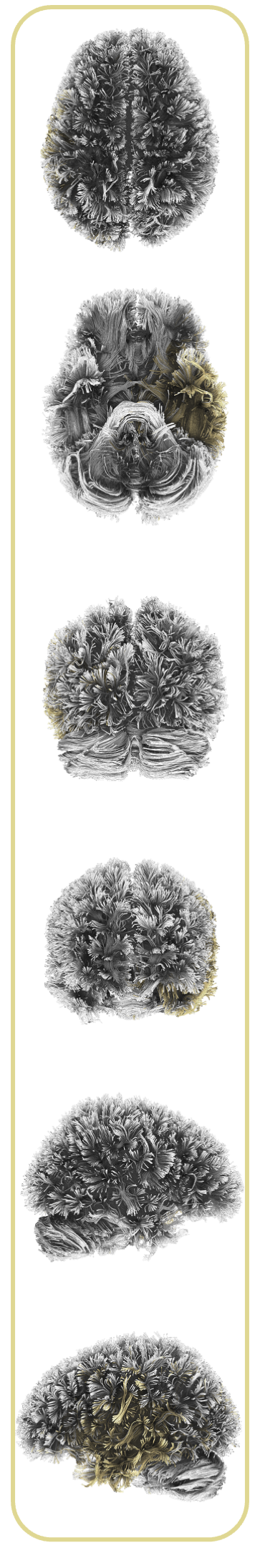}\hspace*{-1em}& 
		  \includegraphics[width=0.09\linewidth,height=10cm]{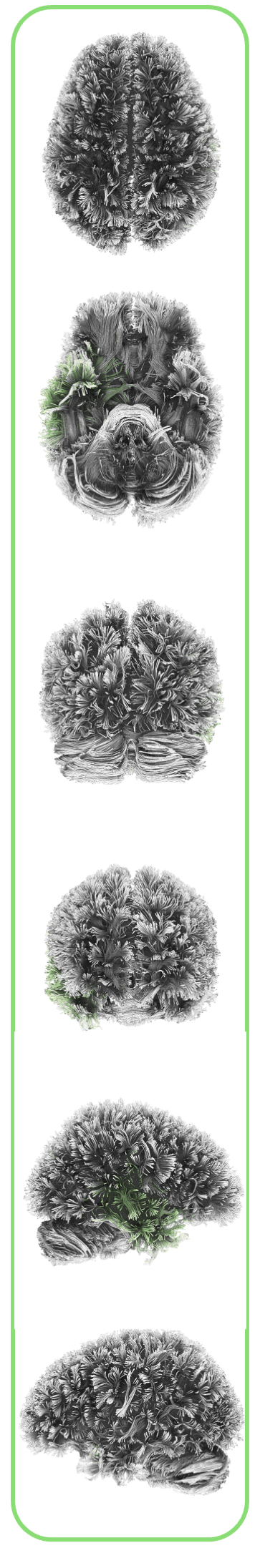}\hspace*{-1em}& 
		  \includegraphics[width=0.09\linewidth,height=10cm]{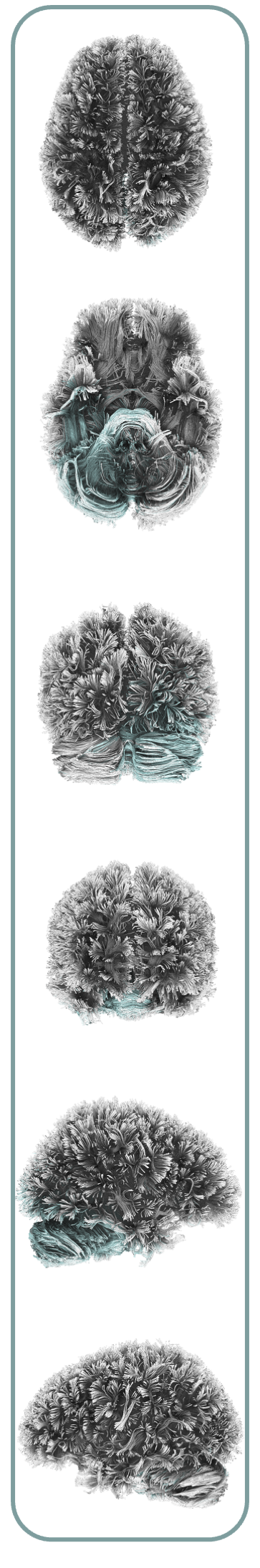}
	\end{tabular}
\caption{An illustration of $K=10$ clusters in one subject at 6 different angles of view (each row denotes one angle of view) where left to right: cluster 1 to cluster 10.}
\label{10clusters} 
\end{figure}

\begin{figure}[h!]
\centering 
	\begin{tabular}{cc}
		\includegraphics[width=0.45\linewidth,height=6cm]{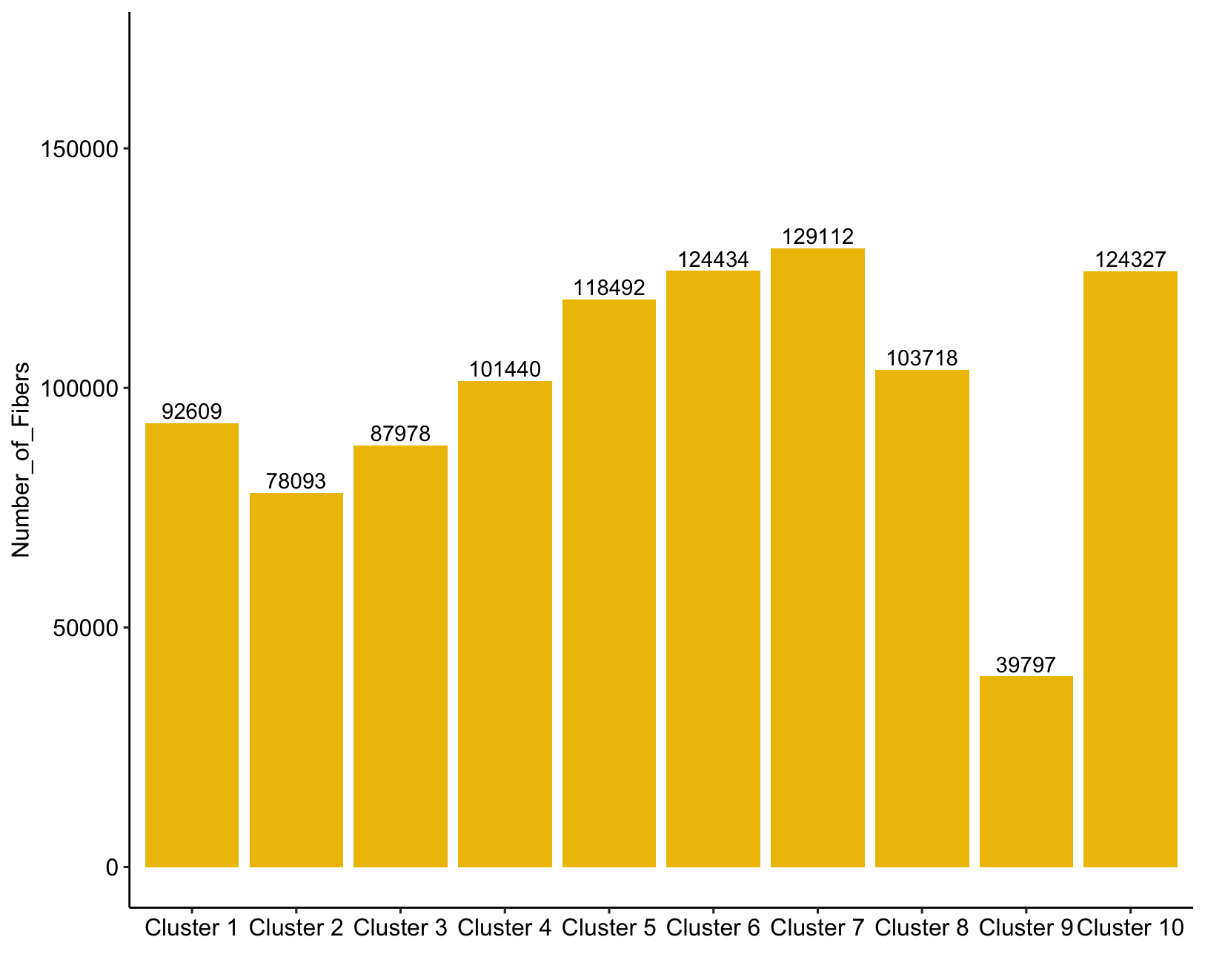}&
		\includegraphics[width=0.45\linewidth,height=6cm]{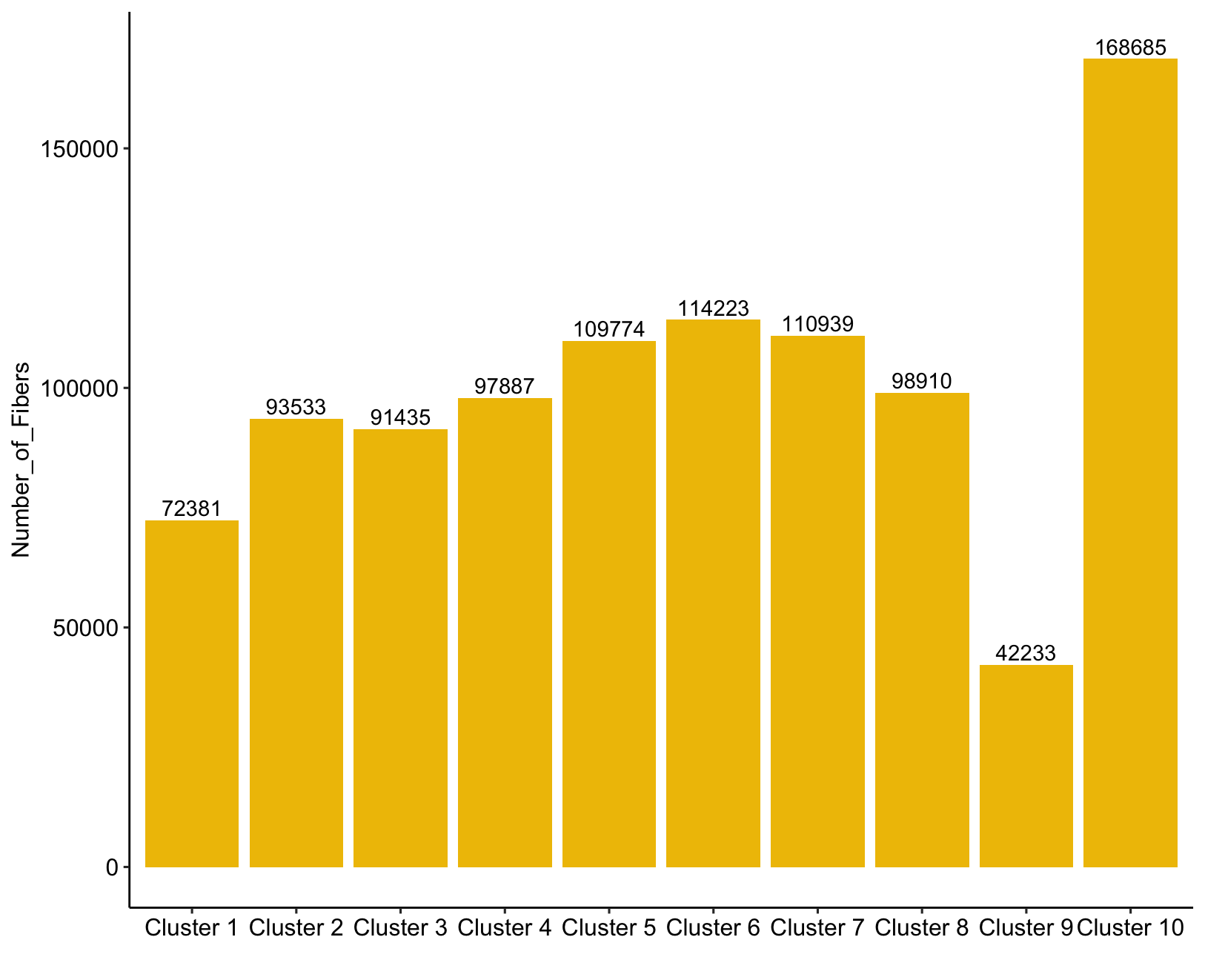}\\
		   {\small (a)  Subject 1} & {\small (b) Subject 2} \\
		  \end{tabular}
\caption{The number of fibers in each cluster for the two subjects from Figure \ref{clusters}.}
\label{sizep} 
\end{figure}

Figures \ref{clusters} and  \ref{10clusters} show examples of the inferred fiber clusters ($K=10$) with each color denoting one cluster. 
The same clusters, and corresponding colors, are used for the different subjects, and some heterogeneity is apparent across subjects. 
The numbers of fibers in each cluster are shown in 
Figure \ref{sizep} for these two subjects. The profile of these counts across clusters is similar for the two subjects, but subject 2 has a considerably greater proportion of fibers in cluster 10. A close inspection of the inferred clusters indicates that the anatomical locations of fiber bundles produced by PPA often correspond to known fascicles in the literature  (\cite{gupta_thomopoulos_rashid_thompson_2017,shin_rowley_chowdhury_jolicoeur_klein_grova_rosa-neto_kobayashi_2019, Friederici2013}).
Taking the 10 clusters in Figure \ref{10clusters} as an example, cluster 2 and cluster 10 cover the {\it inferior longitudinal fasciculus}; cluster 5 overlaps with {\it corticospinal tract}; cluster 3 and cluster 5 contain {\it corpus callosum} and {\it uncinate fasciculus}; cluster 8 and 9 contain {\it superior longitudinal fasciculus}, {\it arcuate fasciculus} and {\it inferior fronto-occipital fasciculus} (\cite{Friederici2013}), which are language function related ROIs.

For APA connectomes, one first chooses brain ROIs according to an atlas template, and then selects a summary of connectivity between each pair of ROIs, such as the number of connections.
 APA-based methods represent the $i$th individual's brain connectome as a $p \times p$ matrix $W_i$. Each cell of this matrix contains a summary of the strength of connection between a pair of brain ROIs; here, we use the number of fibers connecting the regions. Different atlas templates lead to different connectivity matrices having different dimensionality $p$. We used the same fiber tracking method TractoFlow, and chose the HCP842 tractography atlas \citep{yeh2018}, which segments the brain into 80 regions; see Tables S1 
and S2 
in the Appendix for descriptions of these 80 regions. Note that one can also obtain different count values using different tractography algorithms \citep{knosche2015validation} as well as many filtering approaches, e.g., scale-invariant feature transform (SIFT) \citep{burger2016scale} to make streamline counting more quantitative.

For PPA connectomes, we used LASSO to fit Model~\eqref{tpaa_lm}; the hyperparameter in LASSO was selected using cross validation. For APA connectomes, we implemented two recently proposed methods: symmetric bilinear regression (SBL) \citep{wang2019symmetric} and  multi-graph principal component analysis (MultiGraphPCA) \citep{winter2020multiscale}. SBL investigates the relationship between human traits and connectivity matrices through a symmetric bilinear regression model, and MultiGraphPCA proposes a tensor network factorization method that links the scale-specific brain structural connectivity matrices through a common set of individual-specific scores, which are further used for human trait prediction. Tuning in SBL and MultiGraphPCA followed the recommendations by the authors. In particular, for SBL we use K = 14; gamma = 6.9; fullit = 50; maxit = 10000; tol = 1e-6; Replicates = 5. There is a single tuning parameter $K$ in MultiGraphPCA; we compare results for $K = \{2, 10, 20, 50, 70, 200, 400, 500\}$. We also implemented LASSO on the vectorized (only keeping upper-triangular elements) connectivity matrix $W_i$. 
%




In order to assess the robustness of the proposed method, we tested different versions of PPA with respect to the fiber tracking algorithms and regularization strategies. In particular, we adopted another two fiber tracking algorithms in Module (i), Euler Delta Crossings or EuDX \citep{garyfallidis2014dipy}, and local tracking with Sparse Fascicle Model or SFM \citep{rokem2015evaluating}. For the regularization strategy, we also consider elastic net \citep{zou2005regularization},
which combines the ${L_1}$ penalty of LASSO with the 
${L_2}$ penalty of ridge regression  \citep{hoerl1970ridge}, with the goal of simultaneous selection of correlated predictors.

Figures \ref{compare1} and  \ref{compare2} show the comparison of various versions of PPA using different fiber tracking algorithms and regularization strategies, respectively. The results are similar to those shown above, and demonstrate the robustness of the PPA method to the 
 fiber tracking algorithm and regularization approach.

\subsection{Predictive performance \& parsimony}  
We implemented three methods under APA connectomes, which are coded as LASSO, SBL, and MultiGraphPCA, and the version of PPA using Tractoflow fiber tracking algorithm and LASSO regularization to analyze the 1065 HCP subjects. We calculated the 5-fold cross validation mean squared error (MSE) to compare their predictive performance for seven human traits. We also included the performance of a \textit{null model}, which only contains an intercept term and uses the sample average of responses in the training set as the prediction; this model assumes no significance of the connectome in explaining selected traits and serves as a reference model.

\begin{figure}[h!]
\centering 
	\begin{tabular}{ccc}
		 \includegraphics[width=0.3\linewidth, height=4cm]{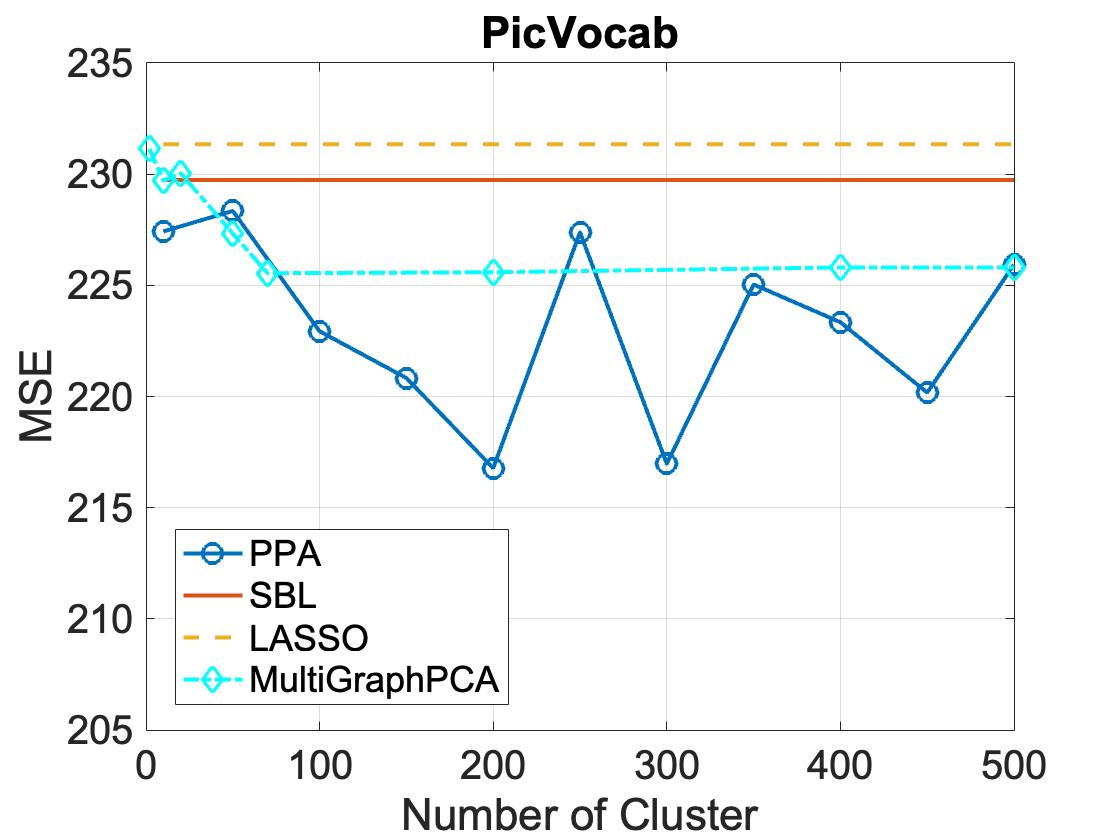} & 
		  \includegraphics[width=0.3\linewidth,height=4cm]{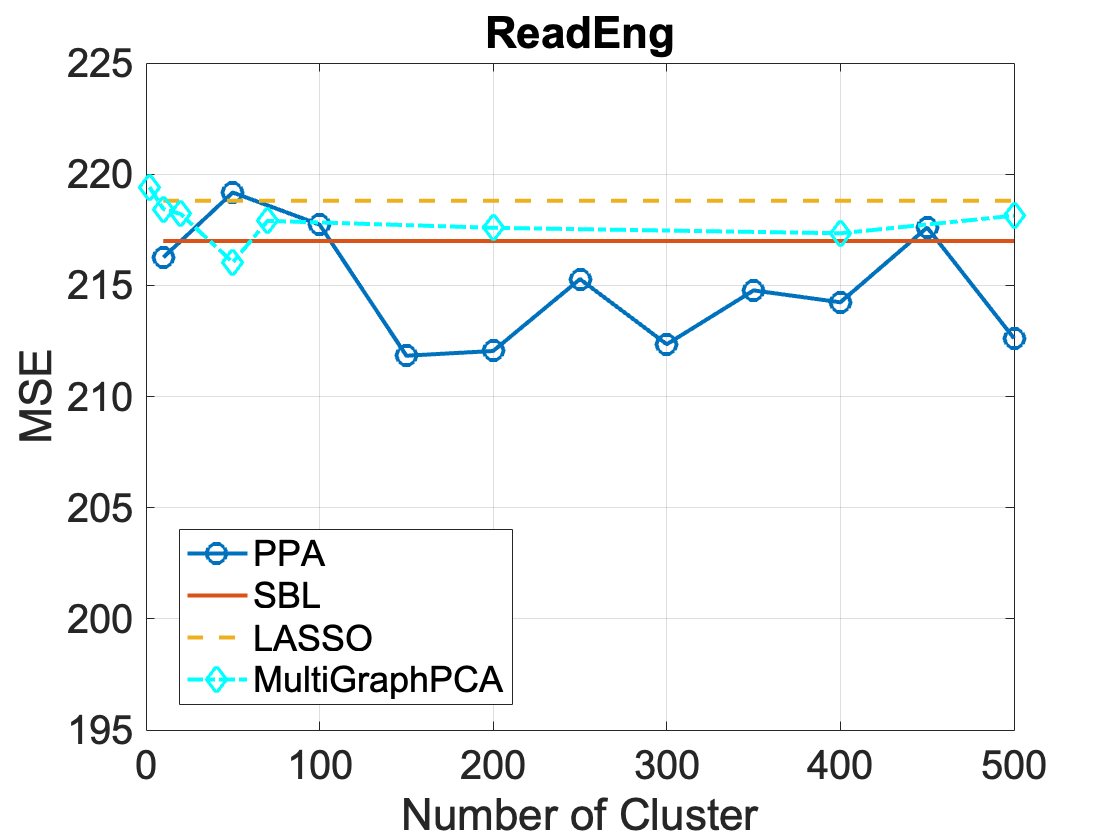}&
		  \includegraphics[width=0.3\linewidth,height=4cm]{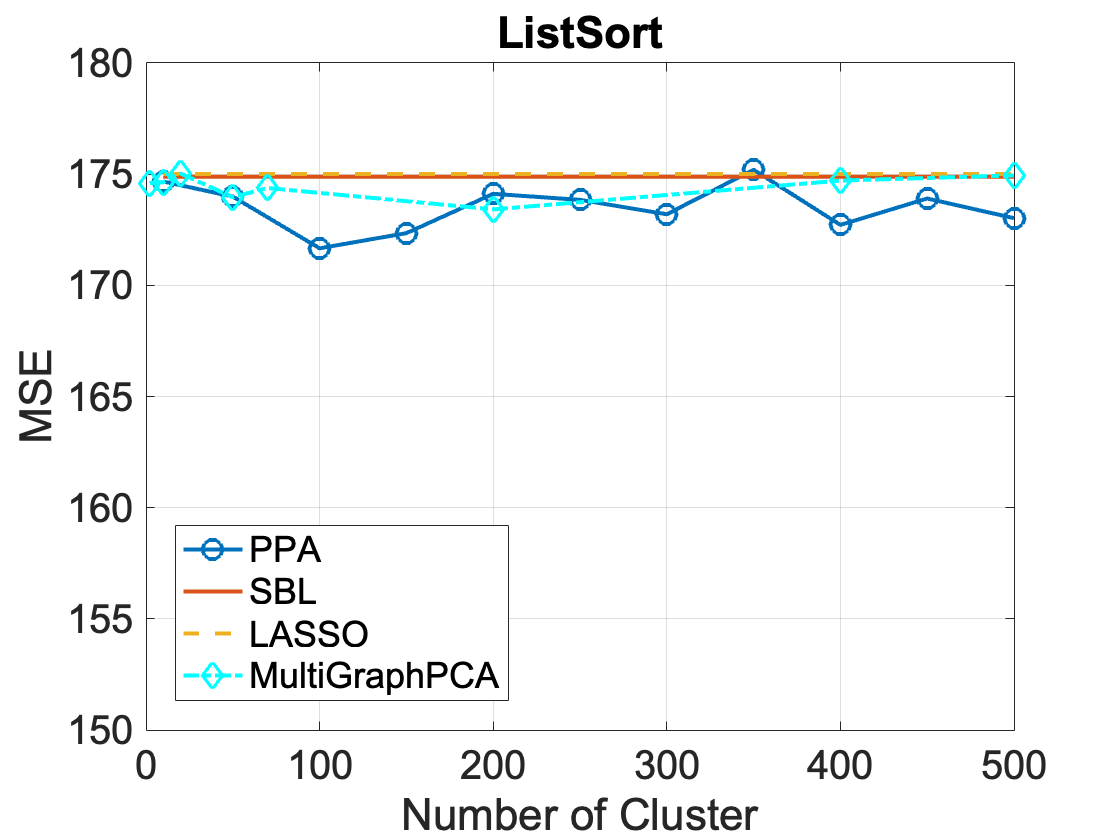}\\
		  \includegraphics[width=0.3\linewidth, height=4cm]{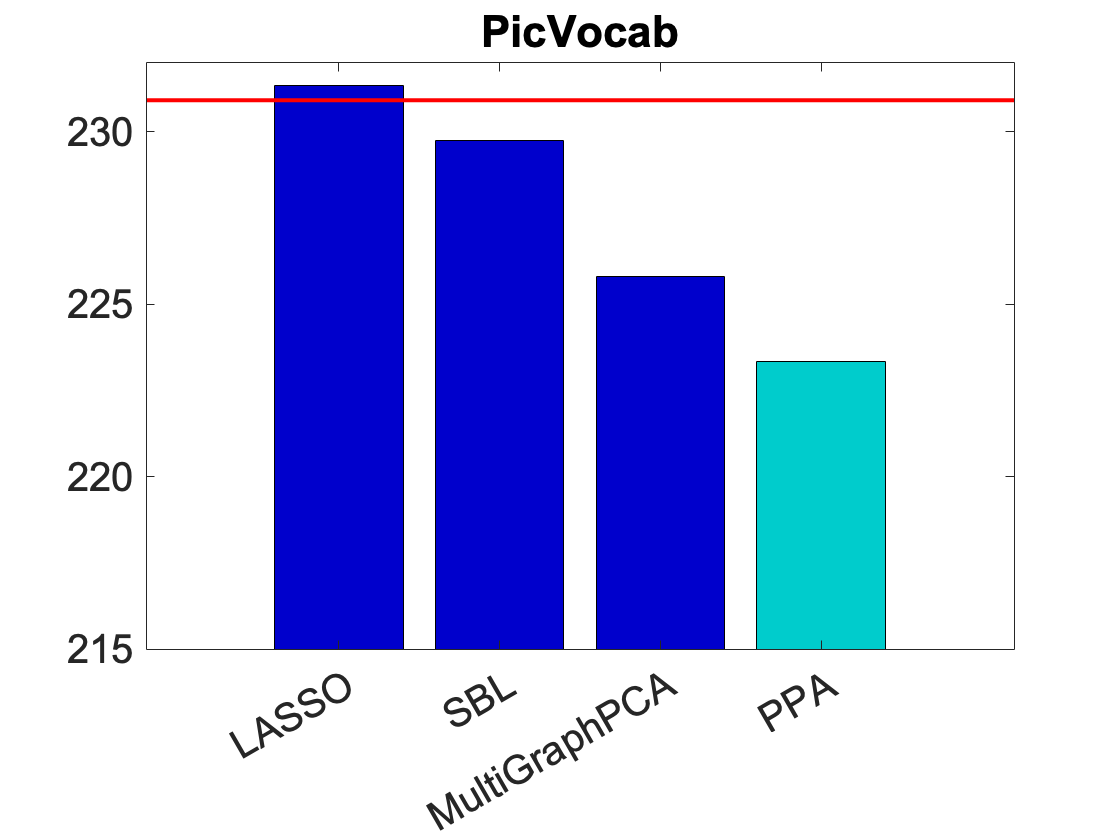} & 
		  \includegraphics[width=0.3\linewidth, height=4cm]{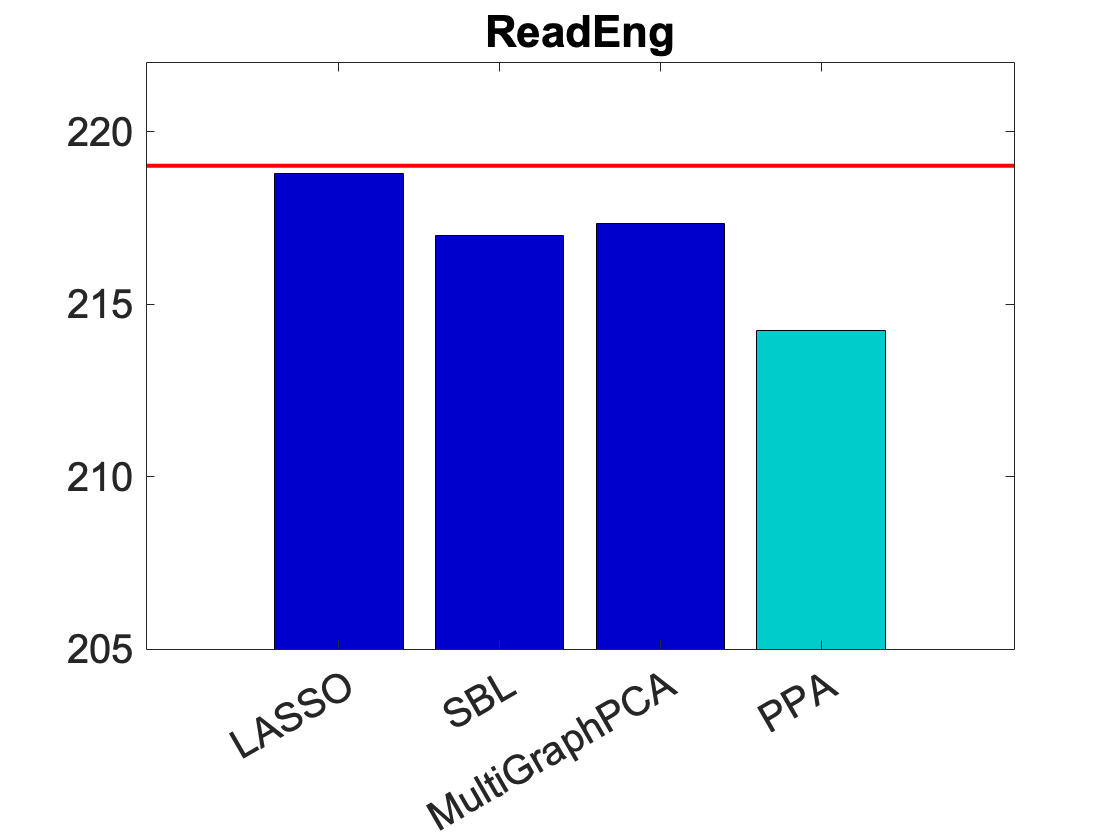}&
		  \includegraphics[width=0.3\linewidth, height=4cm]{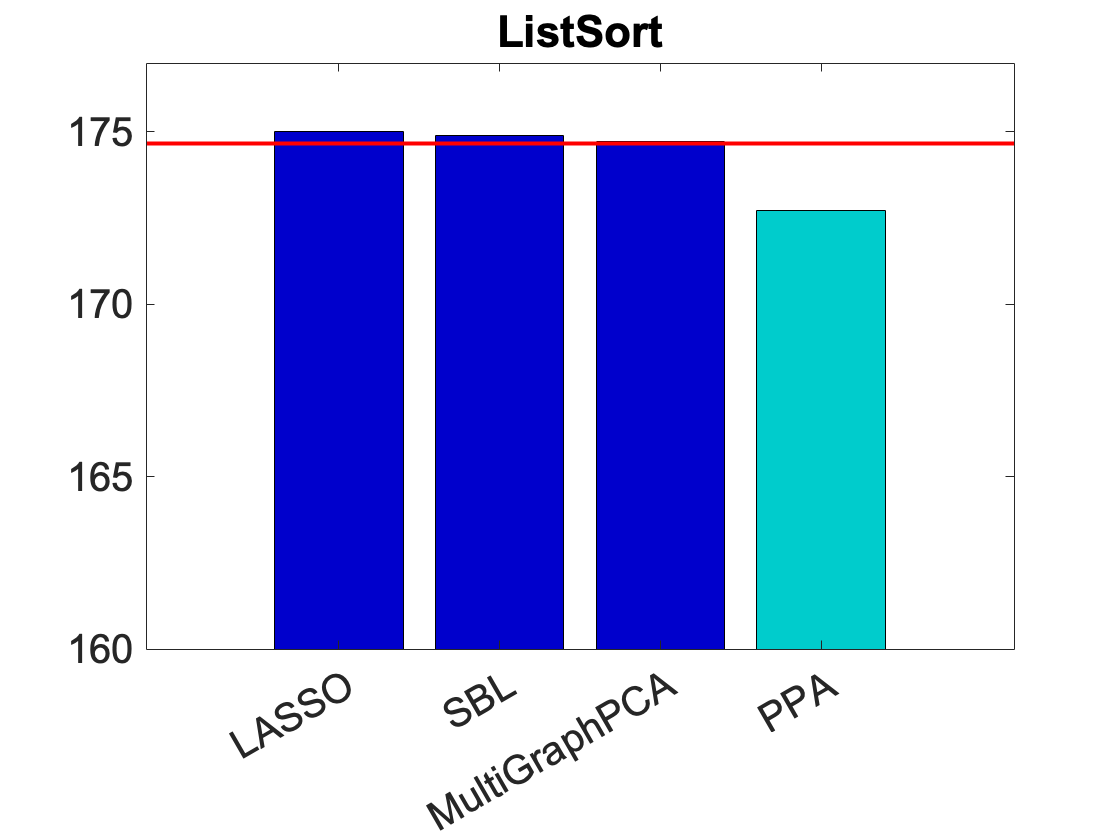}\\
		   {\small (a)  Receptive Vocabulary} & {\small (b) Oral Reading} & {\small (c) List Sorting} \\
		  \end{tabular}
\caption{Comparison of 5-fold cross validation MSE of trait prediction based on PPA and three APA-based methods (LASSO, SBL, and MultiGraphPCA) for three traits: (a) PicVocab, (b) ReadEng, and (c) ListSort. Second row is the bar-plot of MSE for APA and PPA based method for $K=400$. APA methods (SBL, LASSO, and MultiGraphPCA) are in blue and PPA method in cyan. The red horizontal line indicates the MSE of the null model, which is 230.92 in PicVocab; 219.02 in ReadEng; 174.66 in ListSort.}
\label{mse1} 
\end{figure}



\begin{figure}[h!]
\centering 
	\begin{tabular}{cc}
		  \includegraphics[width=0.45\linewidth,height=5cm]{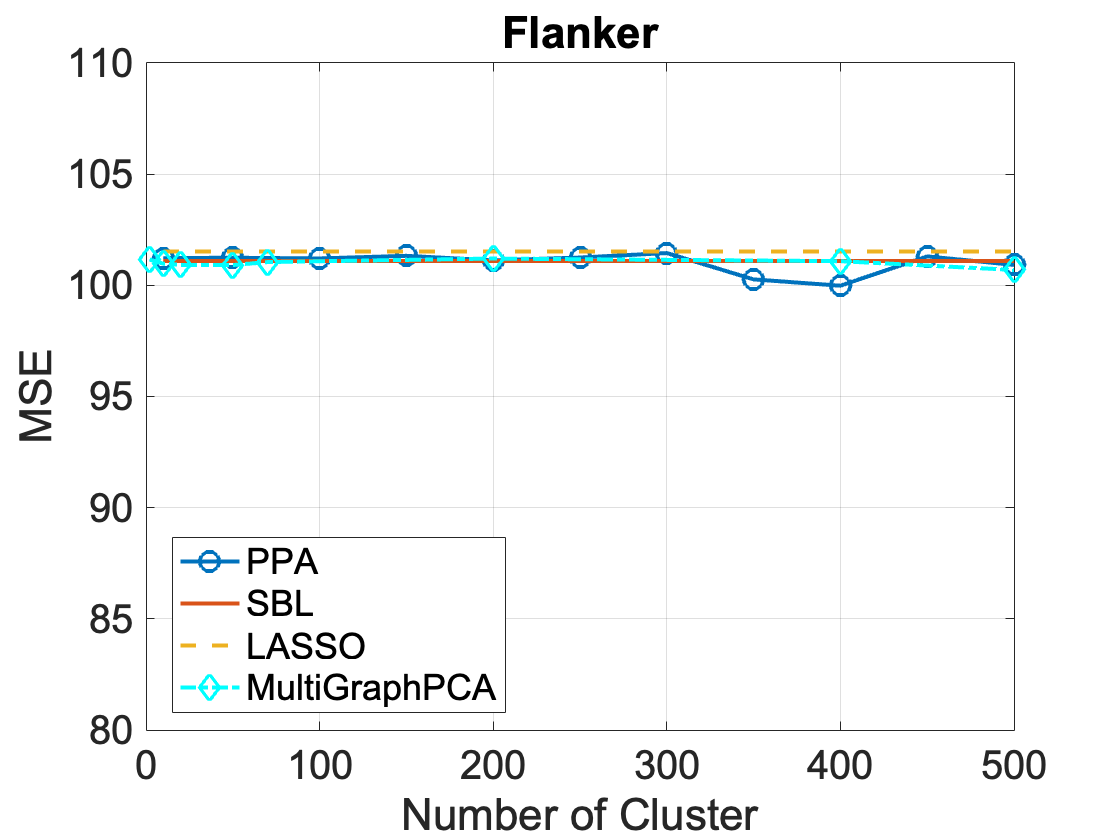} & 
		  \includegraphics[width=0.45\linewidth,height=5cm]{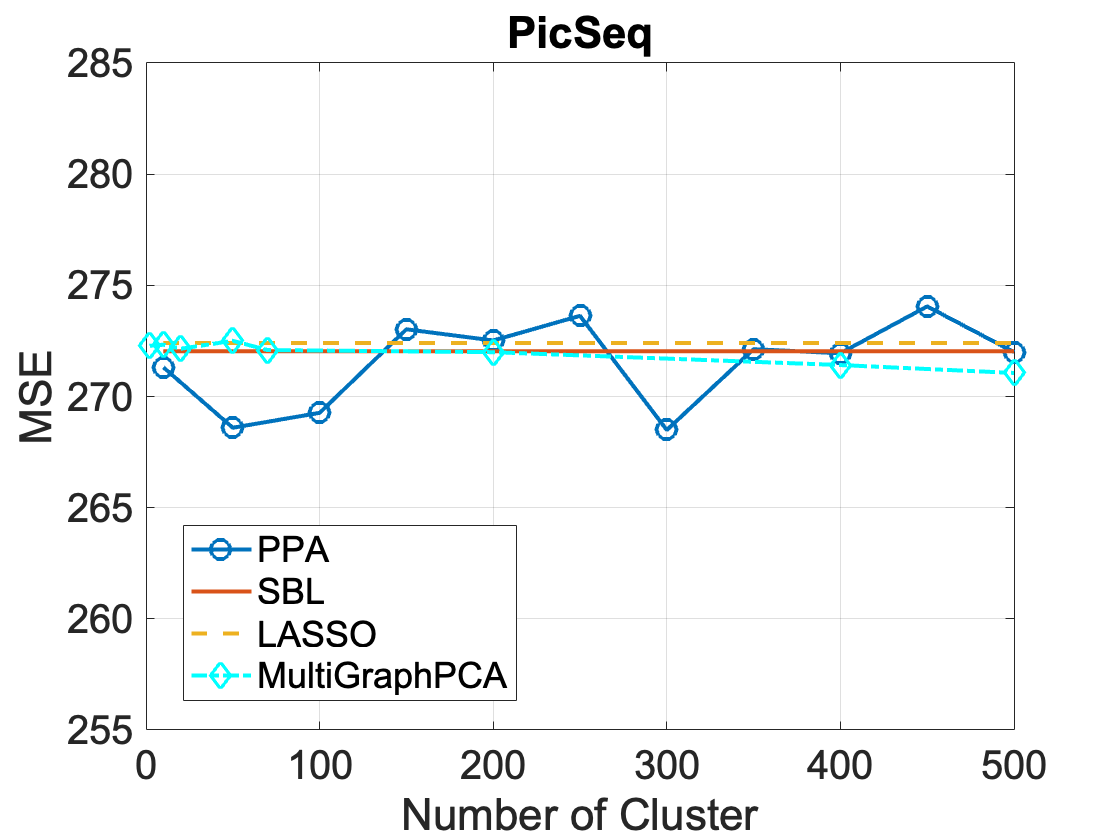} \\
		   {\small (a) Flanker} & {\small (b) Picture Sequence Memory} \\
		  \includegraphics[width=0.45\linewidth,height=5cm]{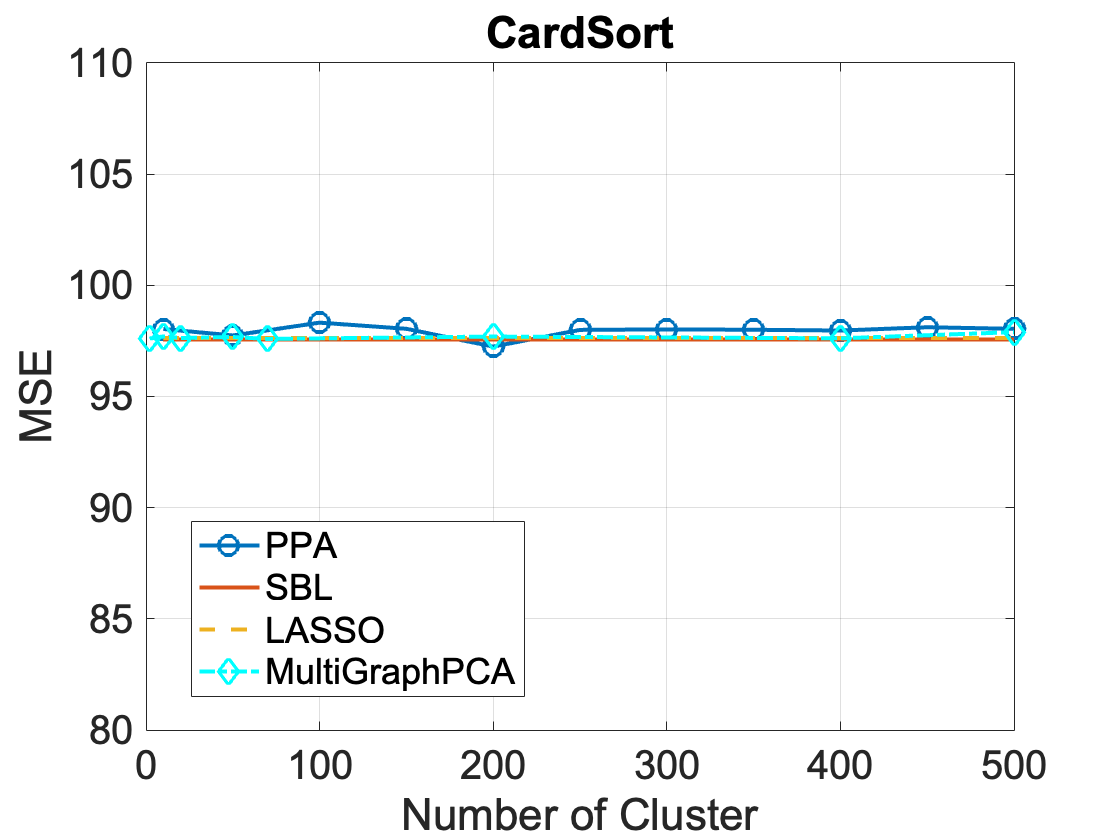} & 
		  \includegraphics[width=0.45\linewidth,height=5cm]{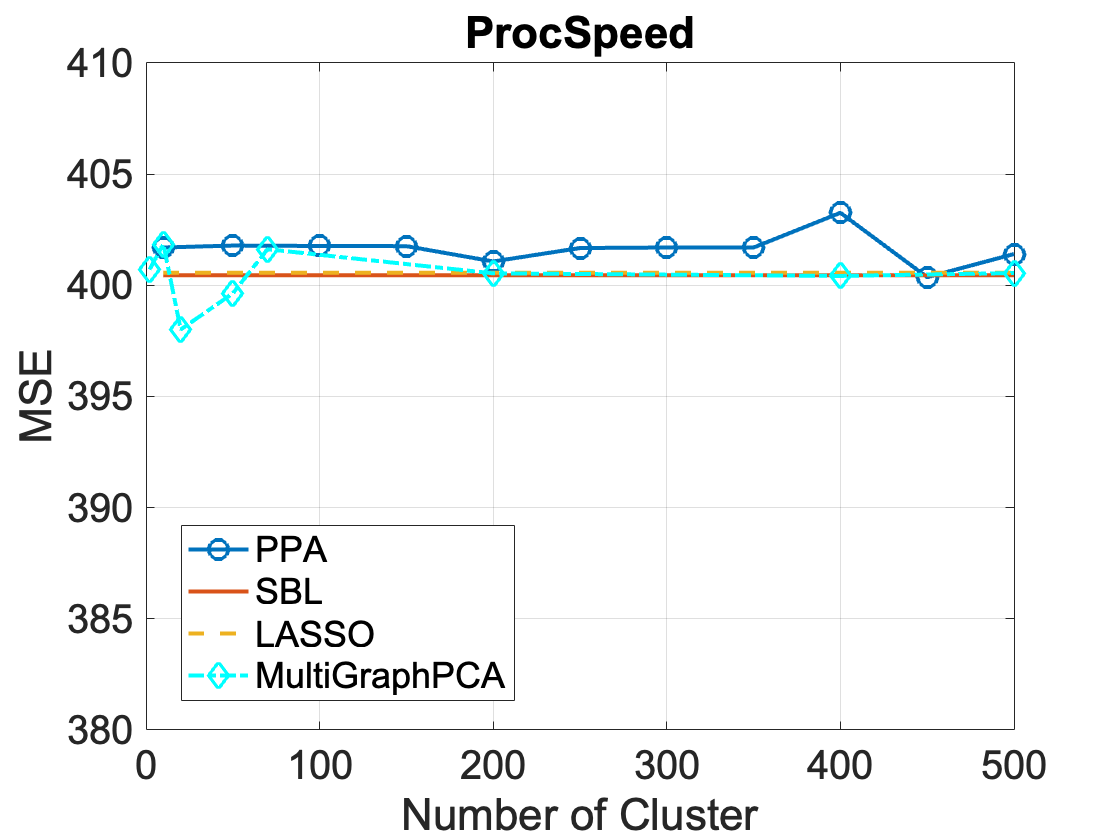} \\
		  {\small (c) Card Sort} & {\small (d) Processing Speed}\\
	\end{tabular}
\caption{Comparison of 5-fold cross validation MSE of trait prediction based on PPA and three APA methods (LASSO, SBL, and MultiGraphPCA) for four traits: (a) Flanker, (b) PicSeq, (c) CardSort, and (d) ProSpeed. Moreover, 5-fold cross validation MSE under the null model is 101.0769 in Flanker; 272.0460 in PicSeq; 97.7893 in CardSort; 402.0294 in ProcSpeed.}
\label{mse2} 
\end{figure}

Figure \ref{mse1} plots the MSEs for three traits. The upper row of Figure \ref{mse1} shows that the proposed PPA, using the simple LASSO method,
clearly outperforms SBL, LASSO, and MultiGraphPCA in most scenarios especially as the number of clusters $K$ increases. 
For the connectivity matrix-based methods, e.g, SBL, LASSO, and MultiGraphPCA, the performance does not depend on the number of clusters $K$, which is a PPA-specific tuning parameter. The bottom row of Figure \ref{mse1} compares the MSEs of our PPA-based methods at $K = 400$ to the selected three APA methods, while the red horizontal line represents the performance of the null model. Since all APA-based methods use TractoFlow, we first focus on the PPA method using the same fiber tracking algorithm. In this case, MSEs of PPA are smaller than the three APA-based methods, uniformly across the three traits. 
In sharp contrast to the excellent performance of LASSO for the PPA connectomes, LASSO predictions based on vectorized APA connectomes did no better than the null model.
SBL and MultiGraphPCA improve the MSEs over LASSO, as a result of a better utilization of the network structure of APA connectomes. The comparison of MSEs suggests that fundamentally changing the connectome representation based on defining population fiber bundles can perhaps lead to even bigger gains.
Figure \ref{mse1} shows that defining large numbers of fiber bundles may lead to predictive gains.

For the other four traits (Flanker, PicSeq, CardSort, ProcSpeed), all methods tend to give a MSE close to the null model (Figure \ref{mse2}), indicating limited predictive power of structural connectivity for these traits. It is reassuring that the proposed method is consistent with APA-based methods in these cases. We remark that the lack of predictive power might be caused in part by a weak relationship between these measured traits and actual innate abilities in the test subjects.

\begin{figure}[ht!]
\centering
	\begin{tabular}{ccc}
		 \includegraphics[width=0.3\linewidth]{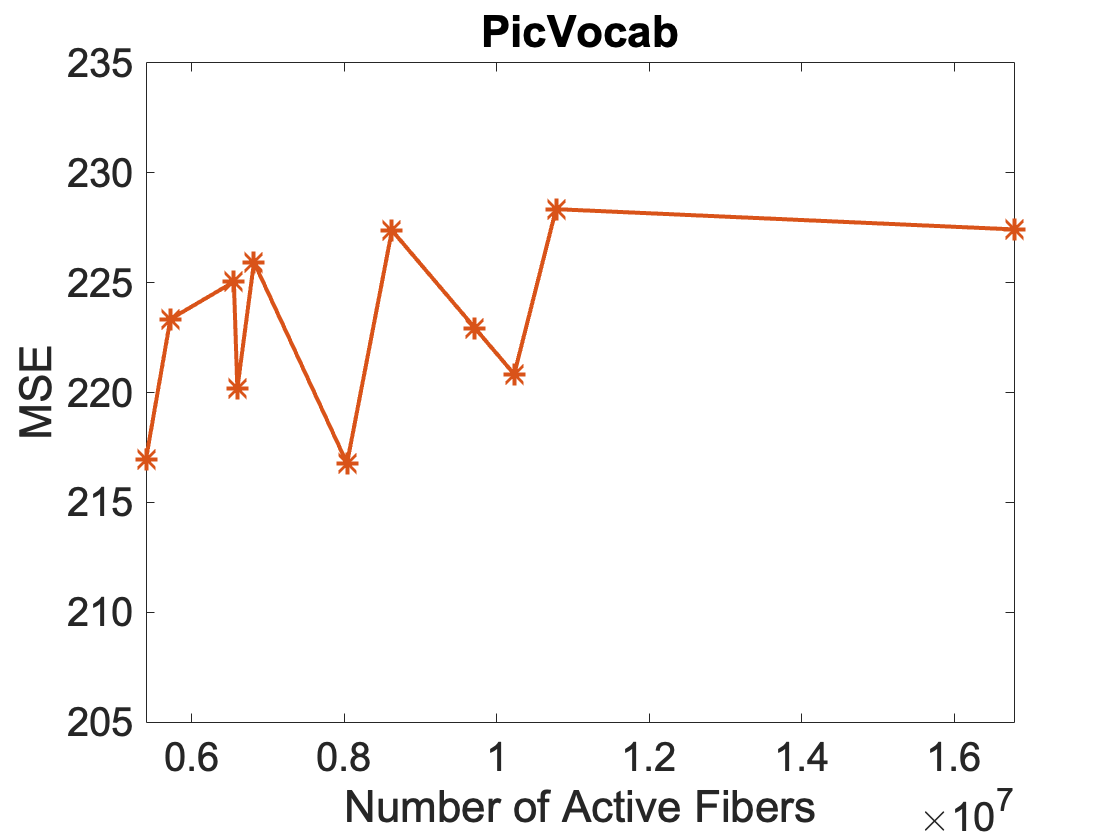} & 
		  \includegraphics[width=0.3\linewidth]{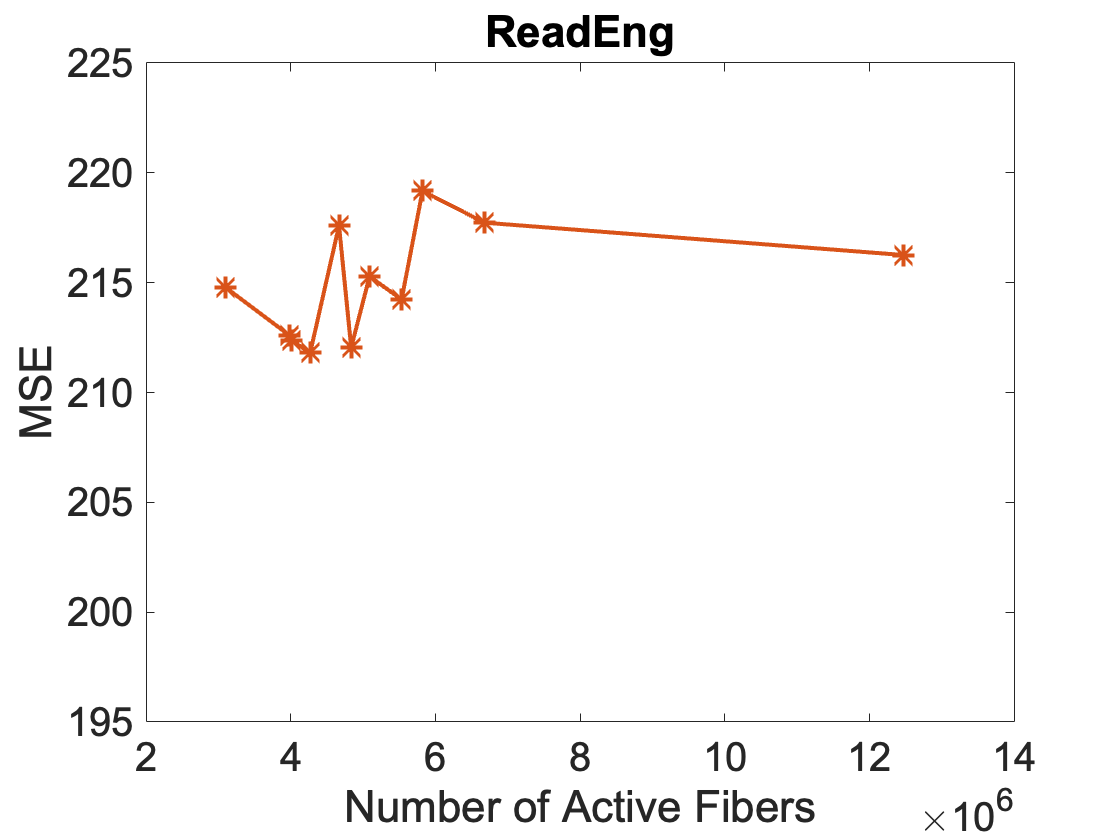}&
		  \includegraphics[width=0.3\linewidth]{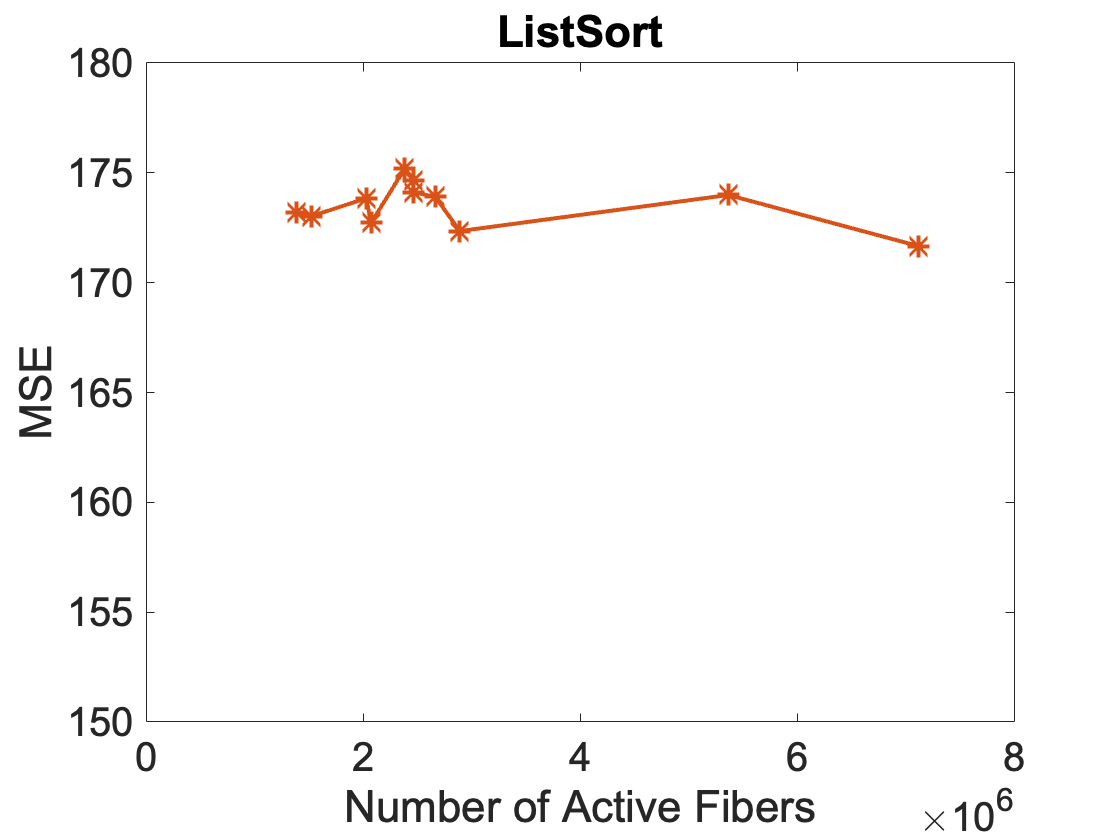}
		  \end{tabular}
\caption{Comparison of 5-fold cross validation MSE of trait prediction based on PPA for three traits: (a) PicVocab, (b) ReadEng, and (c) ListSort.} 
\label{kselect.3traits} 
\end{figure}

While Figure \ref{mse1} shows how MSE varies with $K$ for the three different fiber tracking algorithms, TractoFlow, EuDX and SFM, Figure~\ref{kselect.3traits} plots the MSEs against the number of active fibers to provide additional insight into the impact of $K$. The number of active fibers is defined as the total number of fibers that belong to the active fiber bundles selected by LASSO, i.e., it is 
$\sum_{k \in \mathcal{K}(s)} |A_K^{(k)}|,$ where $|A_K^{(k)}|$ is the number of fibers in the bundle $A_K^{(k)}$.   
Taking picture vocabulary test (PicVocab) as an example, the best MSE is achieved when the number of active fibers is around $0.8*10^7$. This optimal number varies from trait to trait, but a U-shaped curve typically emerges with $K$ varied up to 500. For the other traits in Figure \ref{mse2}, MSEs do not change much as we vary the number of active fibers, which is expected as the MSE curve is flat when plotted against $K$; we omit these curves here as they are redundant. 

Table \ref{fp} reports the number of selected parameters in PPA and APA-based methods, which shows  parsimony and effectiveness of PPA-based methods compared to LASSO applied to APA connectomes and SBL. In particular, LASSO for APA connectomes selects nearly zero active connections, which explains its poor predictive performance in Figure~\ref{mse1}. PPA selects substantially fewer non-zero parameters than SBL; this combined with the better MSEs in Figure~\ref{mse1} shows the effectiveness of PPA connectomes in representing key features of brain networks predictive of traits.  The last four traits show little to no signal for any of the methods and selecting few if any features for these traits seems appropriate.

\begin{table}[h!]
    \centering
    \begin{tabular}{cccccccc}
     \hline \hline
         & \small PicVocab & \small ReadEng & \small ListSort & \small Flanker & \small PicSeq &\small CardSort & \small ProcSpeed \\
         \small PPA (K=50) & \small 23   & \small  20 & \small 12 & \small 11 & \small 15 & \small 0 & \small 1 \\
         \small PPA (K=100) & \small 38   & \small  33 & \small 19 & \small 3 & \small 29 & \small 3 & \small 2 \\
         \small PPA (K=200) & \small 64   & \small  47 & \small 12 & \small 7 & \small 6 & \small 0 & \small 0 \\
         \small PPA (K=300) & \small 53   & \small  31 & \small 2 & \small 3 & \small 1 & \small 0 & \small 0 \\
         \small PPA (K=400) & \small 19   & \small 20 & \small 18 & \small 18 & \small 15 & \small 2 & \small 1 \\
         \small PPA (K=500) & \small 56   & \small 39 & \small 22 & \small 1 & \small 3 & \small 0 & \small 1 \\
         \small LASSO+HCP842 & \small 2   & \small 1 & \small 0 & \small 0 & \small 11 & \small 0 & \small 1 \\
         \small SBL+HCP842 & \small 1134   & \small 1053 & \small 972 & \small 729 & \small 1134 & \small 810 & \small 1134 \\
         \hline\hline
    \end{tabular}
    \caption{Number of selected parameters in different methods. For MultiGraphPCA, the number of parameters is set to be $K$ used in PPA.}
    \label{fp}
\end{table}

To evaluate generalizability of our comparisons between PPA and APA and assess robustness of APA methods, we tested multiple different versions of APA with respect to the atlases and summary of connectivity used in defining the connectivity matrices. In our analyses as we vary these two aspects when implementing APA methods, the improved predictive performance of PPA over APA that was shown earlier is consistently observed, indicating the way connectomes are represented (PPA vs APA) appears to be a more important factor in explaining the performance gain. 

In particular, for all APA related methods (SBL, LASSO, MultiGraphPCA), we checked three different atlases: HCP842, AAL2, and FreeSurferDKT. HCP842 is a tractography atlas \citep{yeh2018}, and we choose the settings used previously in obtaining the results in  Figures \ref{mse1} and \ref{mse2}. This atlas provides tractography of white matter with expert labeling and examination, complementary to traditional histologically-based and voxel-based white matter atlases. AAL2 stands for automated anatomical labeling atlas 2, providing an alternative parcellation of the orbitofrontal cortex \citep{rolls2015implementation}. Finally, FreeSurferDKT is an atlas manually labeled in the macroscopic anatomy in magnetic resonance images by the Desikan–Killiany–Tourville (DKT) protocol \citep{klein2012101}.
From the results in 
 Figure \ref{compare3}, we can see that changing the atlas has little impact on the predictive performance of APA methods.  

For analyses assessing sensitivity of the APA results to the summary of connectivity between regions, we focused on the HCP842 atlas and tested three different ways of calculating the connectivity matrix: ``count", ``ncount", and ``ncount2". For each entry in the connectivity matrix, these three summaries correspond to counting the number of tracts that pass two ROIs (``count''), which is used in the previous comparison in Figures \ref{mse1} and \ref{mse2}, normalizing the count by the median length of the fibers (``ncount"), and multiplying the count by the sum of the inverse of the length (``ncount2"). Figure \ref{compare4} shows that the performance of APA methods is robust to the summary used in calculating connectivity matrices.

\subsection{Integrating PPA and atlas} 

The proposed PPA connectome does not rely on any tractography atlas in defining the connectome or building a regression model for traits. This reduces subjectivity in choosing ROIs and facilitates statistical analyses with improved prediction and parsimony, as shown in the preceding sections. In this section, we demonstrate another feature of PPA in terms of its compatibility with traditional ROIs---as an \textit{ab initio}, tractography-based representation of connectomes, PPA can be integrated with any existing atlas templates to borrow the ROI information encoded therein in a straightforward manner. Such integration allows us to relate the interpretation of PPA results to traditional ROIs. Through visualization, we find the proposed PPA leads to interesting and interpretable findings.

We align active fibers produced by PPA to an atlas. In particular, based on a selected atlas, we build the connectivity matrix at the population level with each matrix entry generated by counting the number of active fibers passing between each pair of ROIs. Note that similarly to deriving the connectivity matrix in APA, one can obtain many different summaries of connectivity between two regions by using various filtering approaches such as SIFT and different ways of normalization, with the count just one simple choice. 
We 
use the HCP842 tractography atlas \citep{yeh2018} as in our implementation of APA-based methods, which segments the brain into 80 regions. Refer to  \url{http://brain.labsolver.org/diffusion-mri-templates/tractography} for more details about the 80 ROIs. 
For each human trait, we visualize the PPA-induced connectivity matrix and anatomy of connections through DSI Studio (\url{http://dsi-studio.labsolver.org}), a tractography software tool that maps brain connections and correlates findings with traits. We adopt the default setting in DSI Studio by thresholding matrix entries with a small number of connecting tracks relative to the maximum connecting tracks in the connectivity matrix. We use 0.5 as the threshold for this ratio.

\begin{figure}[ht!]
\centering 
	\begin{tabular}{cccc}
		 \includegraphics[width=0.225\linewidth, height=3.5cm]{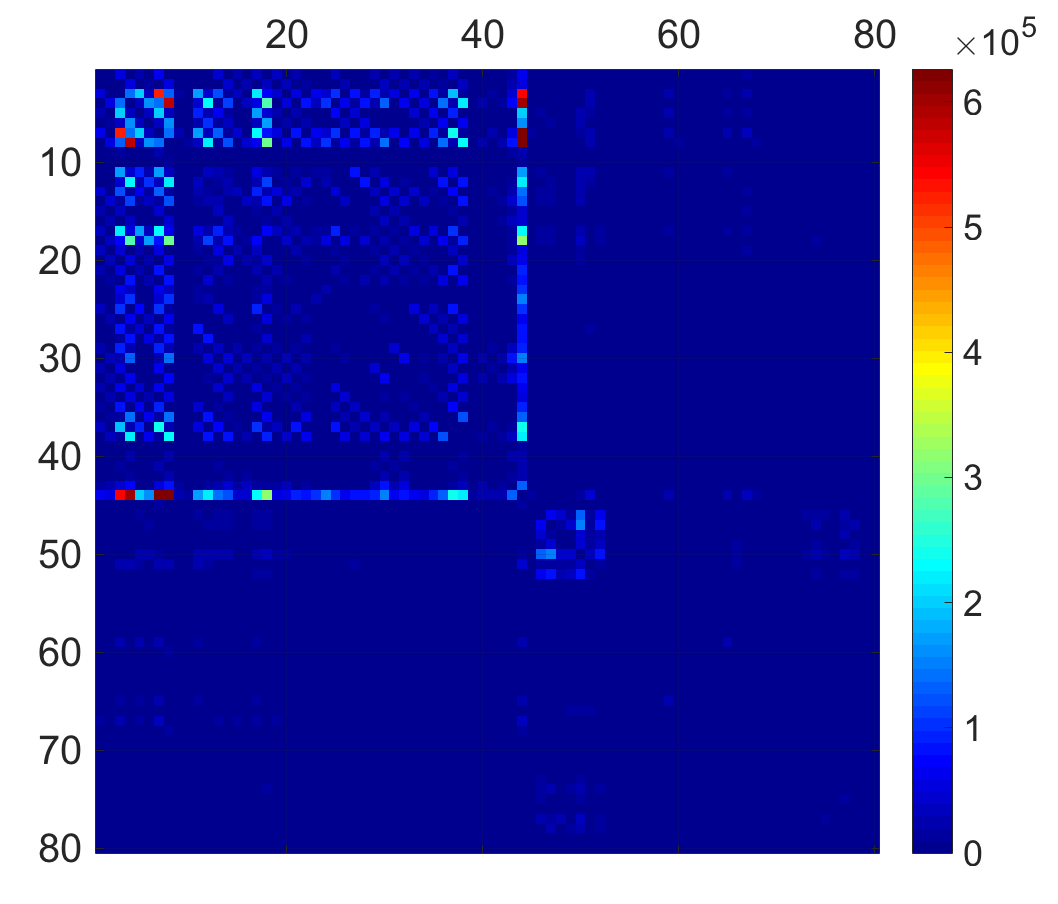} & 
		  \includegraphics[width=0.225\linewidth,height=3.5cm]{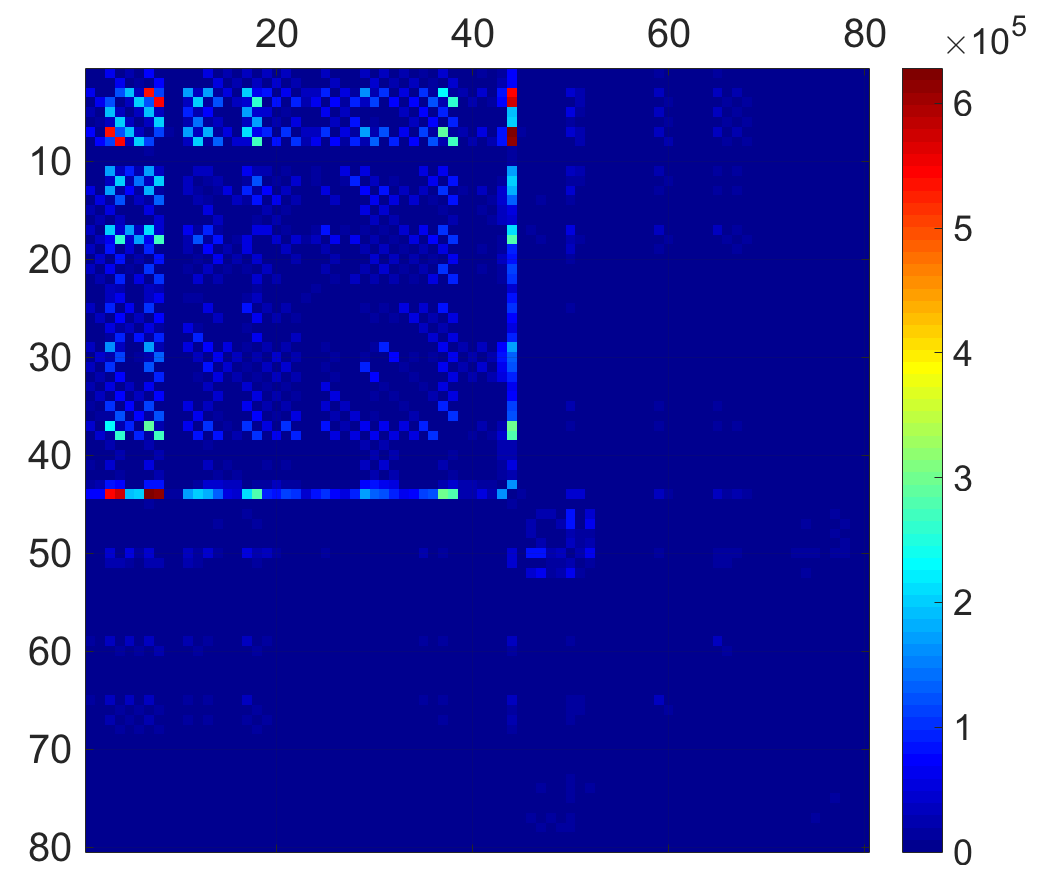} &
		   \includegraphics[width=0.225\linewidth, height=3.5cm]{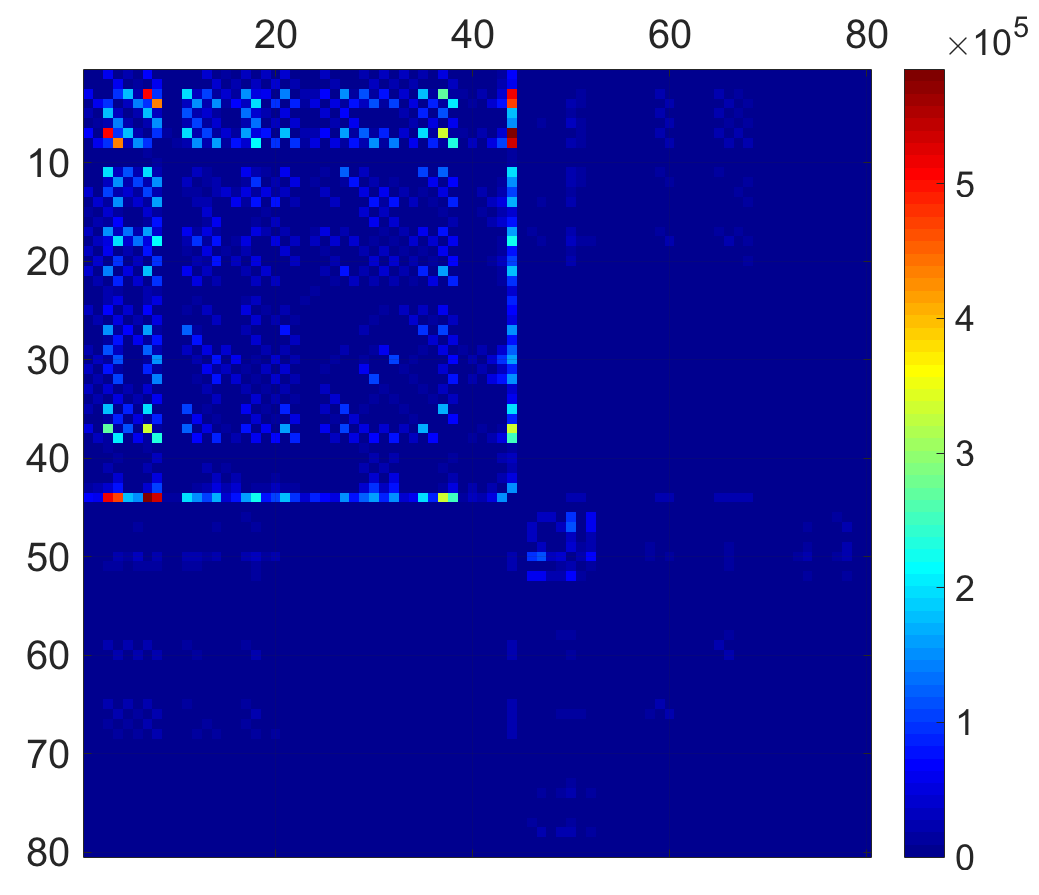} & 
		  \includegraphics[width=0.225\linewidth,height=3.5cm]{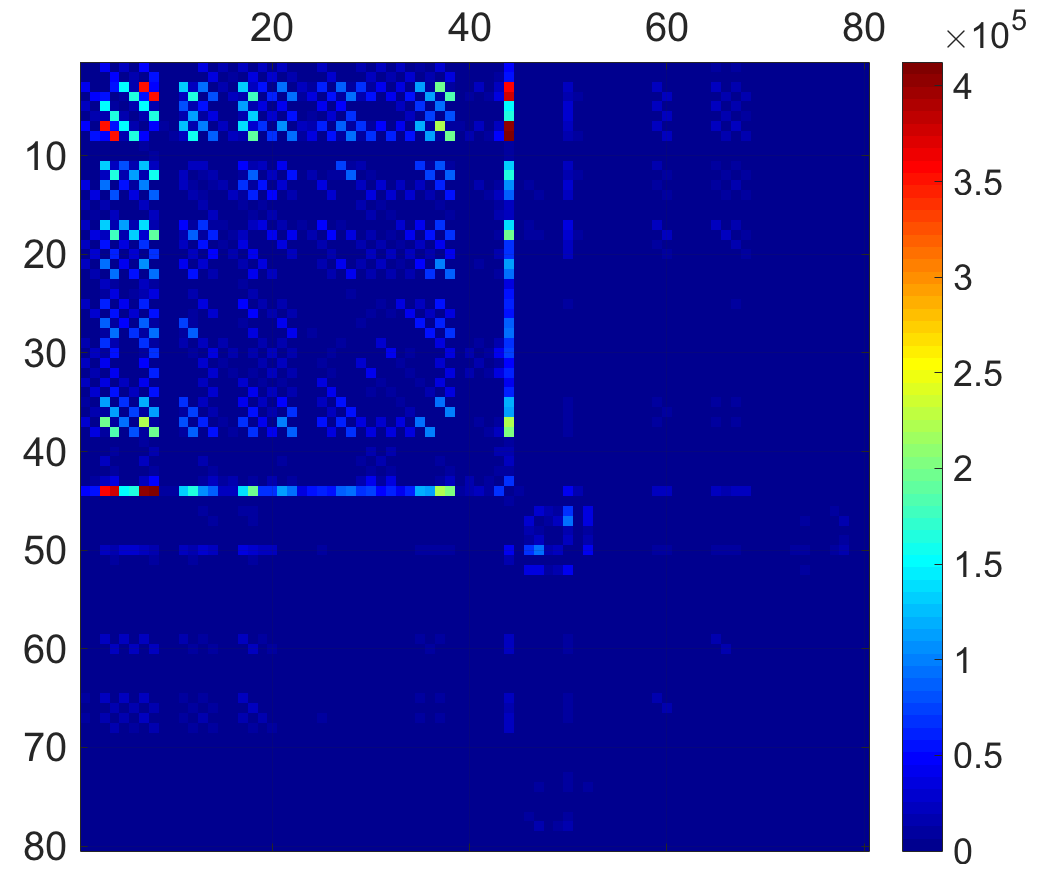} \\
		  \includegraphics[width=0.225\linewidth,height=4cm]{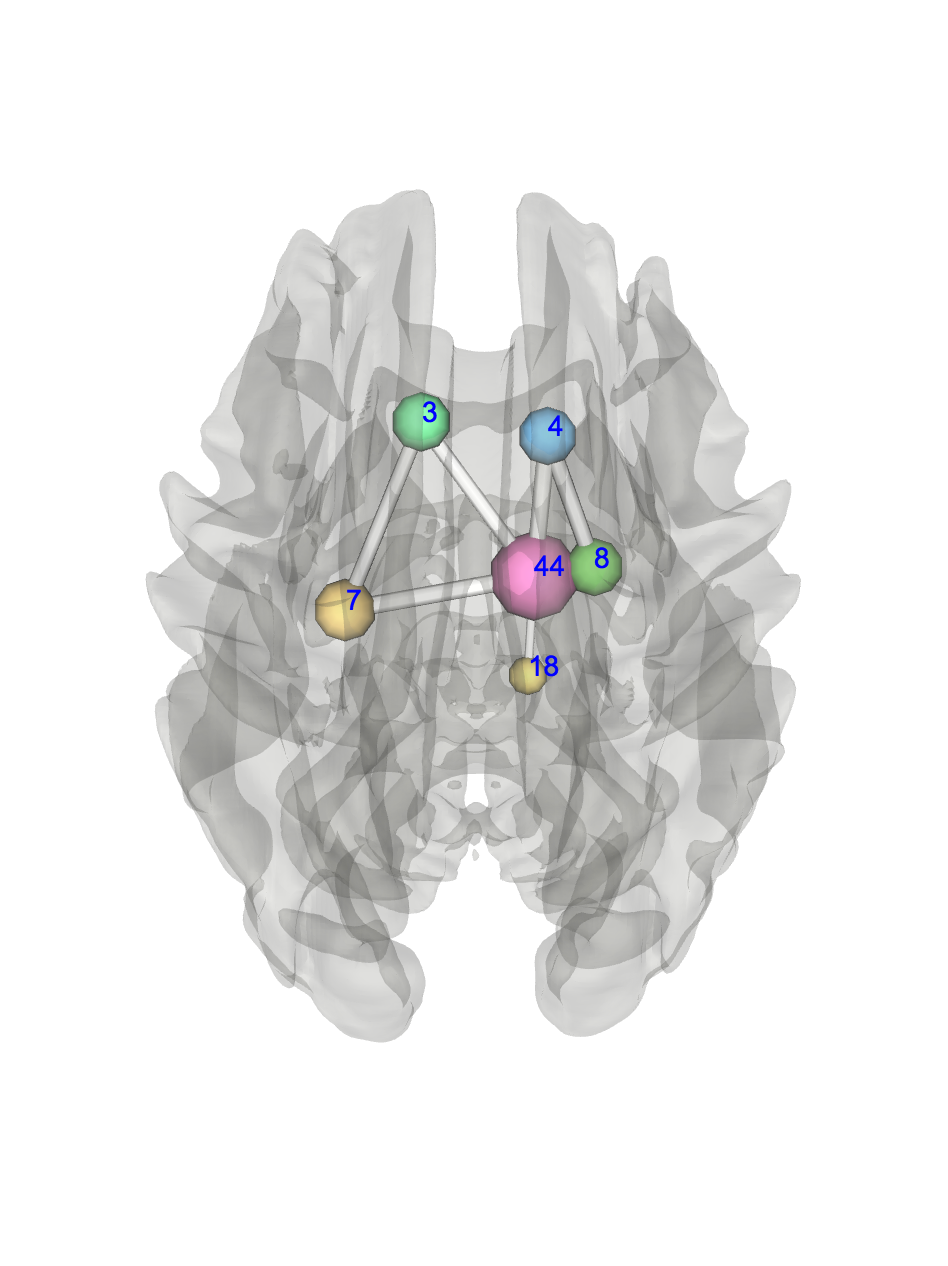} & 
		  \includegraphics[width=0.225\linewidth,height=4cm]{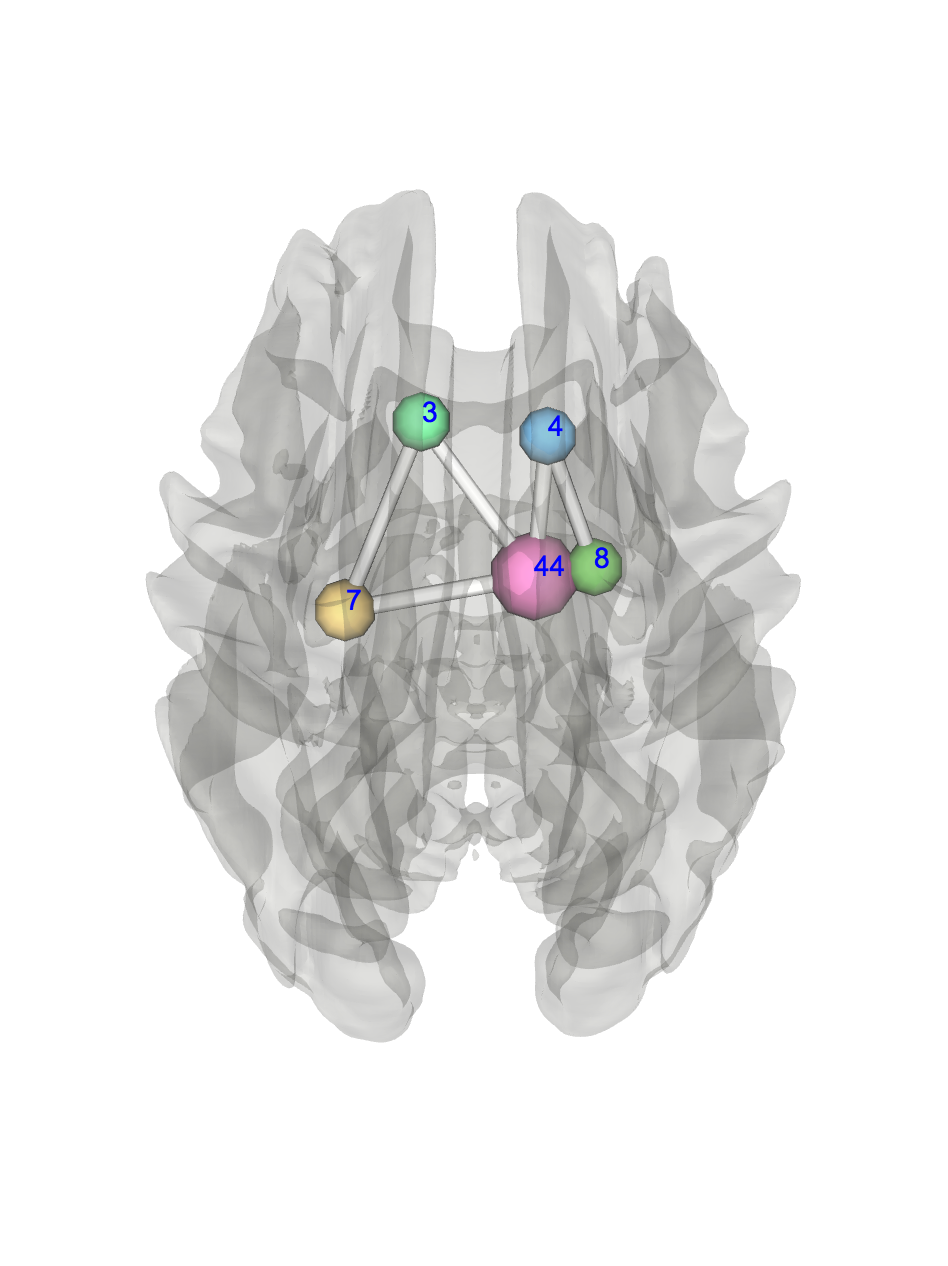} &
		   \includegraphics[width=0.225\linewidth, height=4cm]{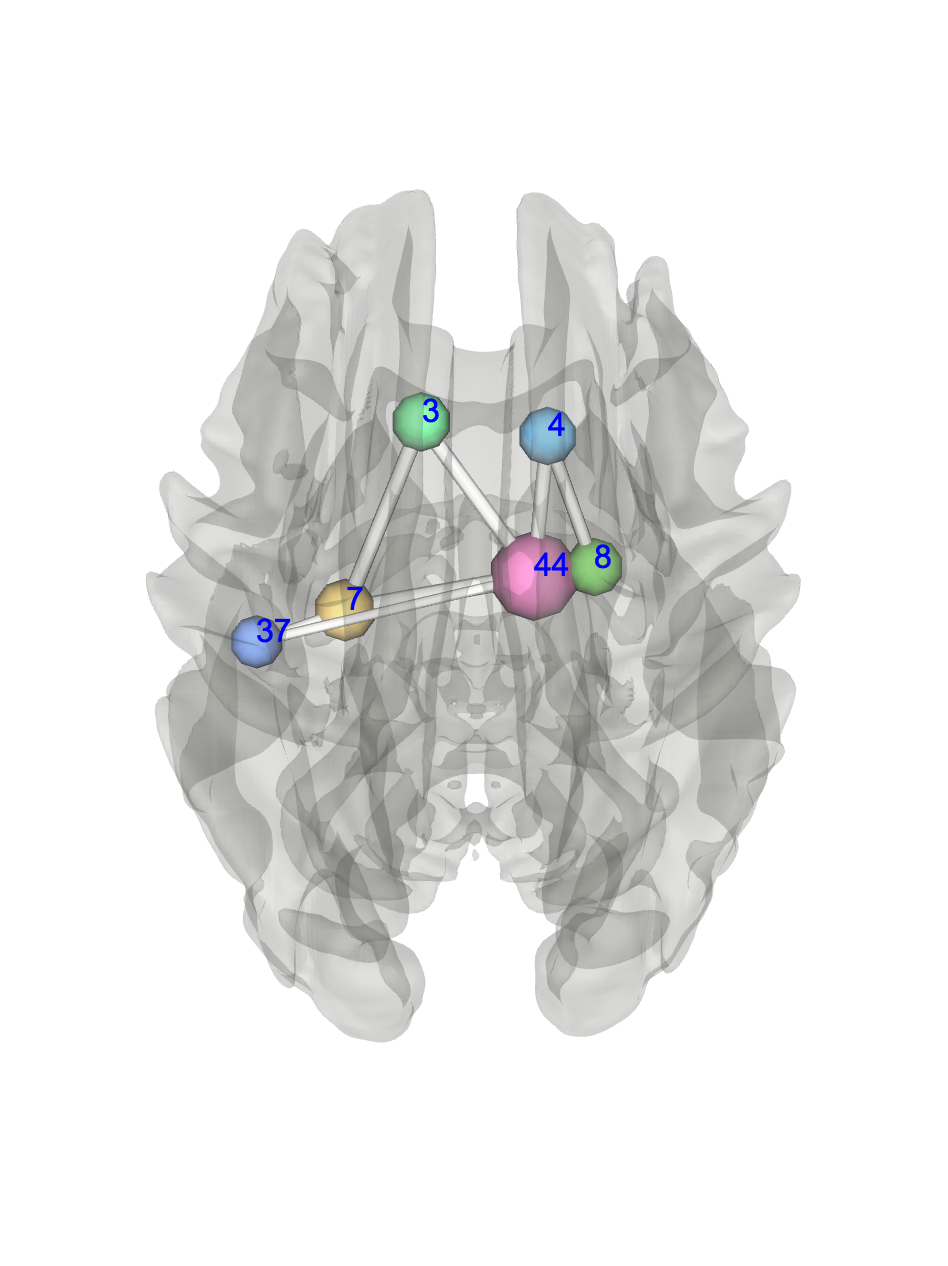} & 
		  \includegraphics[width=0.225\linewidth,height=4cm]{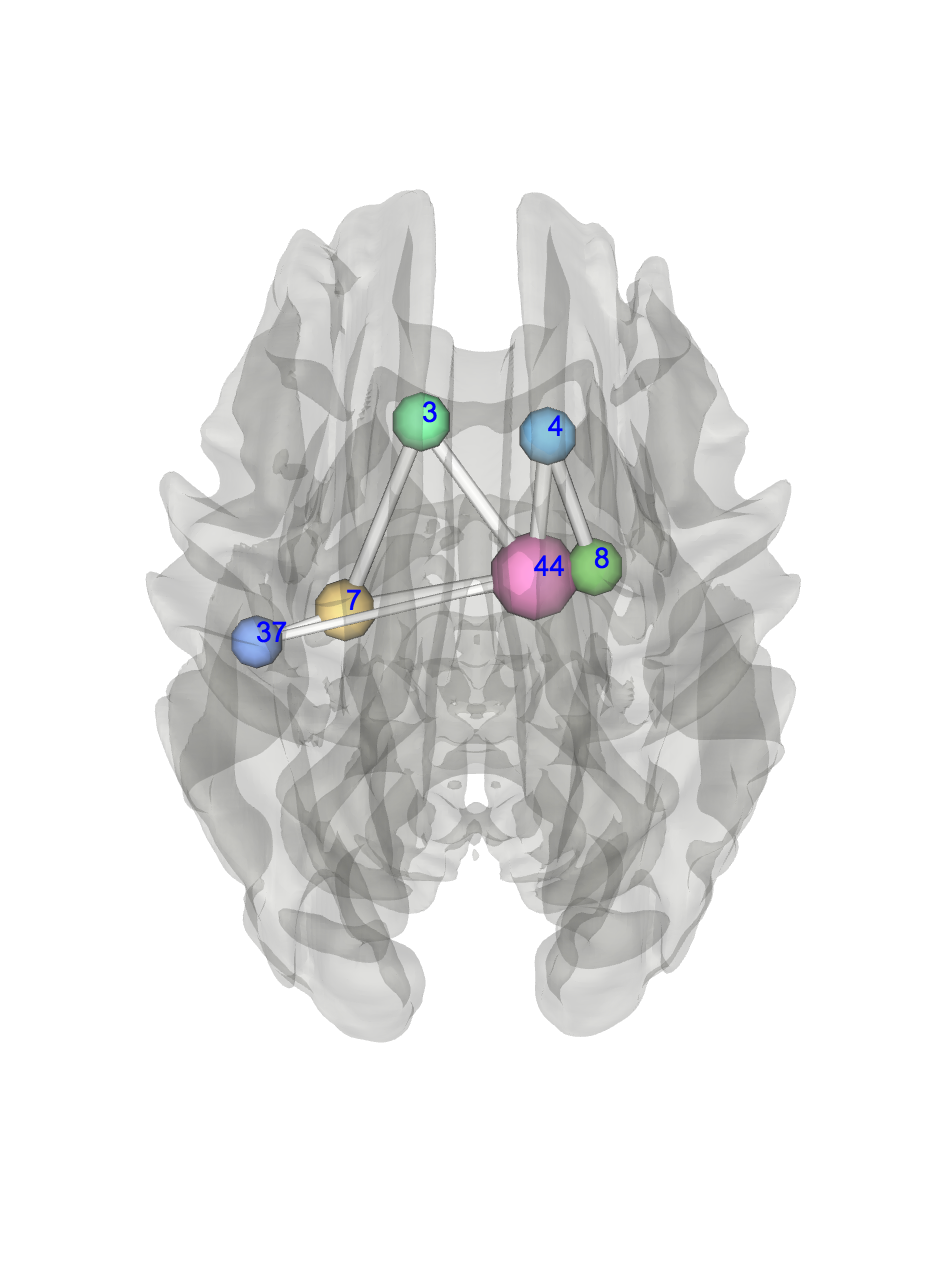} \\
		  \includegraphics[width=0.225\linewidth,height=4cm]{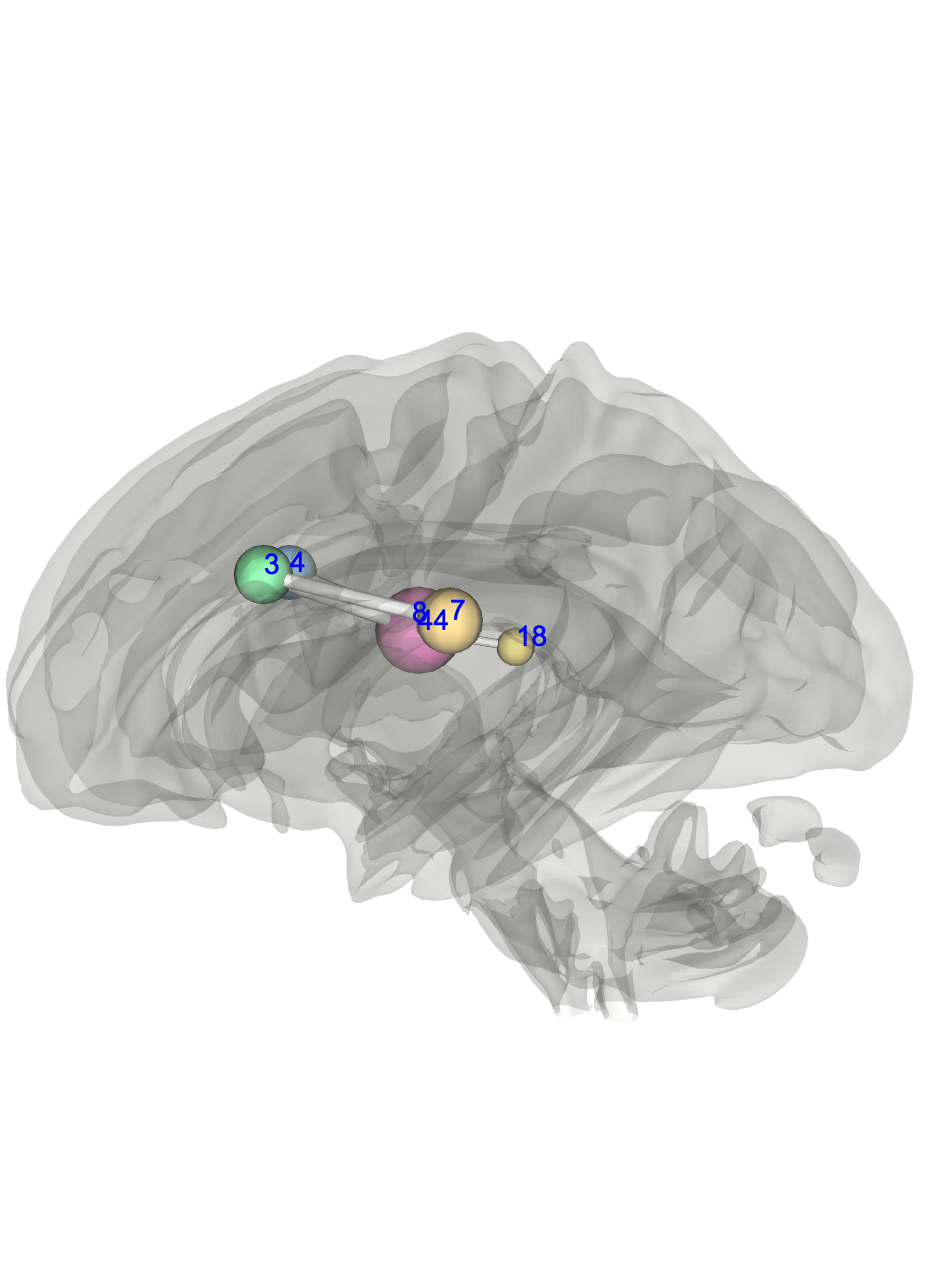} & 
		  \includegraphics[width=0.225\linewidth,height=4cm]{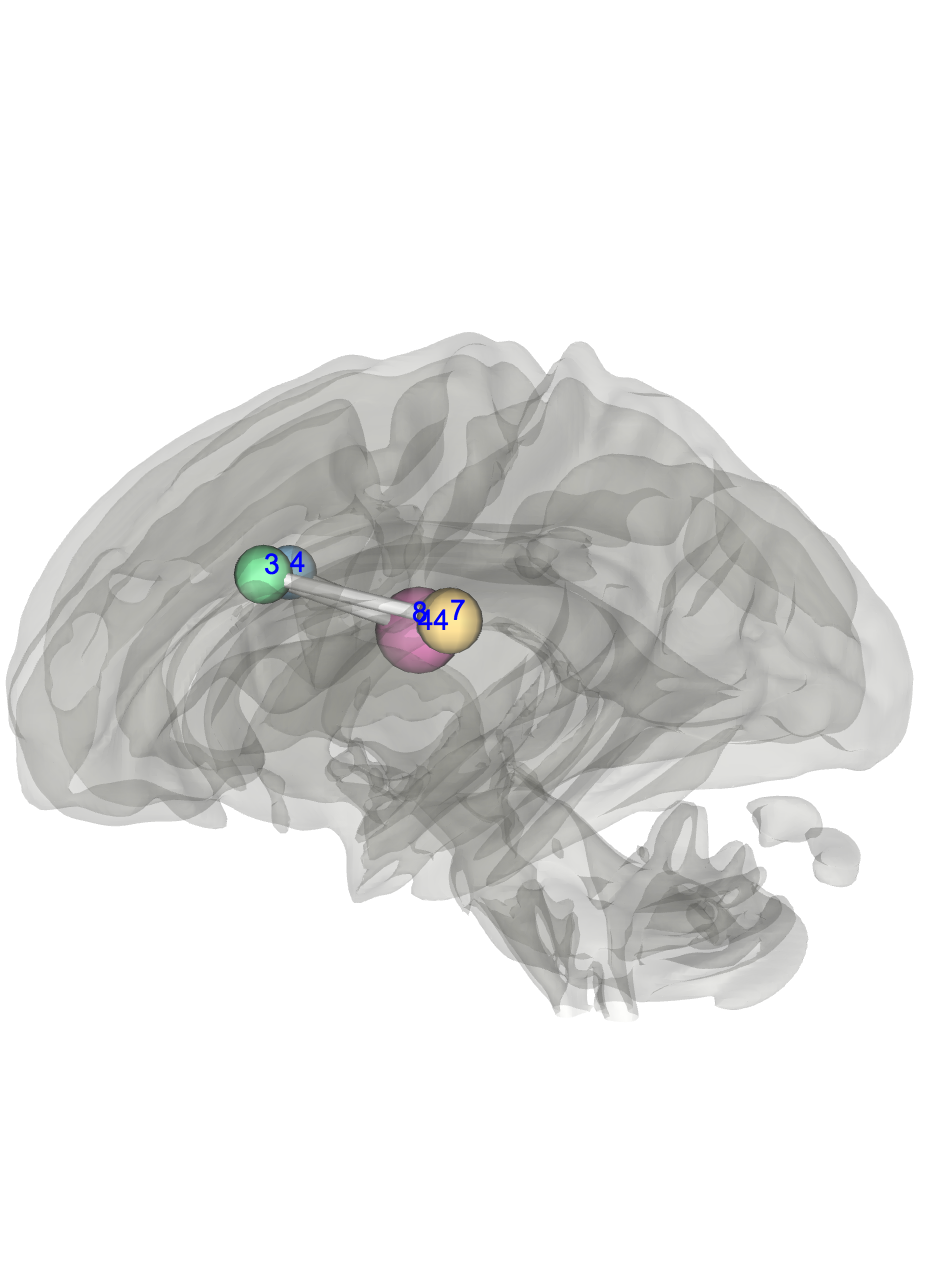} &
		   \includegraphics[width=0.225\linewidth, height=4cm]{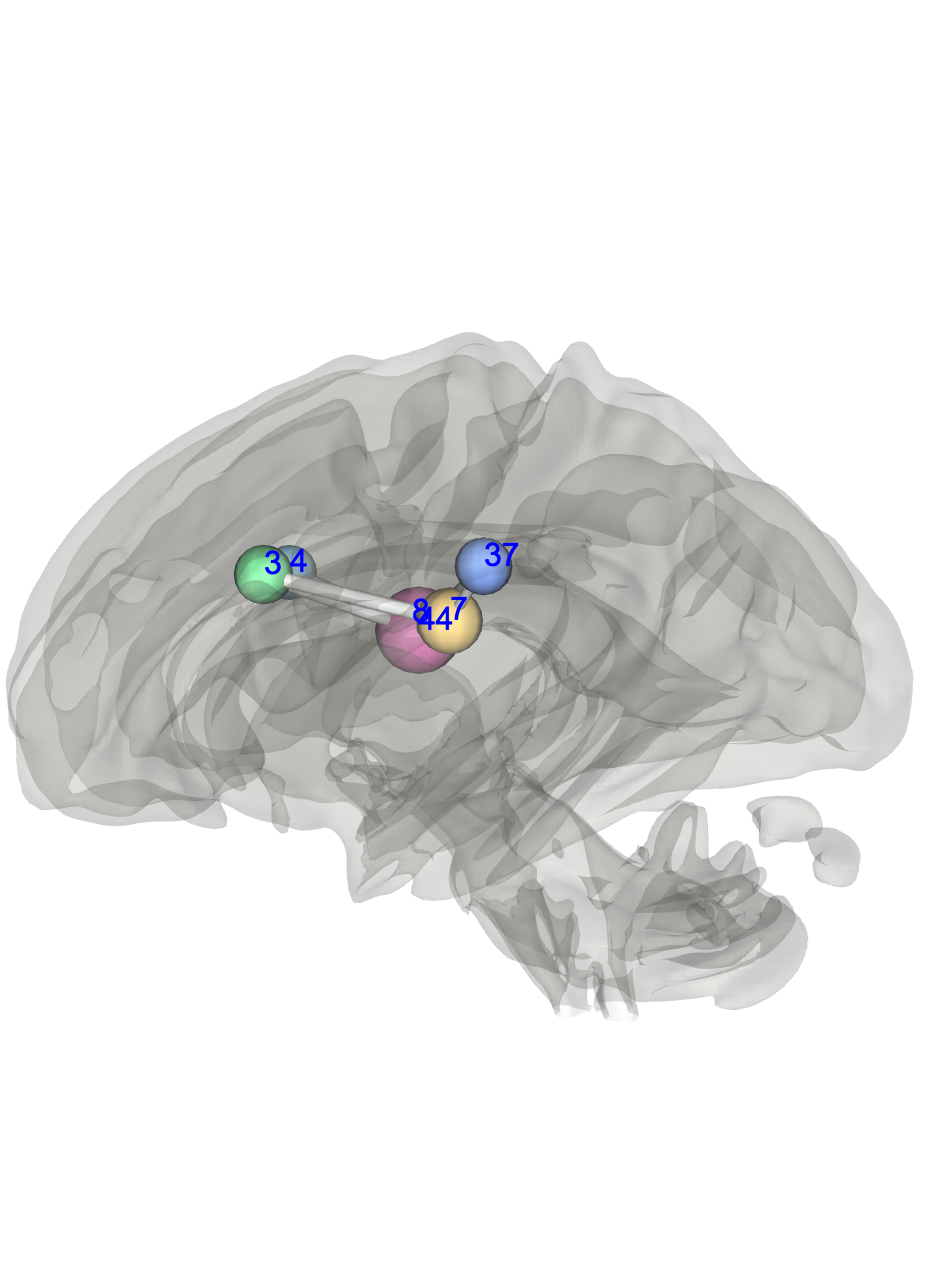} & 
		  \includegraphics[width=0.225\linewidth,height=4cm]{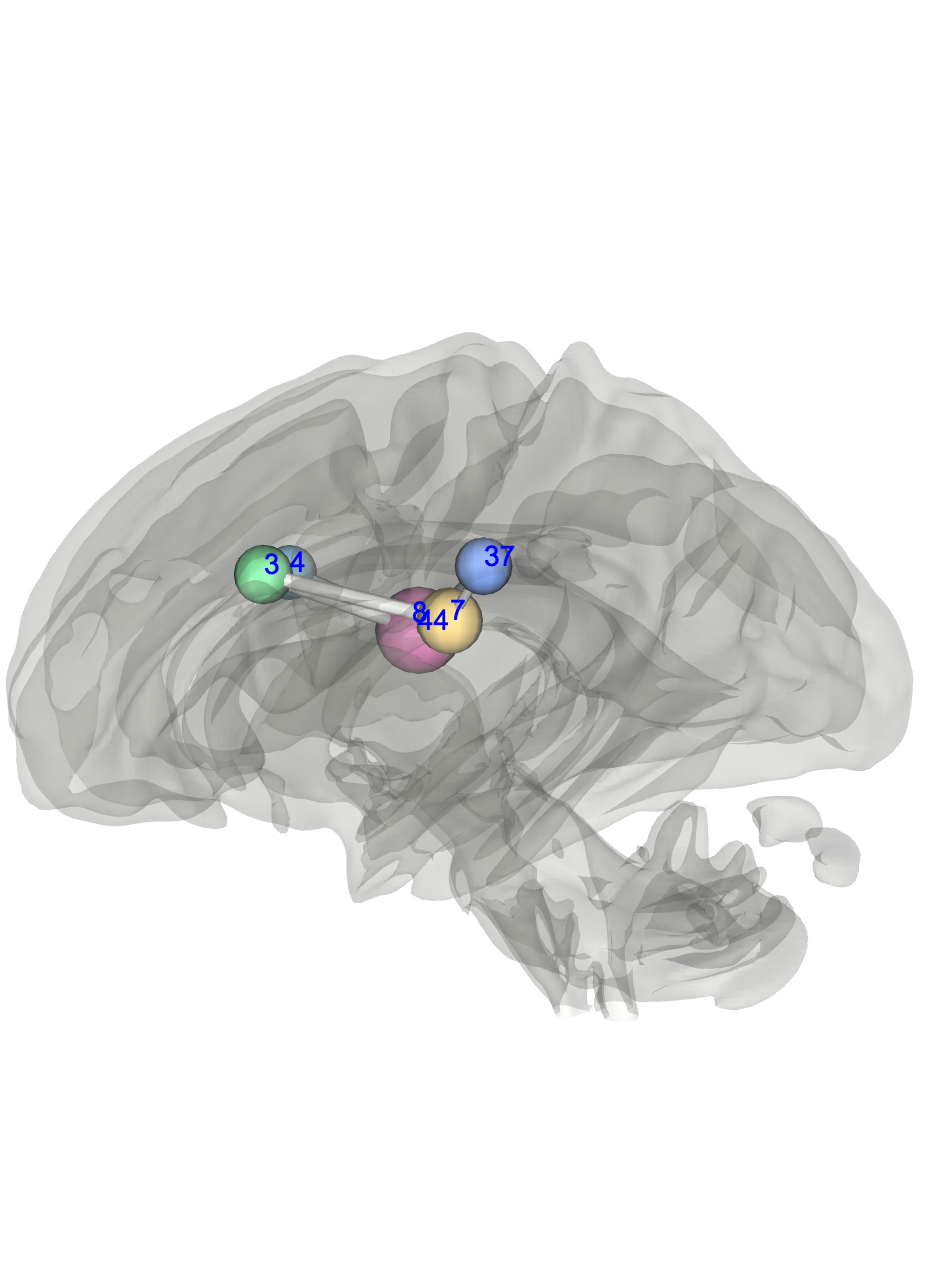} \\
		  {\small (a) $K$=50} & {\small (b) $K$=100} & {\small (c) $K$=200} & {\small (d) $K$=400}
	\end{tabular}
\caption{Visualization for trait PicVocab: each column represents the visualization for a different number of clusters ($K$=50; $K$=100; $K$=200; $K$=400) in PPA; First row represents the connectivity matrix between any two ROI regions in the HCP842 tractography atlas. Second row shows anatomy of connections in an axial view. Third row shows anatomy of connections in a sagittal view.} 
\label{fig:pv} 
\end{figure}

\begin{figure}[ht!]
\centering 
	\begin{tabular}{cccc}
		 \includegraphics[width=0.225\linewidth, height=3.5cm]{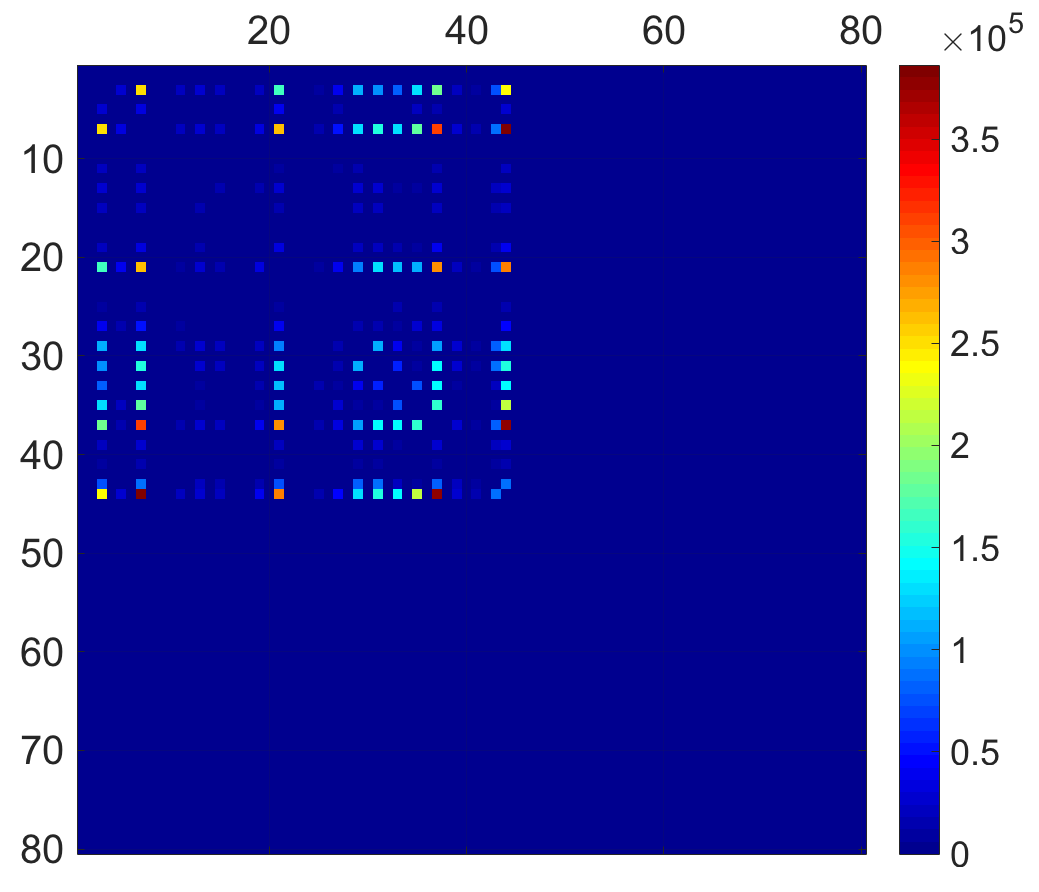} & 
		  \includegraphics[width=0.225\linewidth,height=3.5cm]{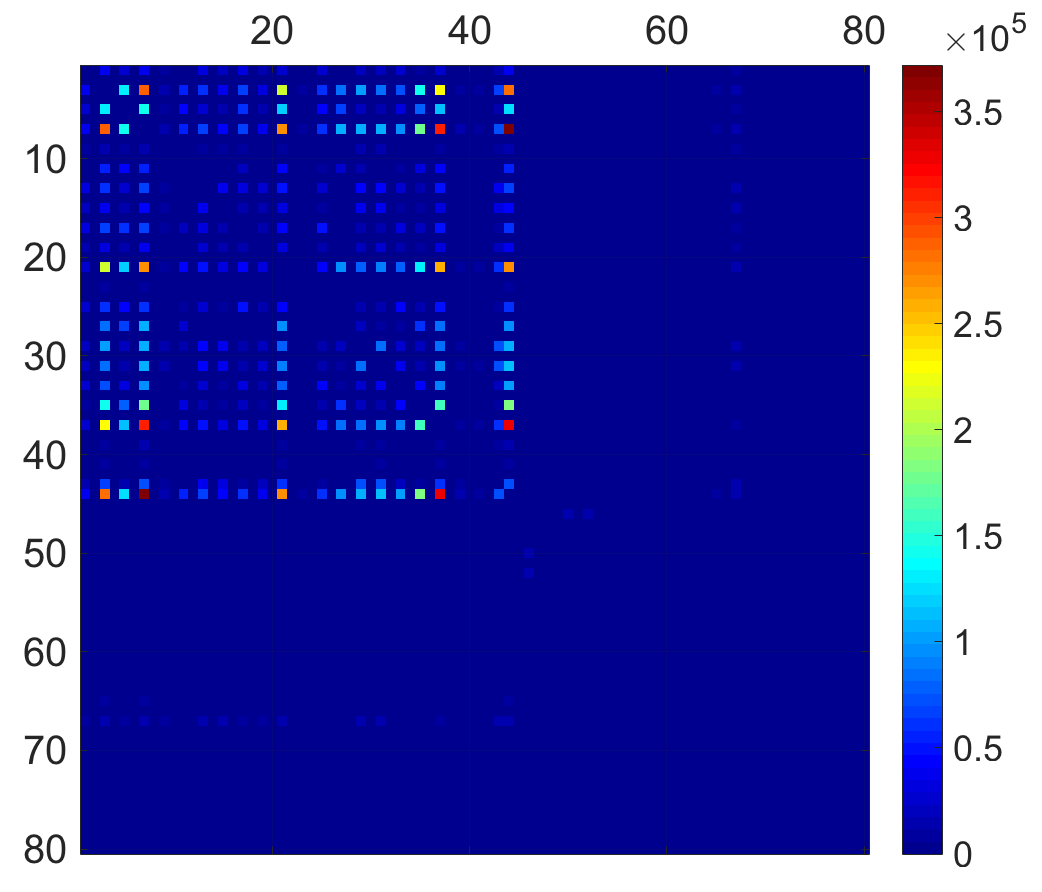} &
		   \includegraphics[width=0.225\linewidth, height=3.5cm]{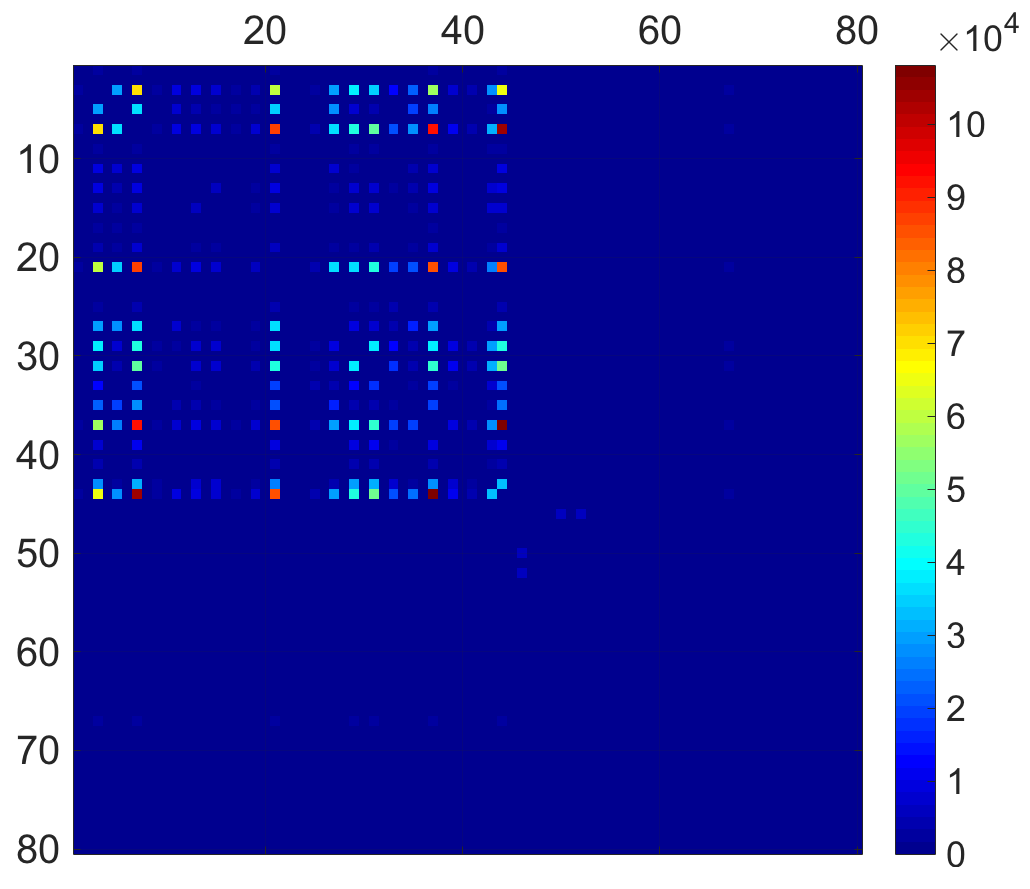} & 
		  \includegraphics[width=0.225\linewidth,height=3.5cm]{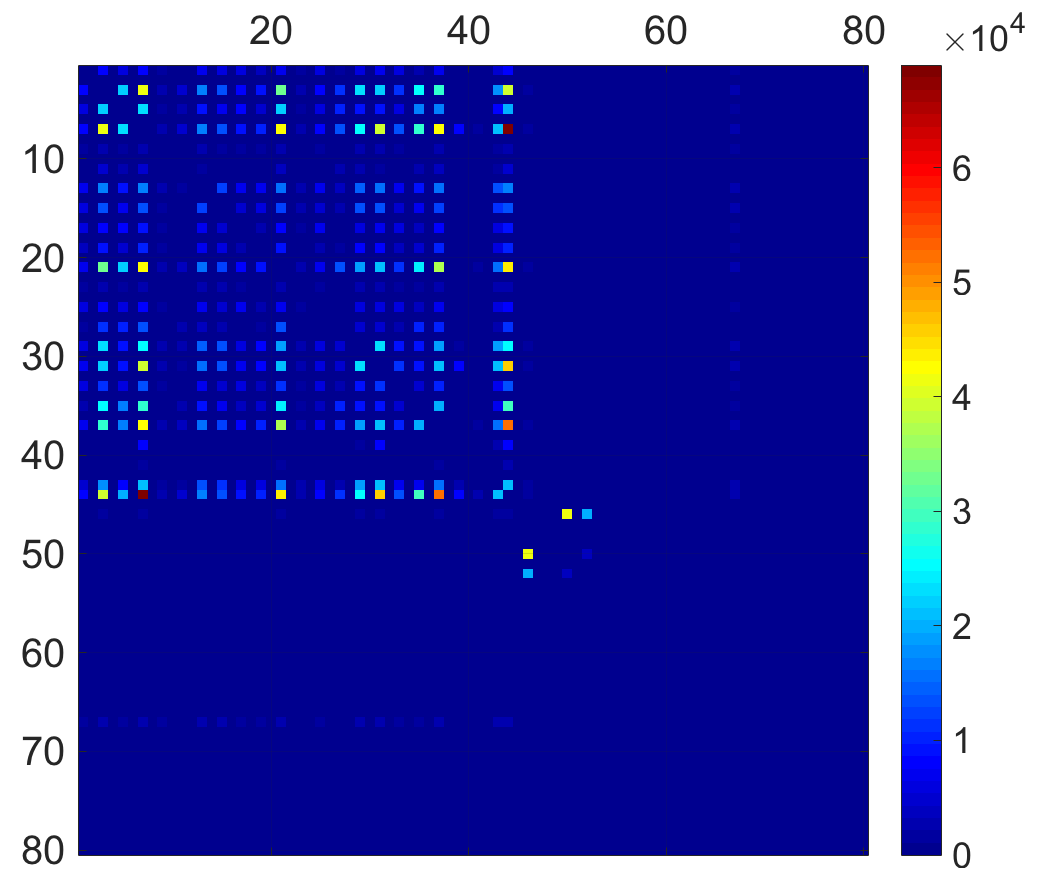} \\
		  \includegraphics[width=0.225\linewidth,height=4cm]{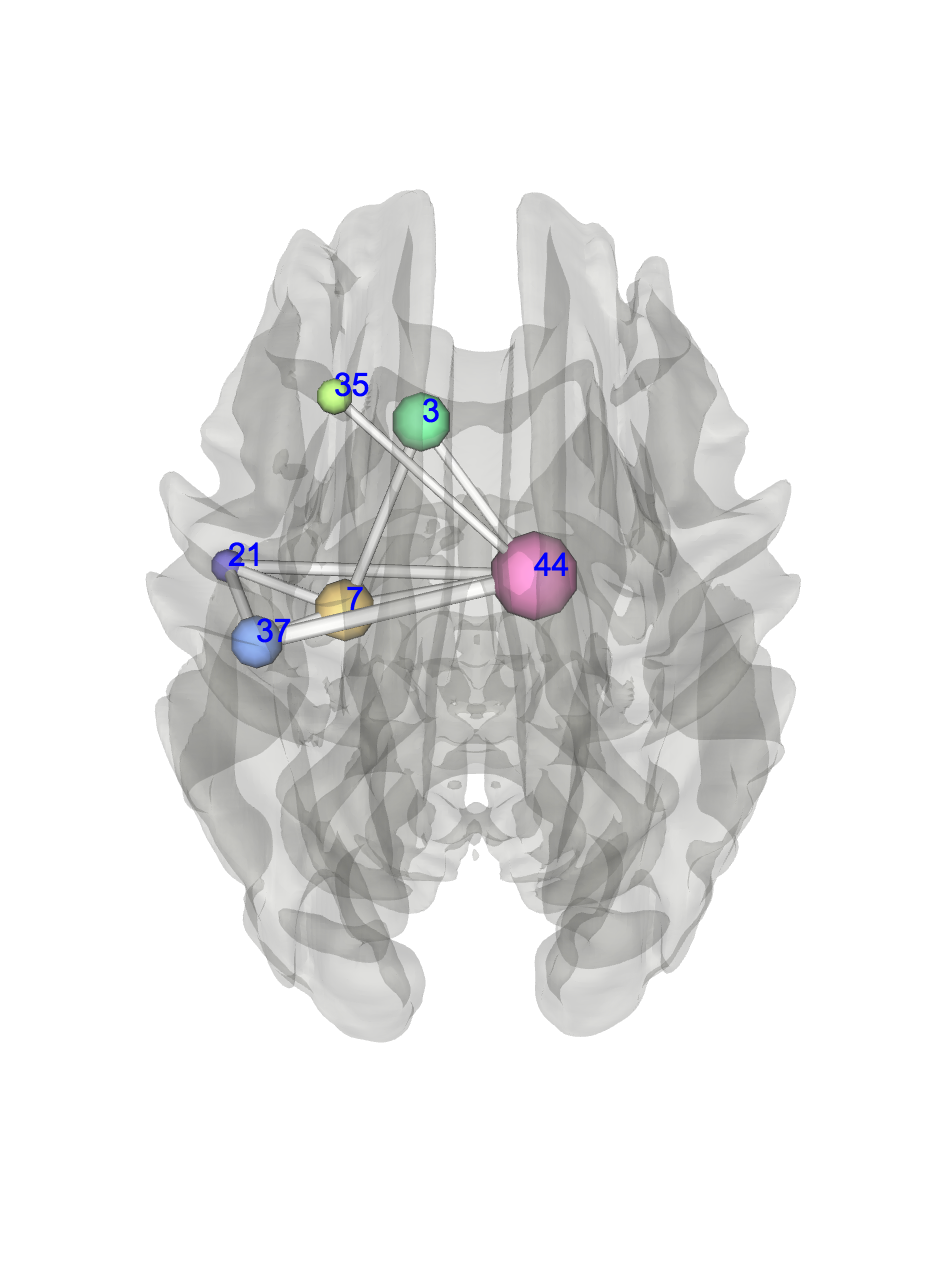} & 
		  \includegraphics[width=0.225\linewidth,height=4cm]{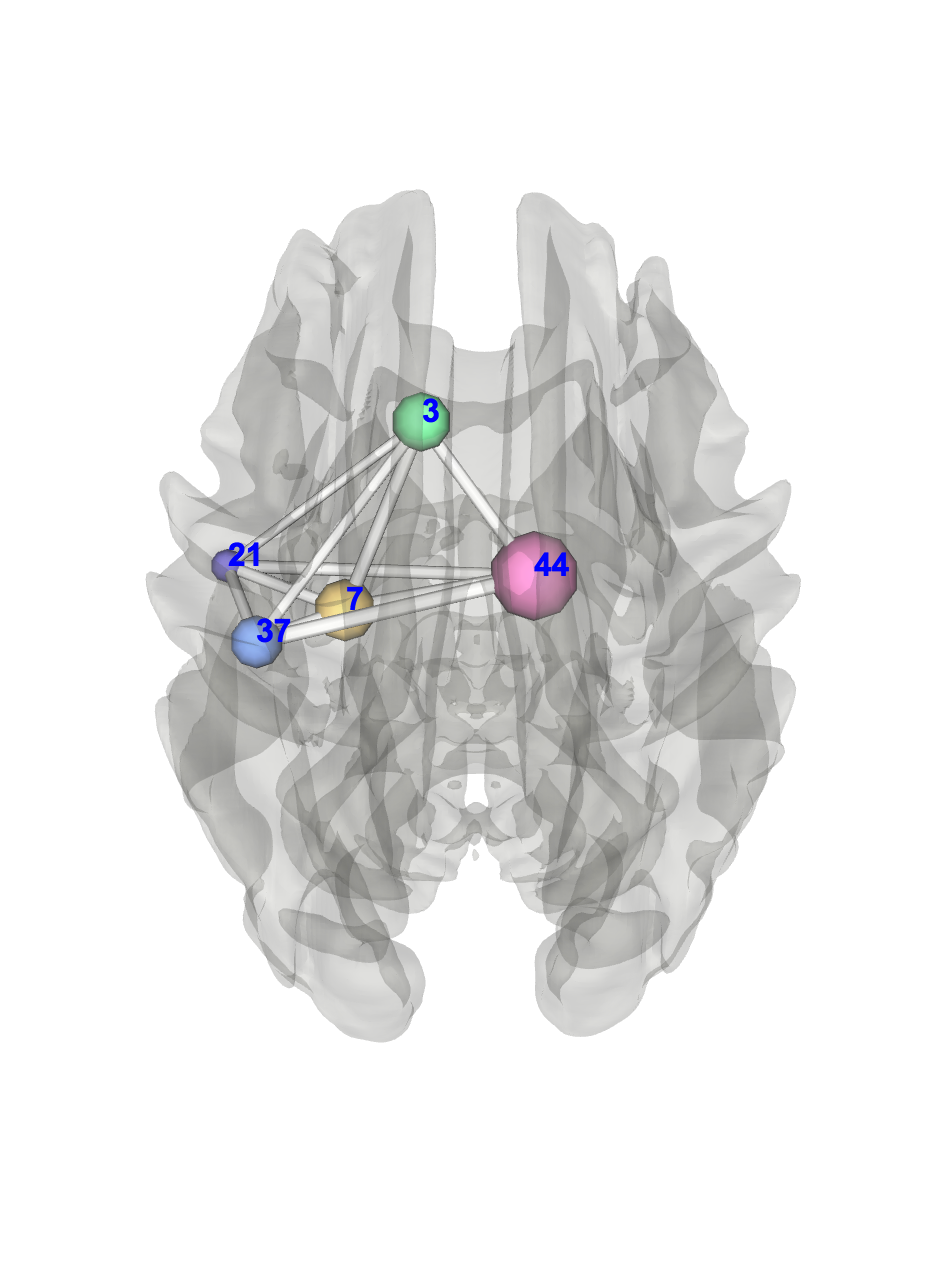} &
		   \includegraphics[width=0.225\linewidth, height=4cm]{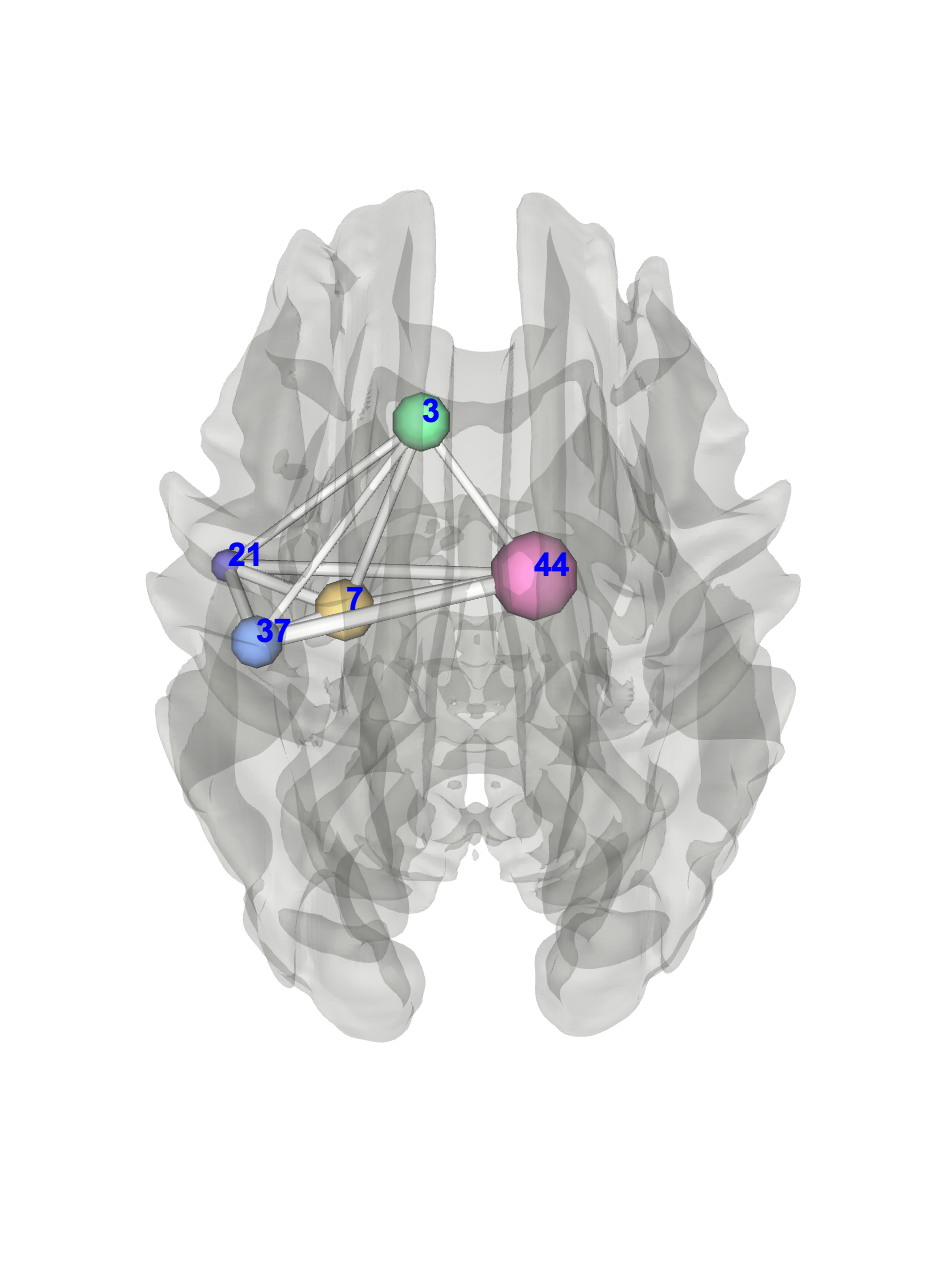} & 
		  \includegraphics[width=0.225\linewidth,height=4cm]{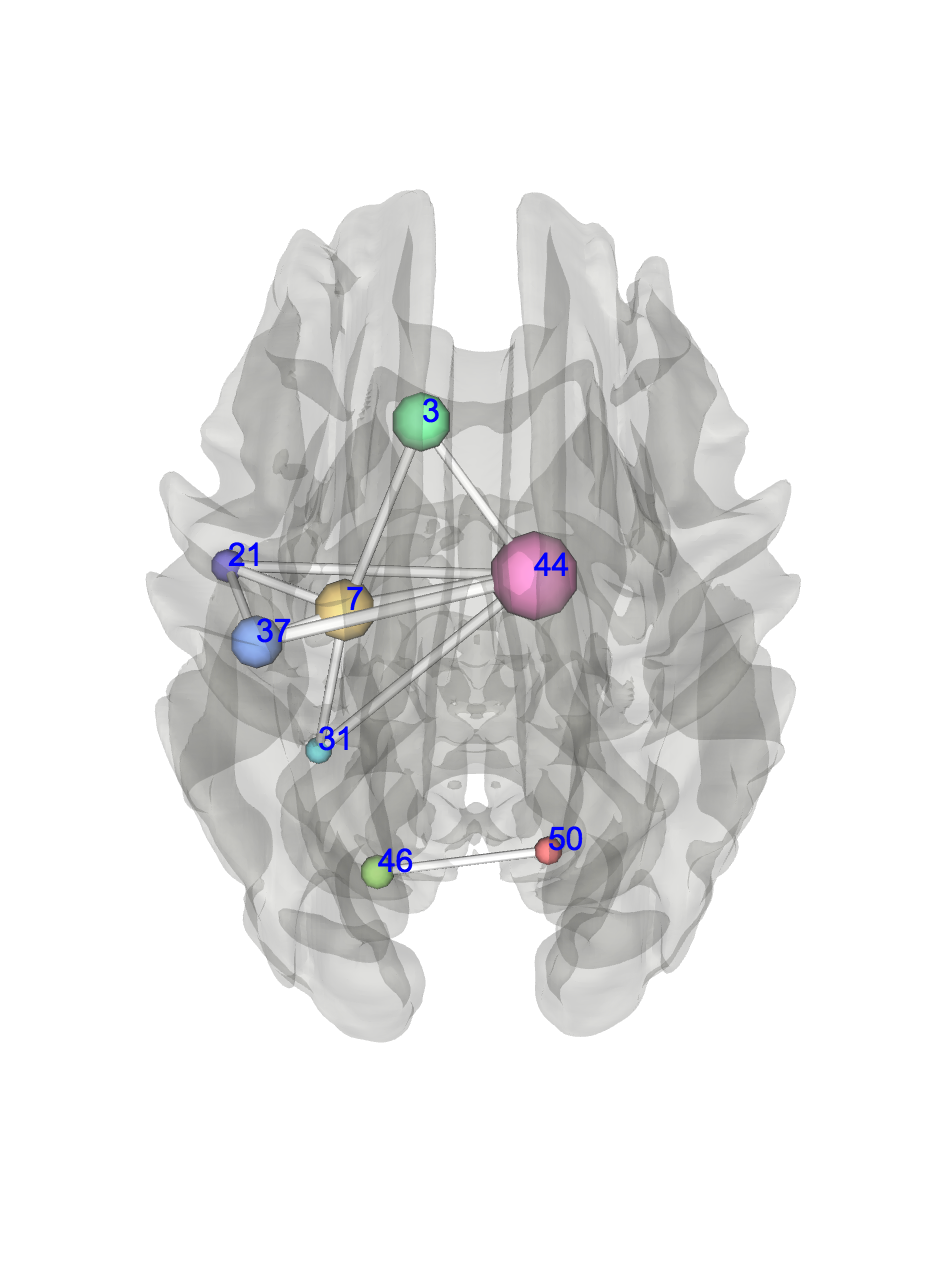} \\
		  \includegraphics[width=0.225\linewidth,height=4cm]{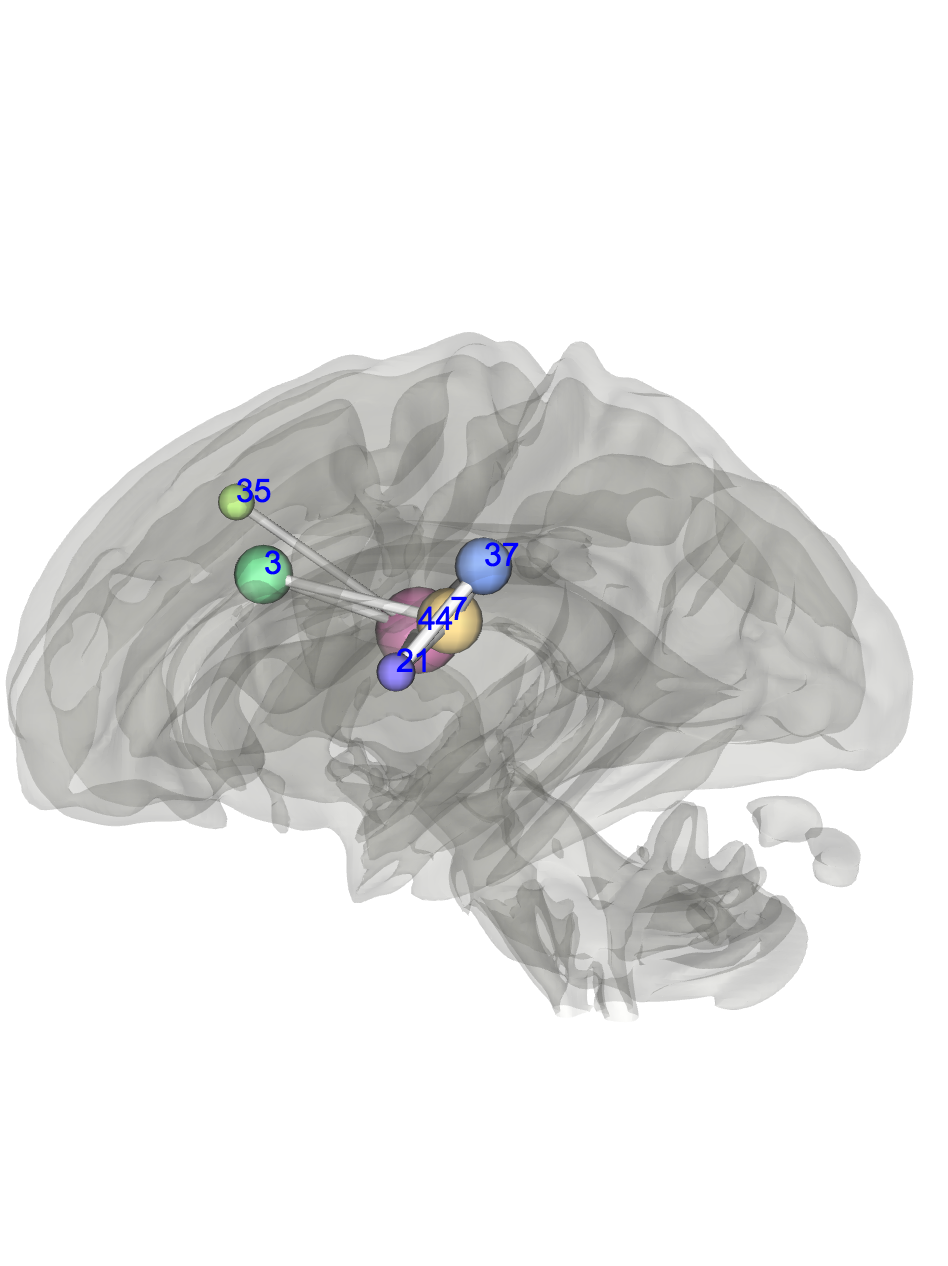} & 
		  \includegraphics[width=0.225\linewidth,height=4cm]{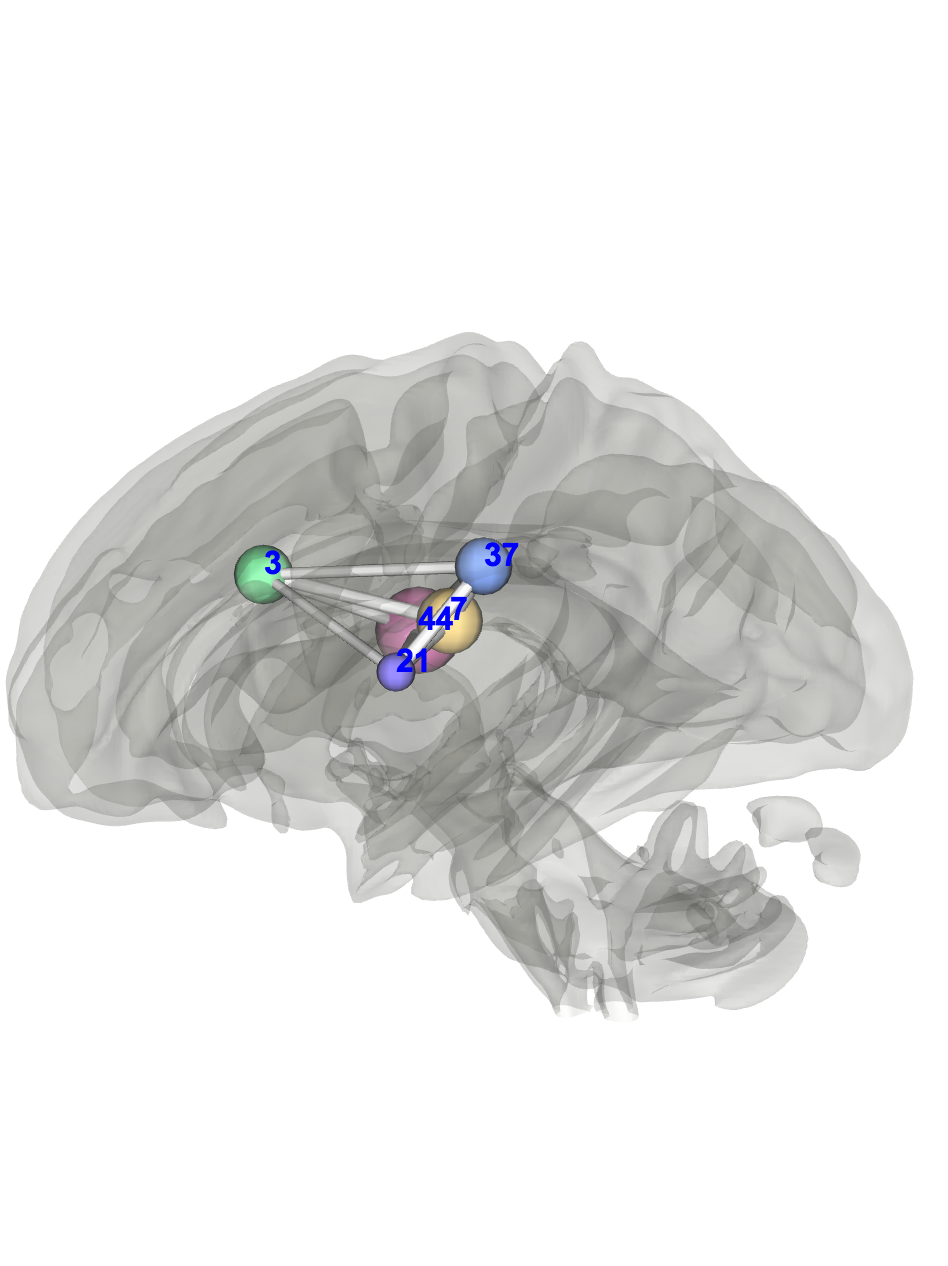} &
		   \includegraphics[width=0.225\linewidth, height=4cm]{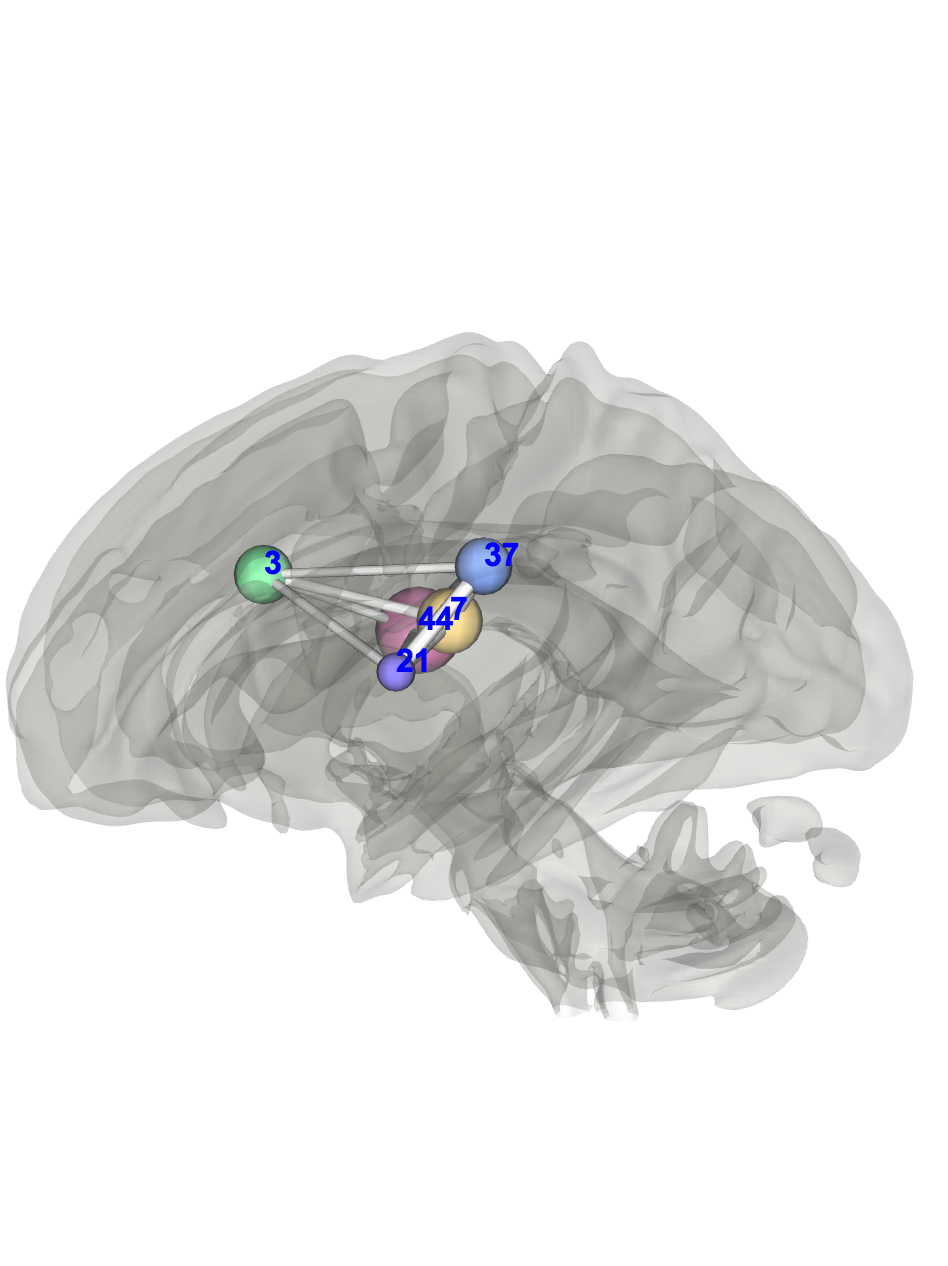} & 
		  \includegraphics[width=0.225\linewidth,height=4cm]{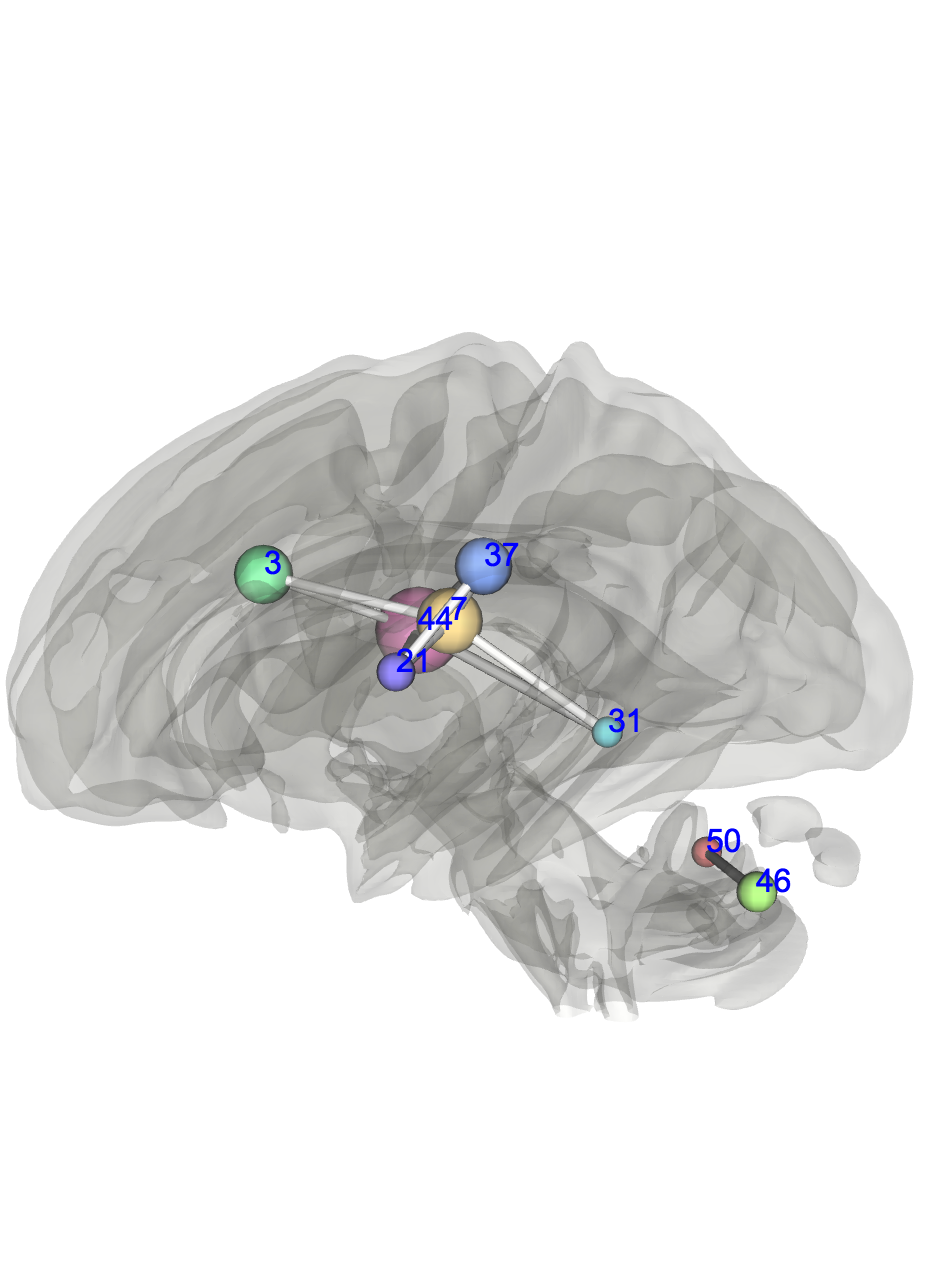} \\
		  {\small (a) $K$=50} & {\small (b) $K$=100} & {\small (c) $K$=200} & {\small (d) $K$=400}
	\end{tabular}
\caption{Visualization for trait ListSort: each column represents the visualization under different $K$ ($K$=50; $K$=100; $K$=200; $K$=400) in PPA; First row represents the connectivity matrix between any two ROI regions in HCP842 tractography atlas. Second row shows anatomy of connections in an axial view. Third row shows anatomy of connections in a sagittal view.}
\label{fig:ls} 
\end{figure}

\begin{figure}[ht!]
\centering 
	\begin{tabular}{cccc}
		 \includegraphics[width=0.225\linewidth, height=3.5cm]{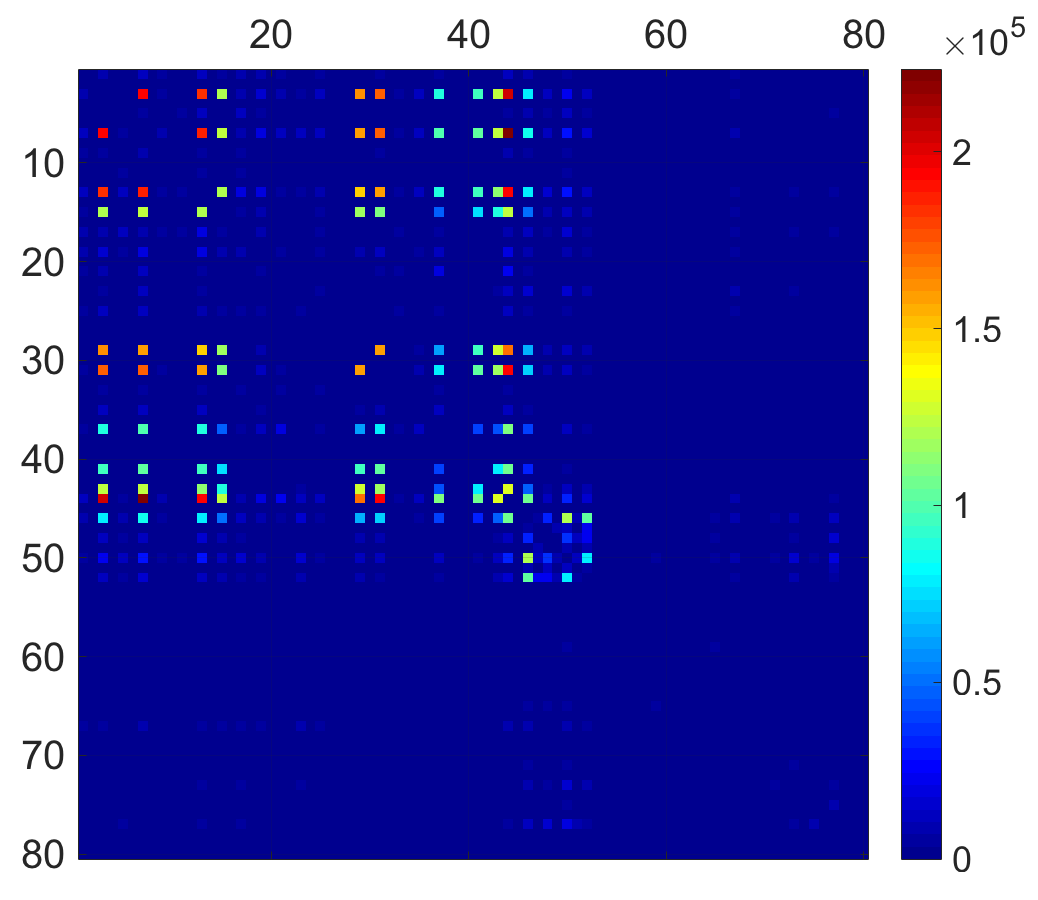} & 
		  \includegraphics[width=0.225\linewidth,height=3.5cm]{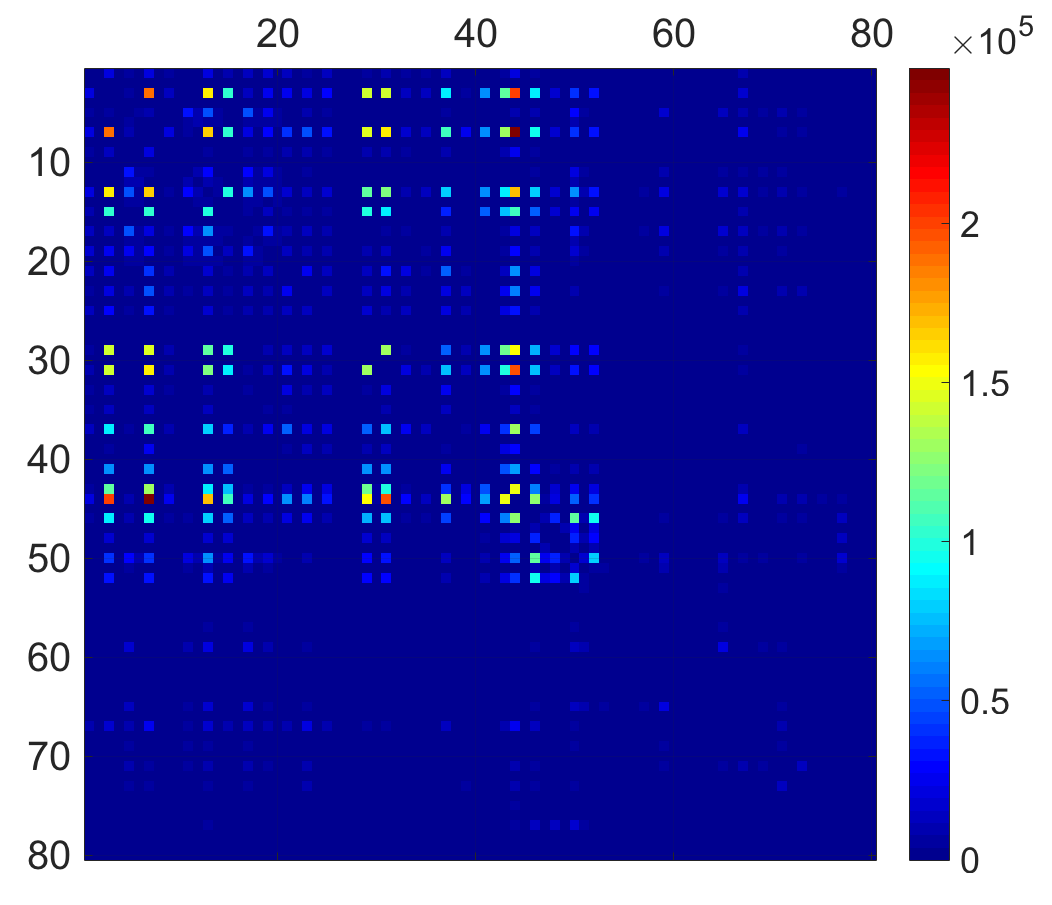} &
		   \includegraphics[width=0.225\linewidth, height=3.5cm]{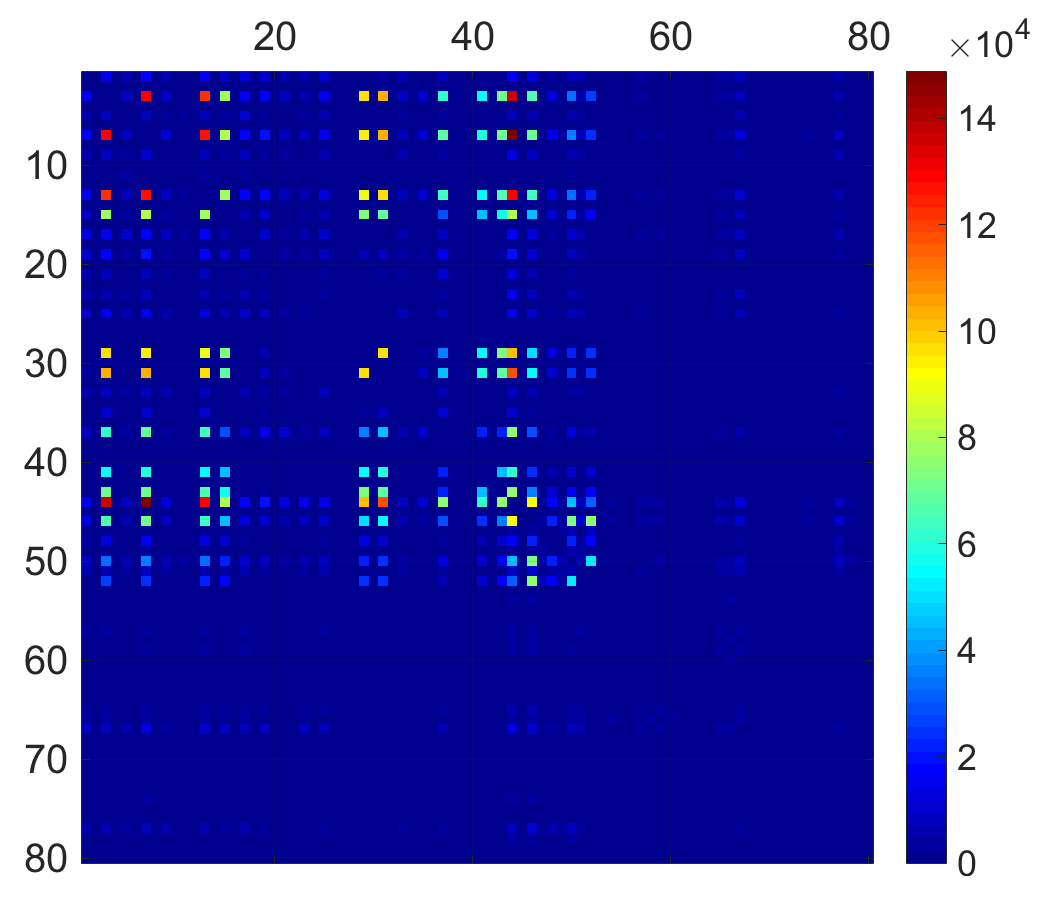} & 
		  \includegraphics[width=0.225\linewidth,height=3.5cm]{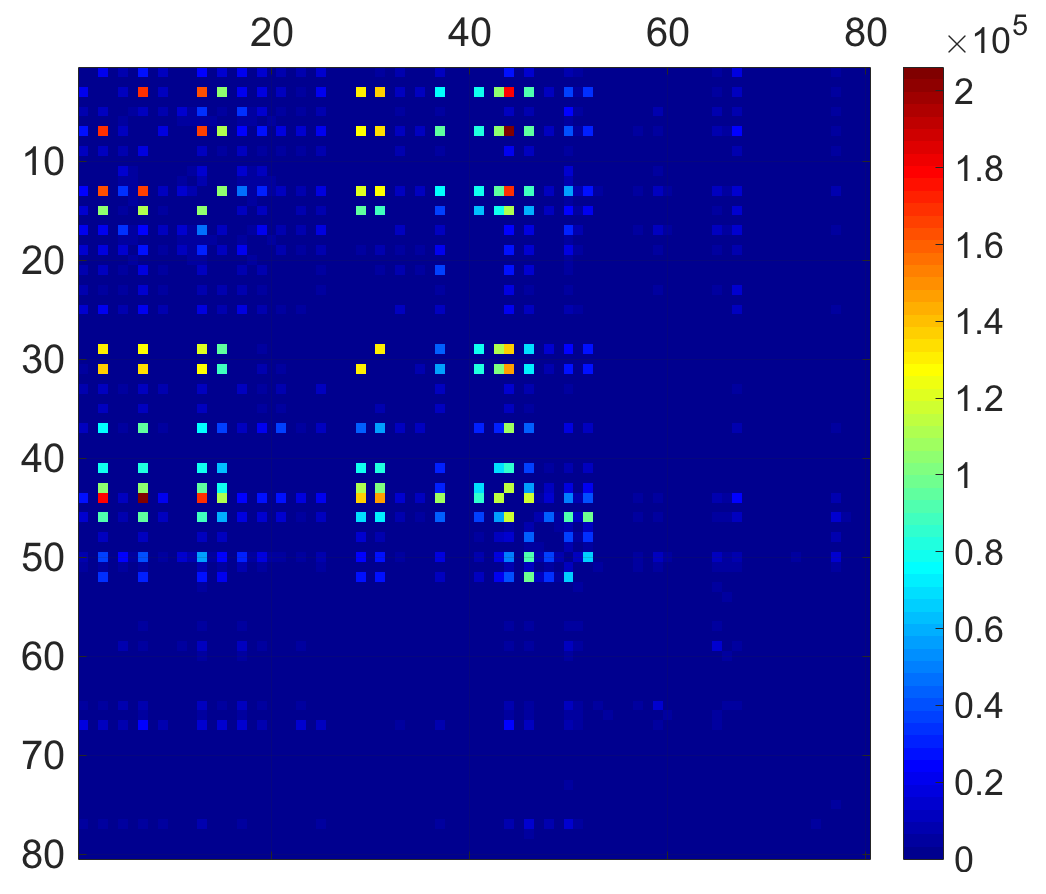} \\
		  \includegraphics[width=0.225\linewidth,height=4cm]{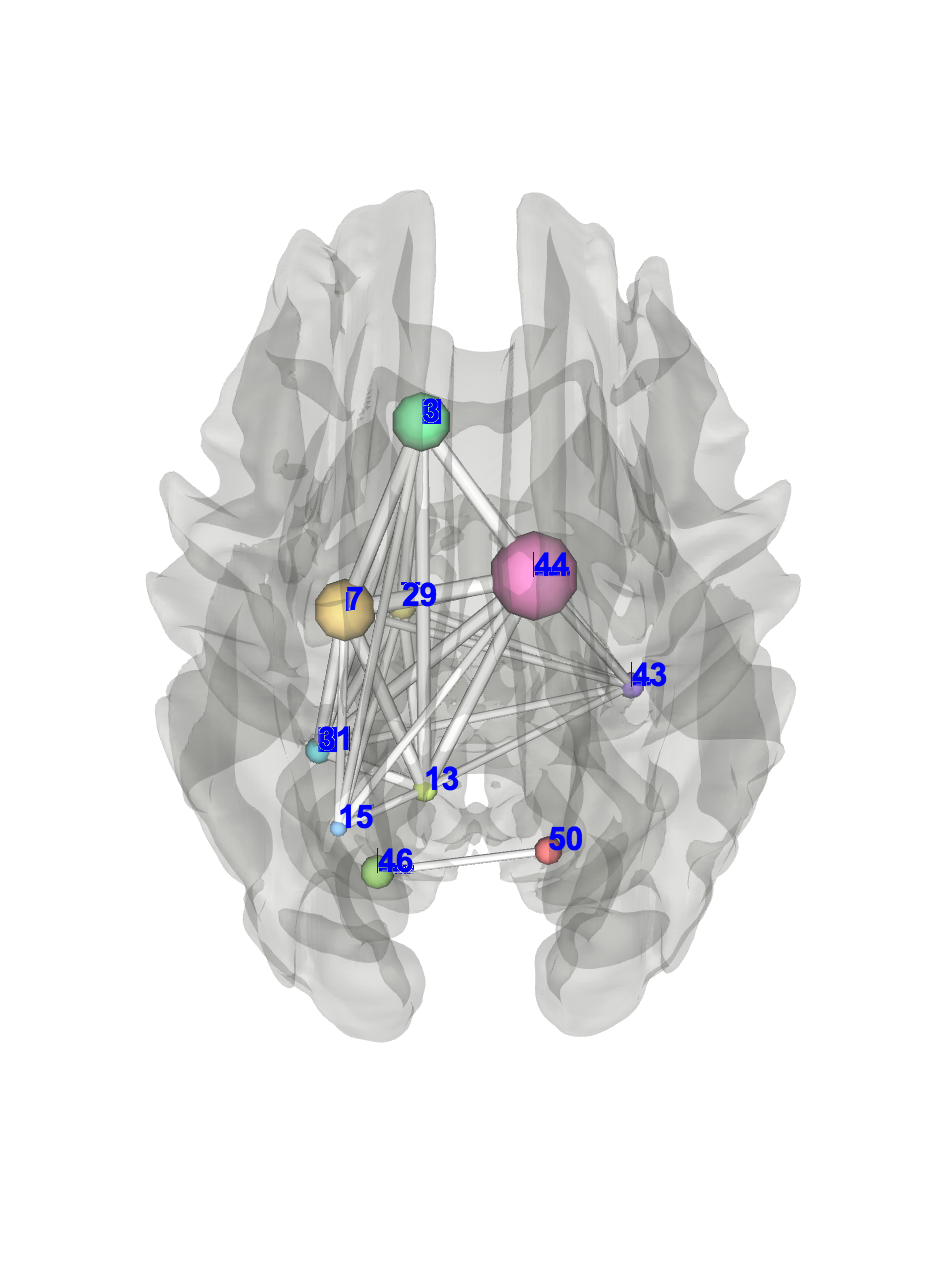} & 
		  \includegraphics[width=0.225\linewidth,height=4cm]{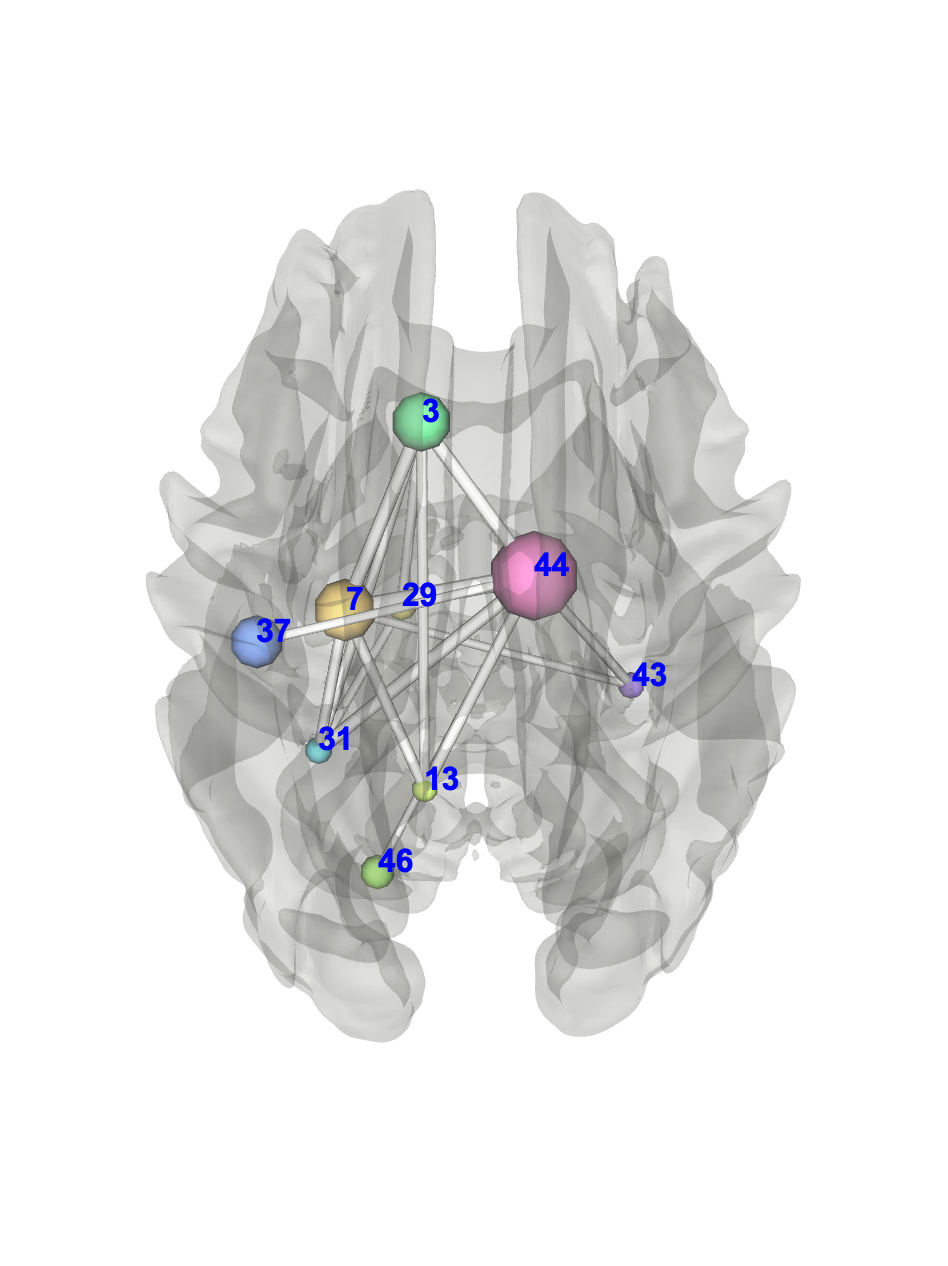} &
		   \includegraphics[width=0.225\linewidth, height=4cm]{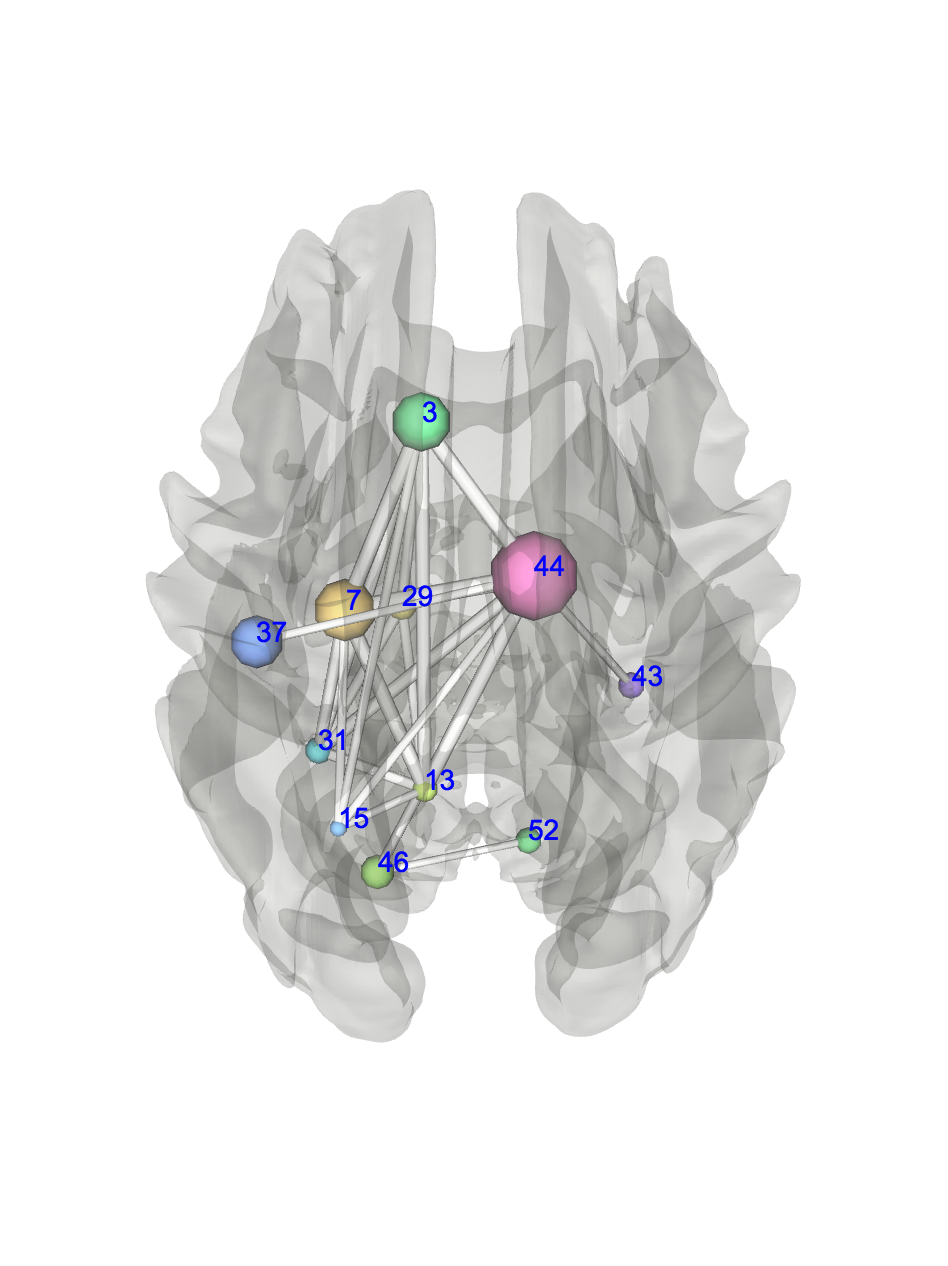} & 
		  \includegraphics[width=0.225\linewidth,height=4cm]{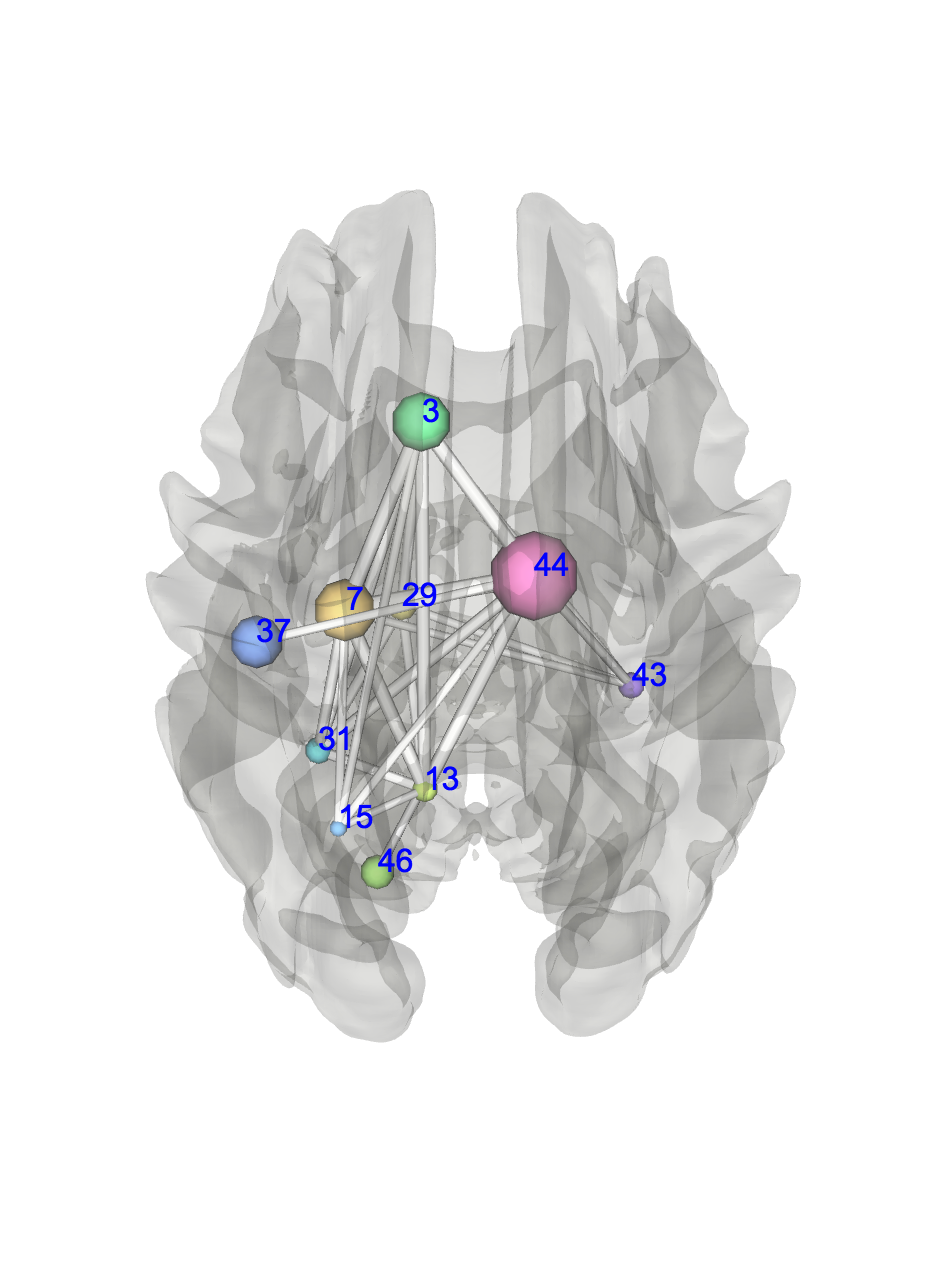} \\
		  \includegraphics[width=0.225\linewidth,height=4cm]{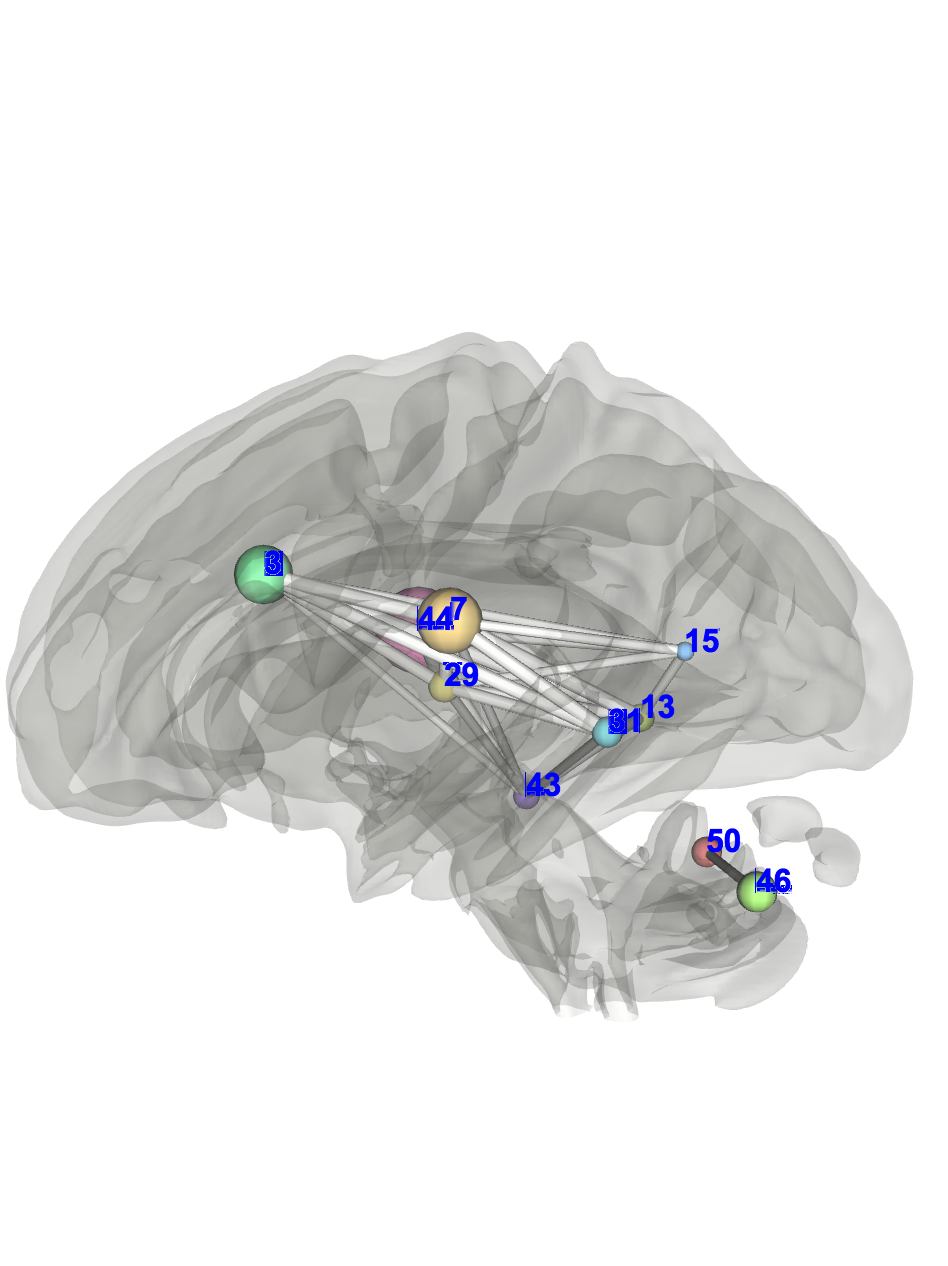} & 
		  \includegraphics[width=0.225\linewidth,height=4cm]{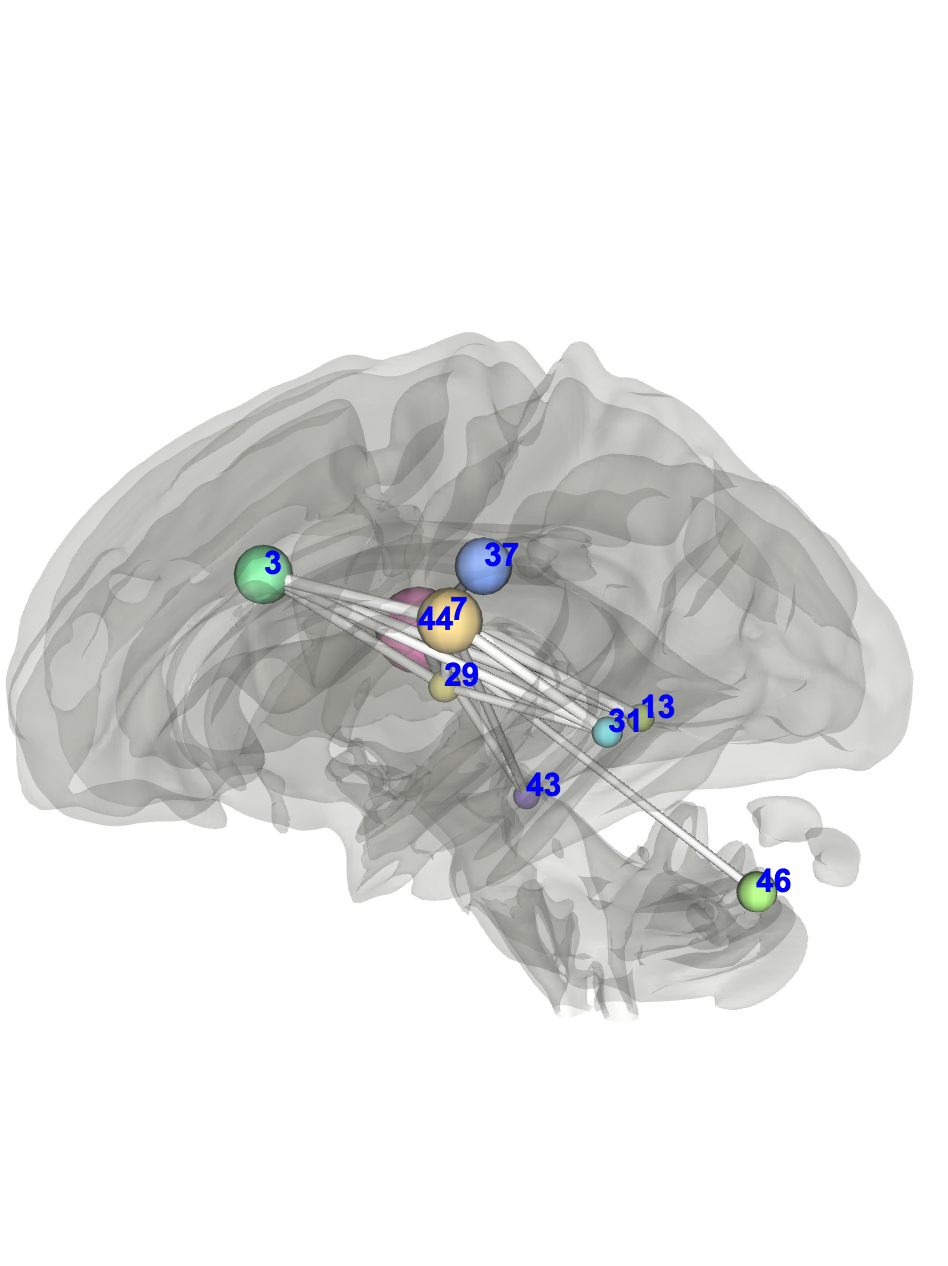} &
		   \includegraphics[width=0.225\linewidth, height=4cm]{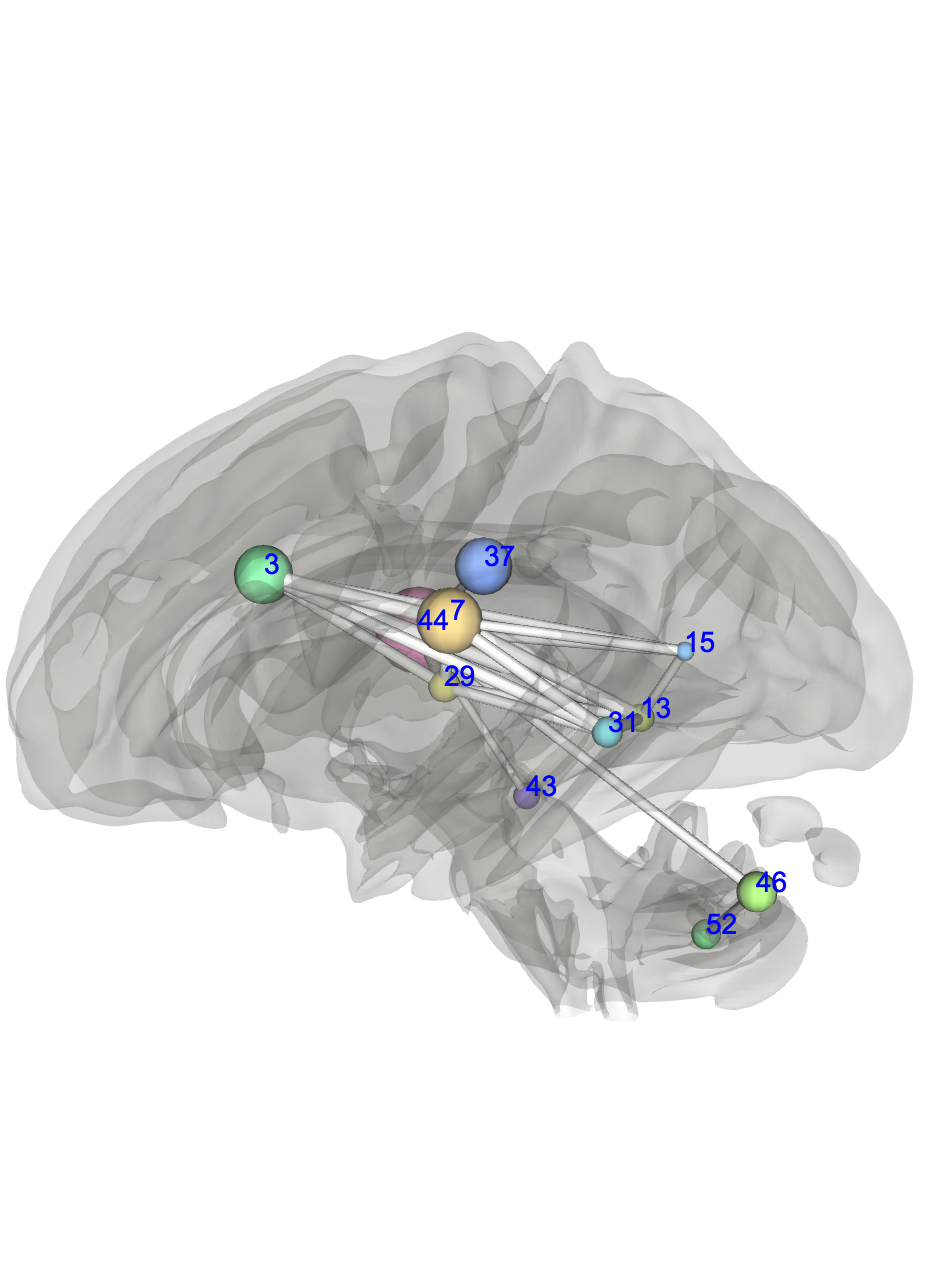} & 
		  \includegraphics[width=0.225\linewidth,height=4cm]{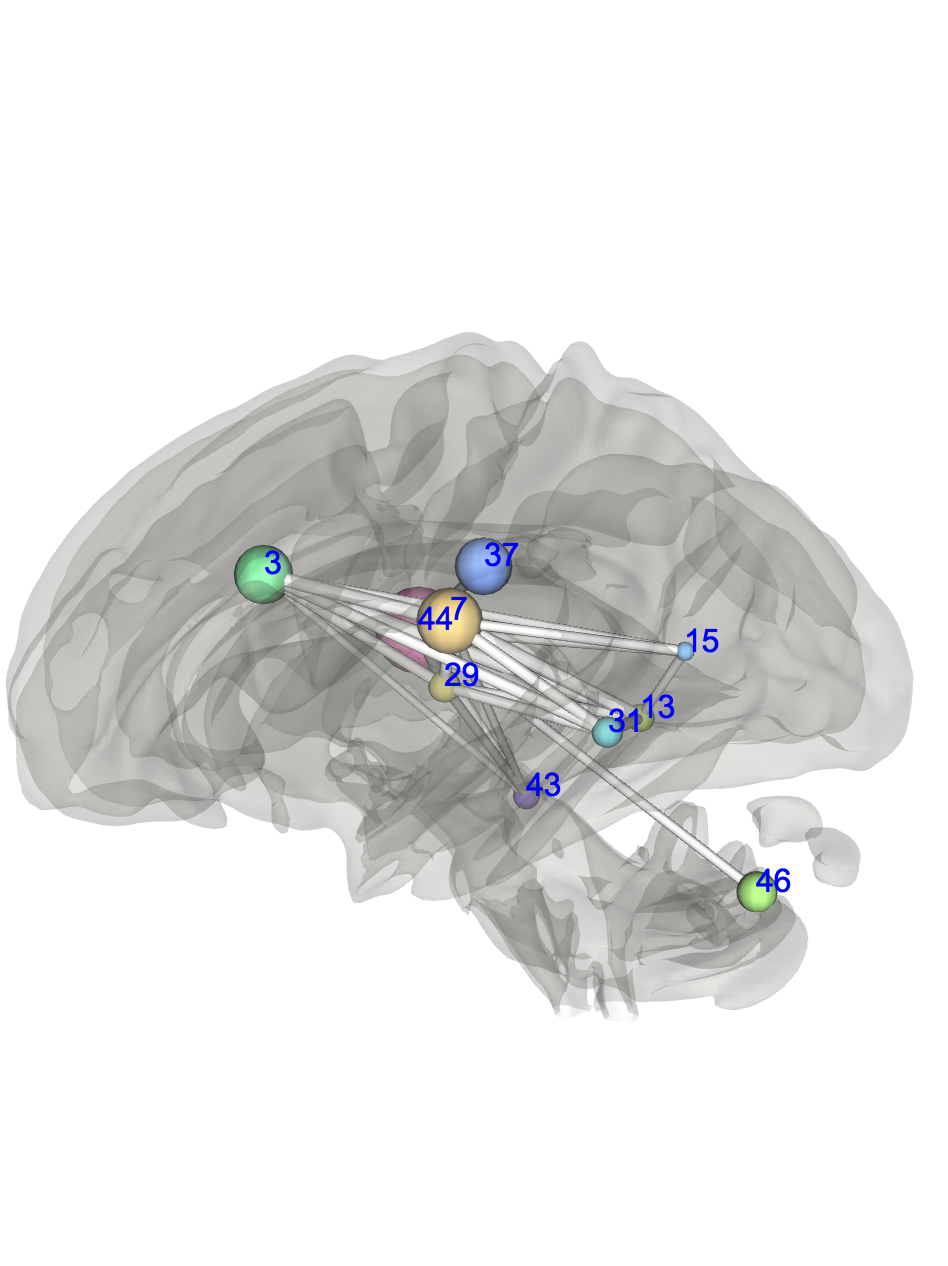} \\
		  {\small (a) $K$=50} & {\small (b) $K$=100} & {\small (c) $K$=200} & {\small (d) $K$=400}
	\end{tabular}
\caption{Visualization for trait ReadEng: each column represents the visualization under different $K$ ($K$=50; $K$=100; $K$=200; $K$=400) in PPA; First row represents the connectivity matrix between any two ROI regions in HCP842 tractography atlas. Second row shows anatomy of connections in an axial view. Third row shows anatomy of connections in a sagittal view.} 
\label{fig:re} 
\end{figure}

\begin{figure}[ht!]
\centering 
	\begin{tabular}{cccc}
		 \includegraphics[width=0.225\linewidth, height=3.5cm]{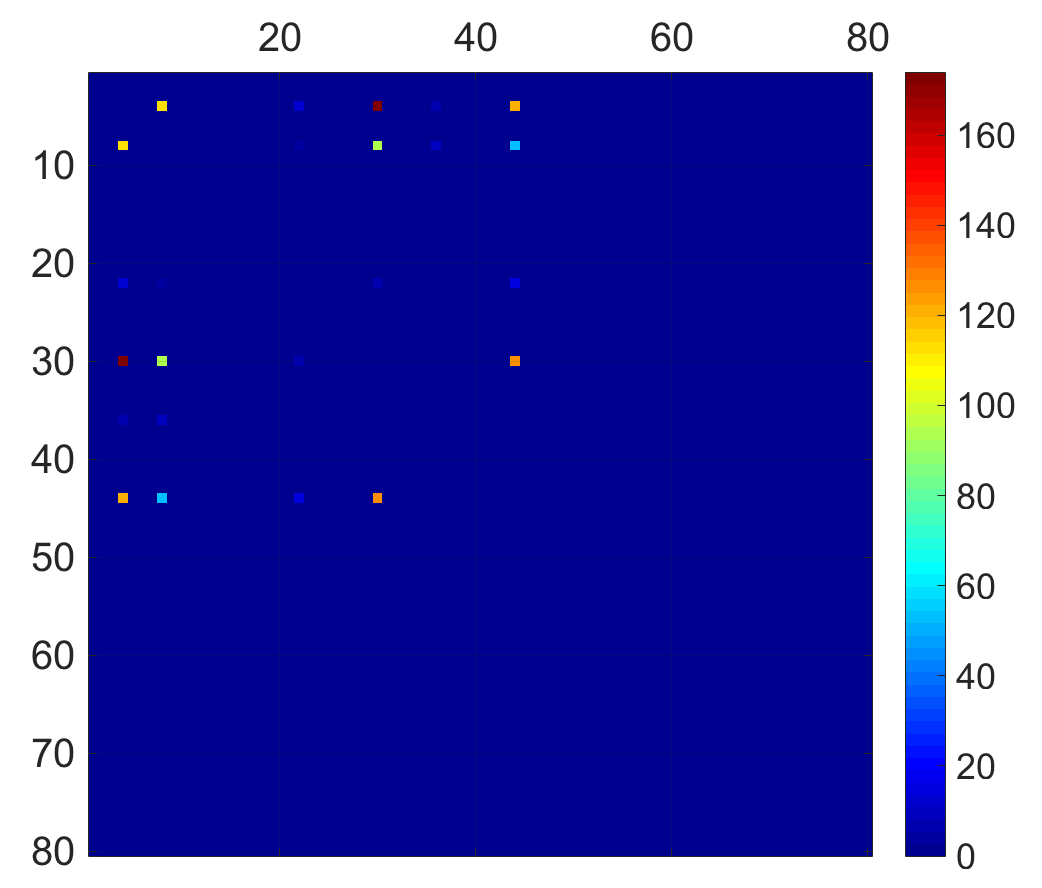} & 
		  \includegraphics[width=0.225\linewidth,height=3.5cm]{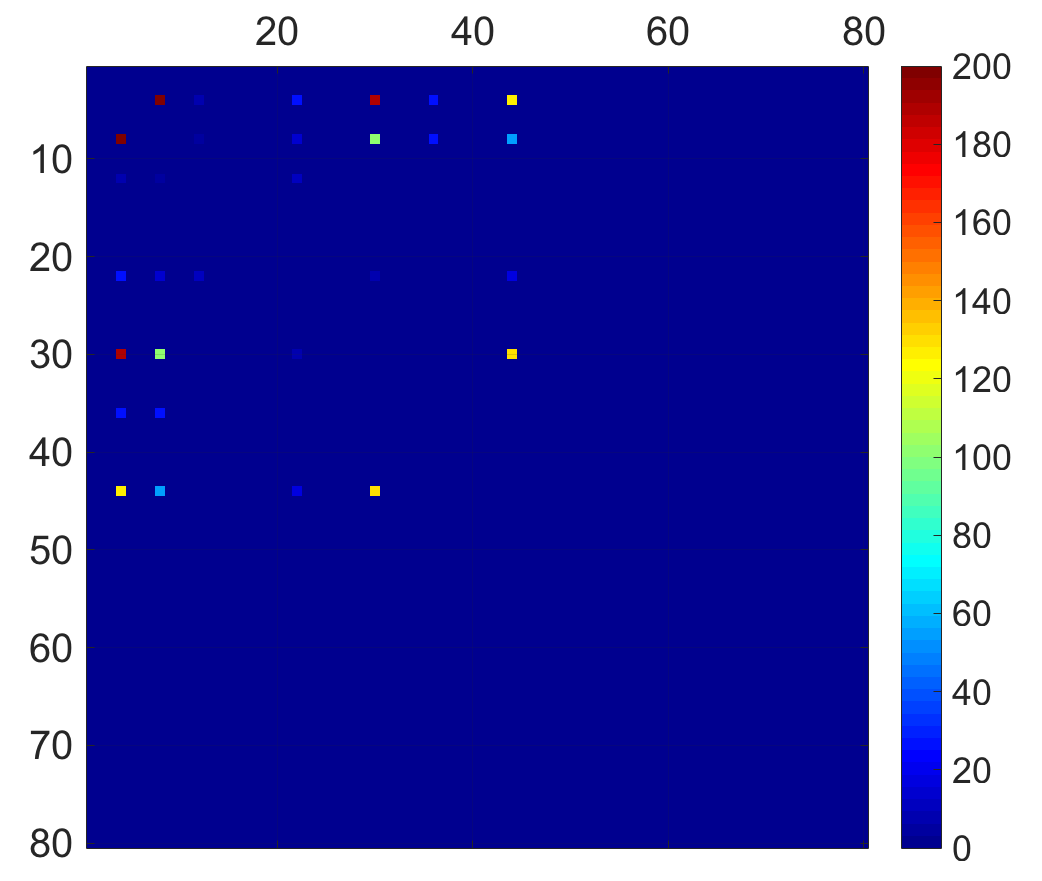} &
		   \includegraphics[width=0.225\linewidth, height=3.5cm]{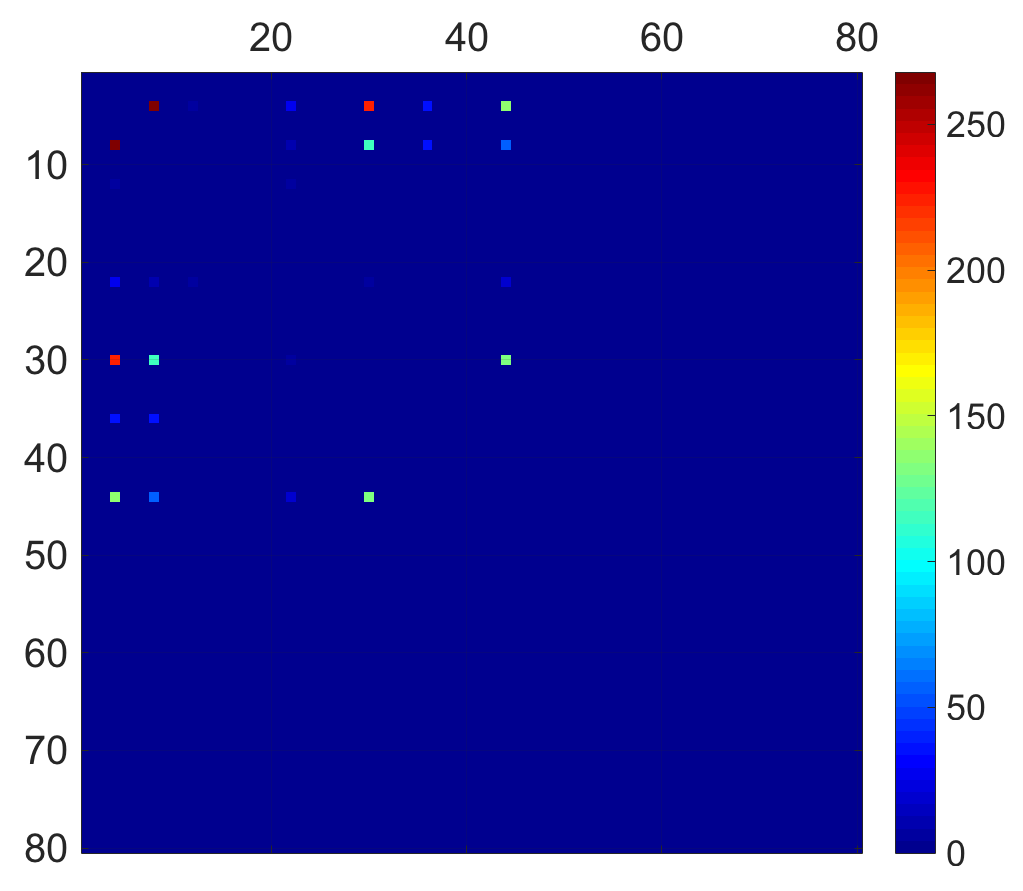} & 
		  \includegraphics[width=0.225\linewidth,height=3.5cm]{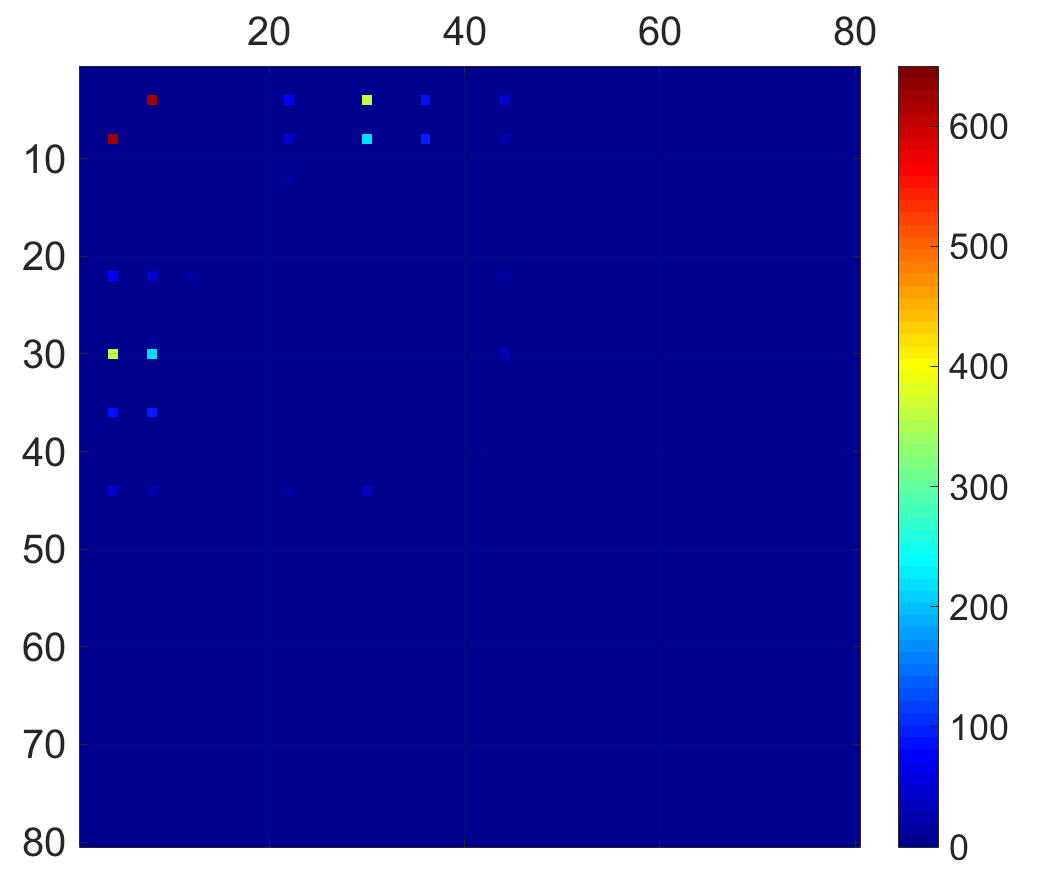} \\
		  \includegraphics[width=0.225\linewidth,height=4cm]{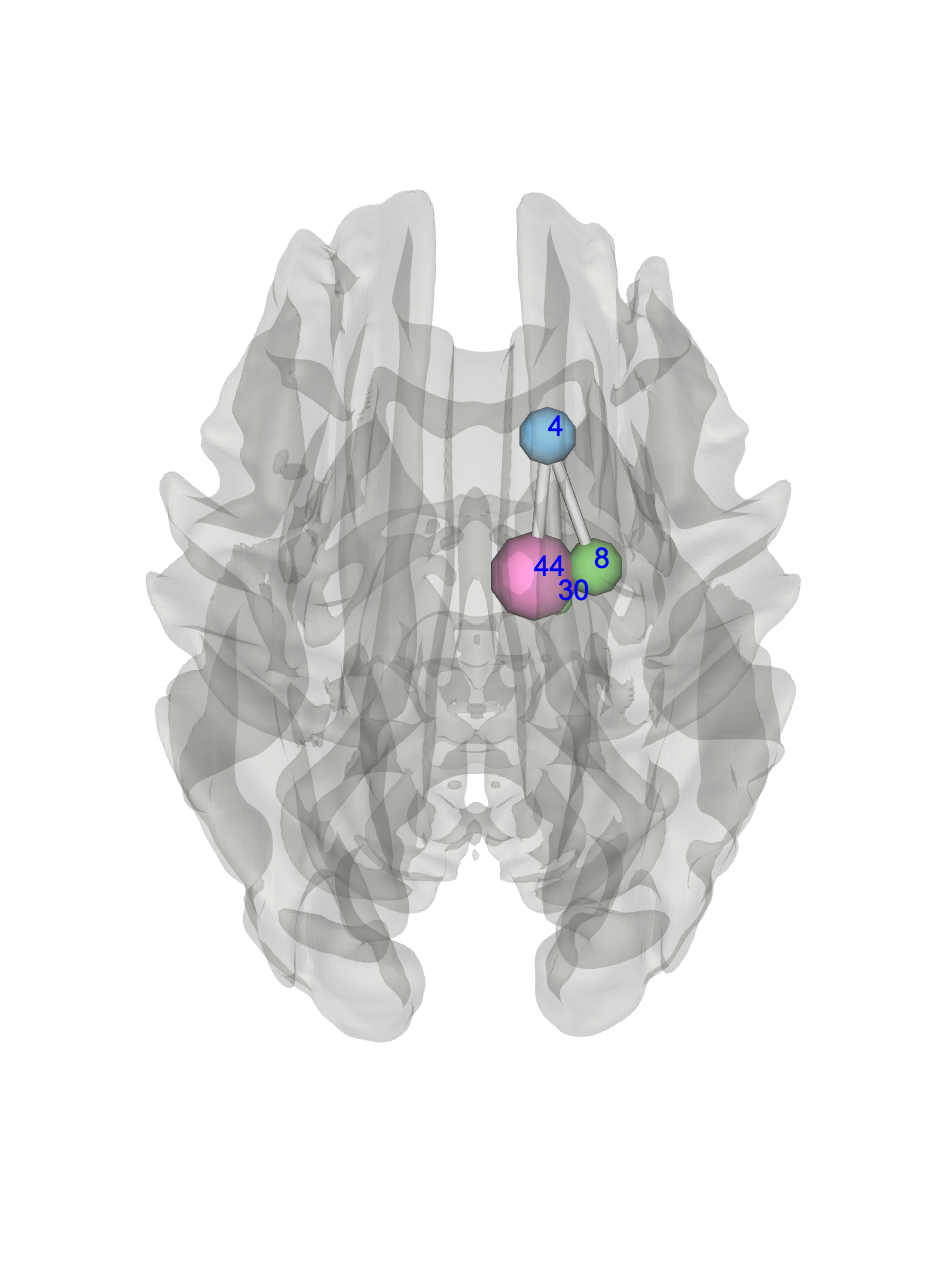} & 
		  \includegraphics[width=0.225\linewidth,height=4cm]{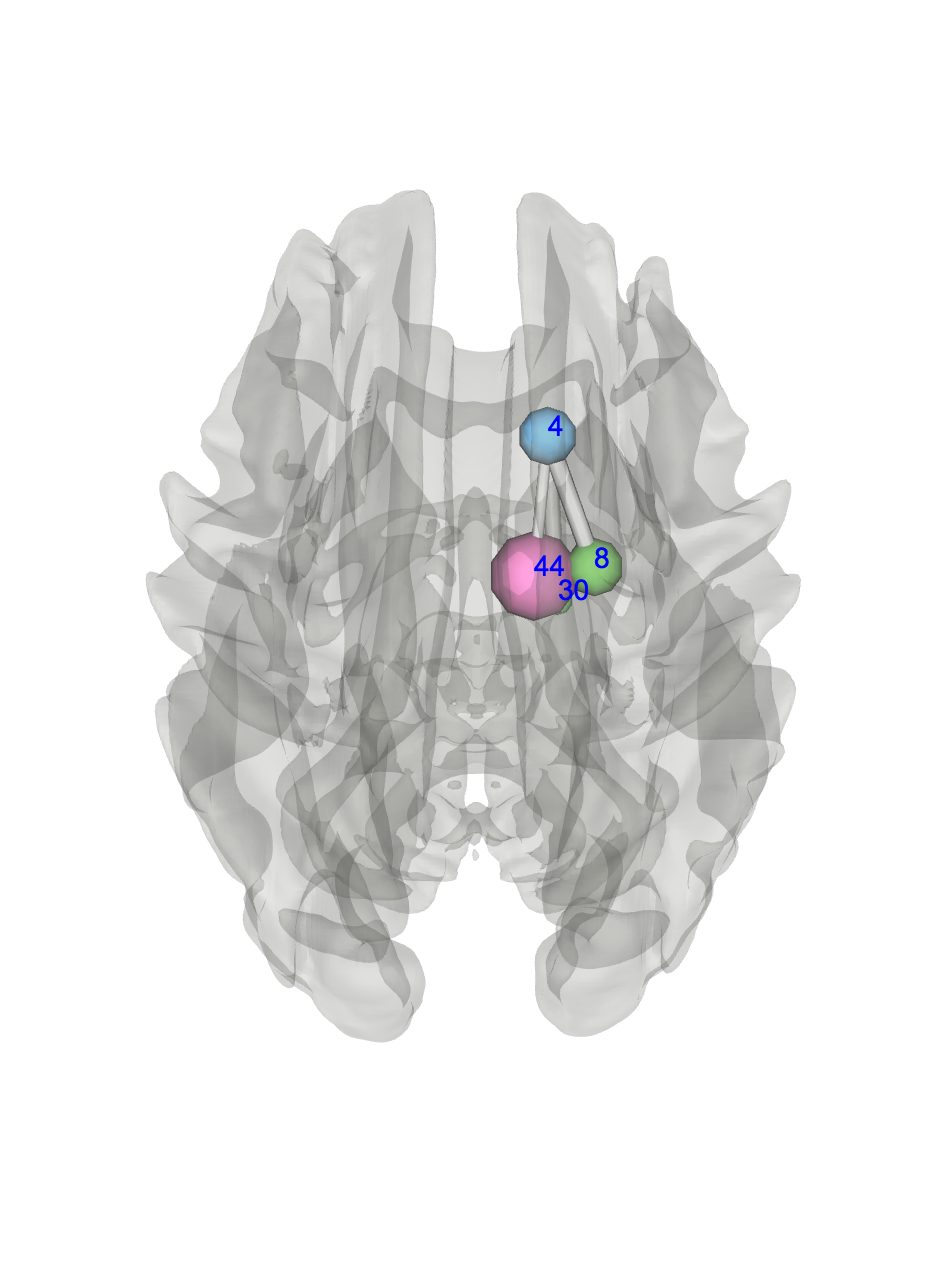} &
		   \includegraphics[width=0.225\linewidth, height=4cm]{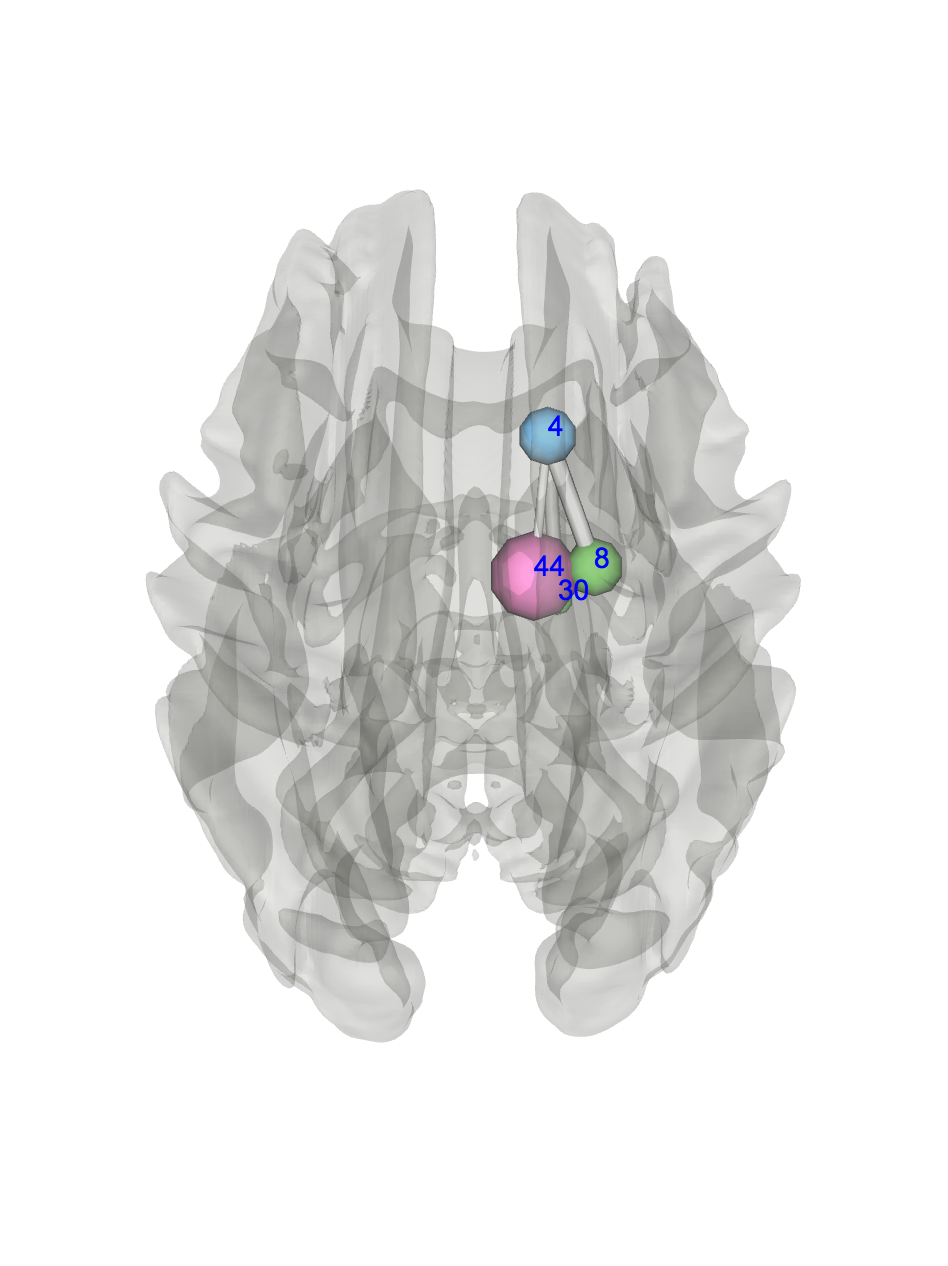} & 
		  \includegraphics[width=0.225\linewidth,height=4cm]{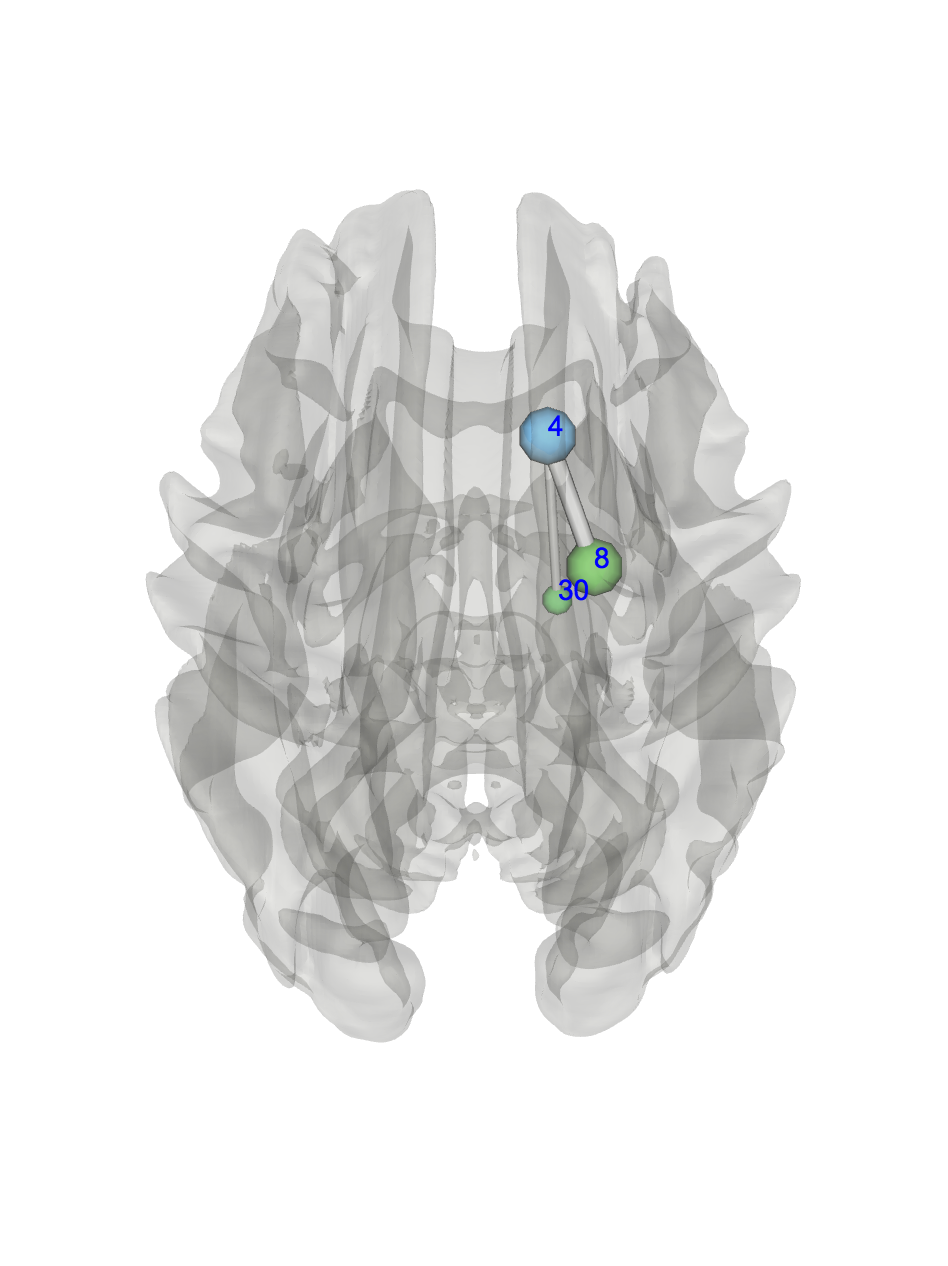} \\
		  \includegraphics[width=0.225\linewidth,height=4cm]{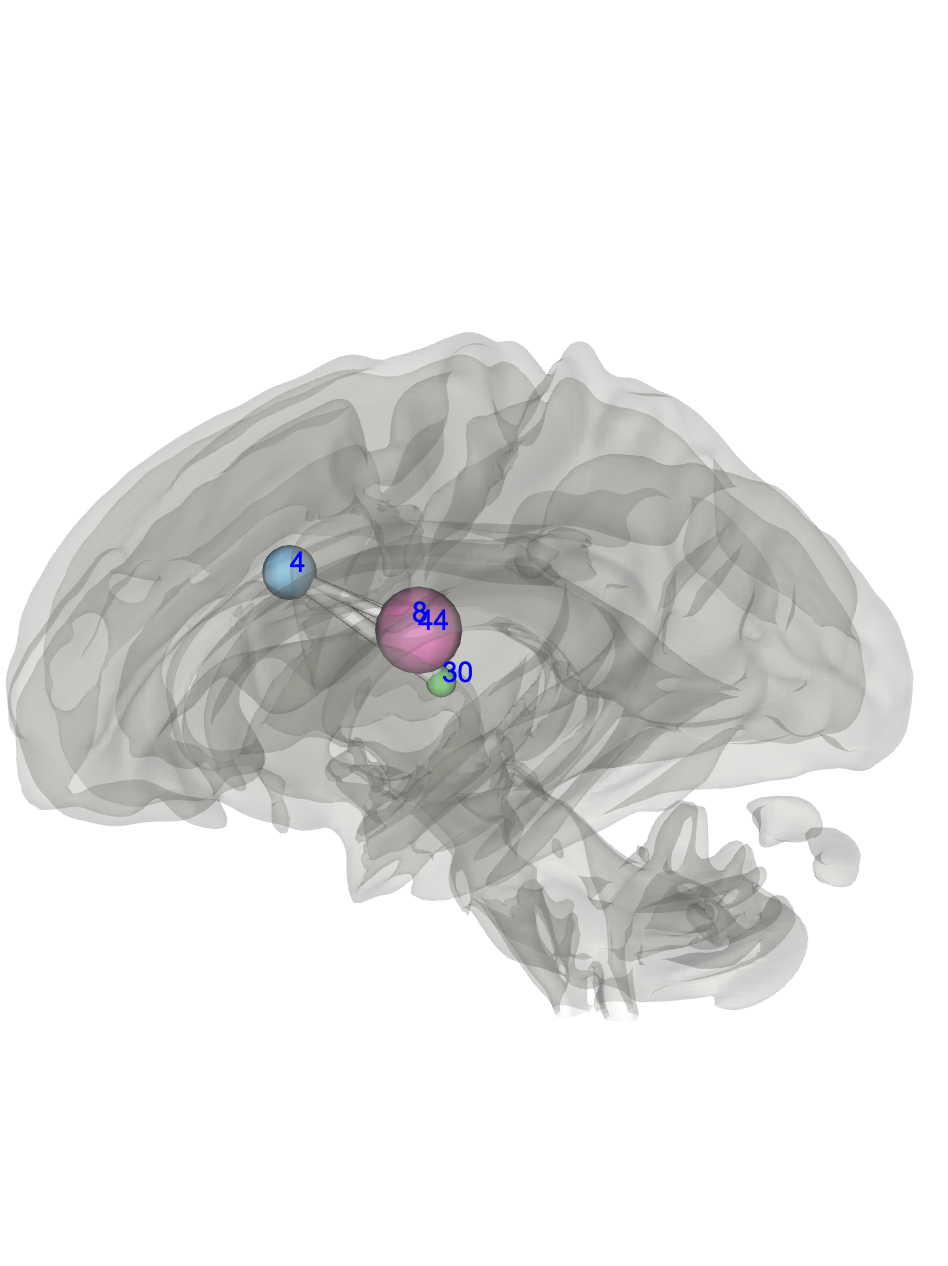} & 
		  \includegraphics[width=0.225\linewidth,height=4cm]{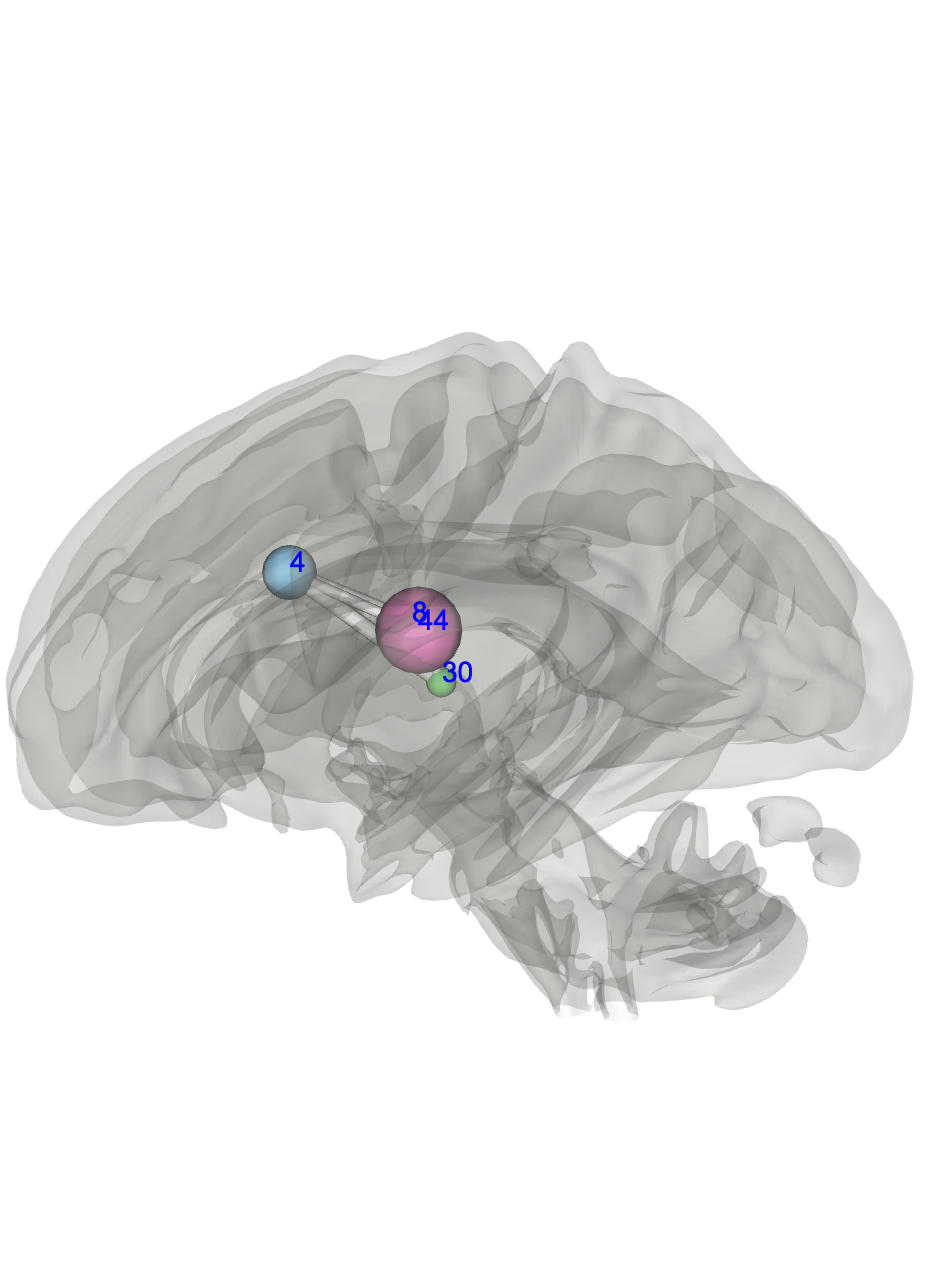} &
		   \includegraphics[width=0.225\linewidth, height=4cm]{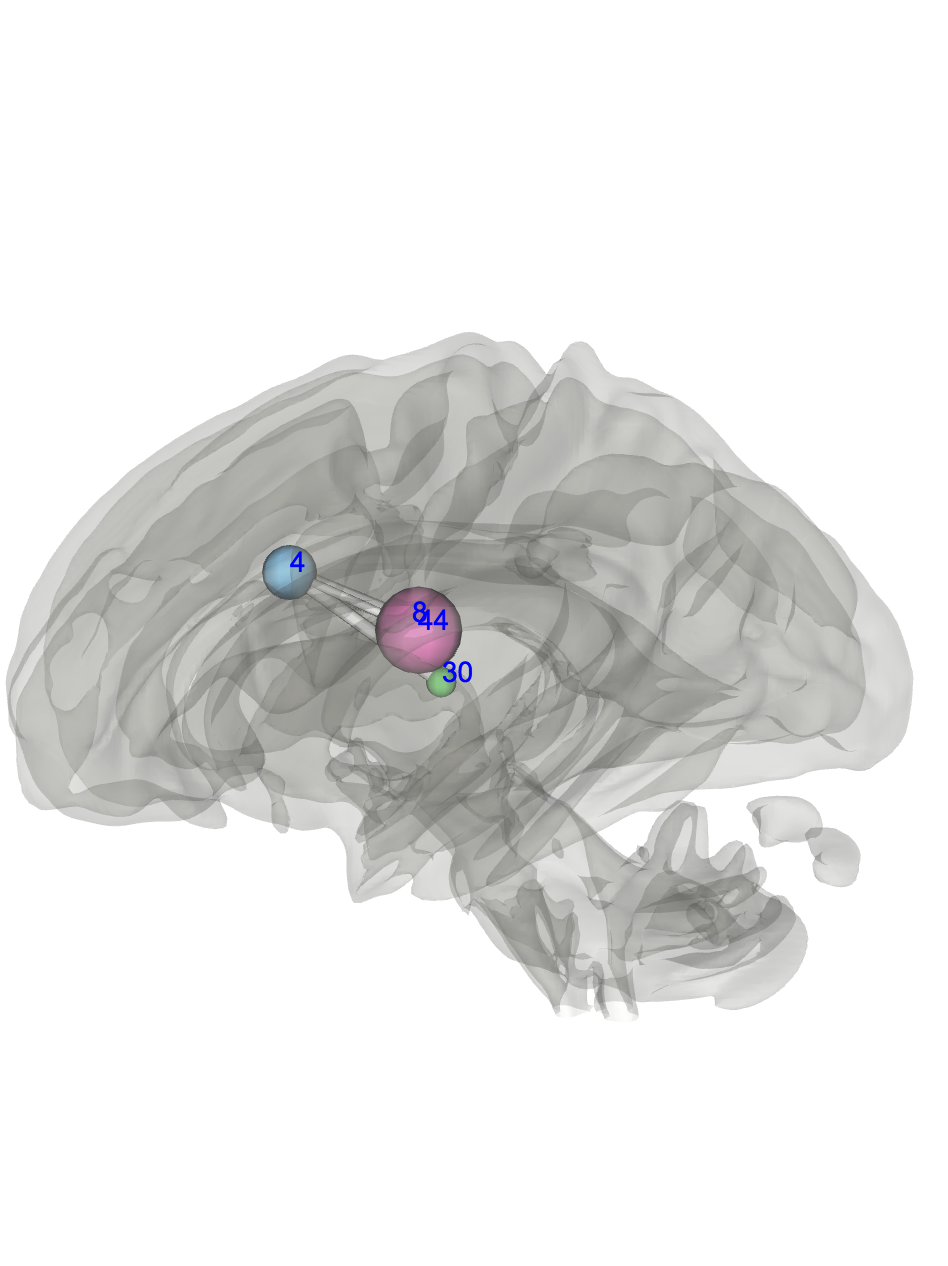} & 
		  \includegraphics[width=0.225\linewidth,height=4cm]{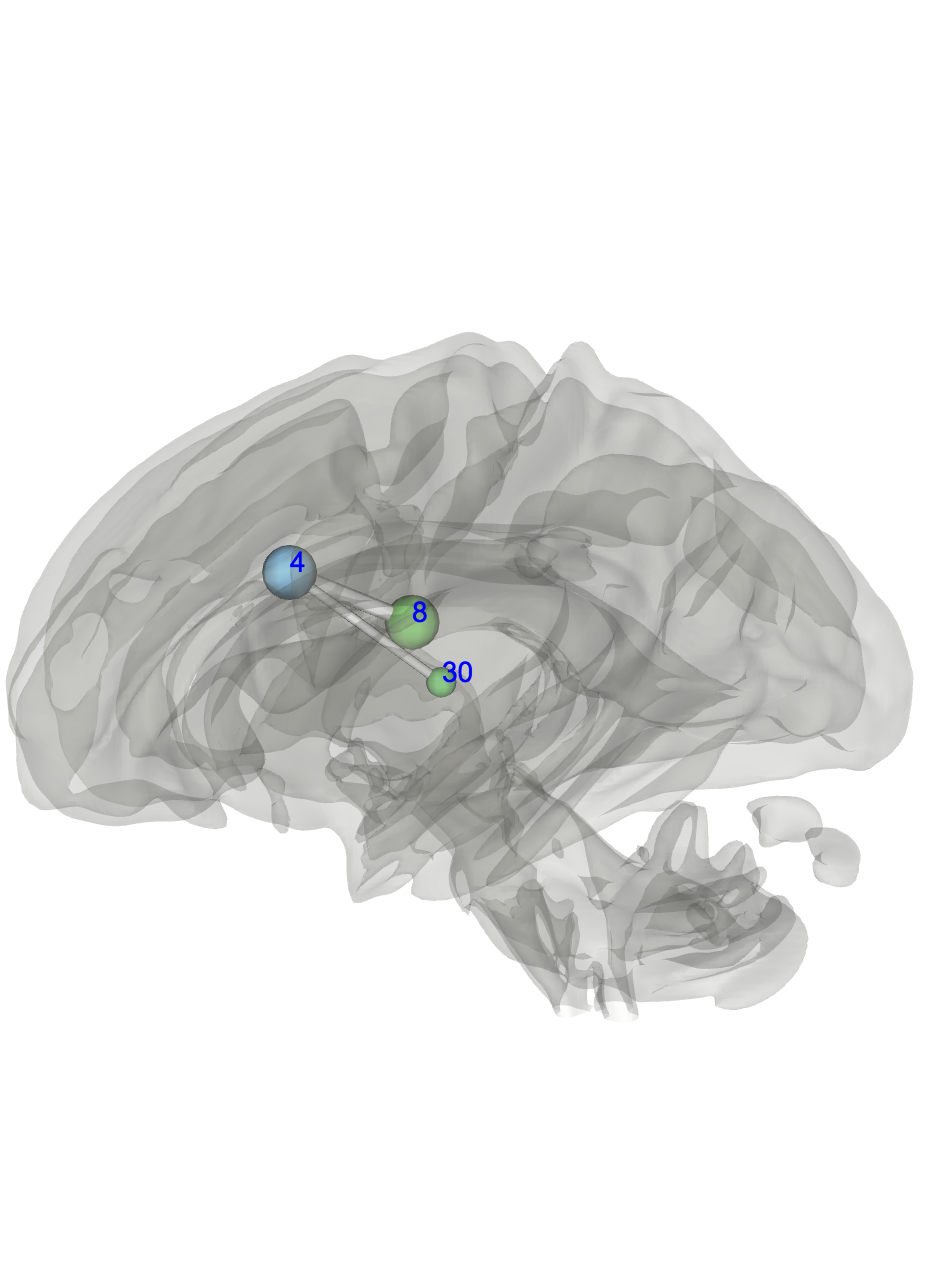} \\
		  {\small (a) $K$=50} & {\small (b) $K$=100} & {\small (c) $K$=200} & {\small (d) $K$=400}
	\end{tabular}
\caption{Visualization for trait Flanker: each column represents the visualization under different $K$ ($K$=50; $K$=100; $K$=200; $K$=400) in PPA; First row represents the connectivity matrix between any two ROI regions in HCP842 tractography atlas. Second row shows anatomy of connections in an axial view. Third row shows anatomy of connections in a sagittal view.} 
\label{fig:f} 
\end{figure}

According to the visualization plots in Figure \ref{fig:pv}-\ref{fig:f}, PPA discovers some insightful connections of various anatomical regions that are related to the human traits. Some interesting findings are listed as follows.
For most human traits, the primary pattern in the connectivity matrix does not vary much as the number of clusters (fiber bundles) increases. When using the trait PicVocab, the subgraph including connections among ROI 3 (Cortico\_Striatal\_Pathway\_L), ROI 4 (Cortico\_Striatal\_Pathway\_R), ROI 7 (Corticothalamic\_Pathway\_L), ROI 8 (Corticothalamic\_Pathway\_R), and ROI 44 (Corpus\_Callosum), can be found in all the 4 different settings of number of clusters, i.e., $K$=50,100,200,400 (see Figure~\ref{fig:pv}). The trait ListSort that is related to human working memory (Figure \ref{fig:ls}), also shows a common pattern in the connectivity matrix across cluster settings, leading to a network with shared nodes containing ROI 3 (Cortico\_Striatal\_Pathway\_L), ROI 7 (Corticothalamic\_Pathway\_L), ROI 21 (Arcuate\_Fasciculus\_L), ROI 37 (U\_Fiber\_L) and ROI 44 (Corpus\_Callosum).

For language associated human traits (e.g., the trait PicVocab is related to language and vocabulary comprehension while the trait ReadEng is related to language and reading decoding), the significant regions are mainly located in the left hemisphere (see Figure \ref{fig:pv} and Figure \ref{fig:re}). This finding indicates that the left hemisphere is particularly important for language, which has been consistently verified in clinical and experimental settings  \citep{ries2016choosing}.


We find that a particularly important ROI 44 (Corpus\_Callosum) is detected for most human traits; ROI 44 is consistently a prominent node in the identified subgroup for most values of $K$ (see the second row in Figure \ref{fig:pv}-\ref{fig:f}).
Corpus Callosum is a large C shape white matter region, forming the floor of the longitudinal fissure that separates the cerebral cortex into the left and right hemispheres \citep{carpenter1985study}. 
This ROI is responsible for transmitting sensory, motor, and cognitive signals between the hemispheres. 

The Flanker task measures both a participant's attention and inhibitory control. According to the visualization plots in Figure \ref{fig:f}, significant regions identified are strongly lateralized to the right hemisphere. Right frontal dominance for inhibitory motor control has become a commonly accepted view \citep{swick2008left, garavan1999right}, indicating that our finding is in agreement with existing literature. The visualization plots for the remaining three human traits that we considered are reported in the appendix; see Figures \ref{fig:ps}-\ref{fig:psd}.

\section{Discussion}
\label{s:discuss}

In this article, we propose a new tractography-based representation of brain connectomes as an alternative to the widely used anatomical connectivity. This representation leads to Principal Parcellation Analysis (PPA), where we represent individual brain connectomes by compositional vectors building on a basis system of fiber bundles that captures the connectivity at the population level.



PPA reduces subjectivity of classical connectomes by eliminating the need to choose atlases and ROIs a priori. Unlike the traditional connectomes where data objects are of complex graph-structure and ultra-dimension, our PPA connectomes can be analyzed using existing statistical tools for high-dimensional vector-valued features.
Our application to HCP data indicates that PPA is robust to the specific choice of fiber tracking algorithm. 
We propose an approach to integrate the parcellation produced by PPA with an atlas, so that the results can be visualized and interpreted using traditional ROIs.

There are several interesting next directions building on our initial PPA approach.  Firstly, the methods used in each of the three modules can be refined.  For example, we can consider different fiber clustering algorithms that take into account more than just the endpoint locations.  Also, instead of just applying LASSO within a linear regression model for the trait responses, we can use more elaborate predictive algorithms and inferential approaches.  Particularly for large datasets, improved predictive accuracy may be possible with flexible non-linear regression methods. 
We have focused on using the proposed connectomes to analyze human traits via a regression model, and it is interesting to consider other inference problems beyond explaining and predicting task-based scores. For example, one may consider both structural and functional connectivity \citep{tian2018characterizing} and utilize other dynamical features \citep{kobeleva2021revealing}, with the proposed PPA connectomes serving as a building block to represent structural connectivity. 


\section*{Data and code availability statement}
The dMRI and human traits data from the original study are available from HCP: \url{https://www.humanconnectome.org/}. Code for implementing PPA is freely available online at \url{https://github.com/xylimeng/PPA}. 

\section*{Declaration of Competing Interest}
The authors declare no conflict of interest.

\section*{Credit authorship contribution statement}

\textbf{Rongjie Liu}: Conceptualization, Methodology, Software, Writing- Original draft preparation. \textbf{Meng Li}: Conceptualization, Methodology, Validation, Supervision, Writing- Reviewing and Editing. \textbf{David Dunson}: Conceptualization, Methodology, Validation, Writing- Reviewing and Editing.

\section*{Acknowledgements}
We thank Lu Wang for helpful discussions on implementing the SBL method, and  Yerong Li for helping with numerical experiments in an earlier version of PPA.  This work was partially funded by grant DMS-2015569 from the National Science Foundation and grant R01MH118927 of the National Institute of Mental Health of the United States National Institutes of Health.

\bibliographystyle{plainnat}
\bibliography{ppa}

%

\newpage
      \setcounter{table}{0}
        \renewcommand{\thetable}{S\arabic{table}}%
        \setcounter{figure}{0}
        \renewcommand{\thefigure}{S\arabic{figure}}%
        \setcounter{page}{1}
        \renewcommand\thepage{S\arabic{page}}
        
\section*{Appendix}
\begin{figure}[h!]
\centering 
\begin{tabular}{ccc}
		  \includegraphics[width=0.3\linewidth,height=4cm]{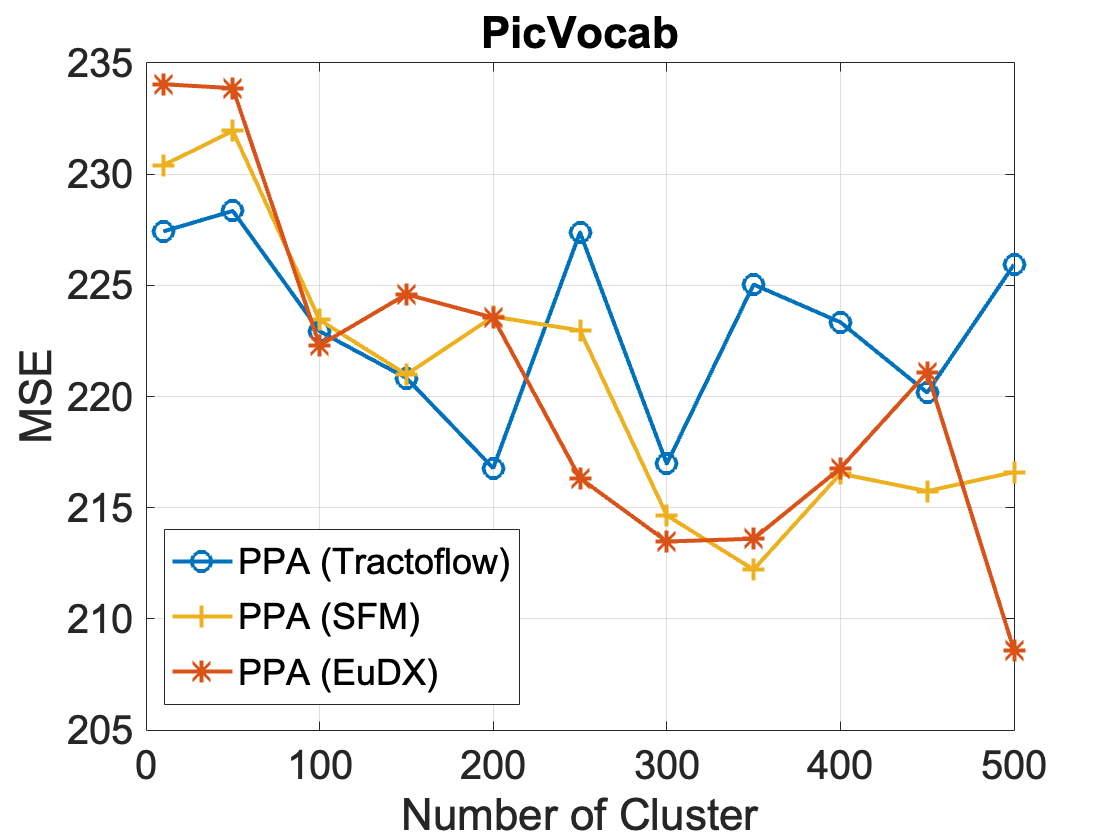} & 
		  \includegraphics[width=0.3\linewidth,height=4cm]{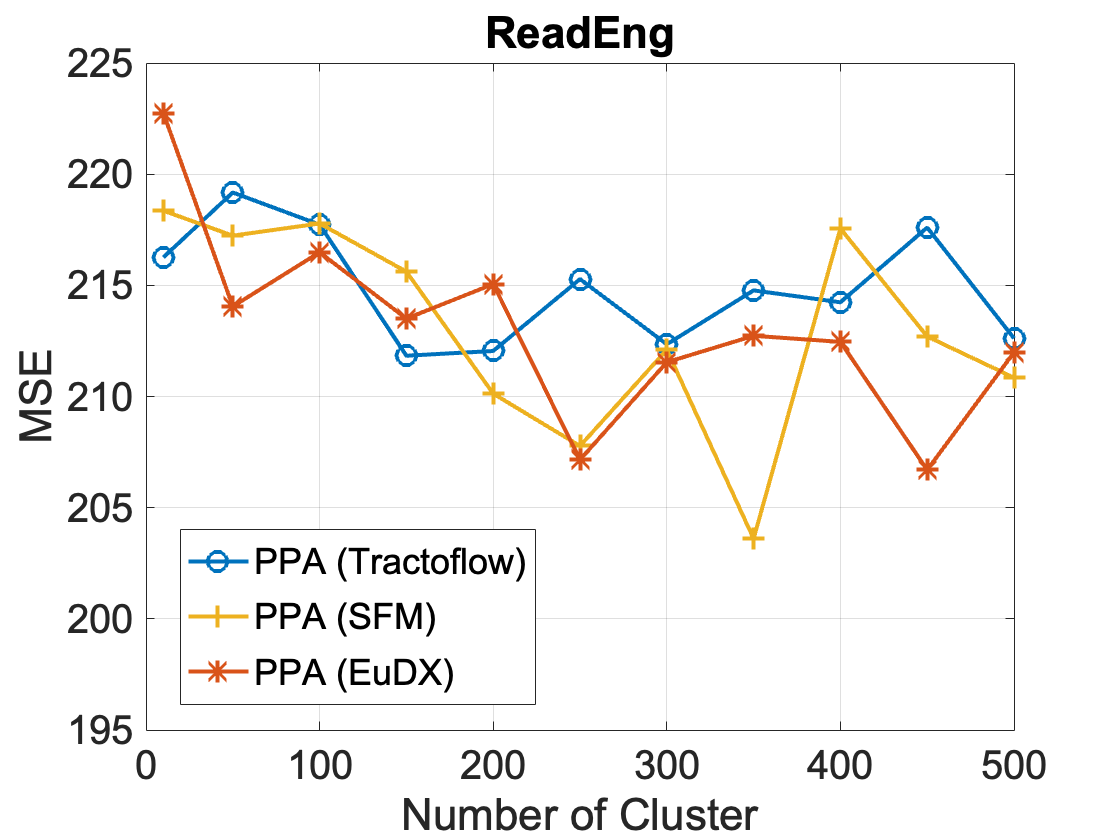} & 
		  \includegraphics[width=0.3\linewidth,height=4cm]{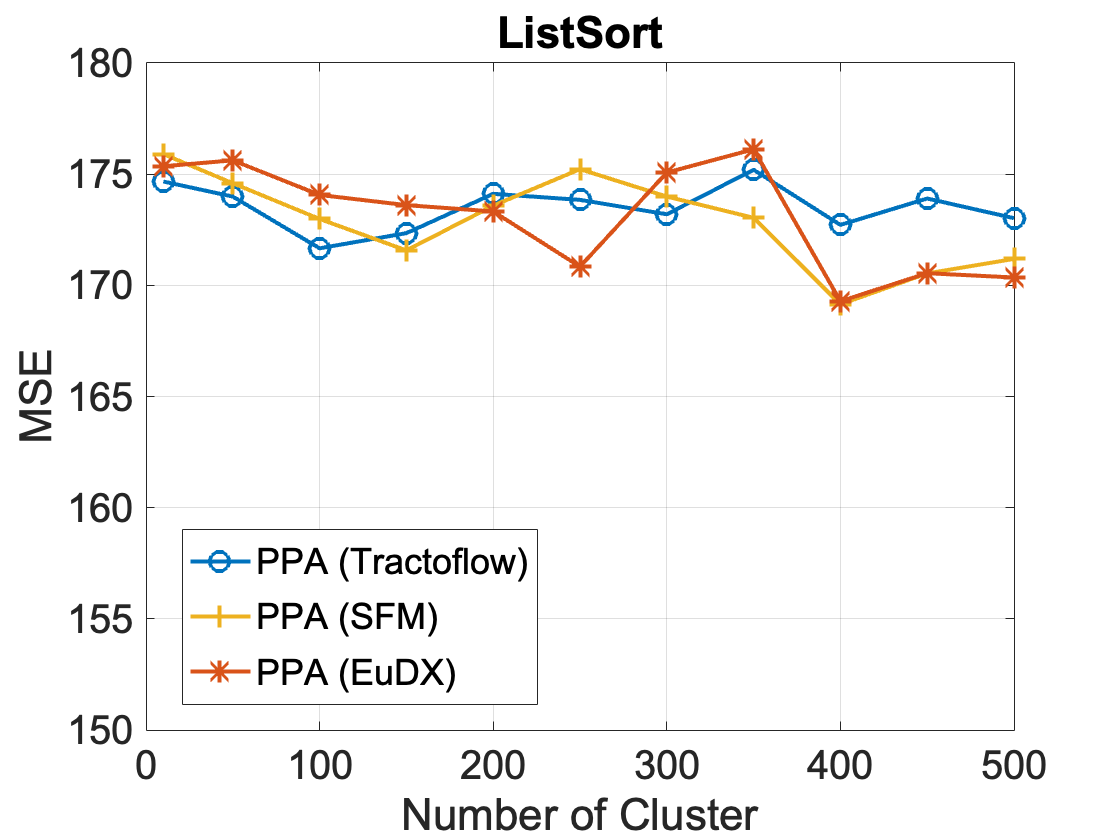} \\
	\end{tabular}
	\begin{tabular}{cc}
		  \includegraphics[width=0.45\linewidth,height=5cm]{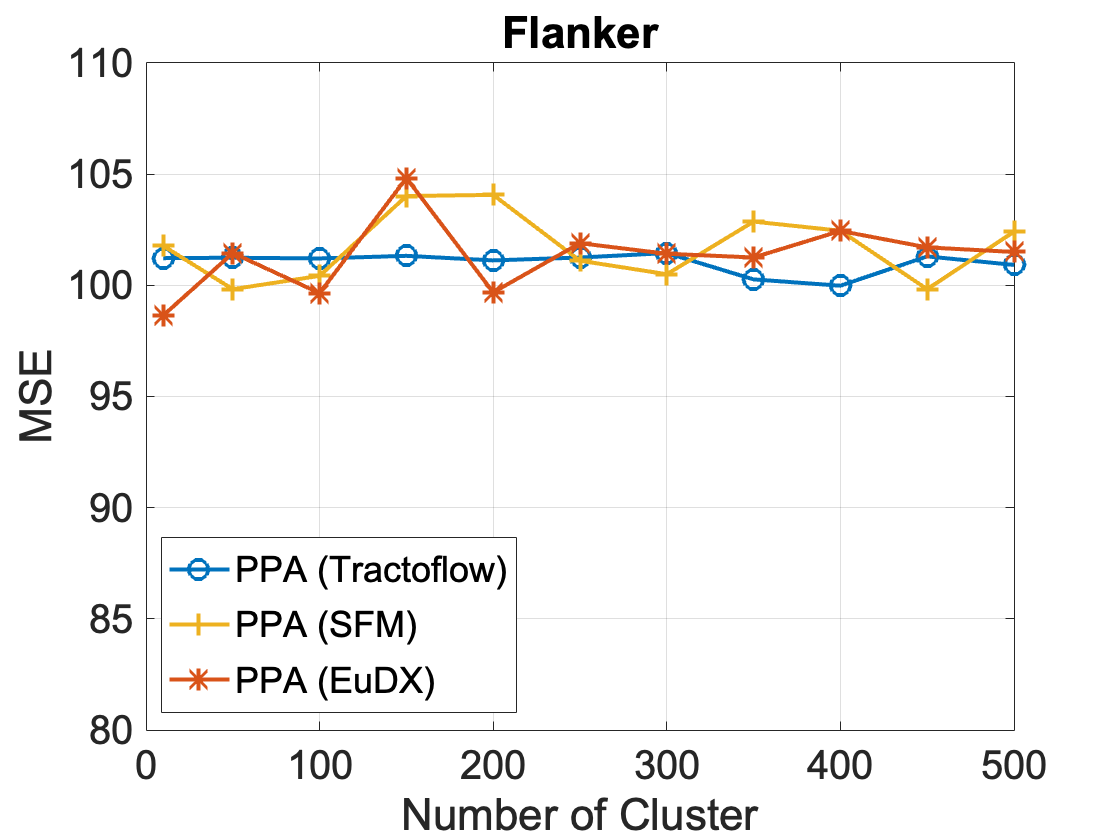} & 
		  \includegraphics[width=0.45\linewidth,height=5cm]{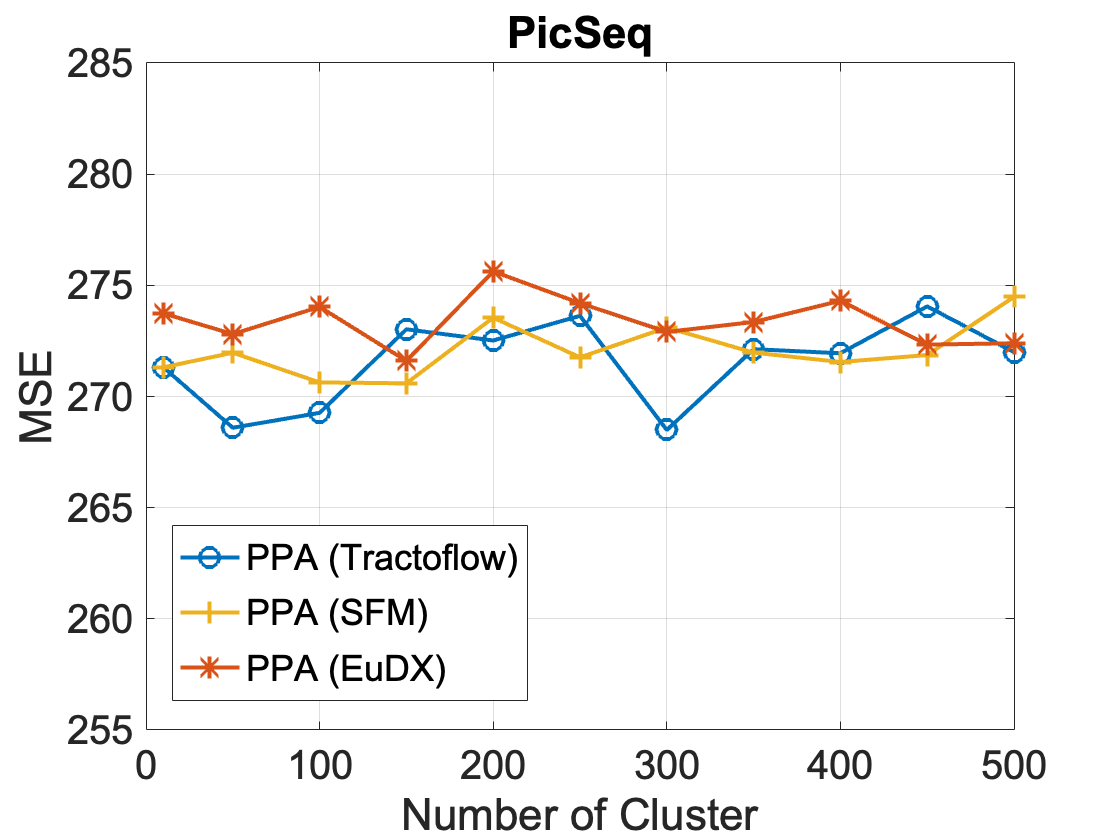} \\
		  \includegraphics[width=0.45\linewidth,height=5cm]{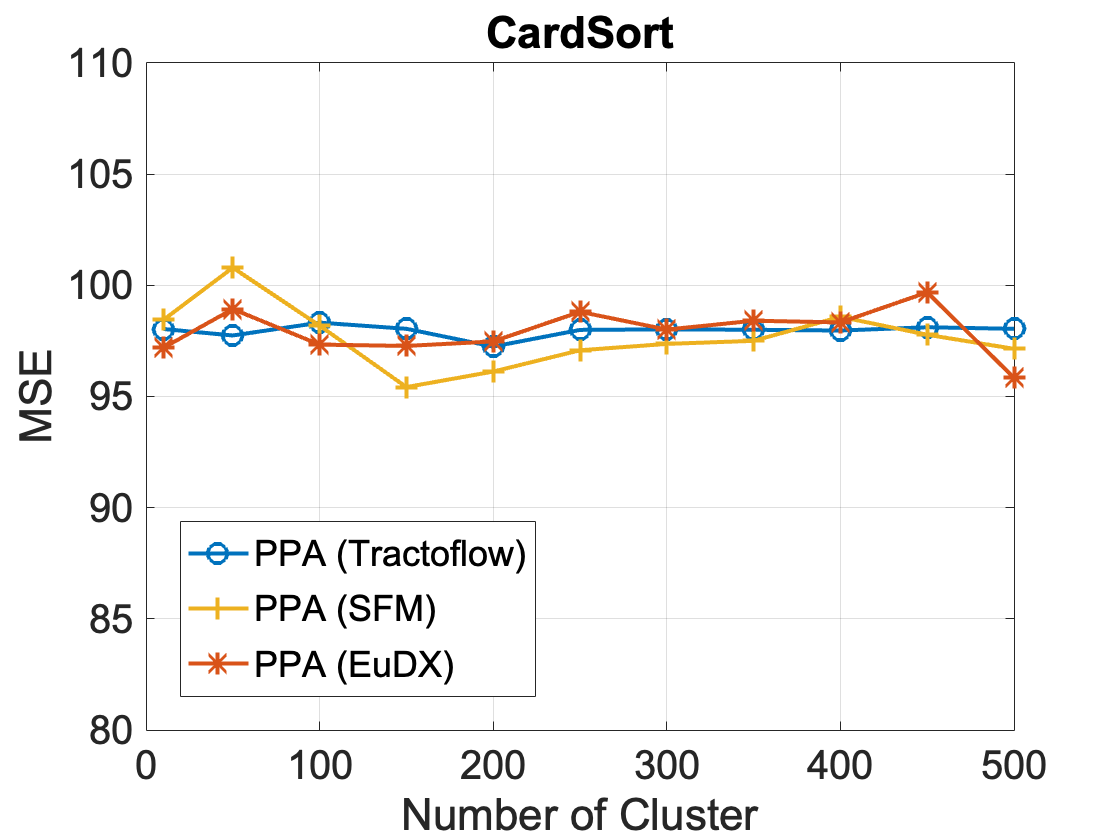} & 
		  \includegraphics[width=0.45\linewidth,height=5cm]{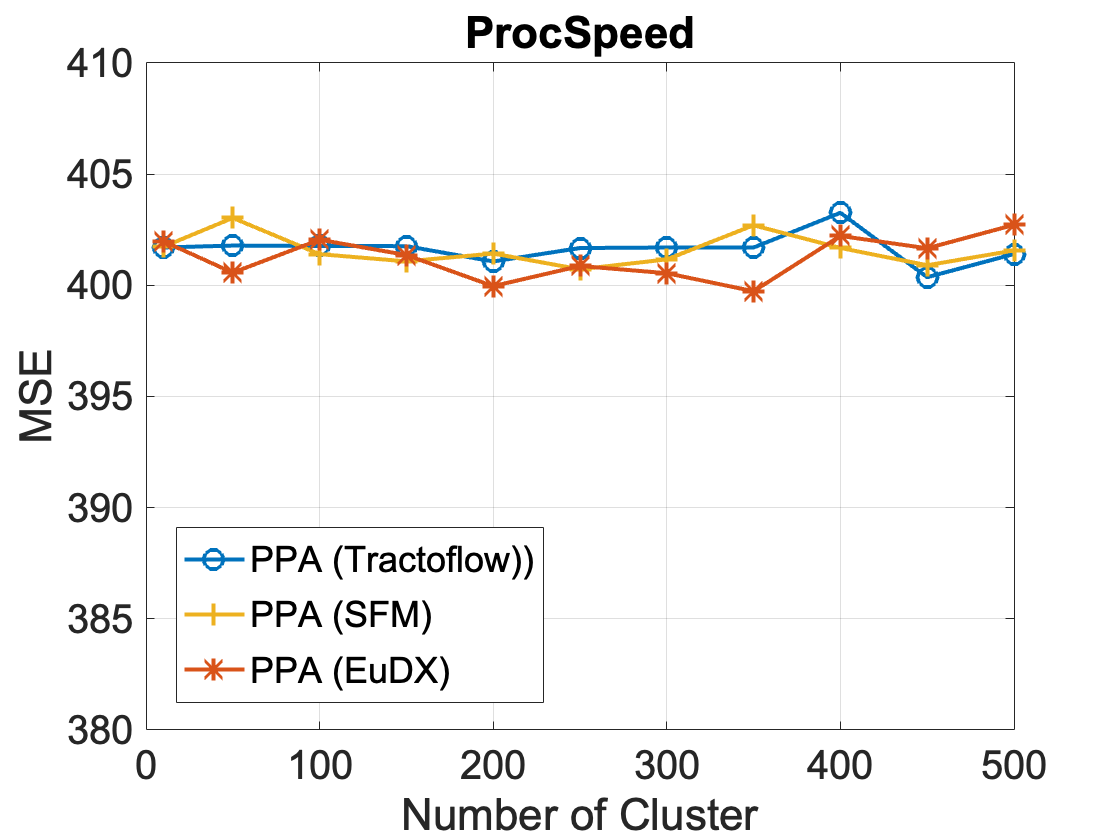} \\
	\end{tabular}
\caption{Comparison of 5-fold cross validation MSE of trait prediction based on PPA with three fiber tracking algorithms: TractoFlow, EuDX, and SFM.}\label{compare1} 
\end{figure}

\begin{figure}[h!]
\centering 
\begin{tabular}{ccc}
		  \includegraphics[width=0.3\linewidth,height=4cm]{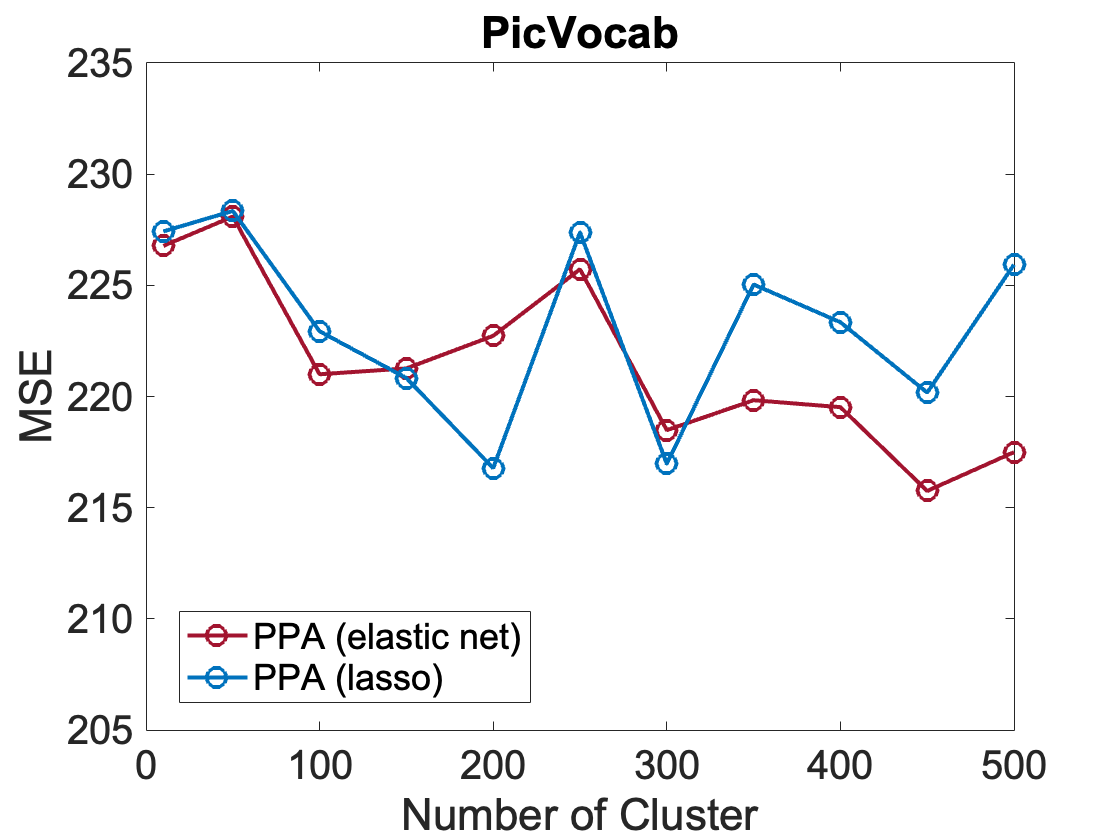} & 
		  \includegraphics[width=0.3\linewidth,height=4cm]{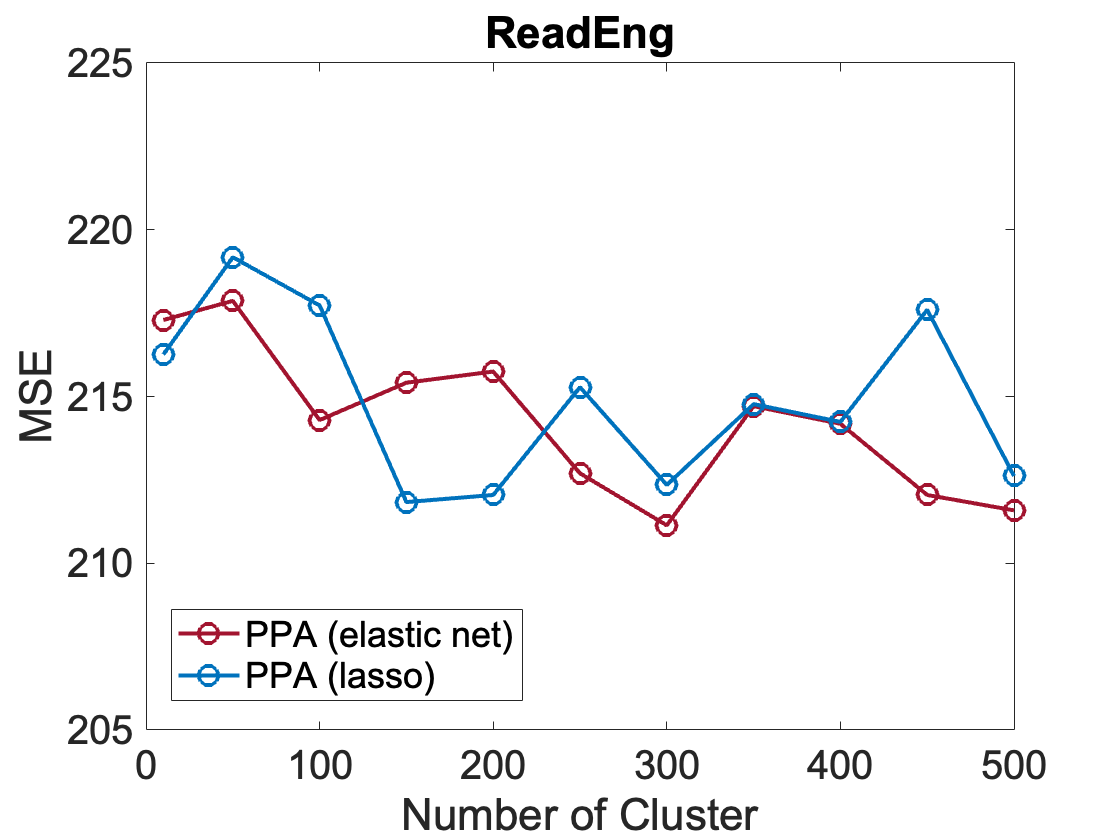} & 
		  \includegraphics[width=0.3\linewidth,height=4cm]{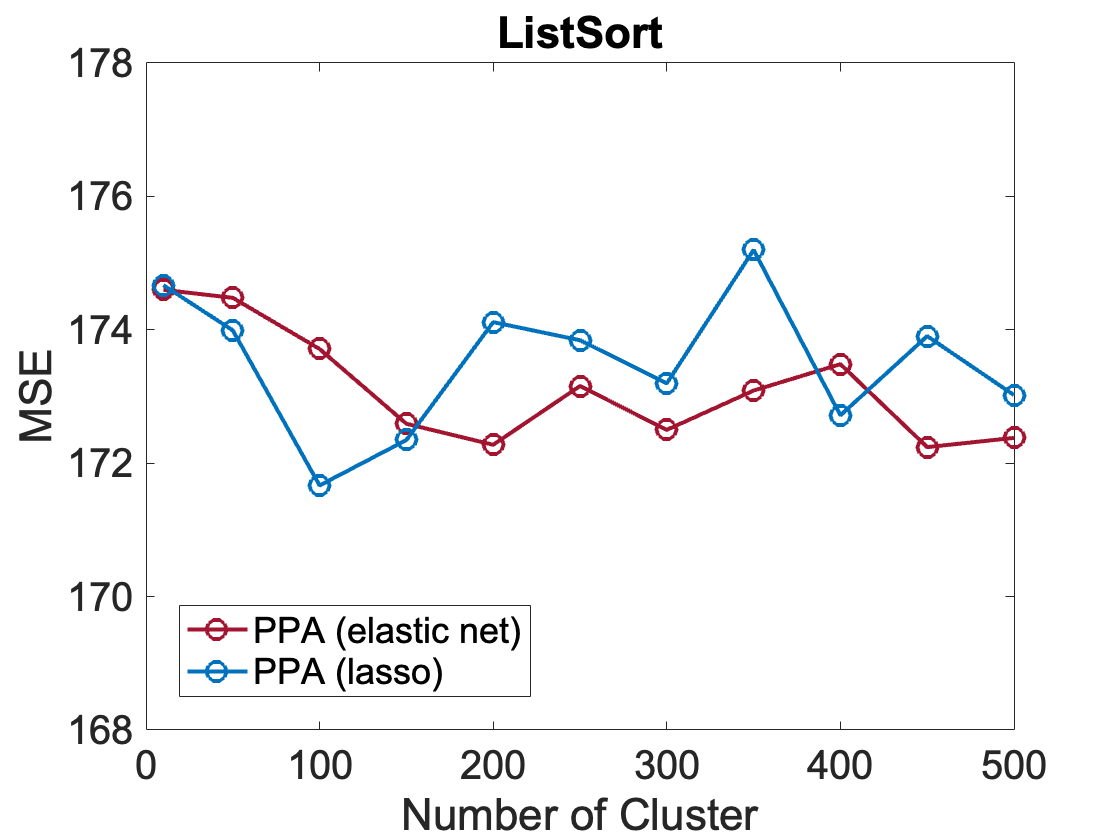} \\
	\end{tabular}
	\begin{tabular}{cc}
		  \includegraphics[width=0.45\linewidth,height=5cm]{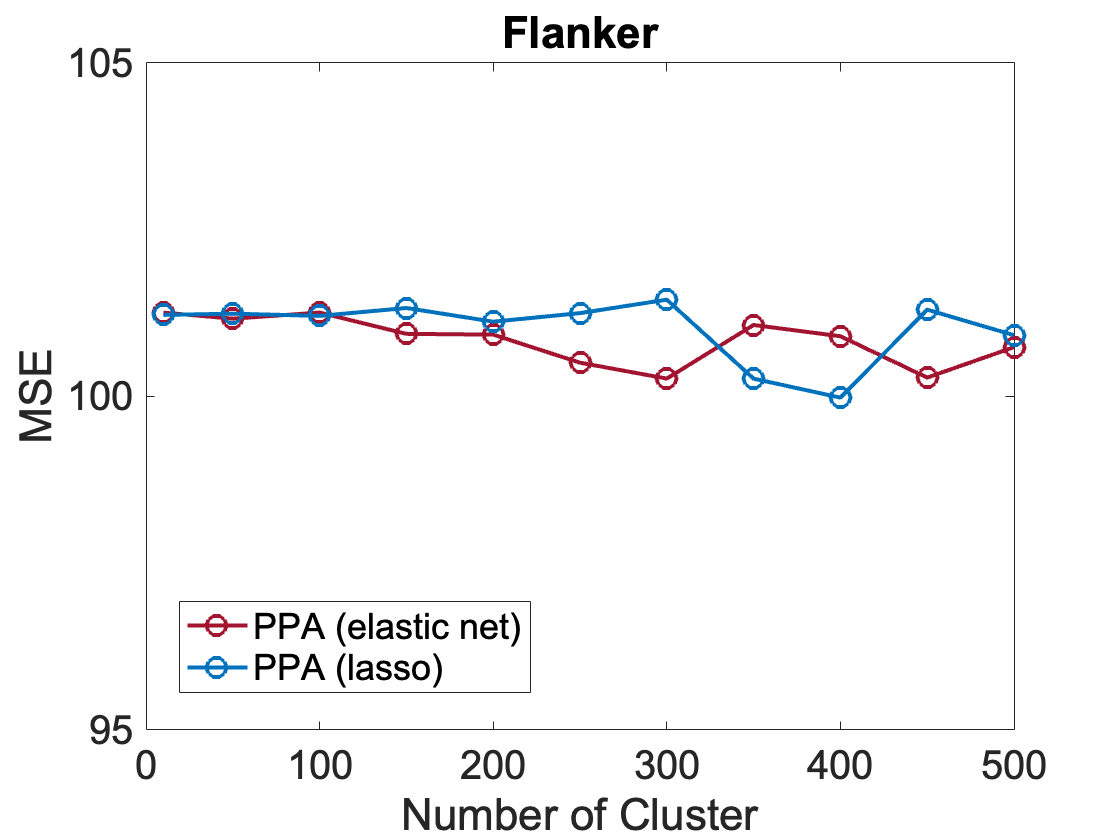} & 
		  \includegraphics[width=0.45\linewidth,height=5cm]{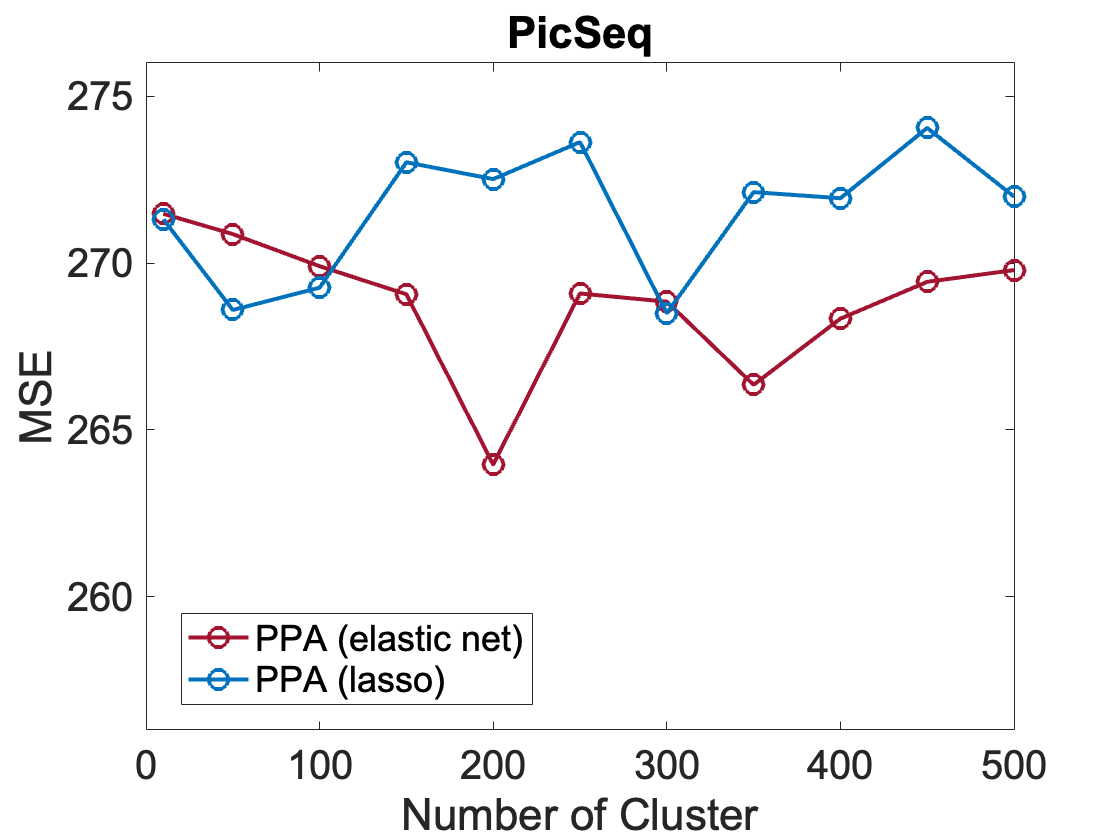} \\
		  \includegraphics[width=0.45\linewidth,height=5cm]{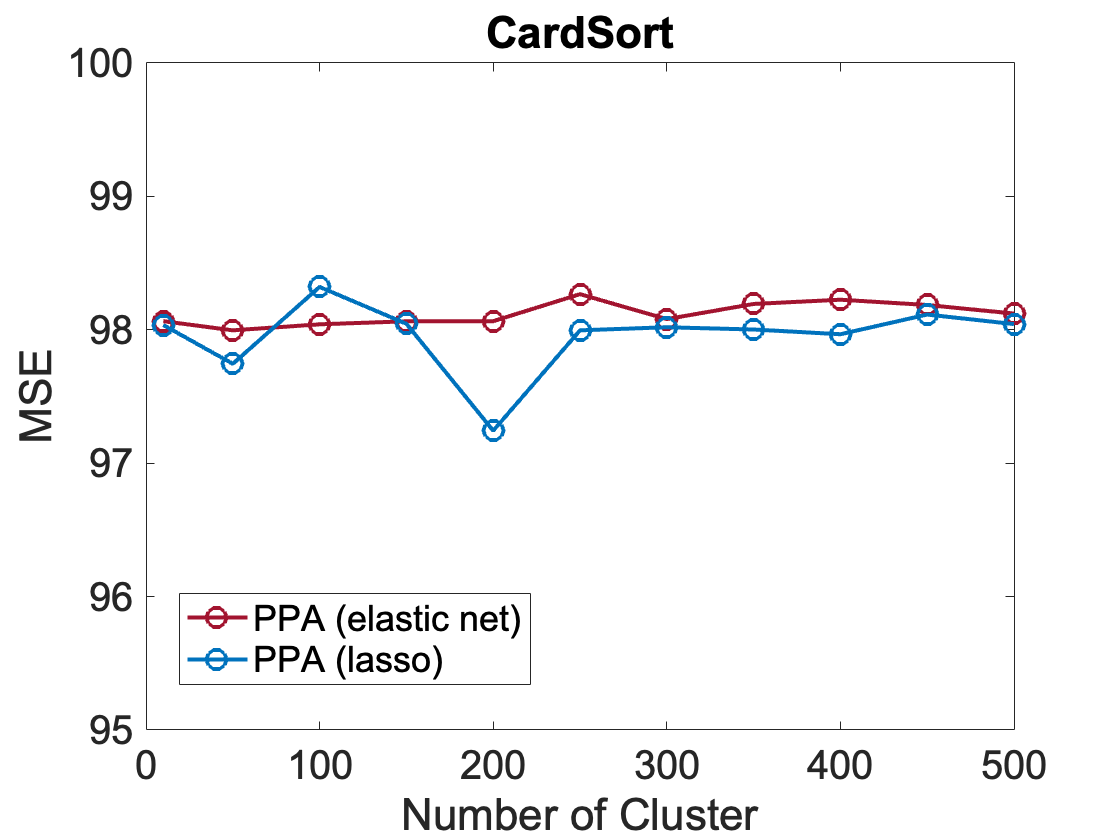} & 
		  \includegraphics[width=0.45\linewidth,height=5cm]{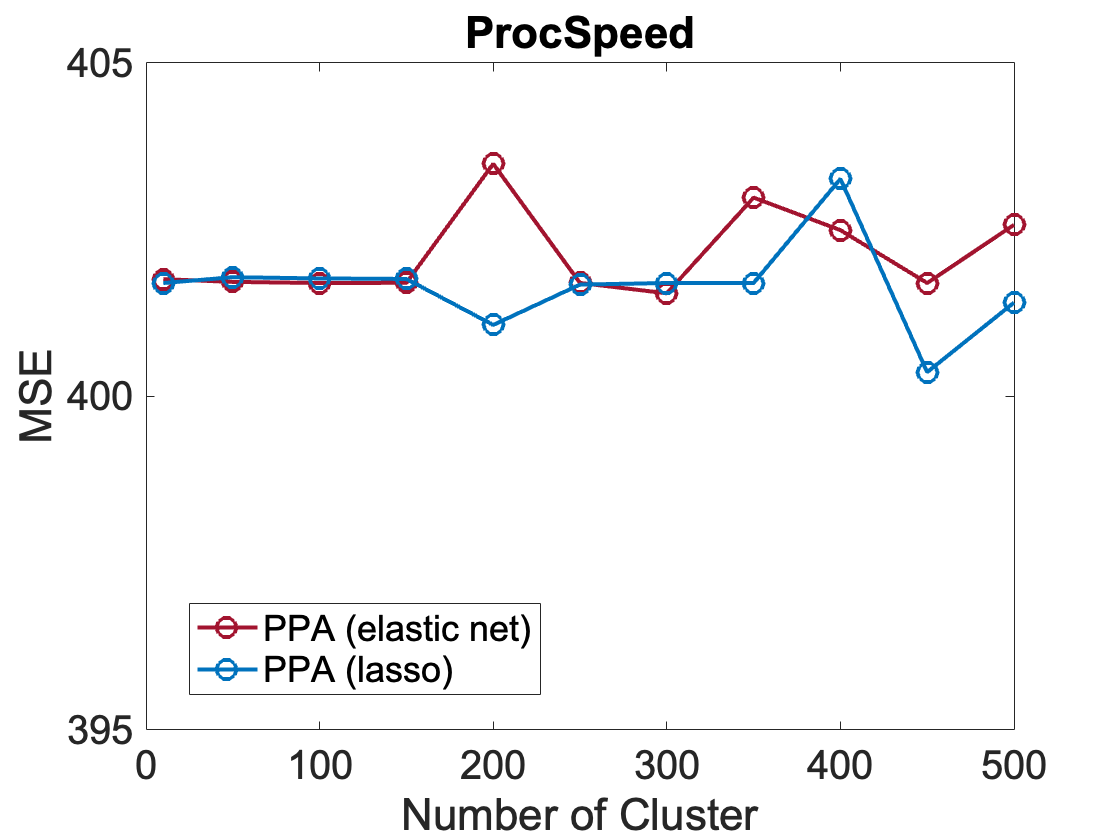} \\
	\end{tabular}
\caption{Comparison of 5-fold cross validation MSE of trait prediction based on PPA with two regularization strategies: LASSO and elastic net.}\label{compare2} 
\end{figure}

\begin{figure}[h!]
\centering 
\begin{tabular}{ccc}
		  \includegraphics[width=0.3\linewidth,height=4cm]{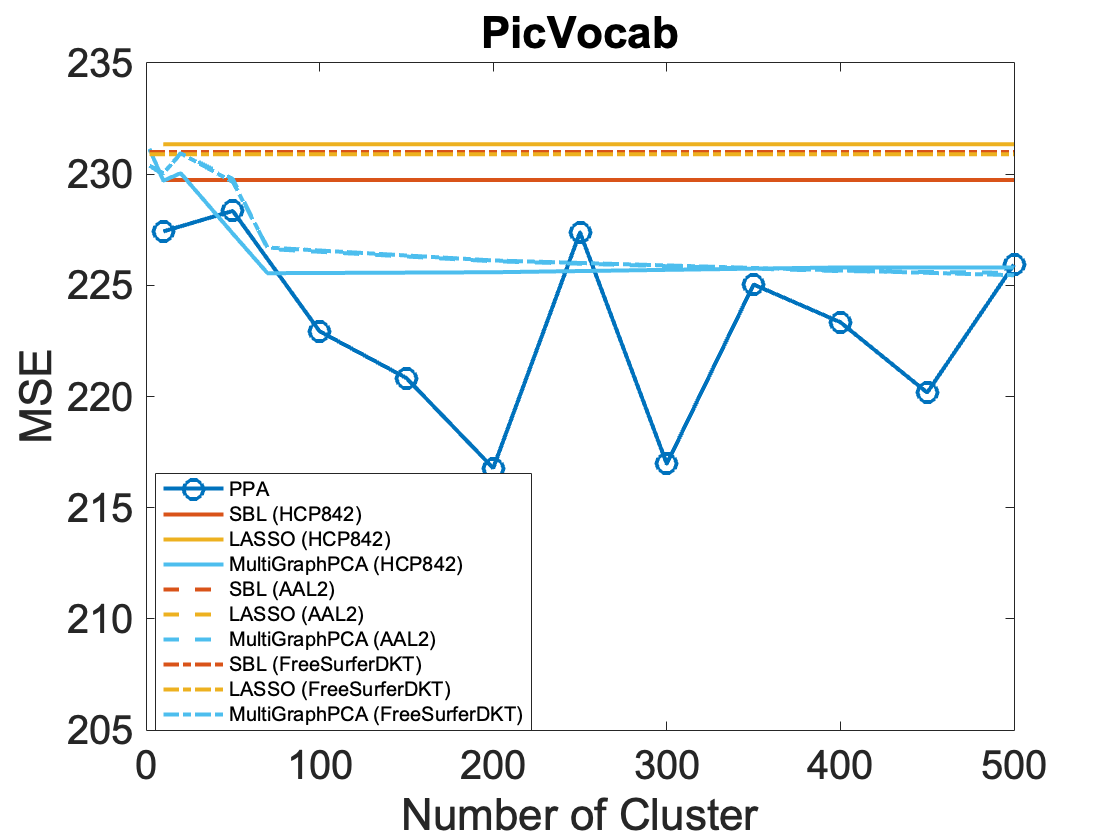} & 
		  \includegraphics[width=0.3\linewidth,height=4cm]{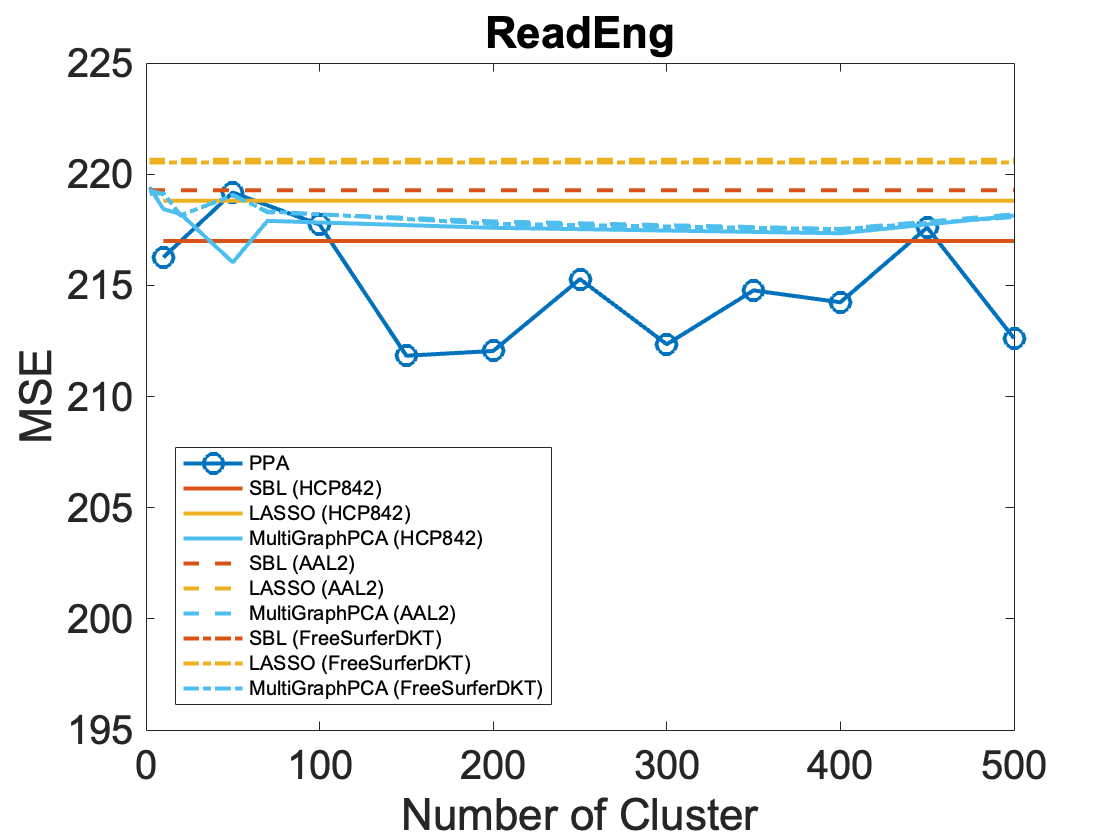} & 
		  \includegraphics[width=0.3\linewidth,height=4cm]{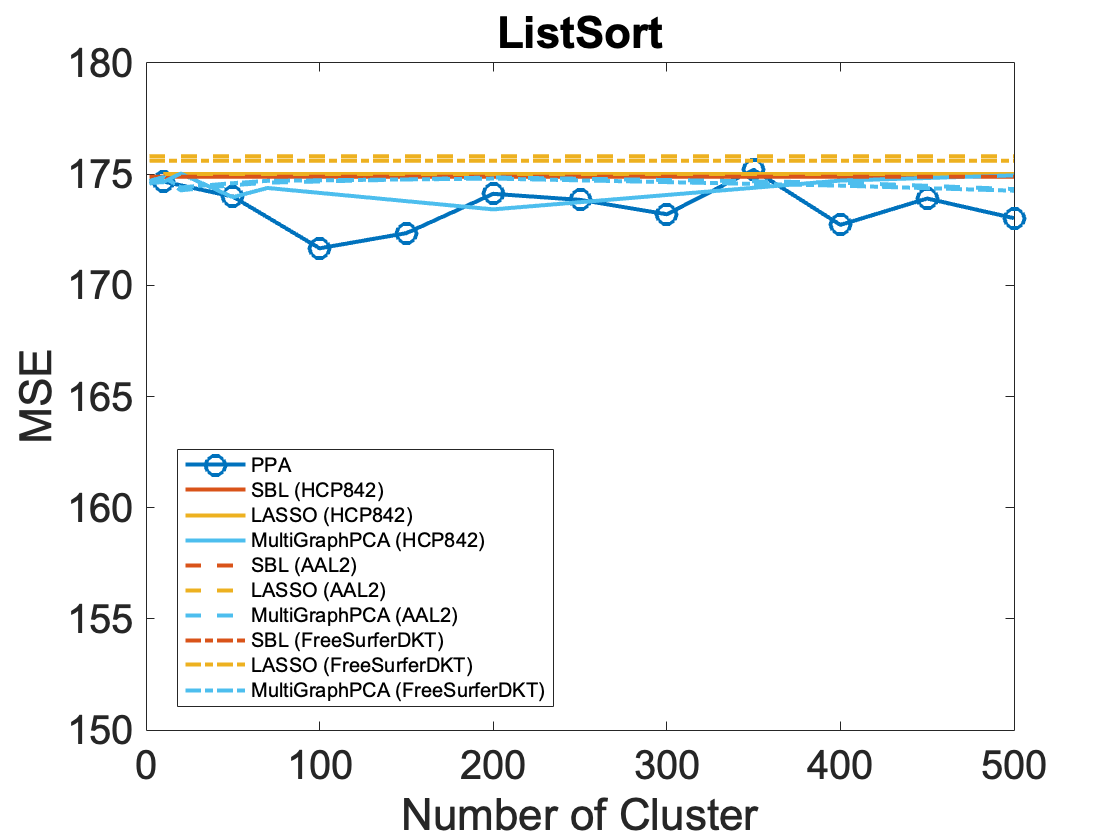} \\
	\end{tabular}
	\begin{tabular}{cc}
		  \includegraphics[width=0.45\linewidth,height=5cm]{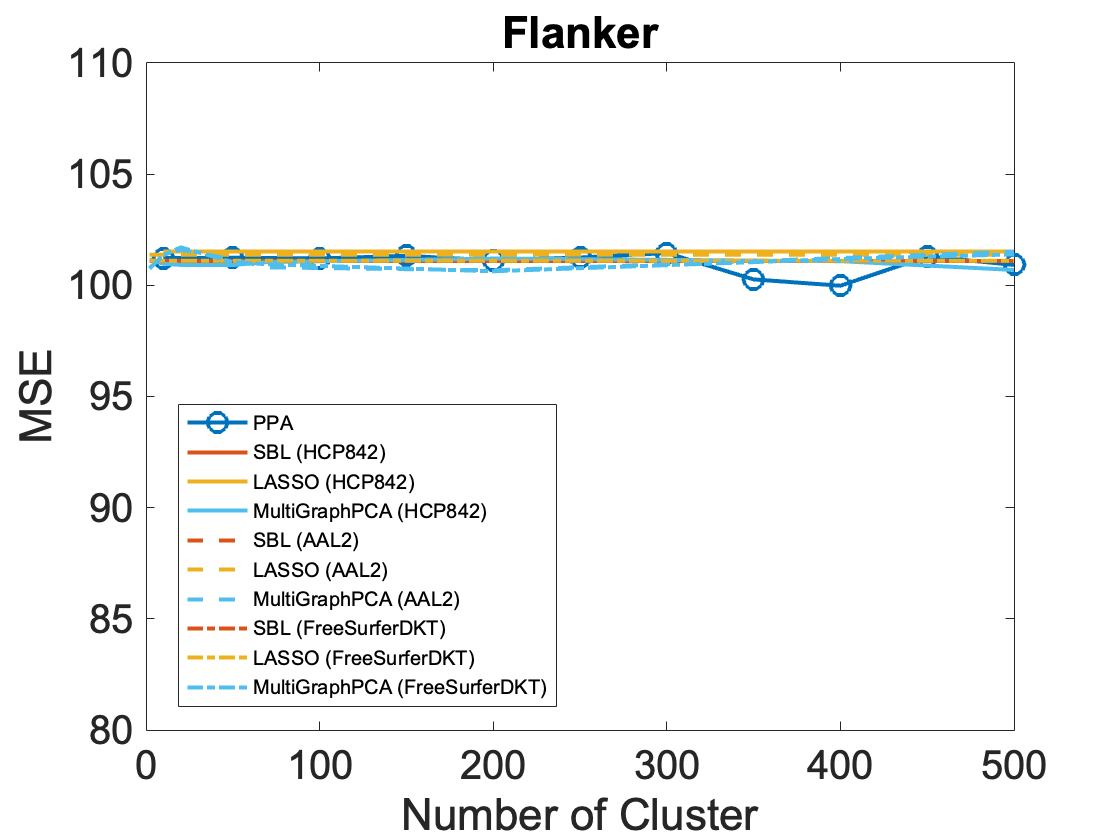} & 
		  \includegraphics[width=0.45\linewidth,height=5cm]{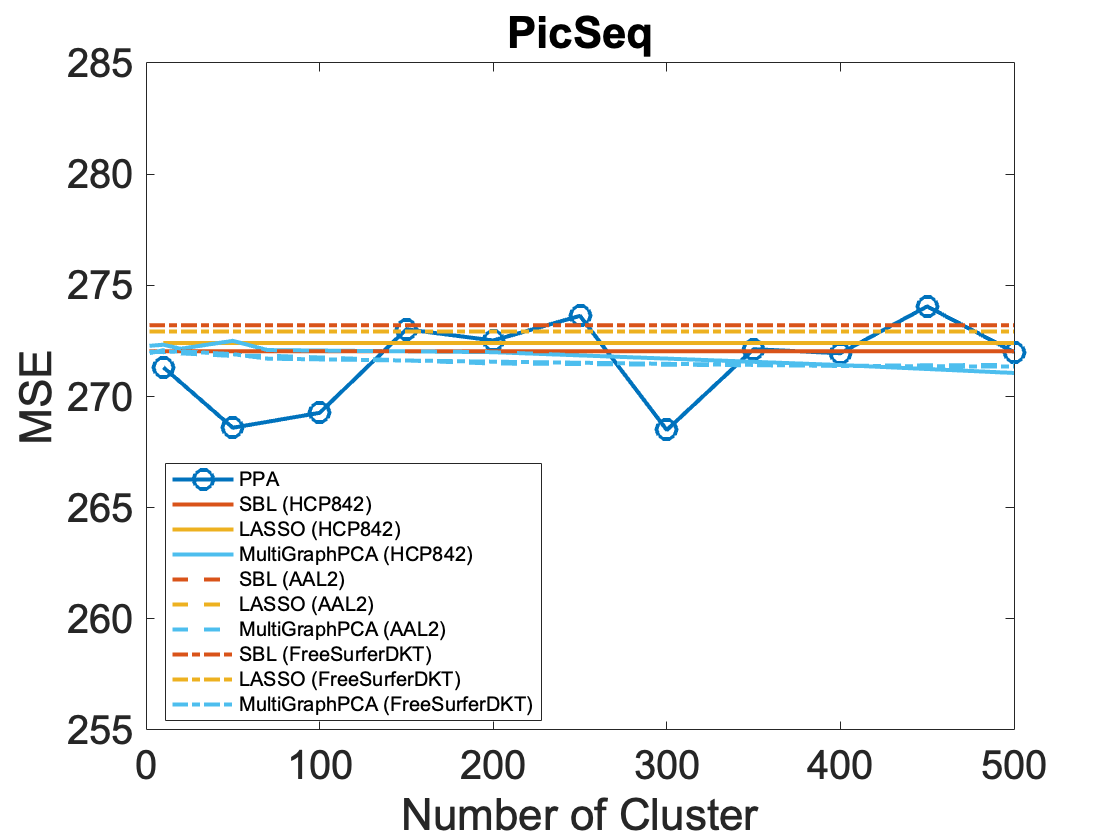} \\
		  \includegraphics[width=0.45\linewidth,height=5cm]{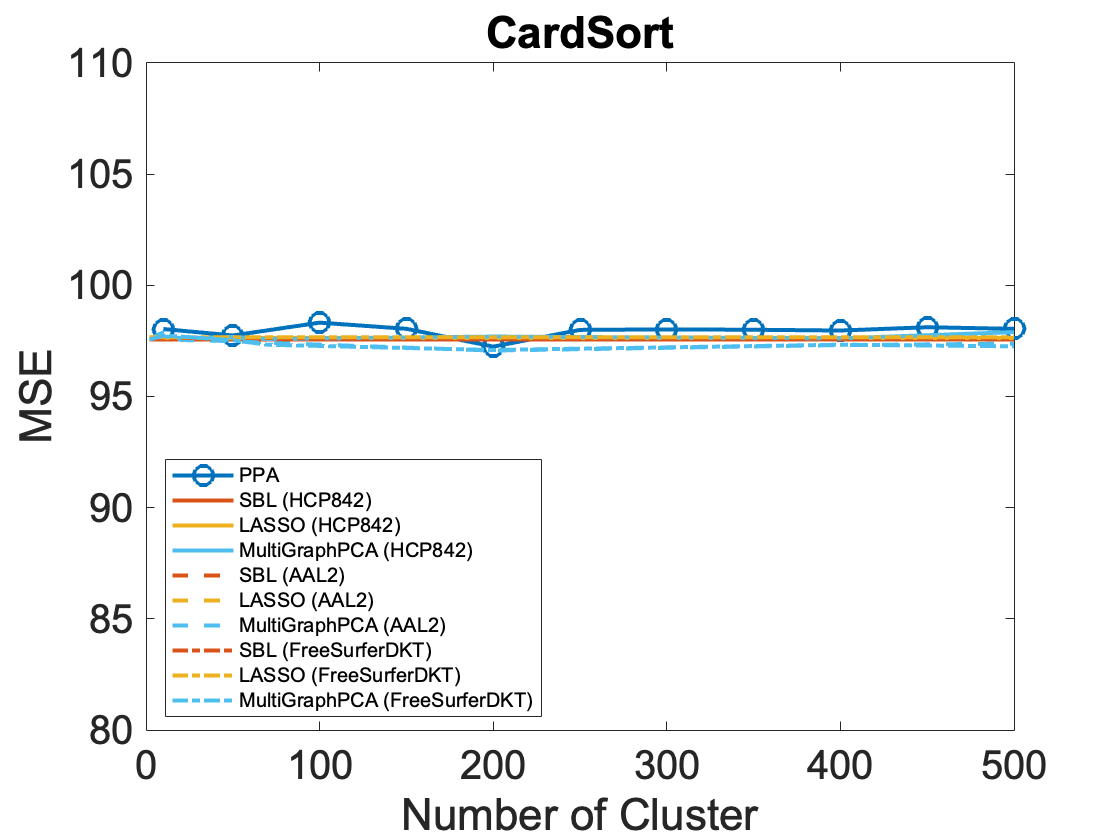} & 
		  \includegraphics[width=0.45\linewidth,height=5cm]{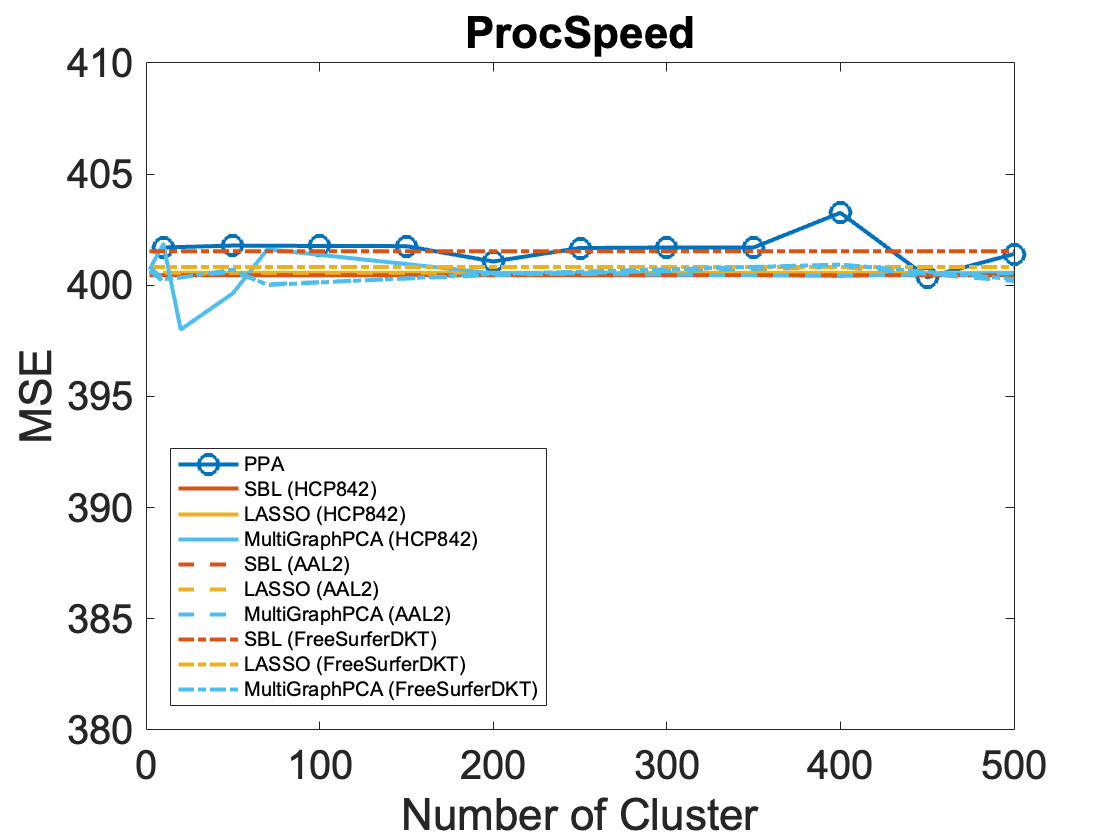} \\
	\end{tabular}
\caption{Comparison of 5-fold cross validation MSE of trait prediction based on APA with three atlases: HCP842, AAL2, and FreeSurferDKT.} \label{compare3} 
\end{figure}

\begin{figure}[h!]
\centering 
\begin{tabular}{ccc}
		  \includegraphics[width=0.3\linewidth,height=4cm]{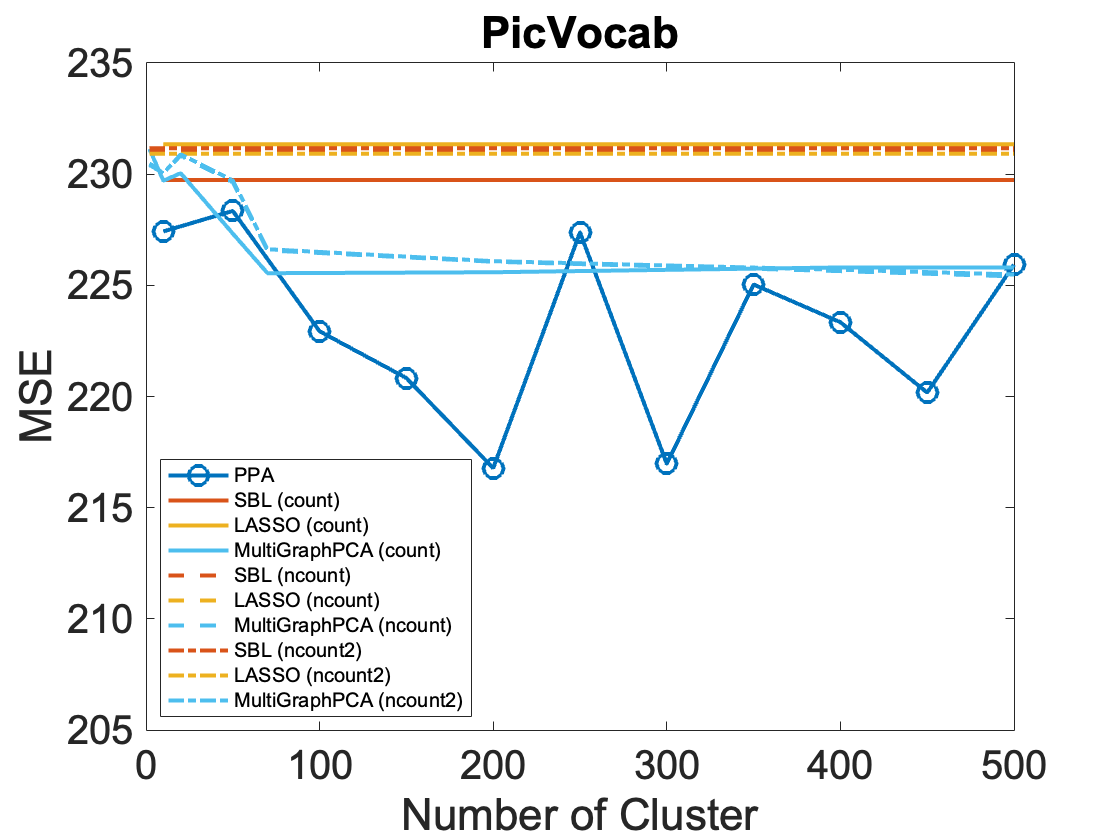} & 
		  \includegraphics[width=0.3\linewidth,height=4cm]{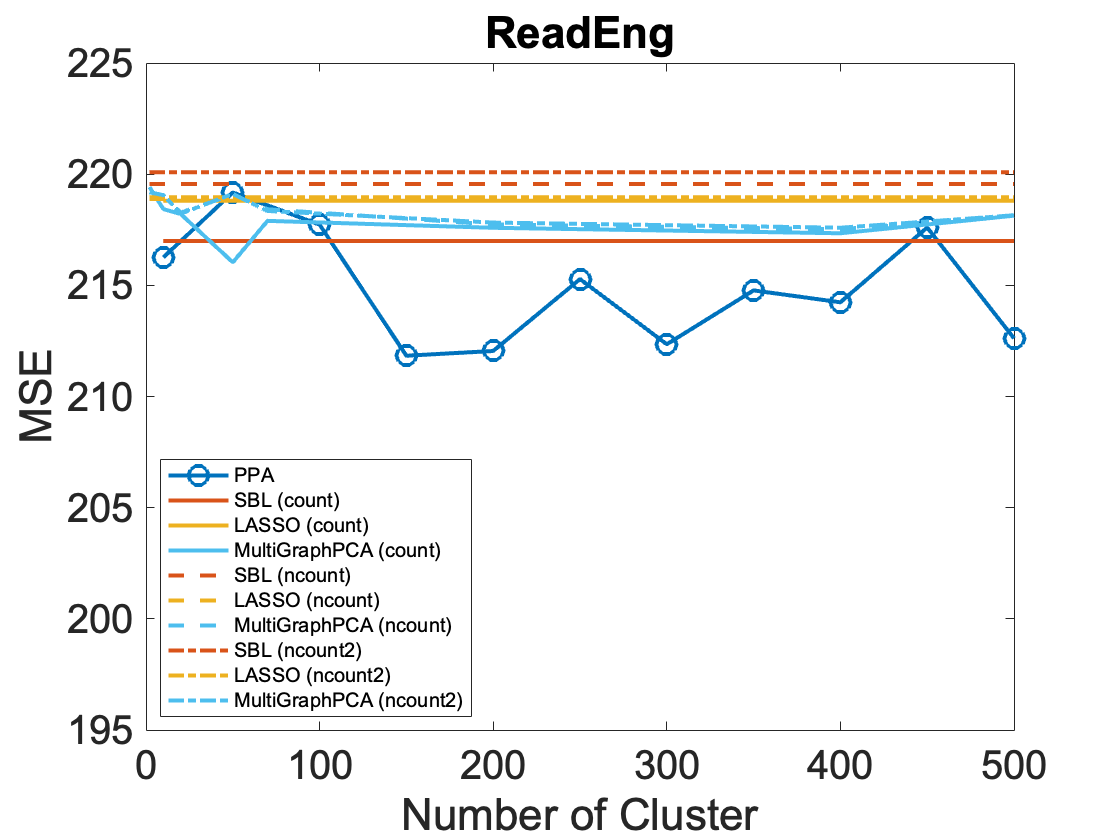} & 
		  \includegraphics[width=0.3\linewidth,height=4cm]{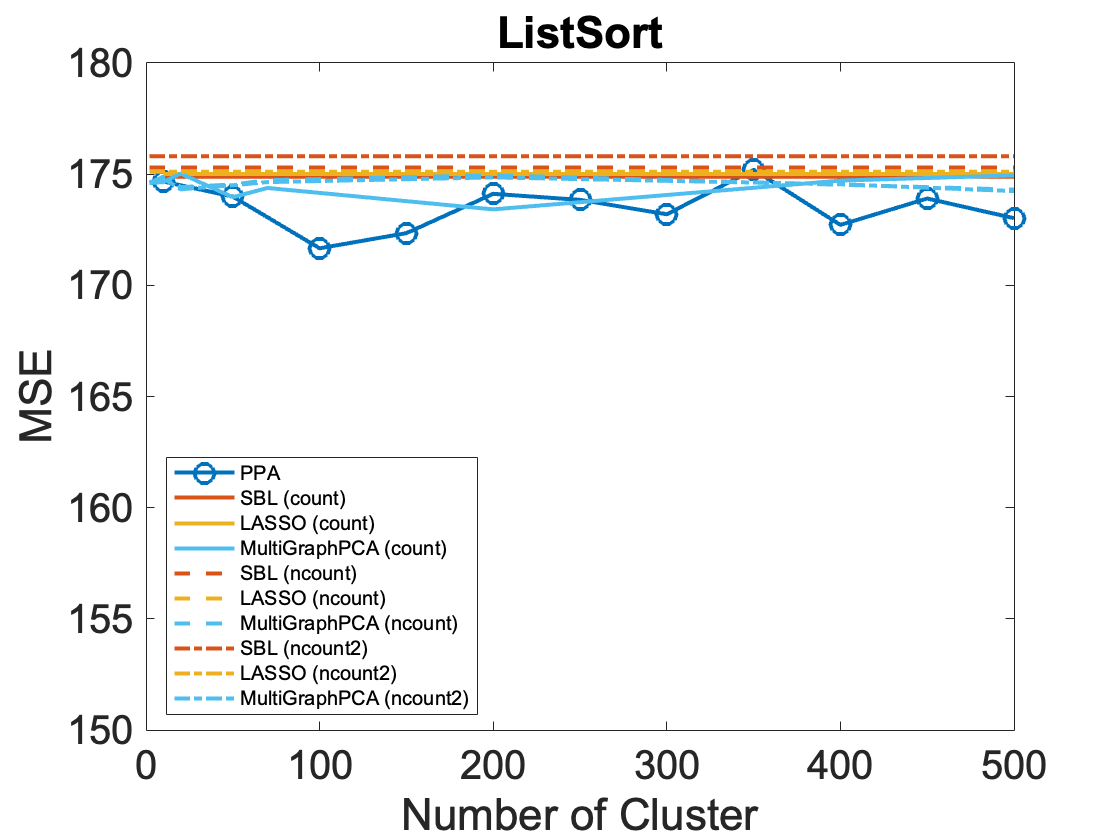} \\
	\end{tabular}
	\begin{tabular}{cc}
		  \includegraphics[width=0.45\linewidth,height=5cm]{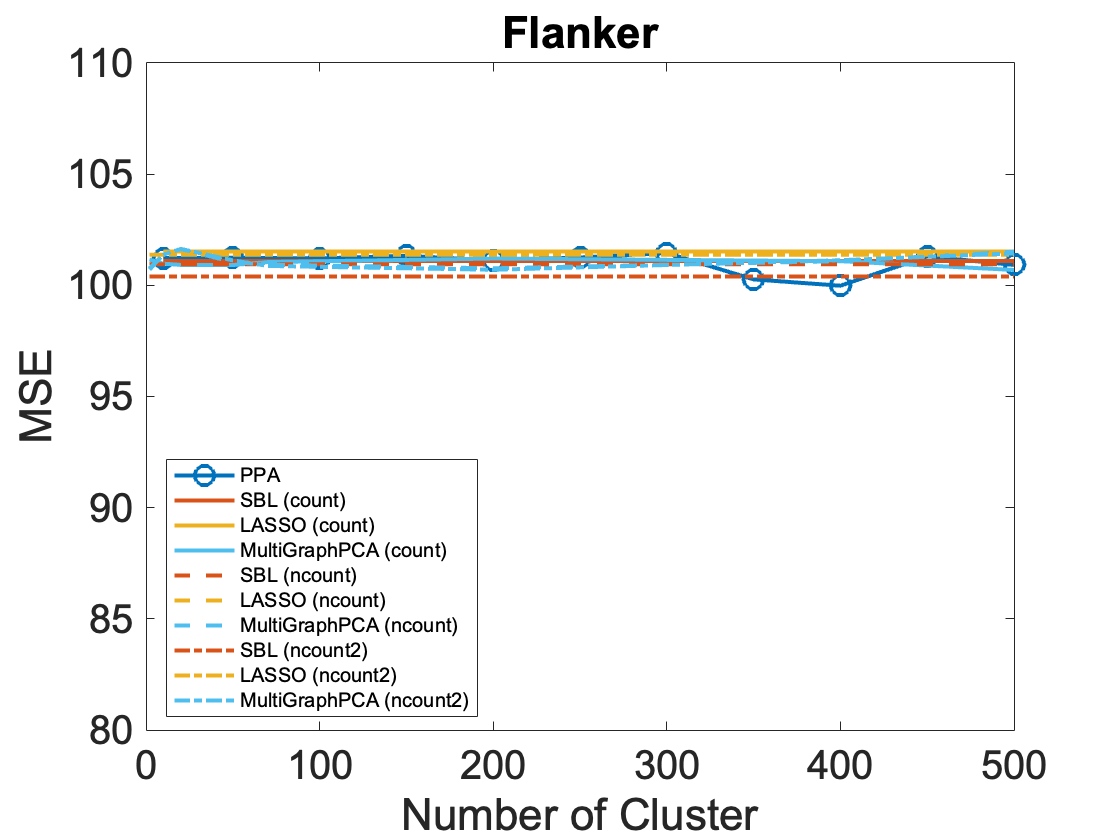} & 
		  \includegraphics[width=0.45\linewidth,height=5cm]{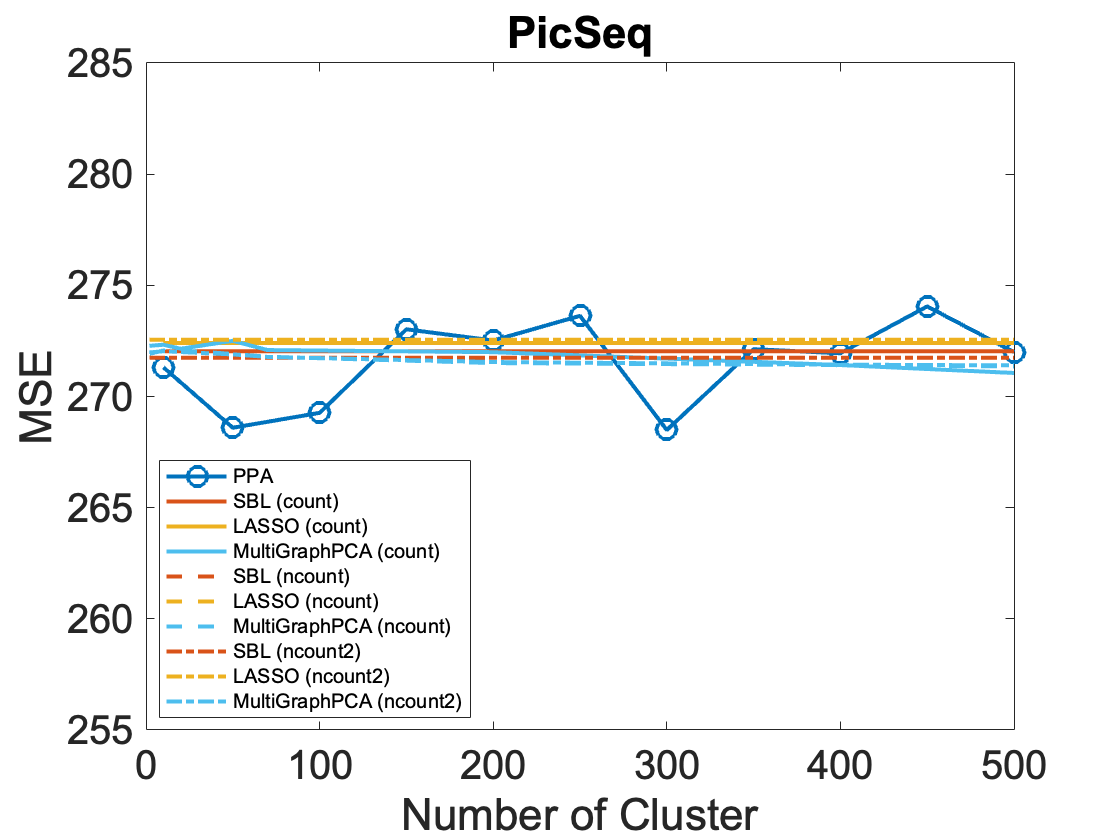} \\
		  \includegraphics[width=0.45\linewidth,height=5cm]{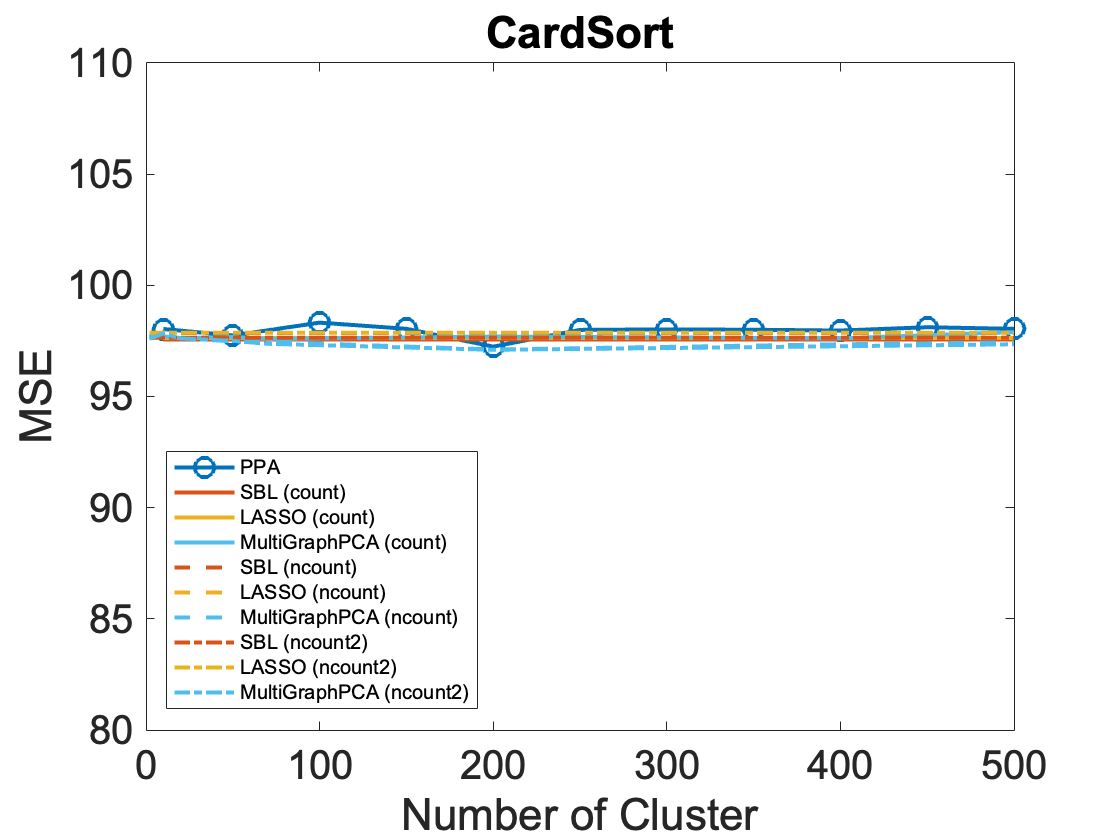} & 
		  \includegraphics[width=0.45\linewidth,height=5cm]{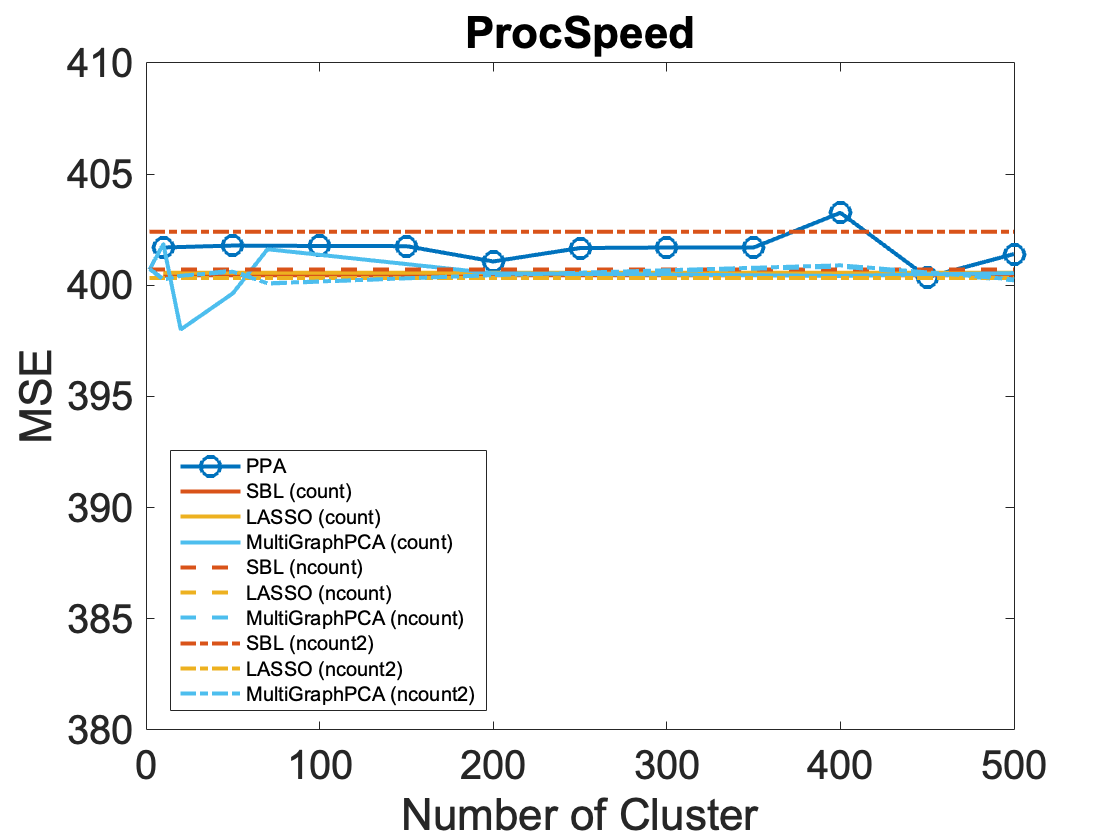} \\
	\end{tabular}
\caption{Comparison of 5-fold cross validation MSE of trait prediction based on APA with three metrics in calculating the connectivity matrix: count, ncount, and ncount2.} \label{compare4} 
\end{figure}

\begin{figure}[h!]
\centering 
	\begin{tabular}{ccc}
		 \includegraphics[width=0.3\linewidth, height=4cm]{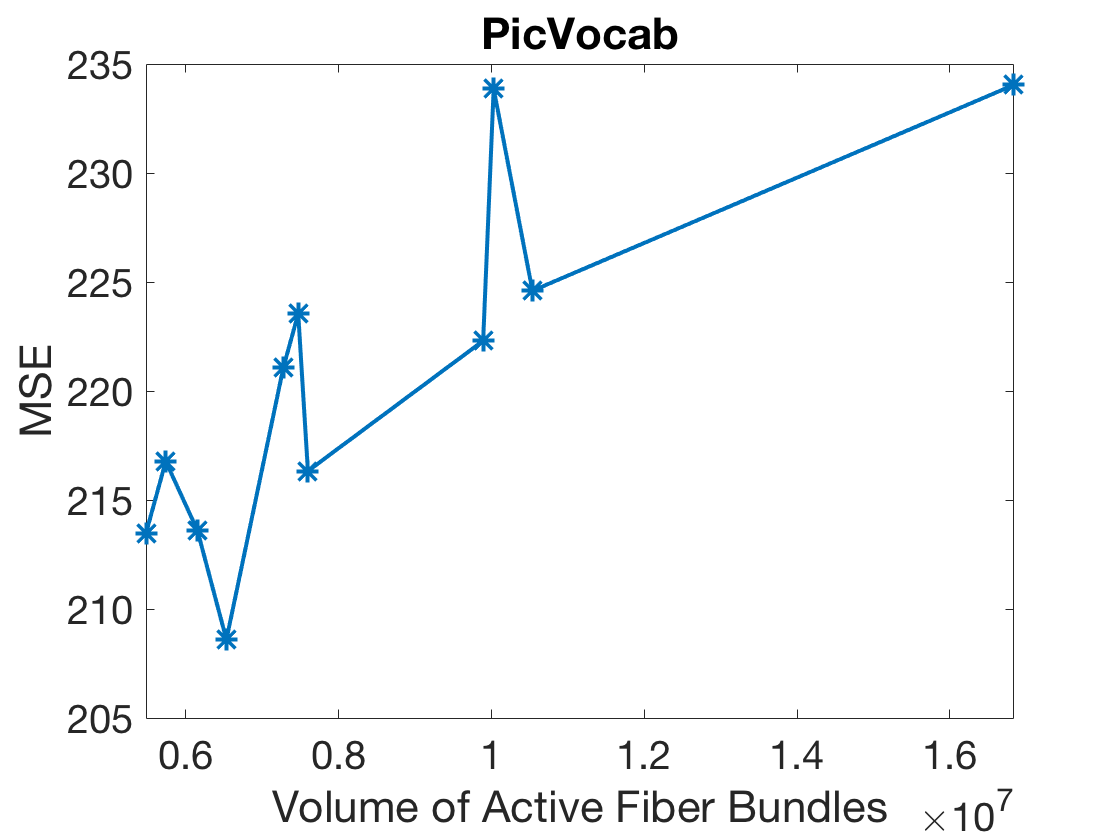} & 
		  \includegraphics[width=0.3\linewidth,height=4cm]{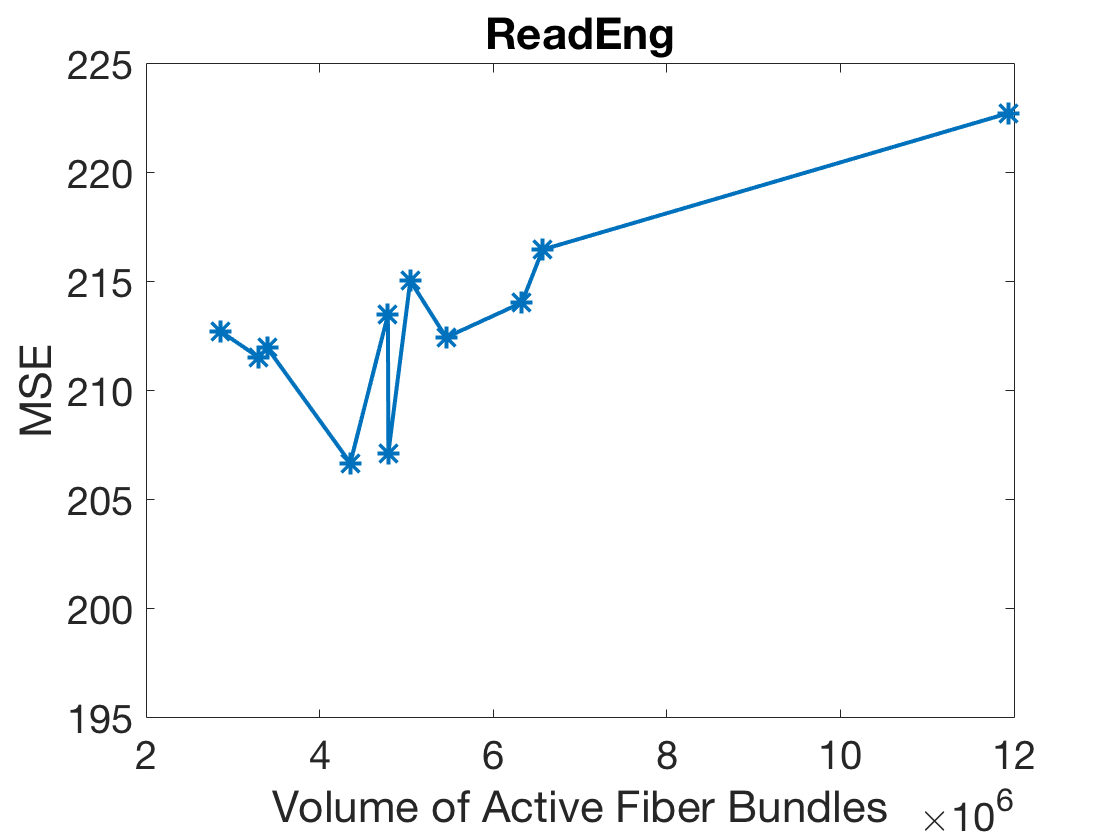}&
		  \includegraphics[width=0.3\linewidth,height=4cm]{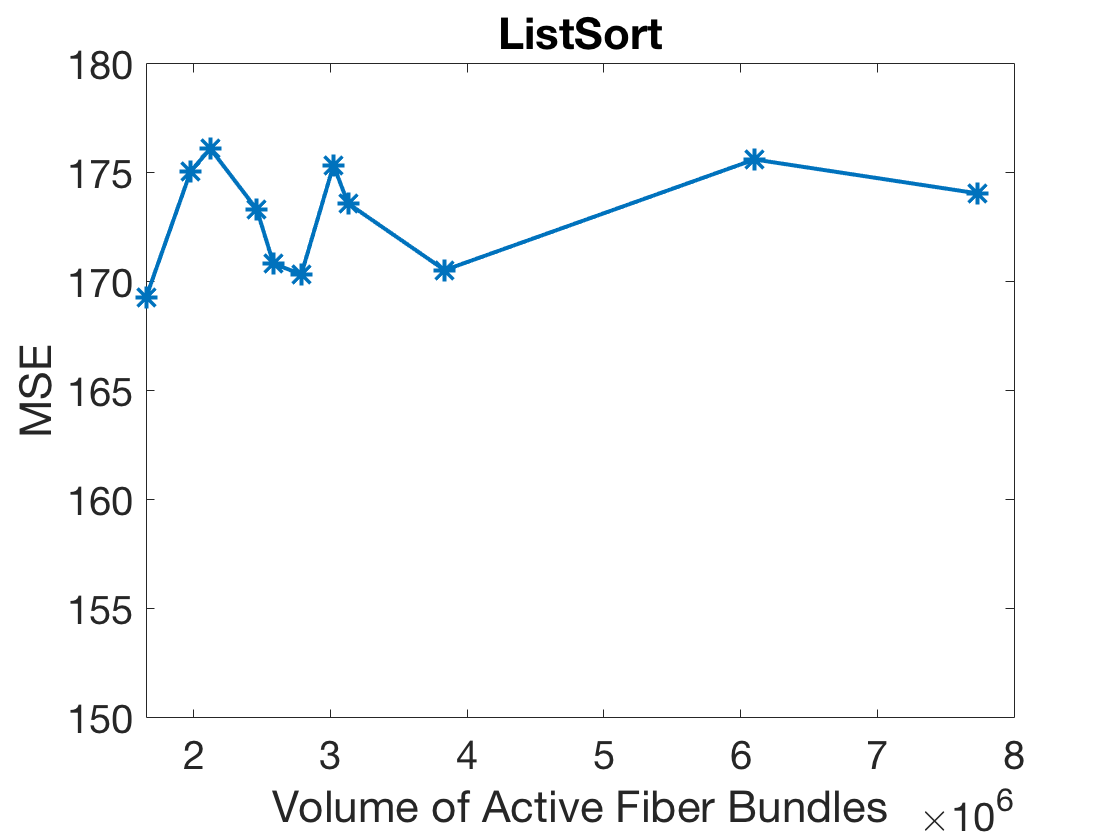}\\
		   {\small (a)  EuDX for Receptive Vocabulary} & {\small (b) EuDX for Oral Reading} & {\small (c) EuDX for List Sorting} \\
		  \includegraphics[width=0.3\linewidth, height=4cm]{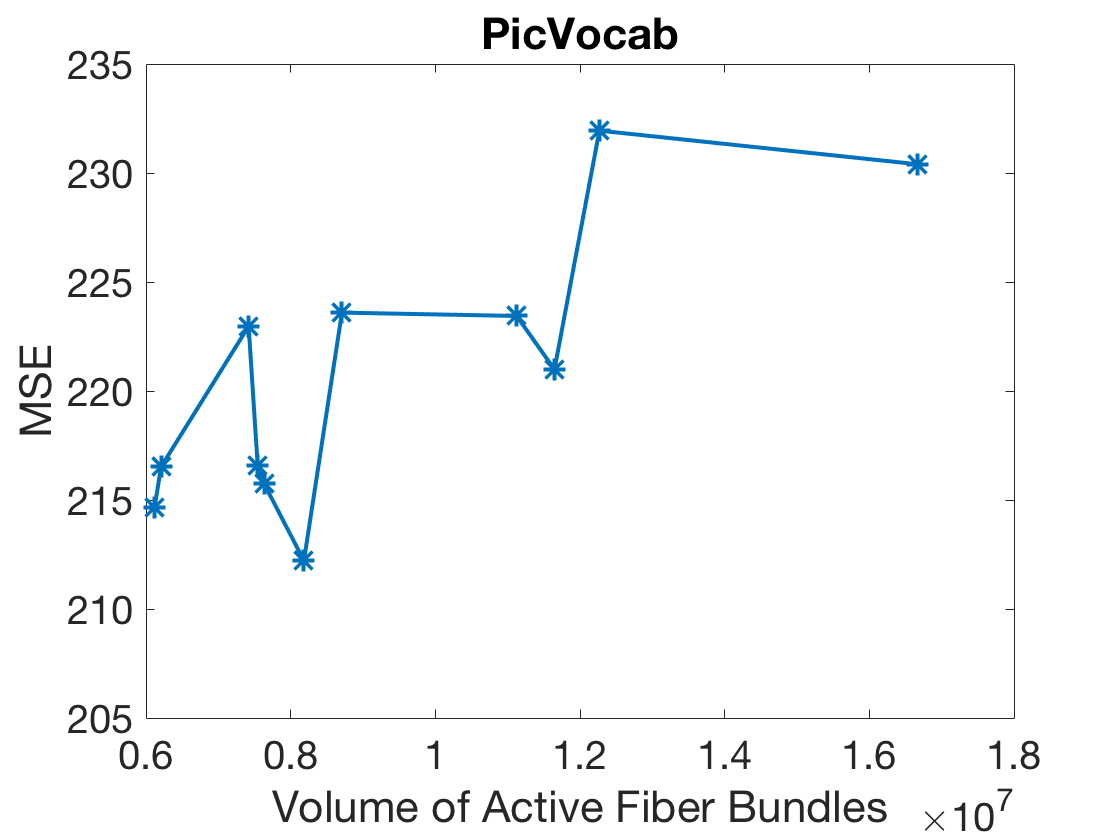} & 
		  \includegraphics[width=0.3\linewidth, height=4cm]{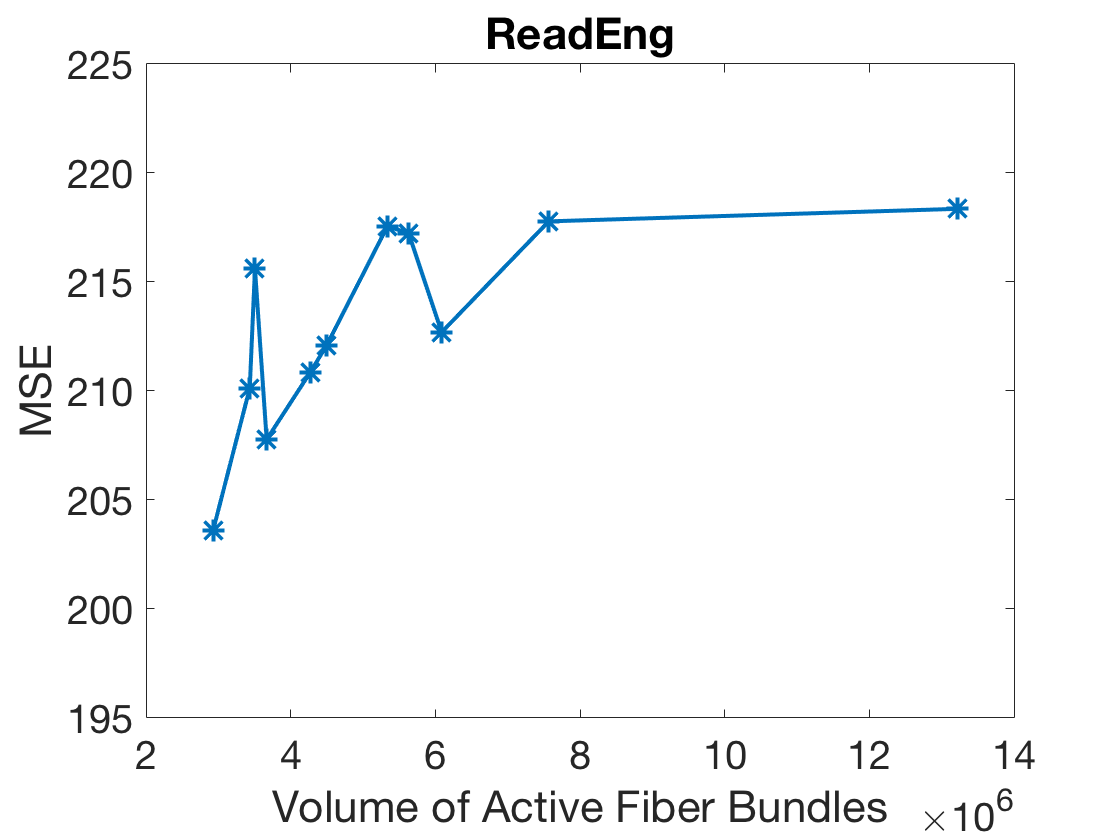}&
		  \includegraphics[width=0.3\linewidth, height=4cm]{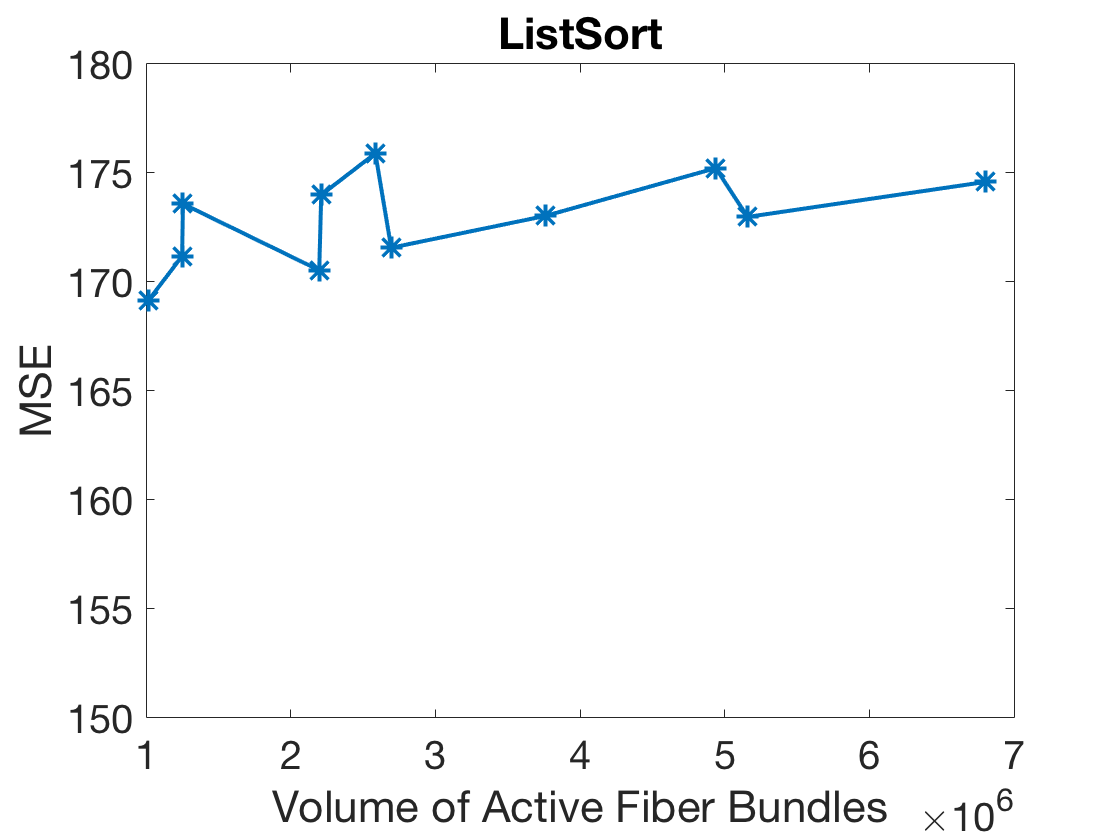}\\
		   {\small (d) SFM for Receptive Vocabulary} & {\small (e) SFM for Oral Reading} & {\small (f) SFM for List Sorting} \\
		  \end{tabular}
\caption{The comparison of 5-fold cross validation MSE of trait prediction along with the number of active fiber bundles based on PPA with the other two fiber tracking options: EuDX ((a)-(c)) and SFM ((d)-(f)) for three traits: PicVocab, ReadEng and ListSort.} 
\label{mse_2trackings} 
\end{figure}

\begin{figure*}
\centering 
	\begin{tabular}{cccc}
		 \includegraphics[width=0.225\linewidth, height=3.5cm]{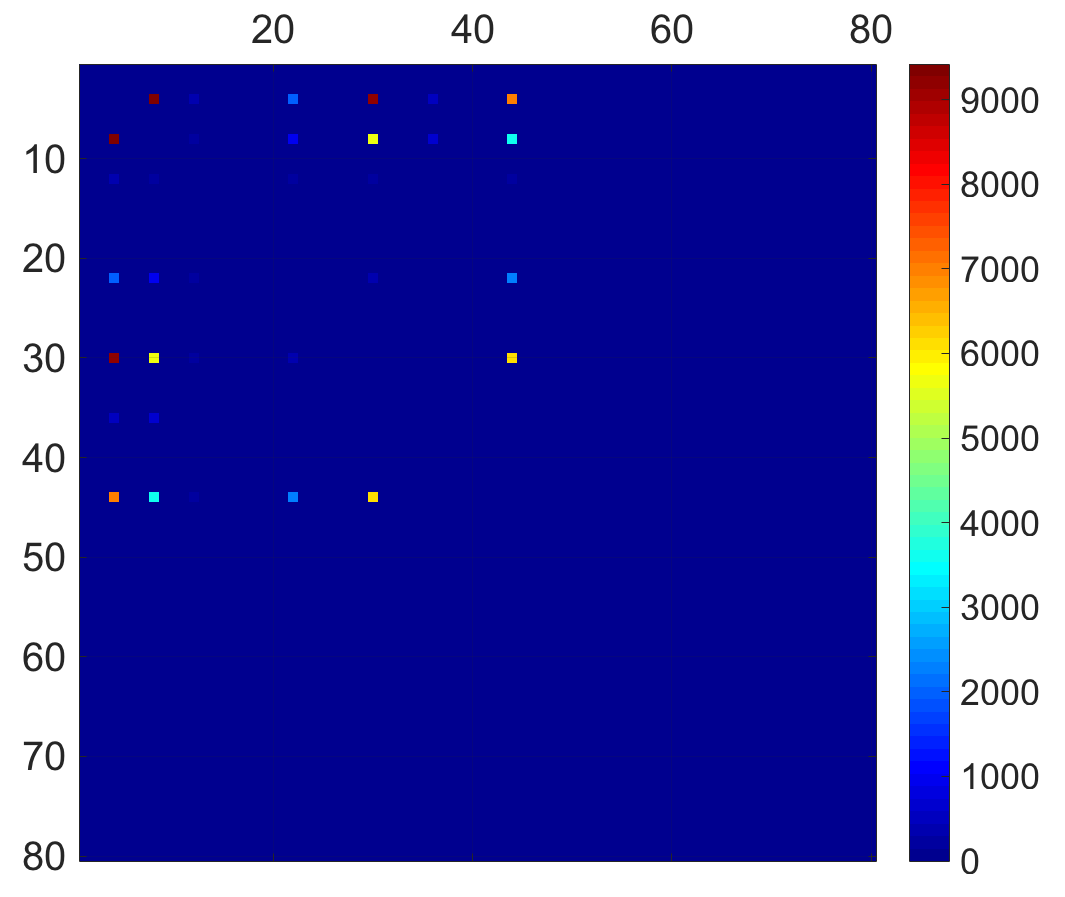} & 
		  \includegraphics[width=0.225\linewidth,height=3.5cm]{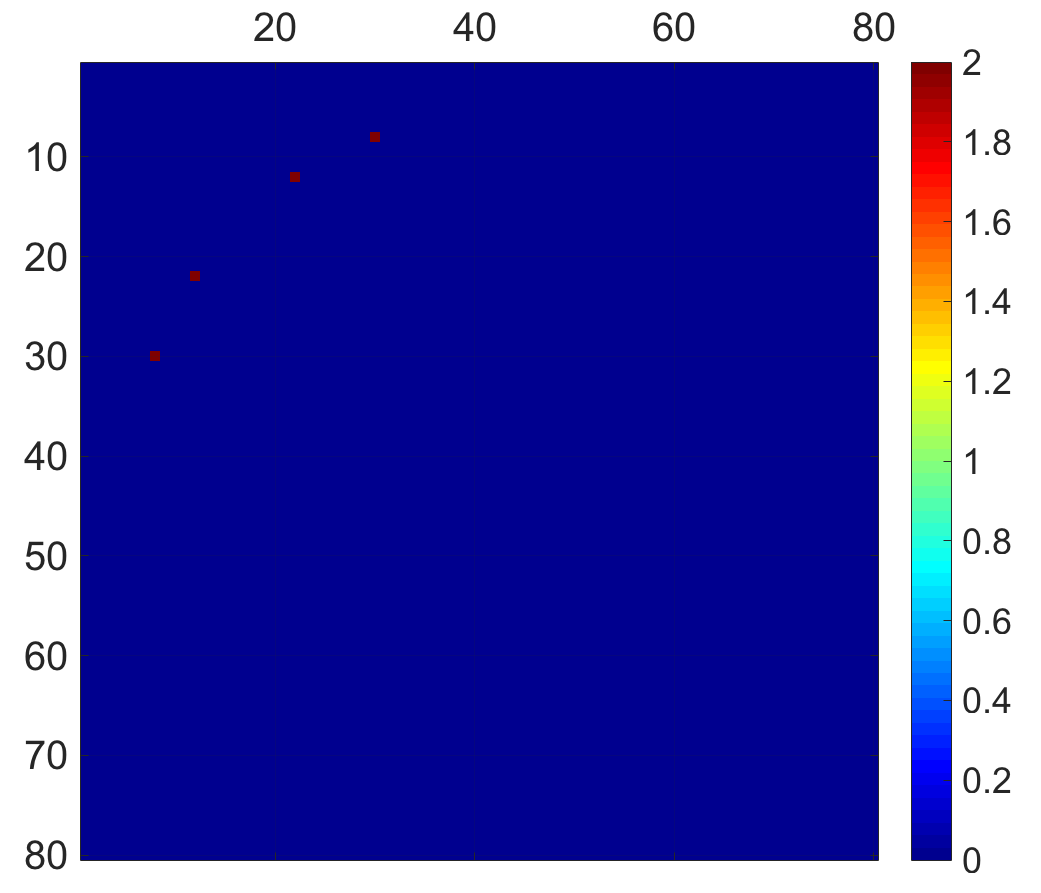} &
		   \includegraphics[width=0.225\linewidth, height=3.5cm]{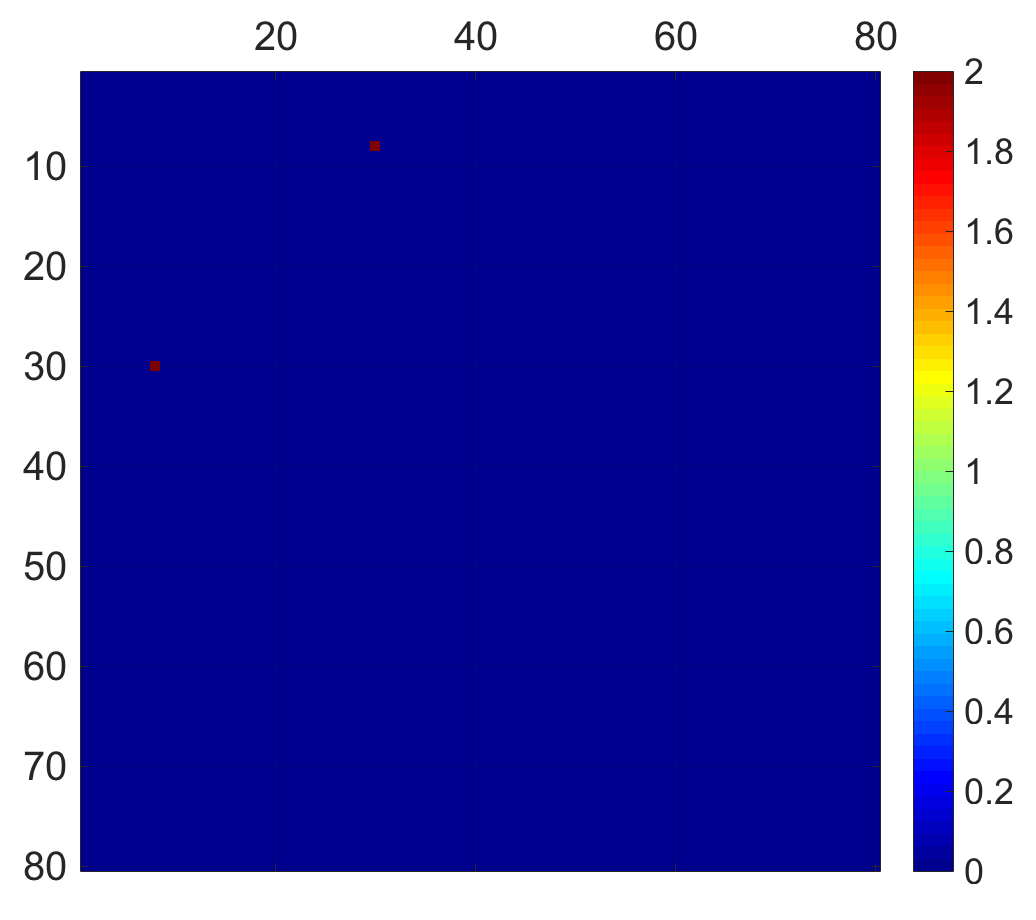} & 
		  \includegraphics[width=0.225\linewidth,height=3.5cm]{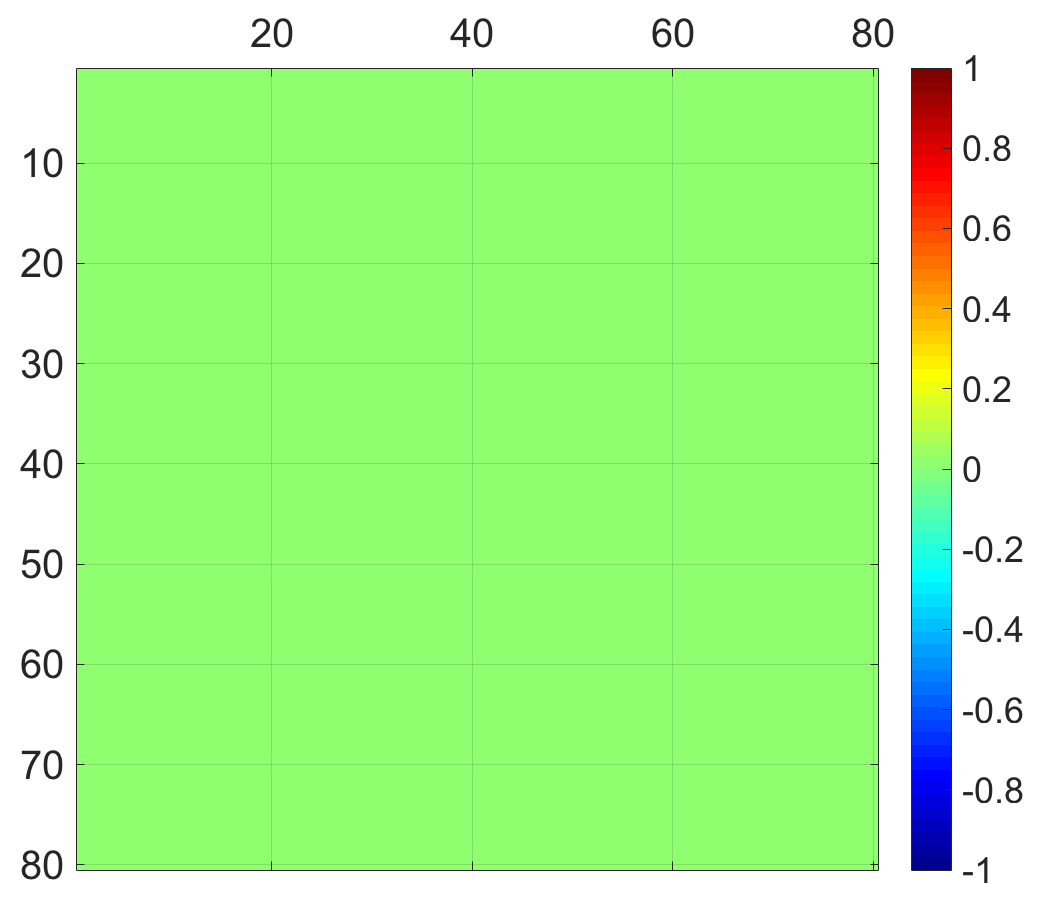} \\
		  \includegraphics[width=0.225\linewidth,height=4cm]{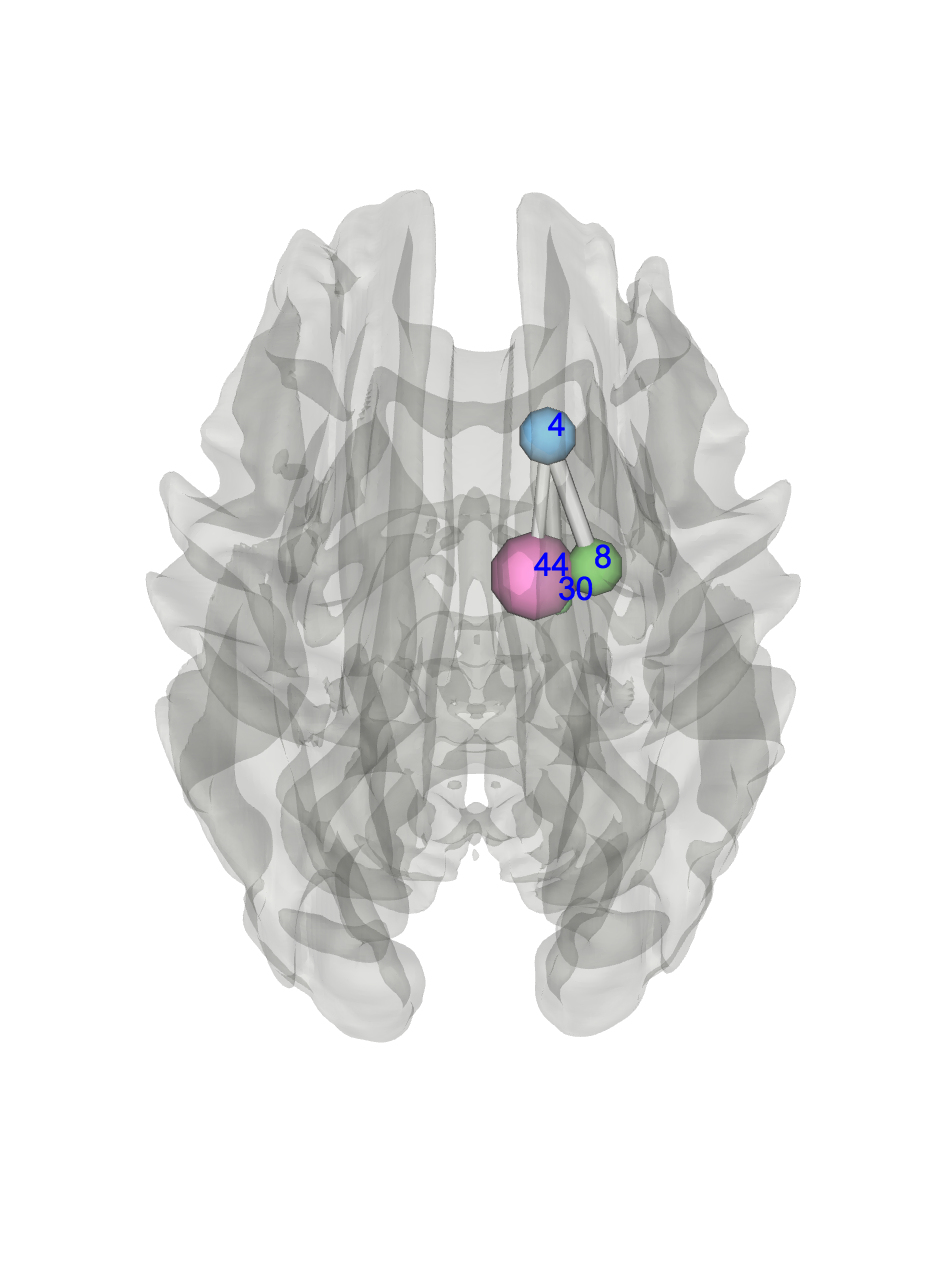} & 
		  \includegraphics[width=0.225\linewidth,height=4cm]{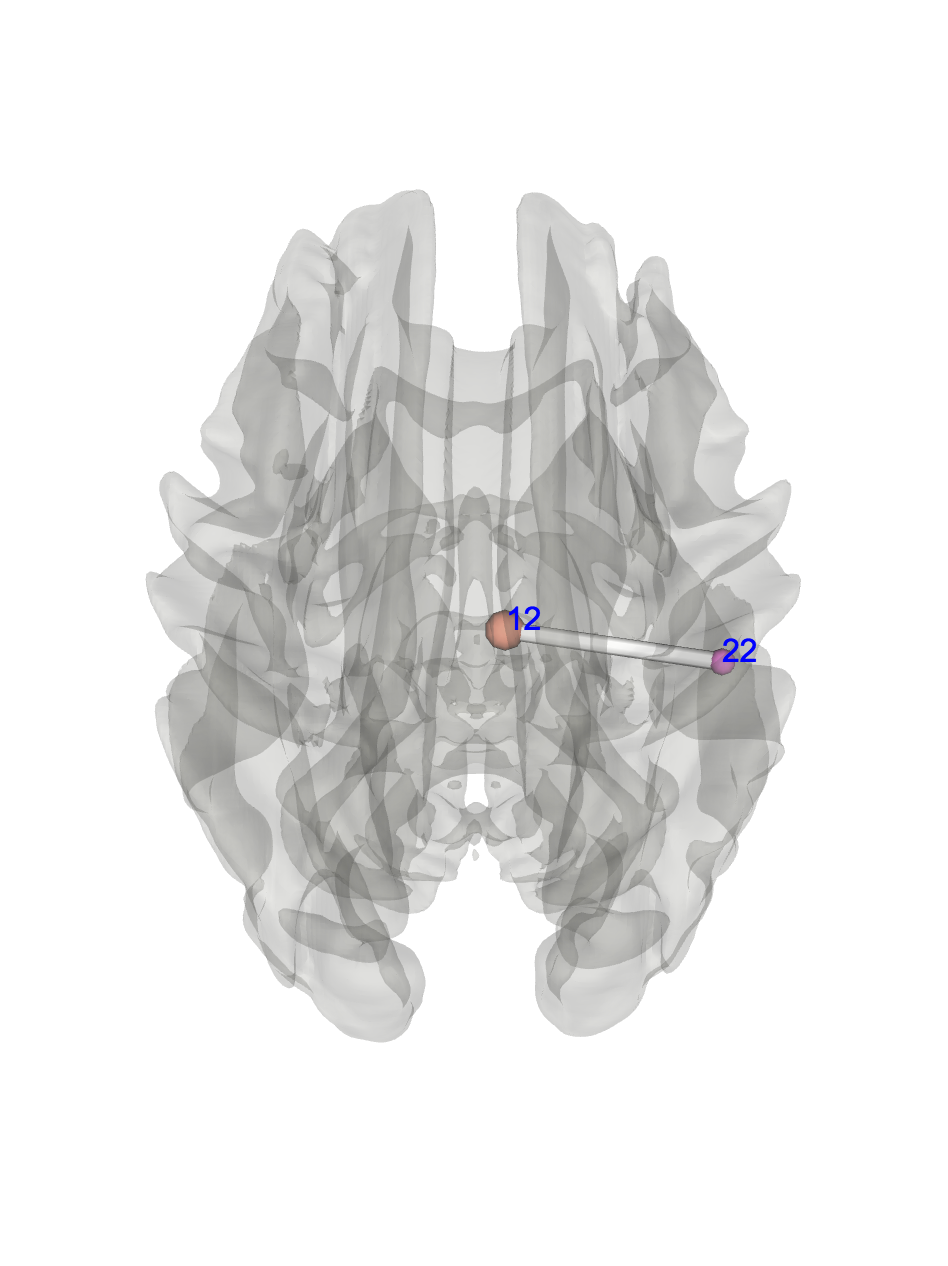} &
		   \includegraphics[width=0.225\linewidth, height=4cm]{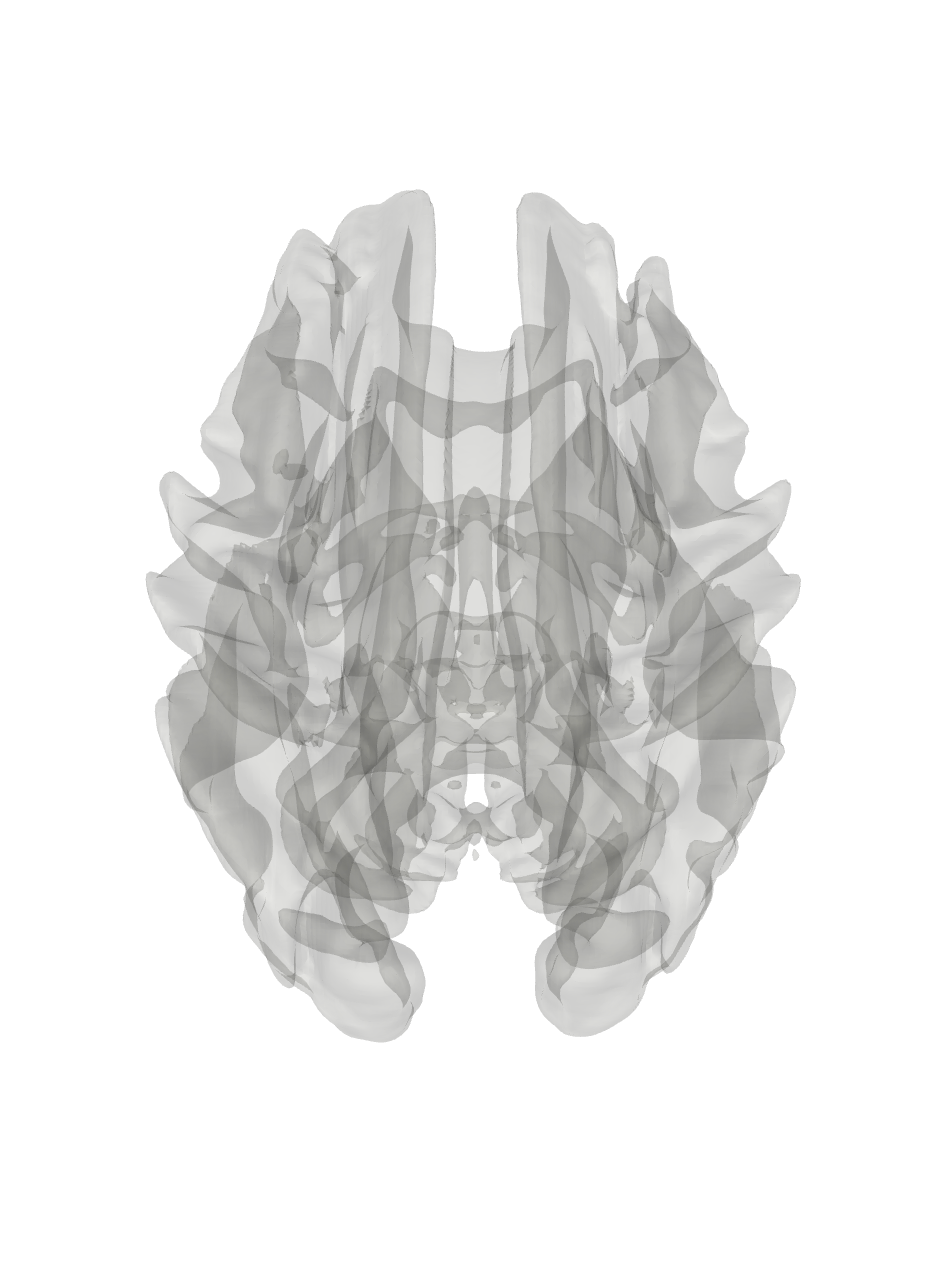} & 
		  \includegraphics[width=0.225\linewidth,height=4cm]{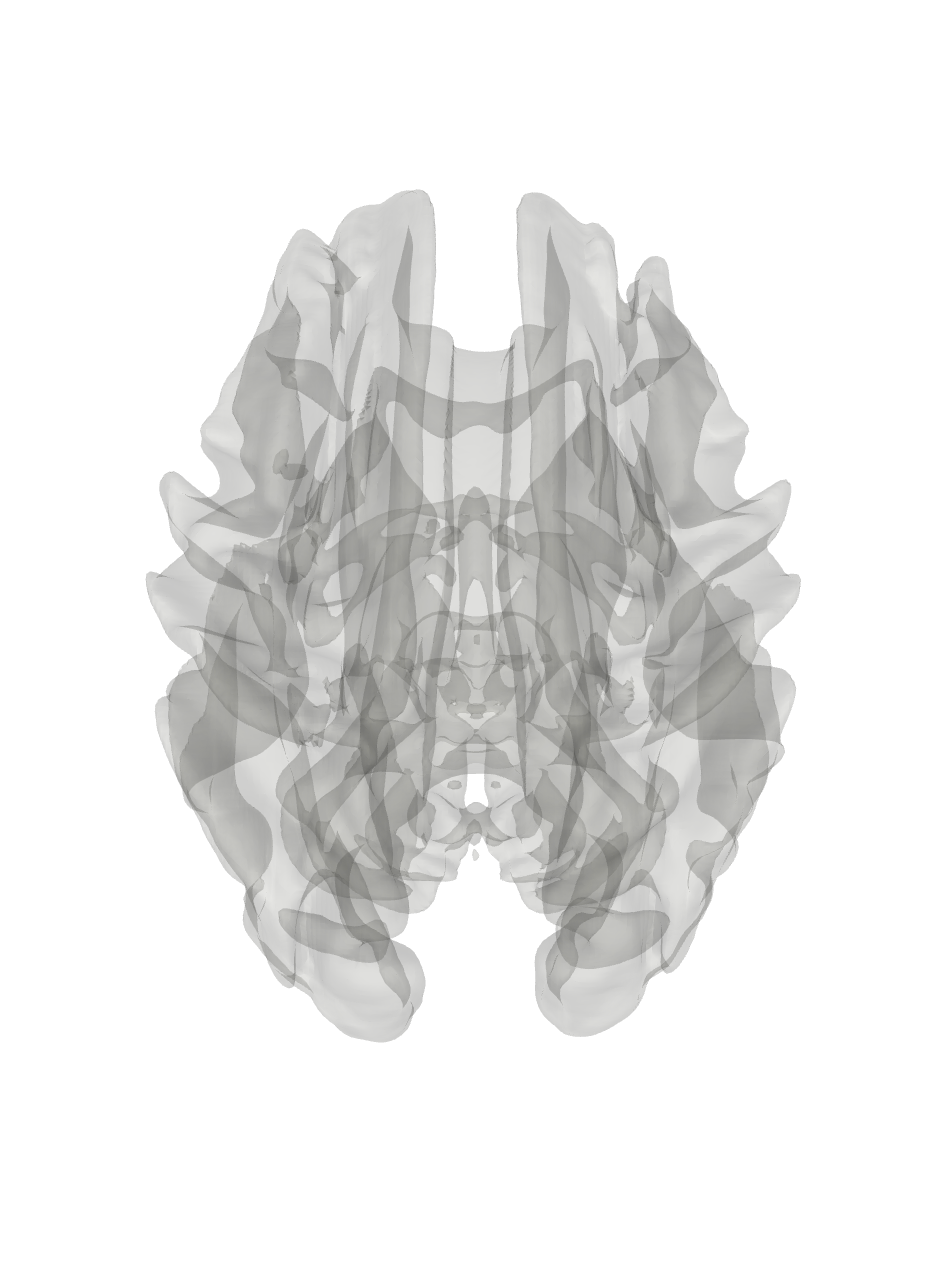} \\
		  \includegraphics[width=0.225\linewidth,height=4cm]{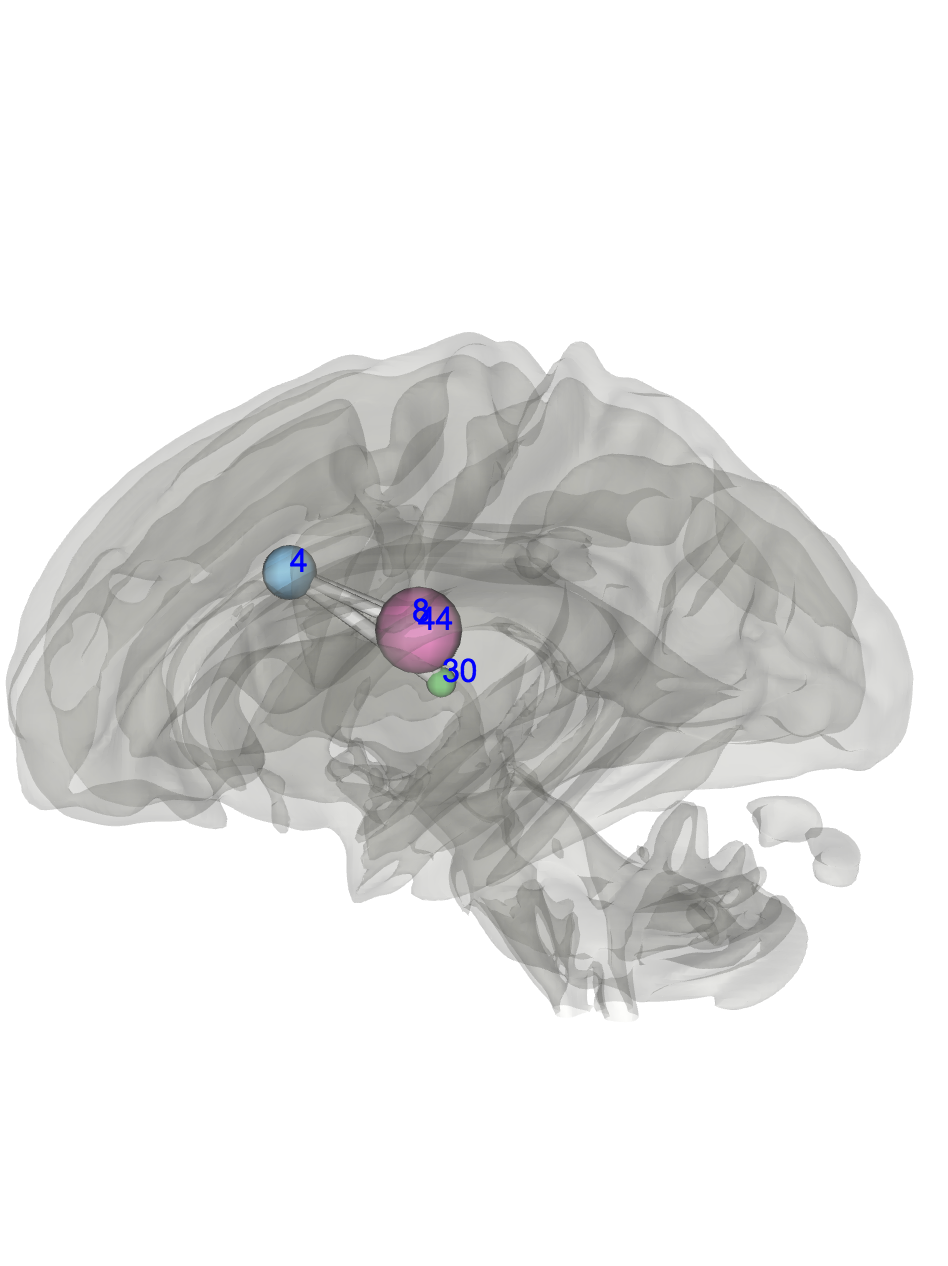} & 
		  \includegraphics[width=0.225\linewidth,height=4cm]{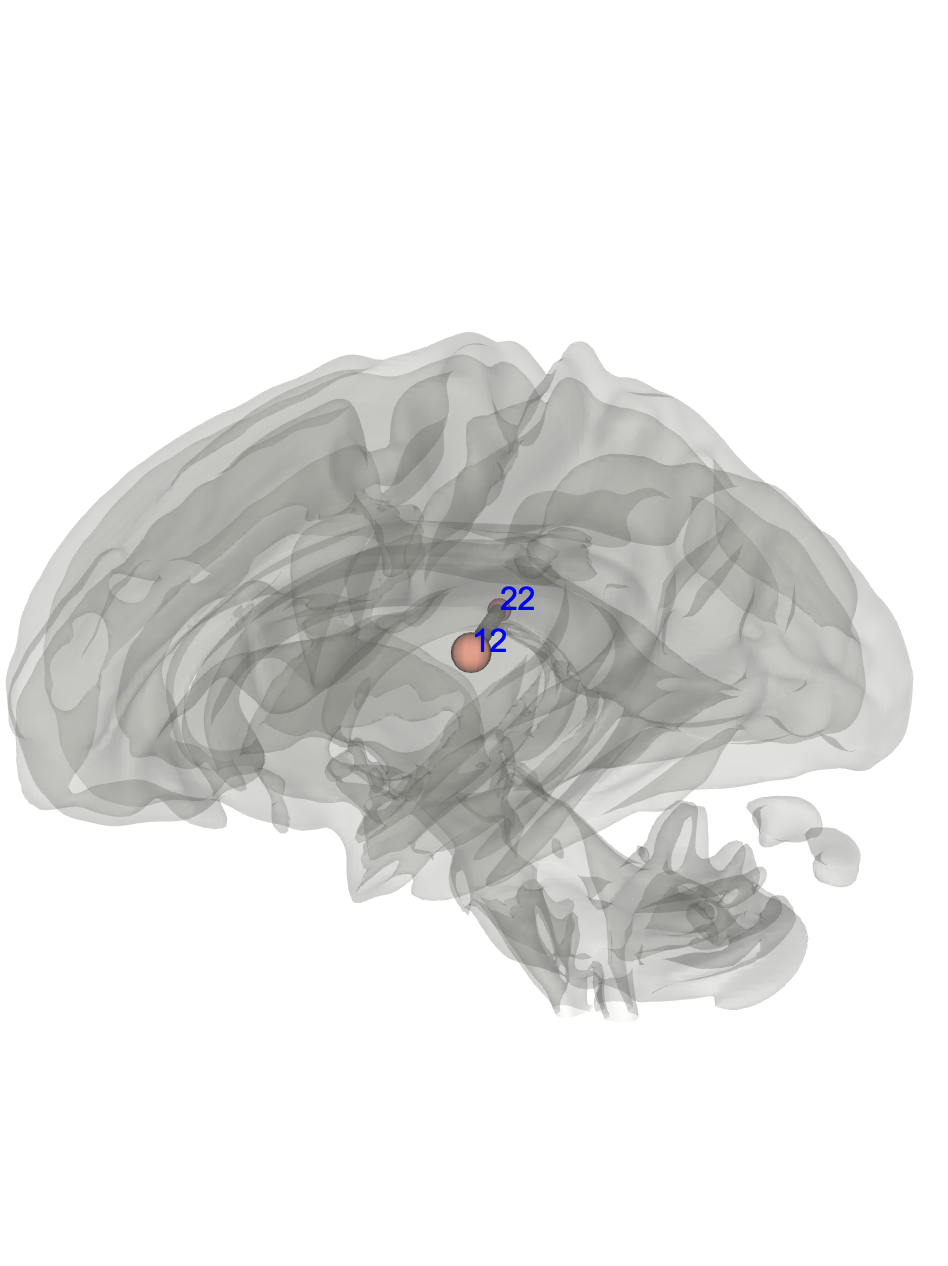} &
		   \includegraphics[width=0.225\linewidth, height=4cm]{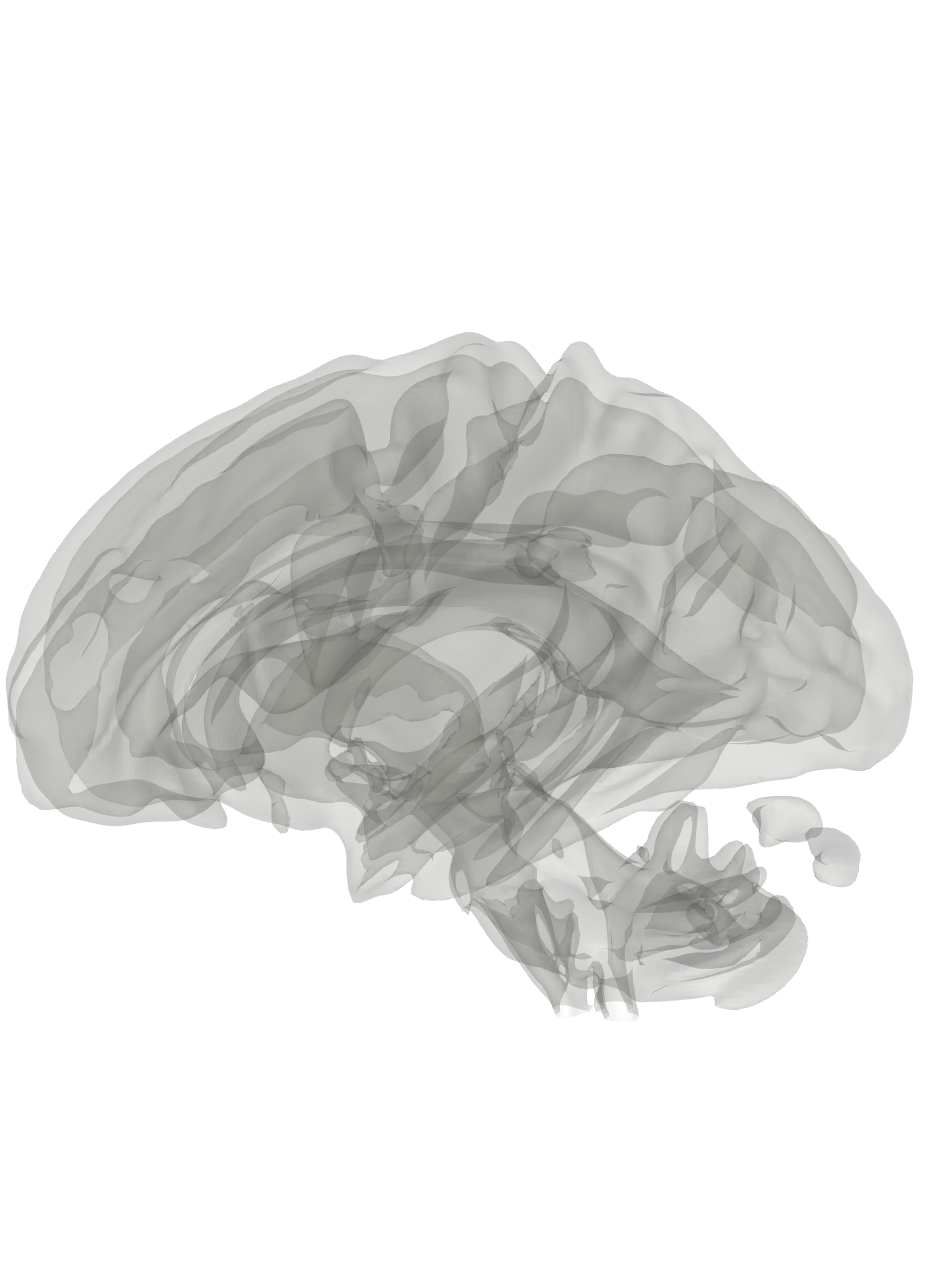} & 
		  \includegraphics[width=0.225\linewidth,height=4cm]{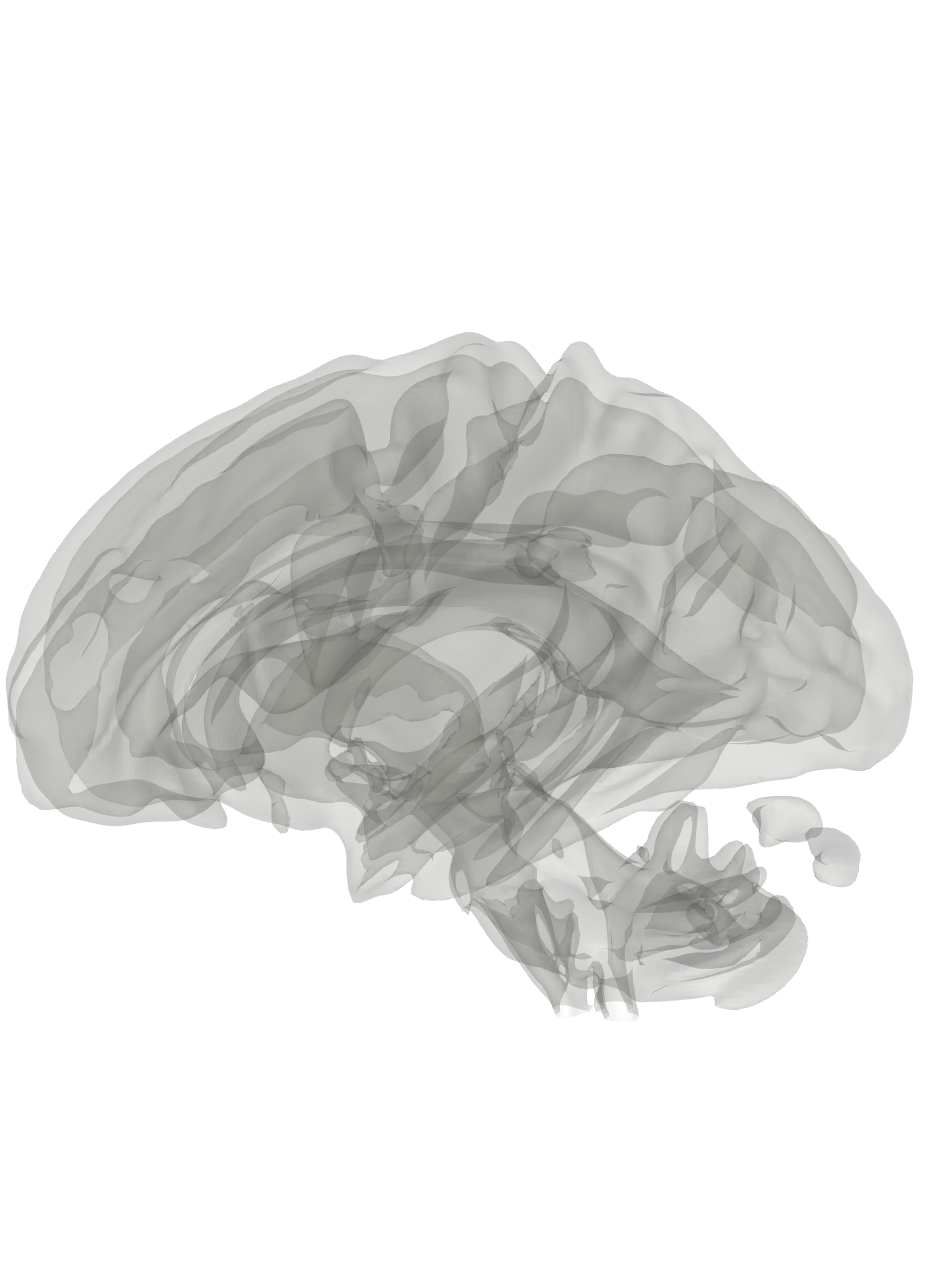} \\
		  {\small (a) $K$=50} & {\small (b) $K$=100} & {\small (c) $K$=200} & {\small (d) $K$=400}
	\end{tabular}
\caption{Visualization for trait PicSeq: each column represents the visualization under different $K$ ($K$=50; $K$=100; $K$=200; $K$=400) in PPA; First row represents the connectivity matrix between any two ROI regions in HCP842 tractography atlas. Second row shows anatomy of connections in an axial view. Third row shows anatomy of connections in a sagittal view.} 
\label{fig:ps} 
\end{figure*}

\begin{figure*}
\centering 
	\begin{tabular}{cccc}
		 \includegraphics[width=0.225\linewidth, height=3.5cm]{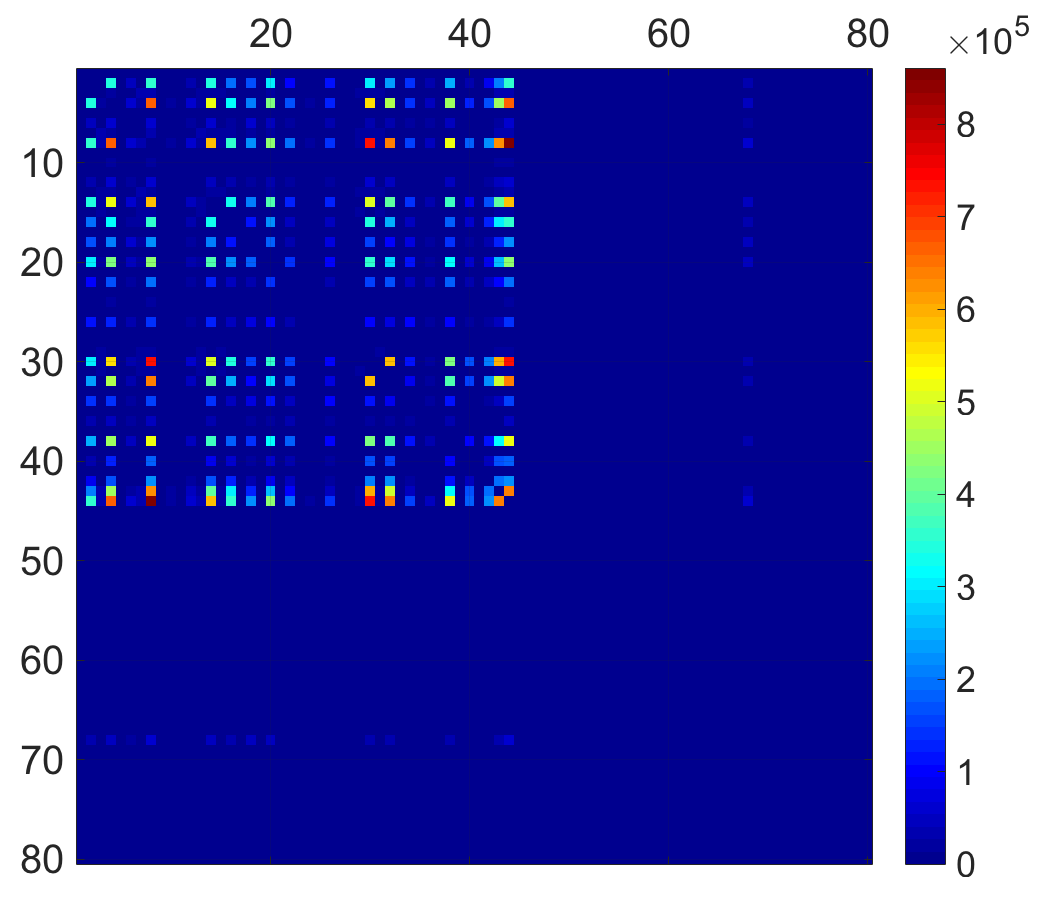} & 
		  \includegraphics[width=0.225\linewidth,height=3.5cm]{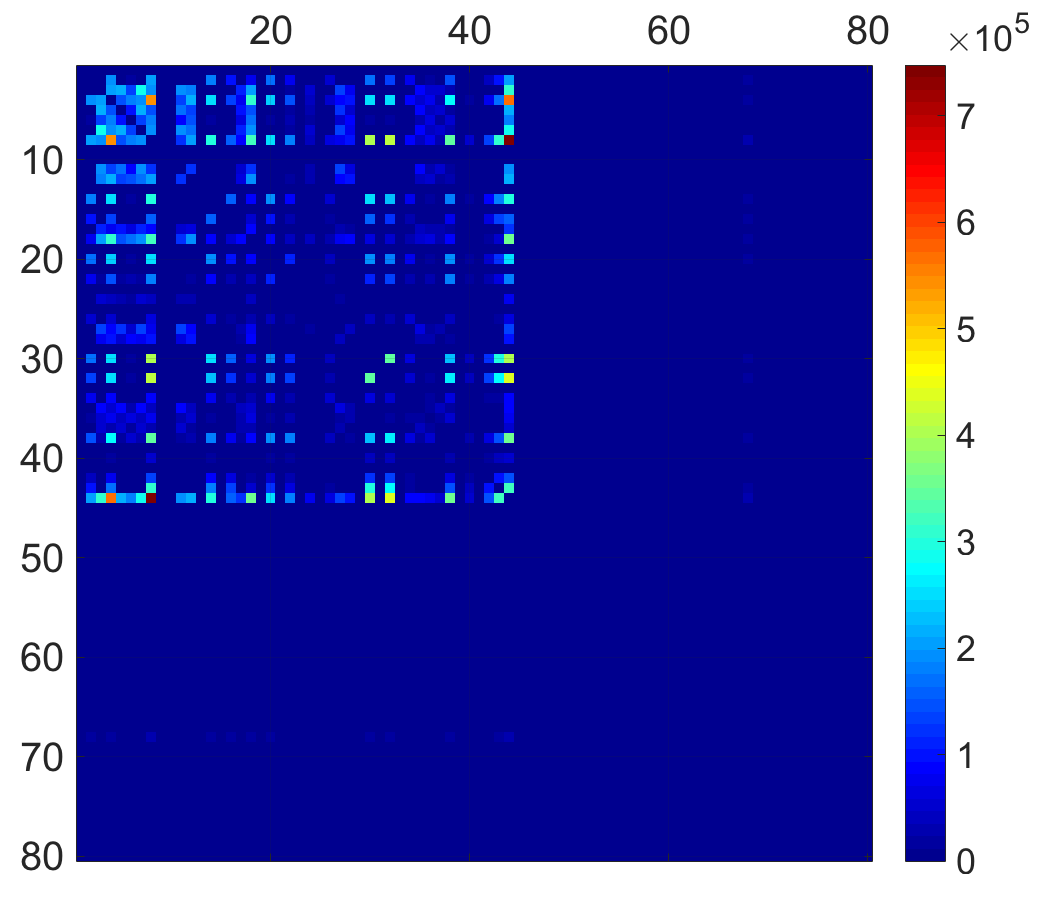} &
		   \includegraphics[width=0.225\linewidth, height=3.5cm]{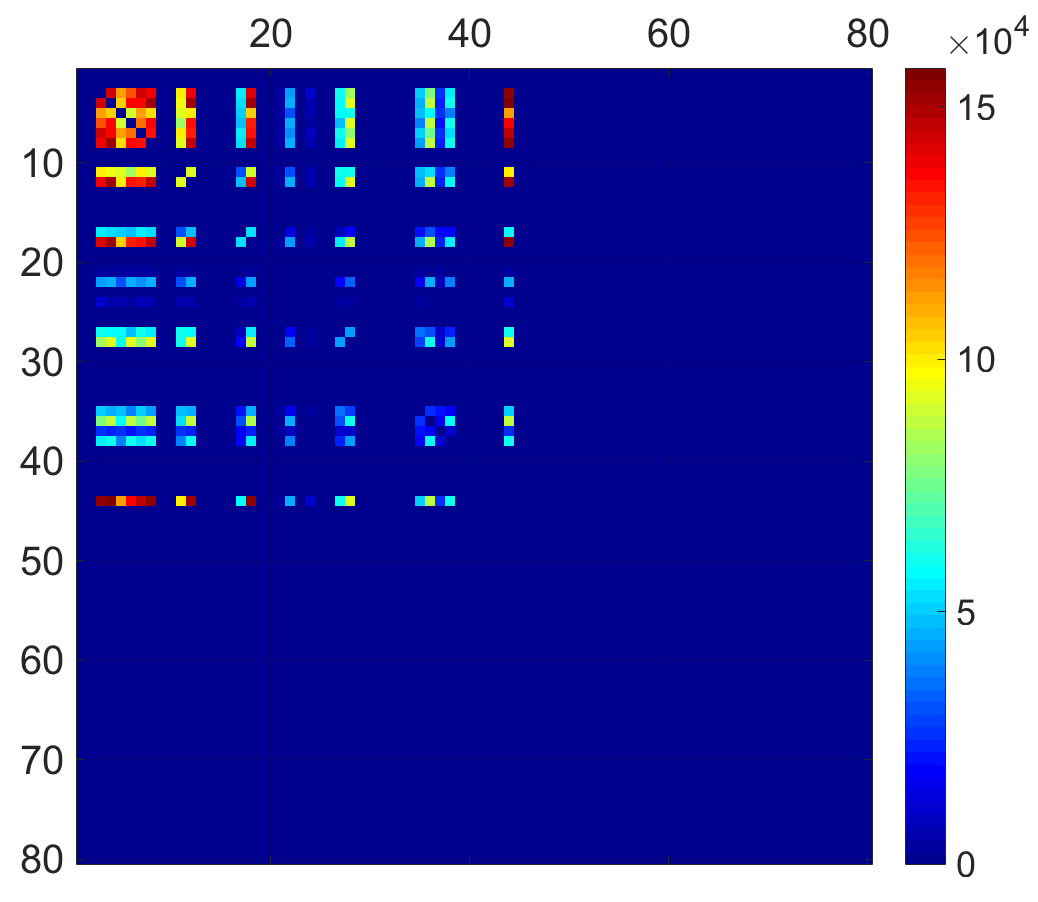} & 
		  \includegraphics[width=0.225\linewidth,height=3.5cm]{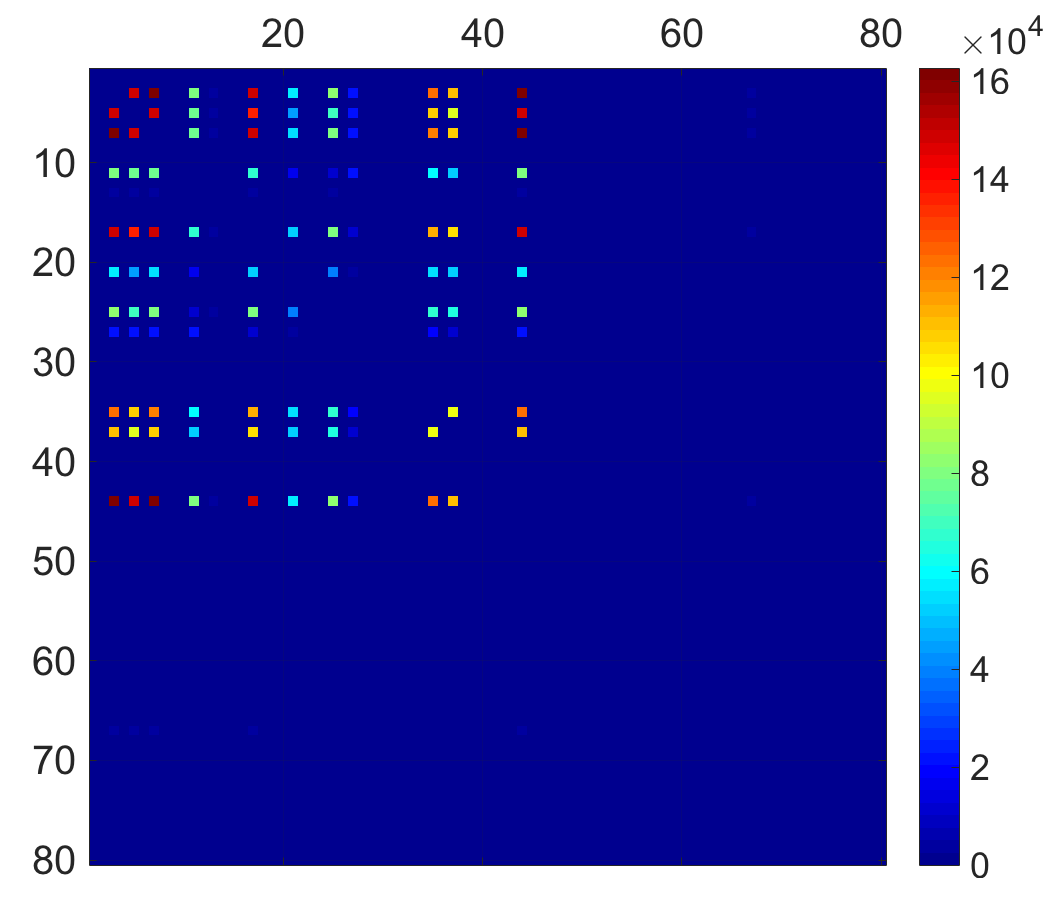} \\
		  \includegraphics[width=0.225\linewidth,height=4cm]{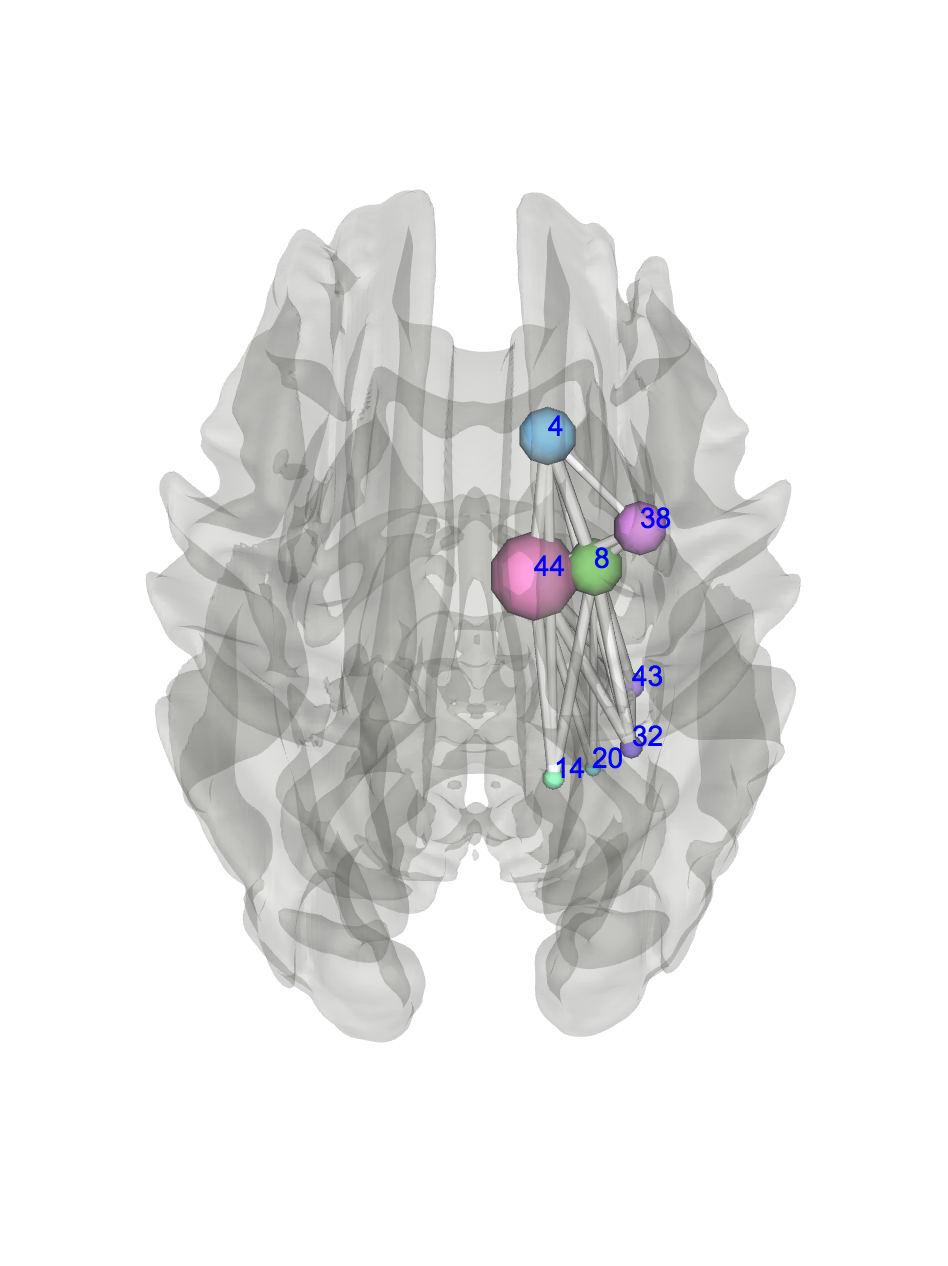} & 
		  \includegraphics[width=0.225\linewidth,height=4cm]{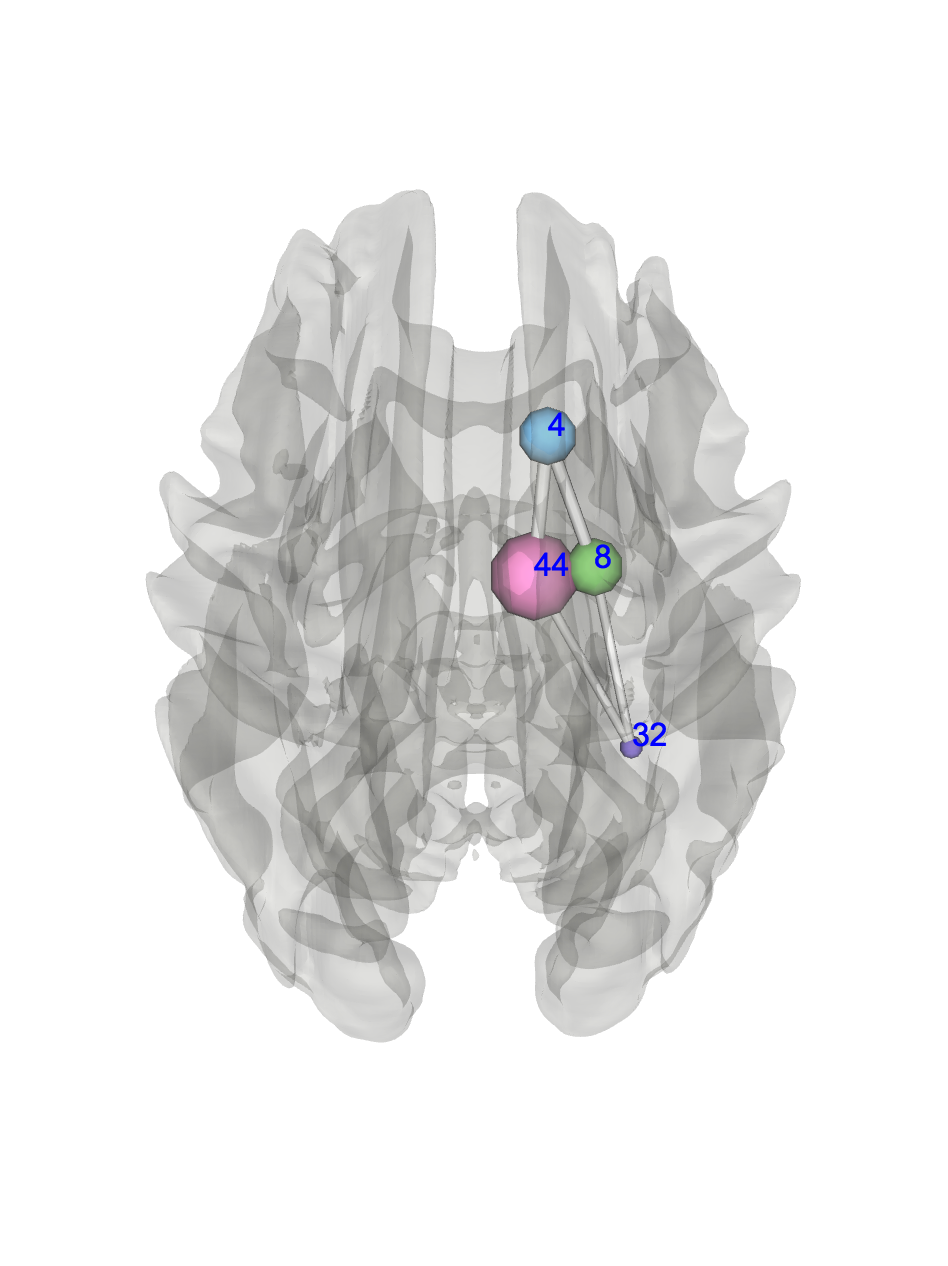} &
		   \includegraphics[width=0.225\linewidth, height=4cm]{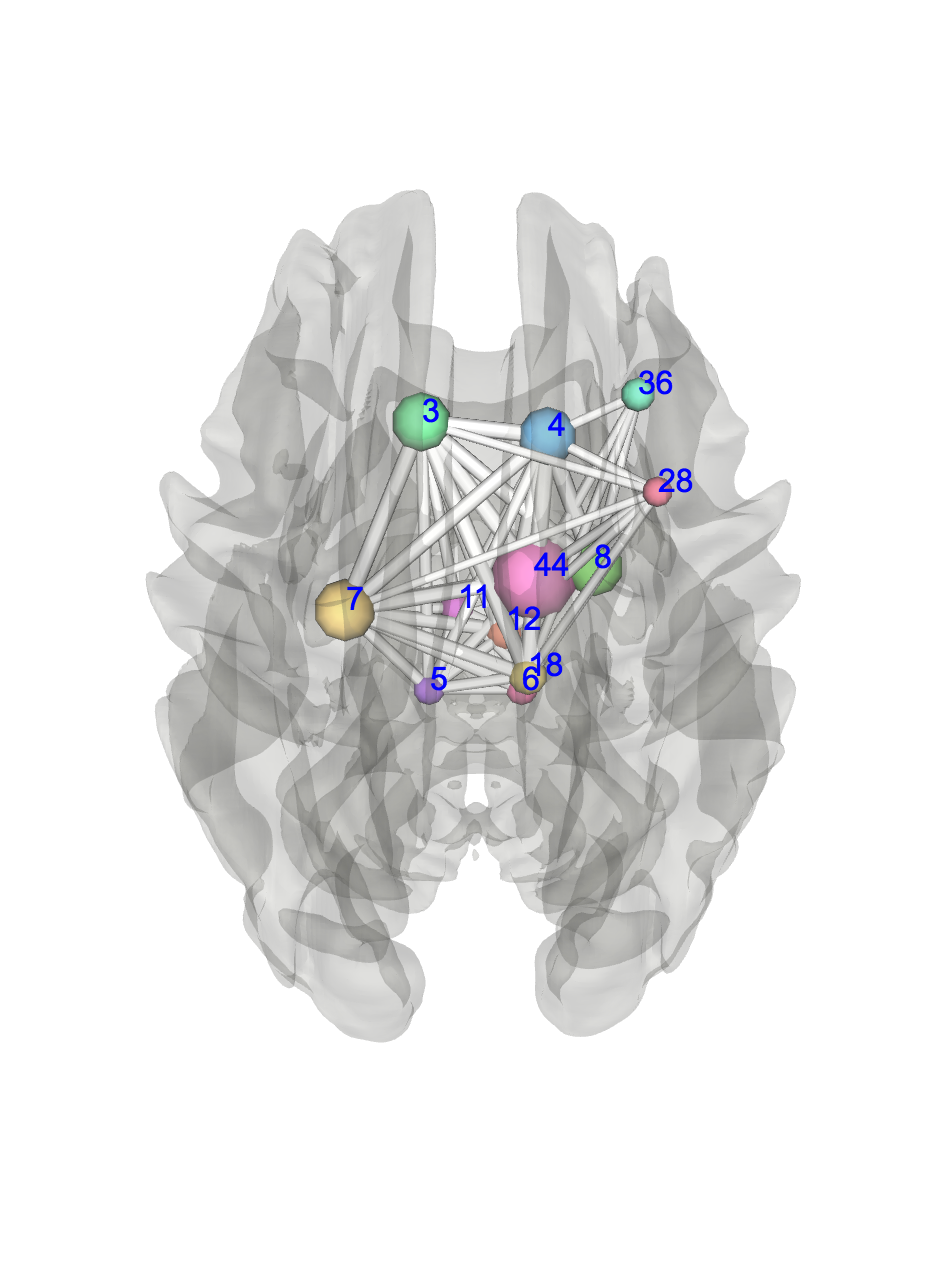} & 
		  \includegraphics[width=0.225\linewidth,height=4cm]{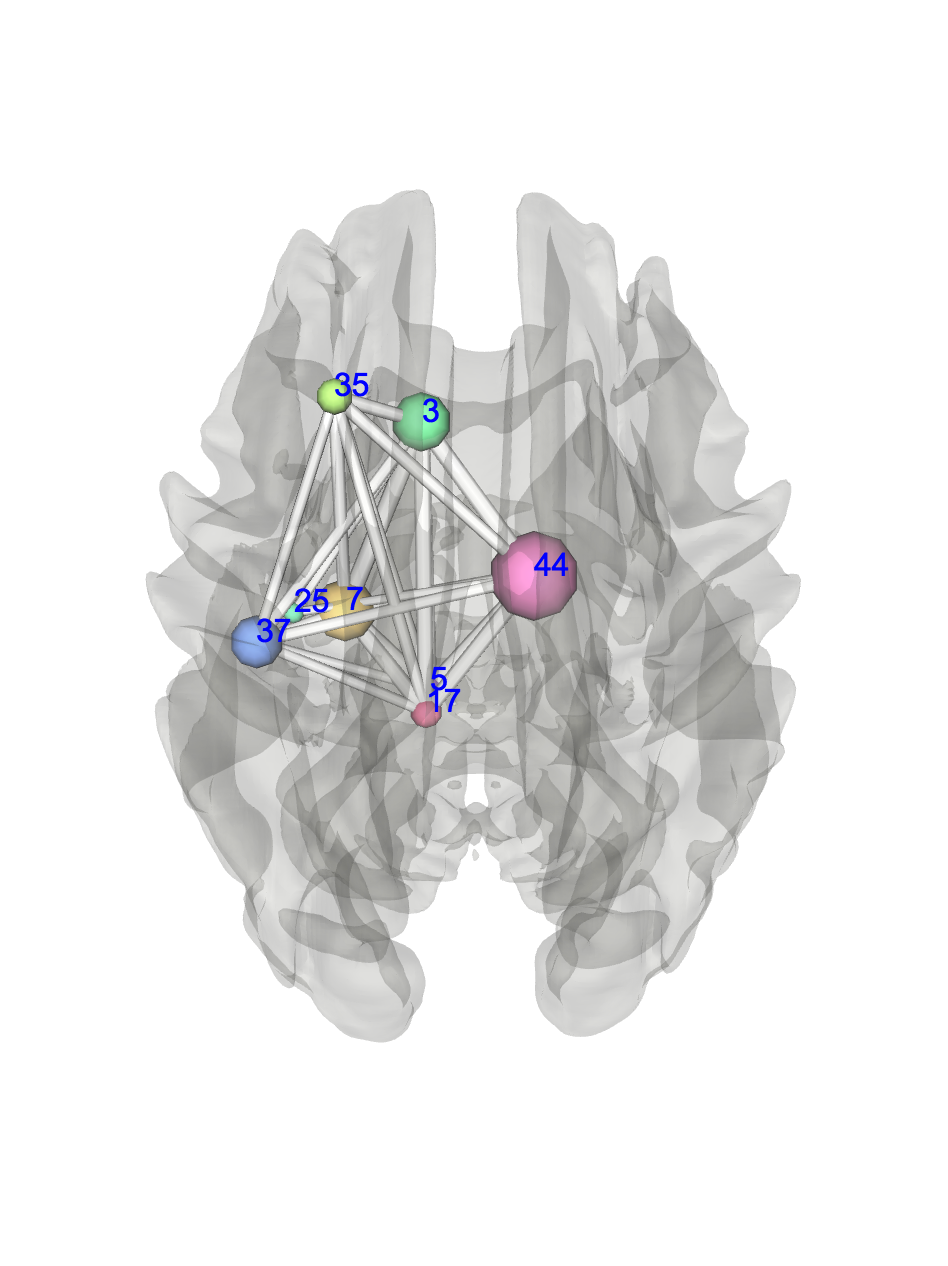} \\
		  \includegraphics[width=0.225\linewidth,height=4cm]{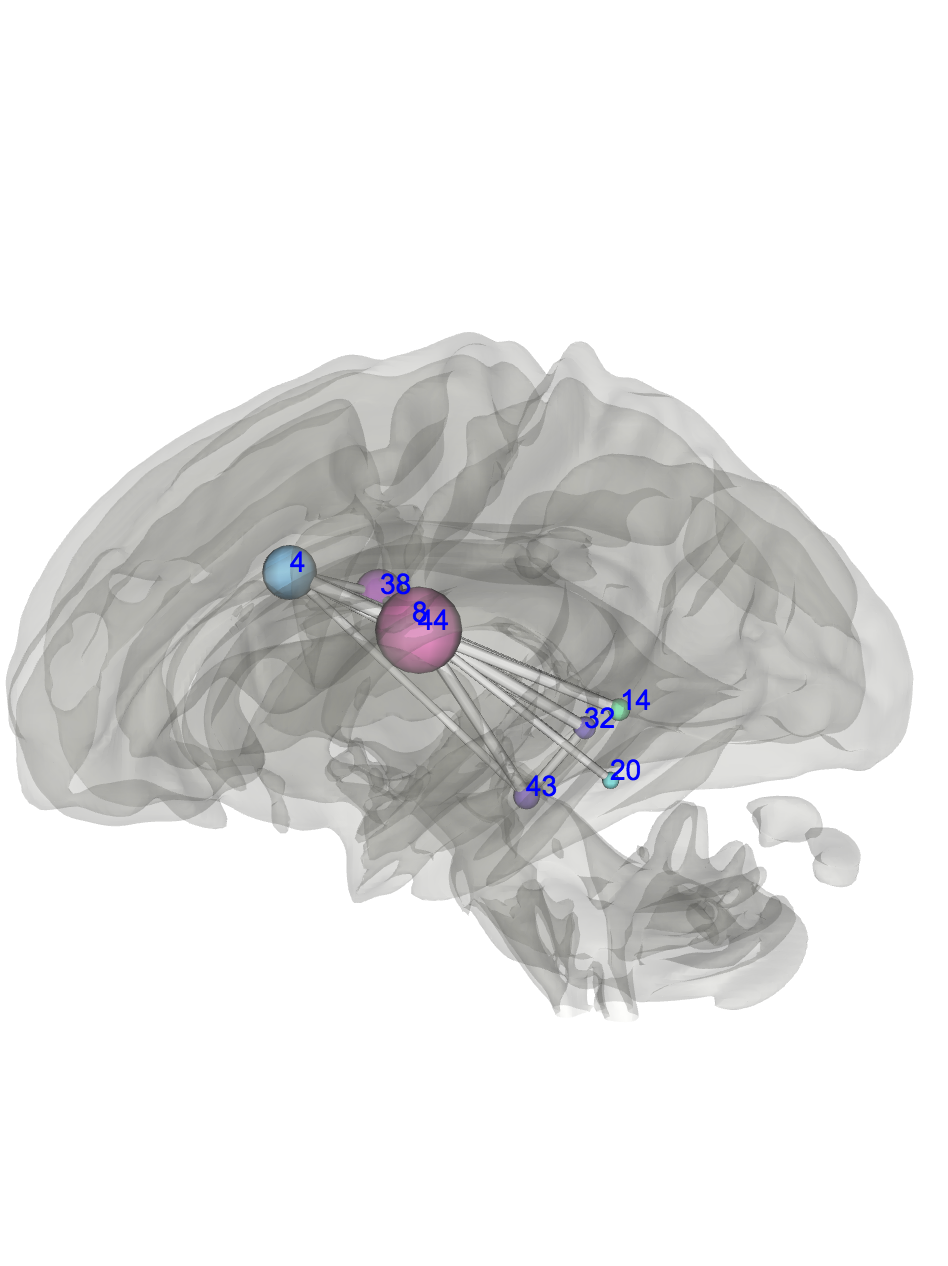} & 
		  \includegraphics[width=0.225\linewidth,height=4cm]{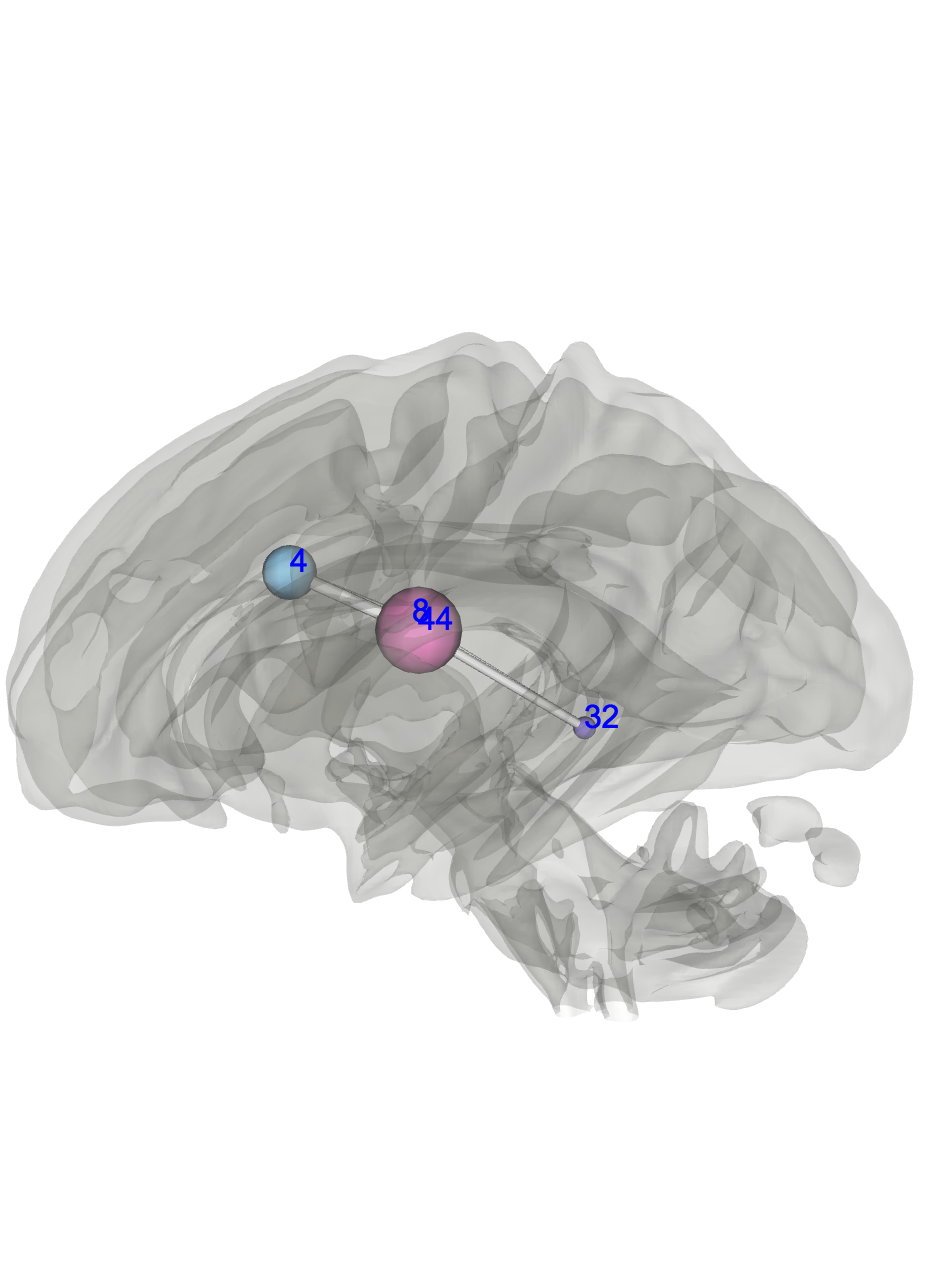} &
		   \includegraphics[width=0.225\linewidth, height=4cm]{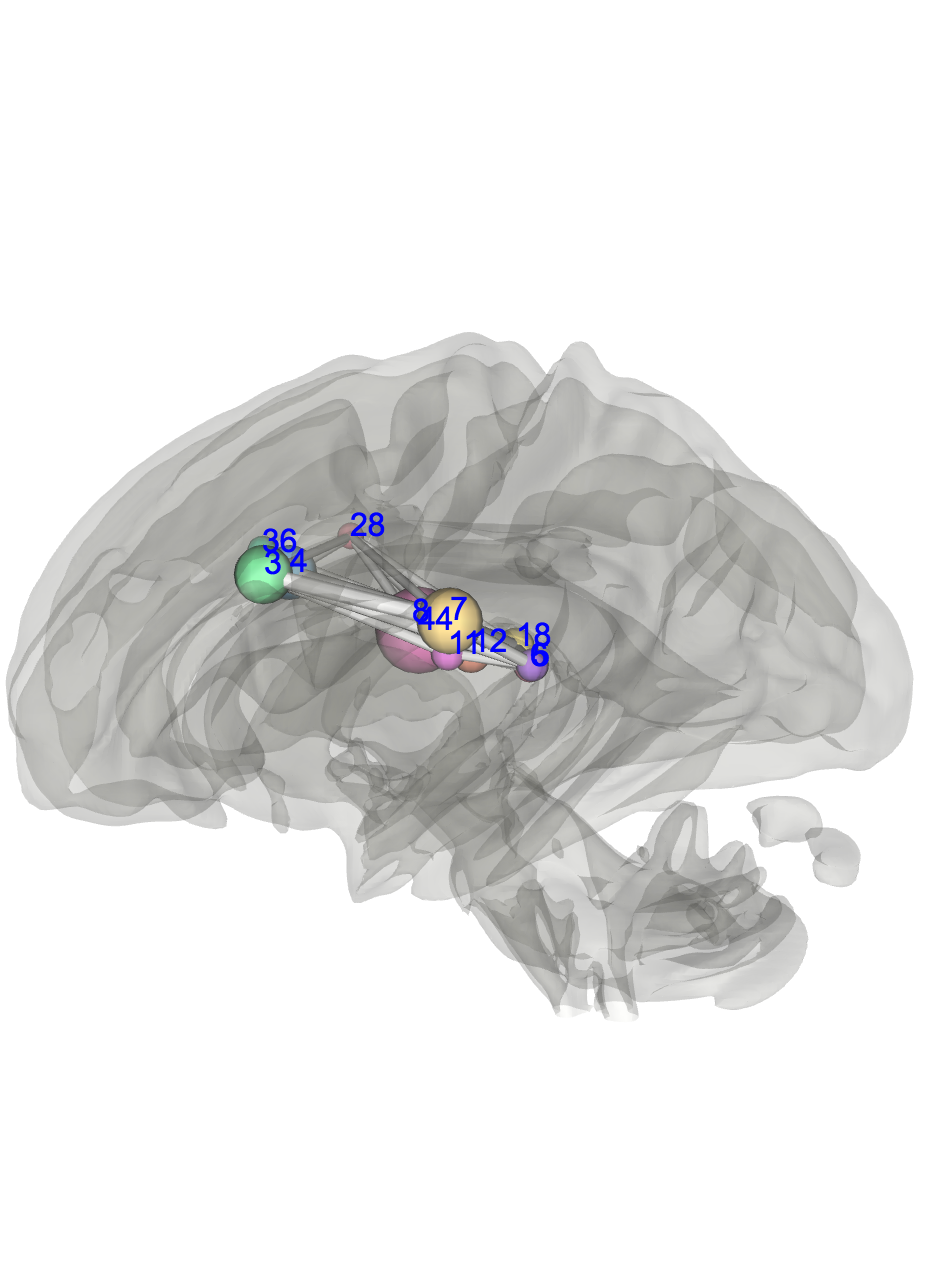} & 
		  \includegraphics[width=0.225\linewidth,height=4cm]{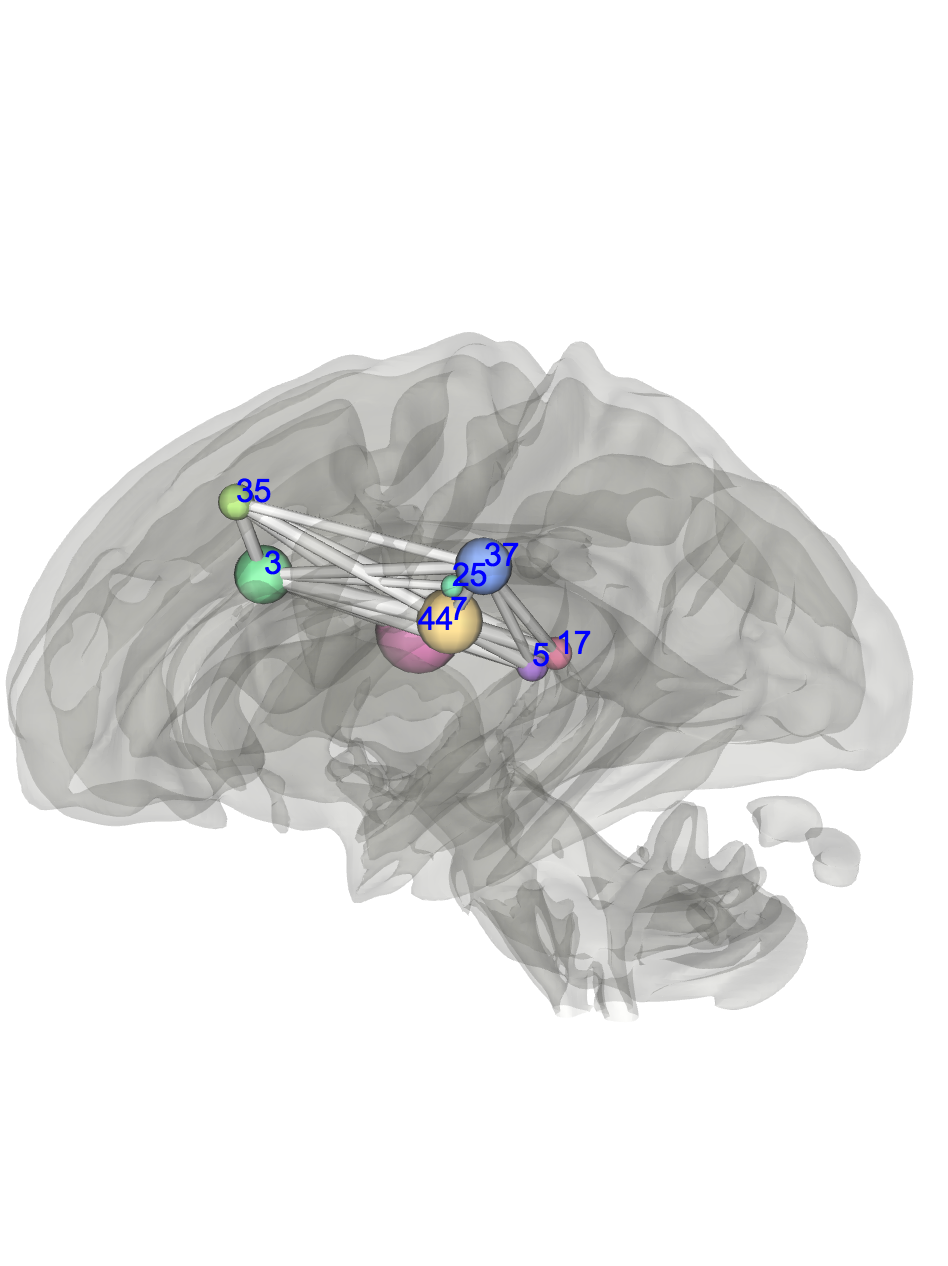} \\
		  {\small (a) $K$=50} & {\small (b) $K$=100} & {\small (c) $K$=200} & {\small (d) $K$=400}
	\end{tabular}
\caption{Visualization for trait CardSort: each column represents the visualization under different $K$ ($K$=50; $K$=100; $K$=200; $K$=400) in PPA; First row represents the connectivity matrix between any two ROI regions in HCP842 tractography atlas. Second row shows anatomy of connections in an axial view. Third row shows anatomy of connections in a sagittal view.
} 
\label{fig:cs} 
\end{figure*}

\begin{figure*}
\centering 
	\begin{tabular}{cccc}
		 \includegraphics[width=0.225\linewidth, height=3.5cm]{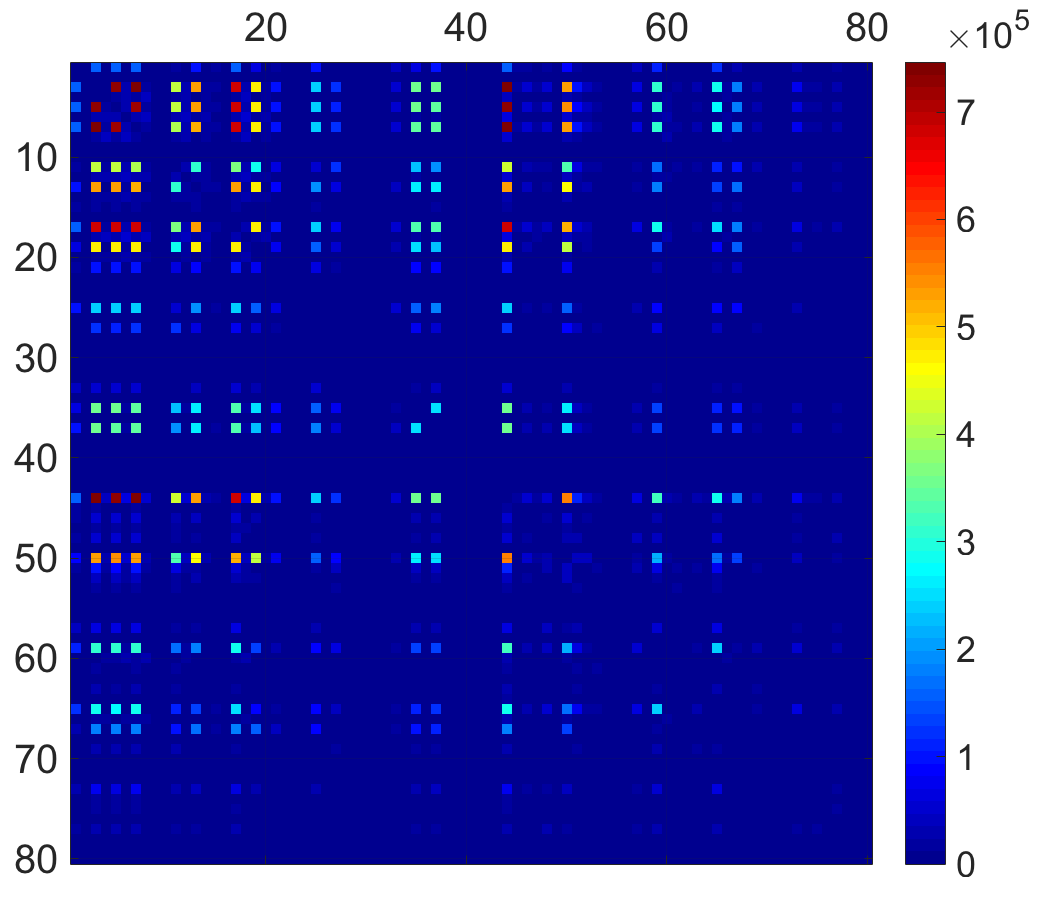} & 
		  \includegraphics[width=0.225\linewidth,height=3.5cm]{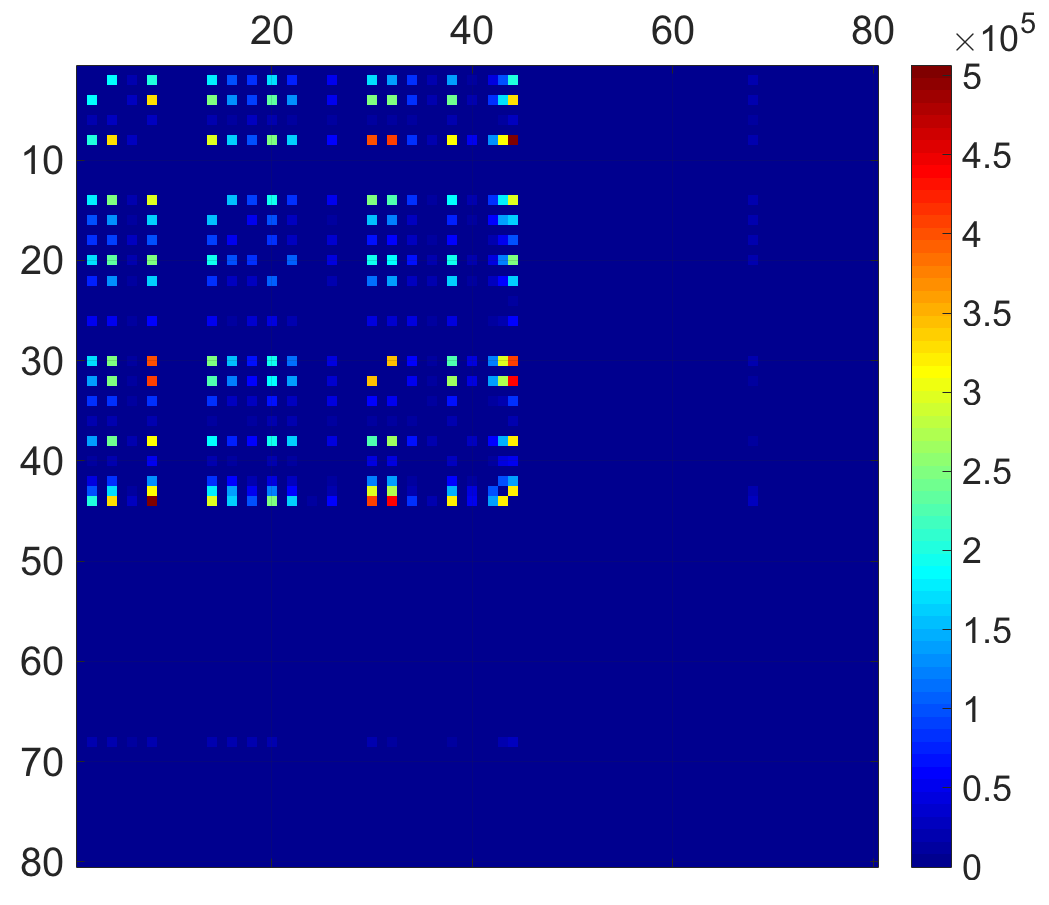} &
		   \includegraphics[width=0.225\linewidth, height=3.5cm]{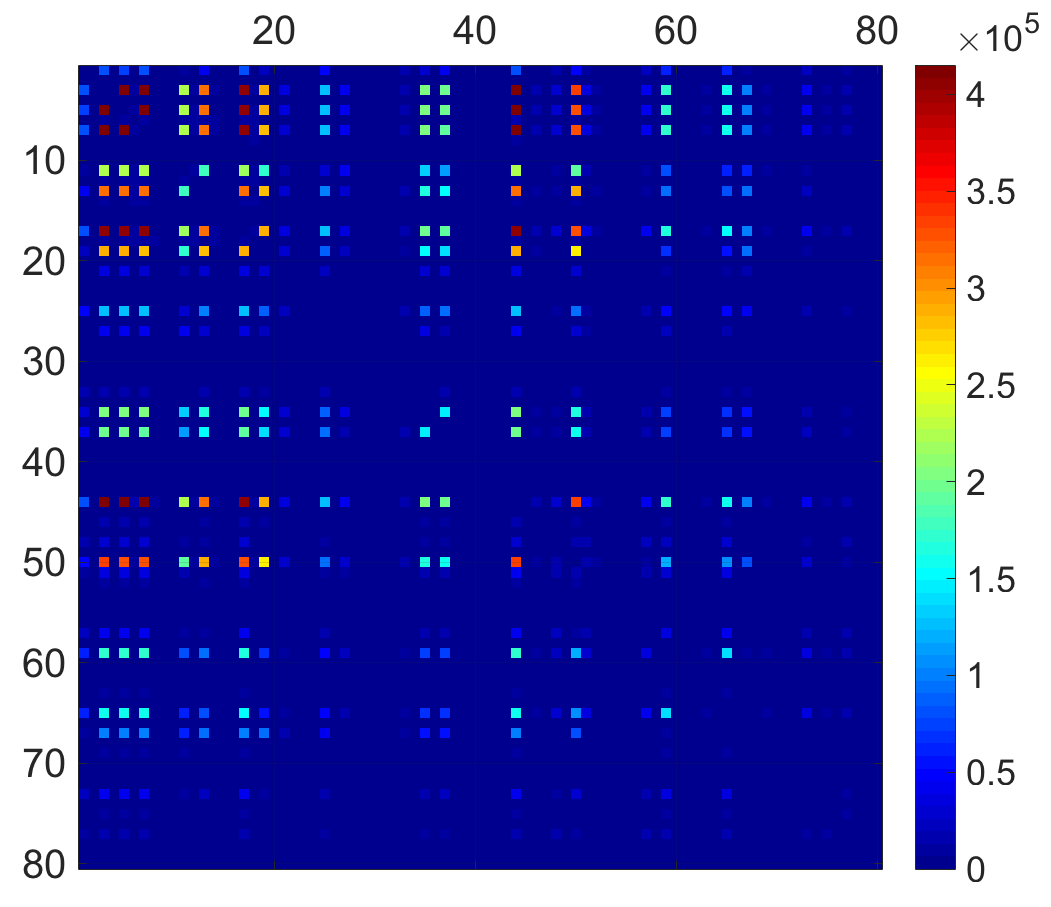} & 
		  \includegraphics[width=0.225\linewidth,height=3.5cm]{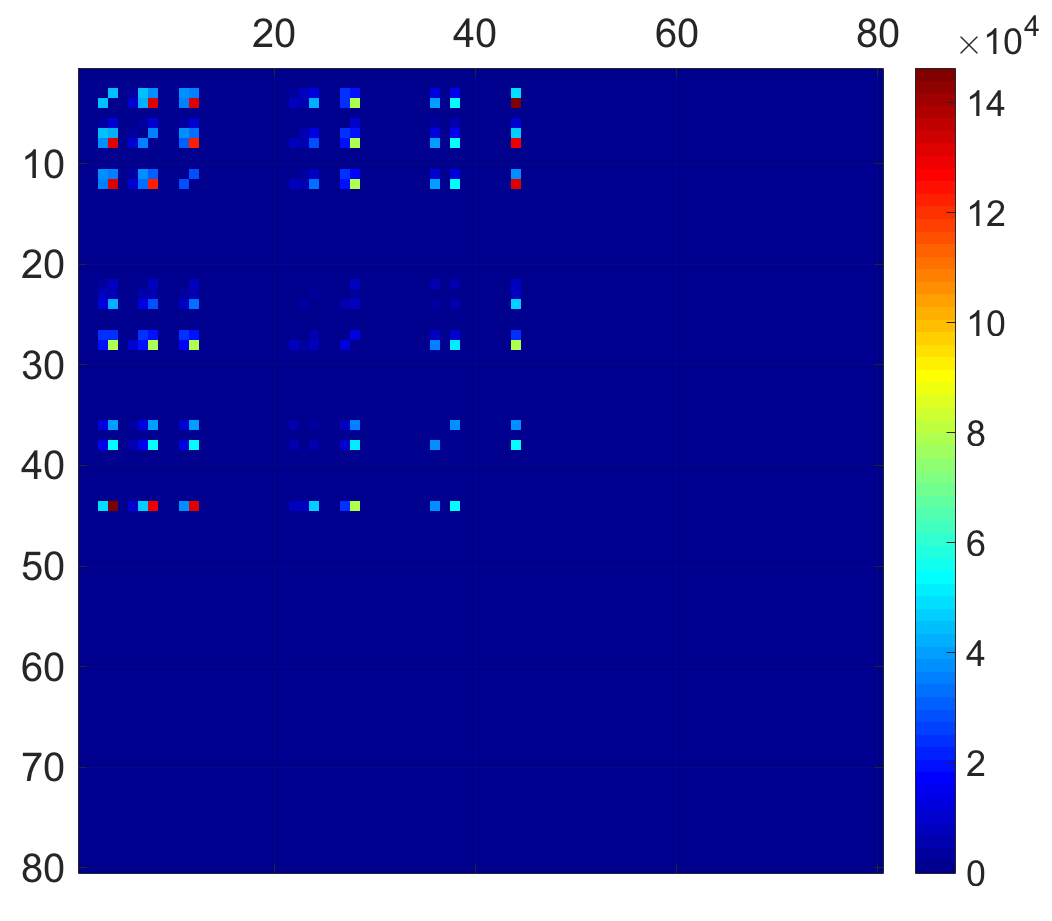} \\
		  \includegraphics[width=0.225\linewidth,height=4cm]{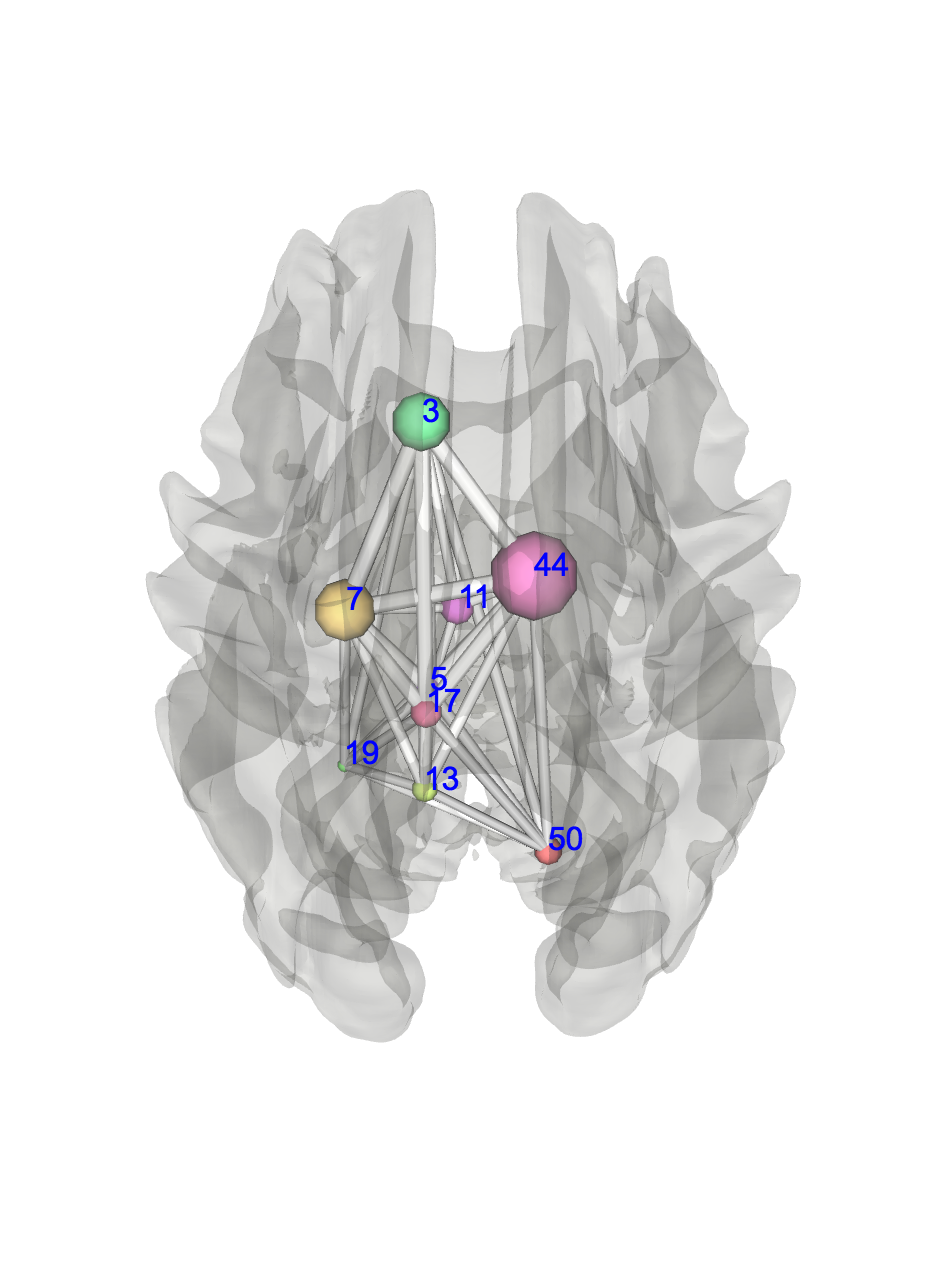} & 
		  \includegraphics[width=0.225\linewidth,height=4cm]{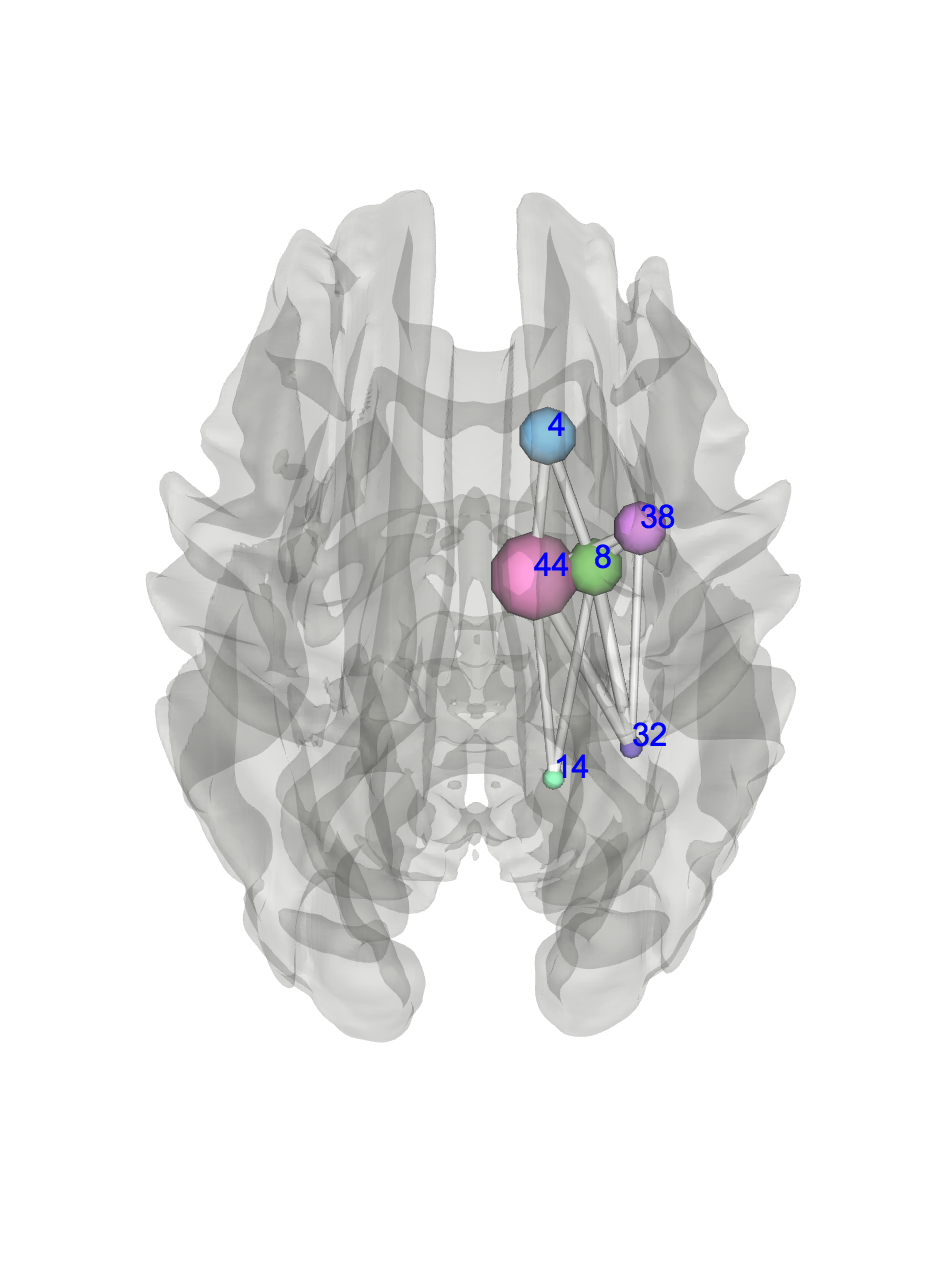} &
		   \includegraphics[width=0.225\linewidth, height=4cm]{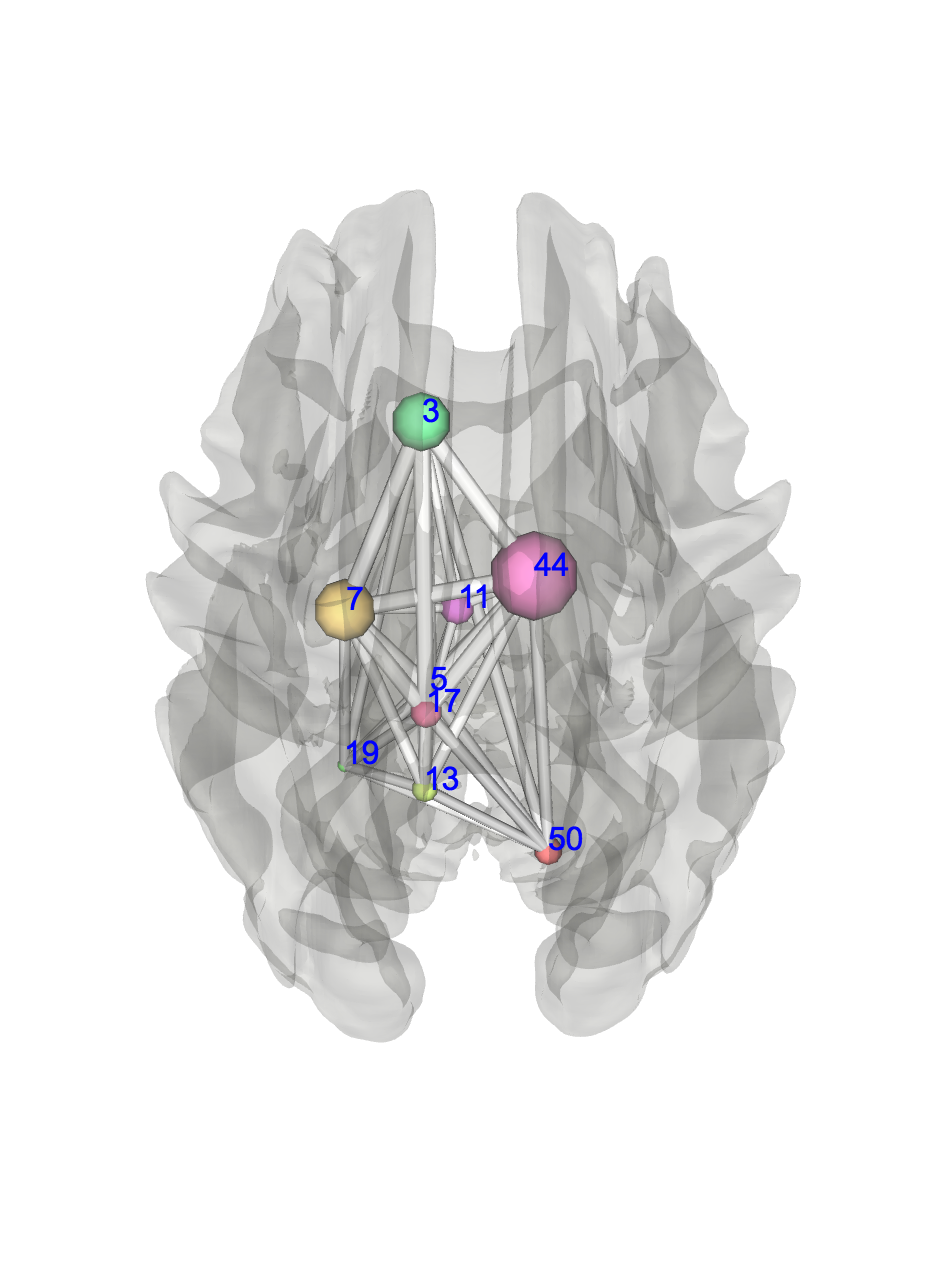} & 
		  \includegraphics[width=0.225\linewidth,height=4cm]{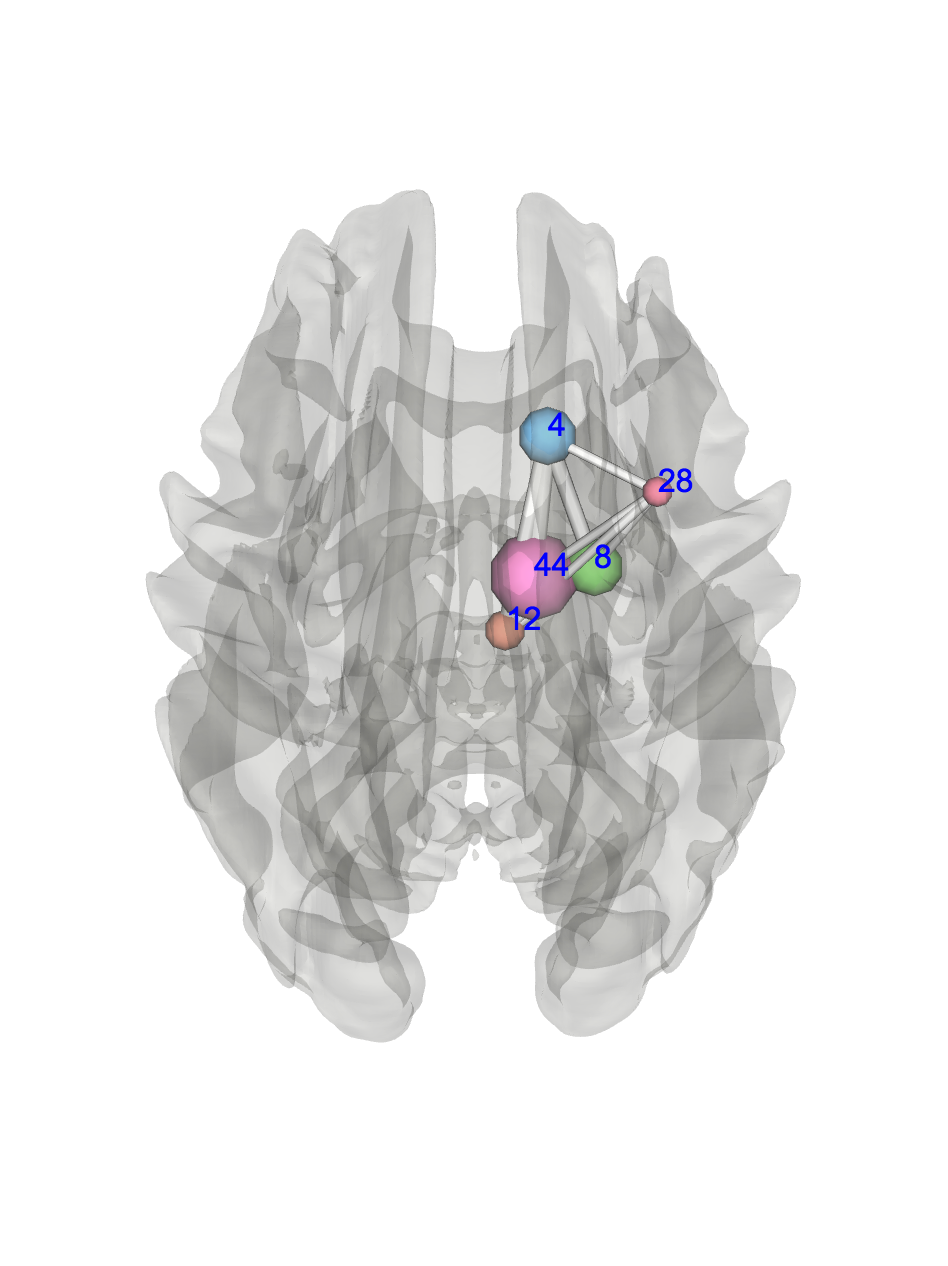} \\
		  \includegraphics[width=0.225\linewidth,height=4cm]{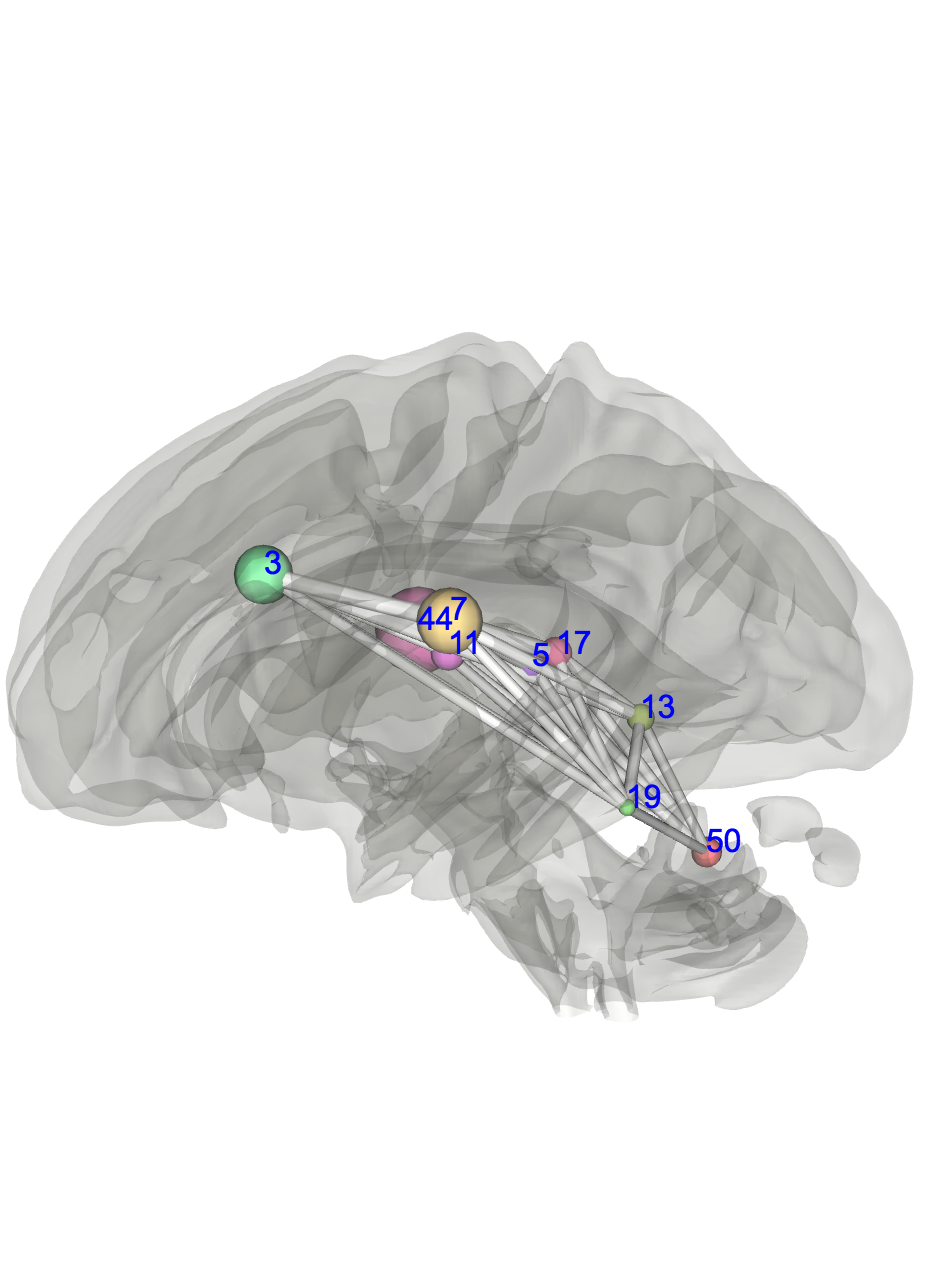} & 
		  \includegraphics[width=0.225\linewidth,height=4cm]{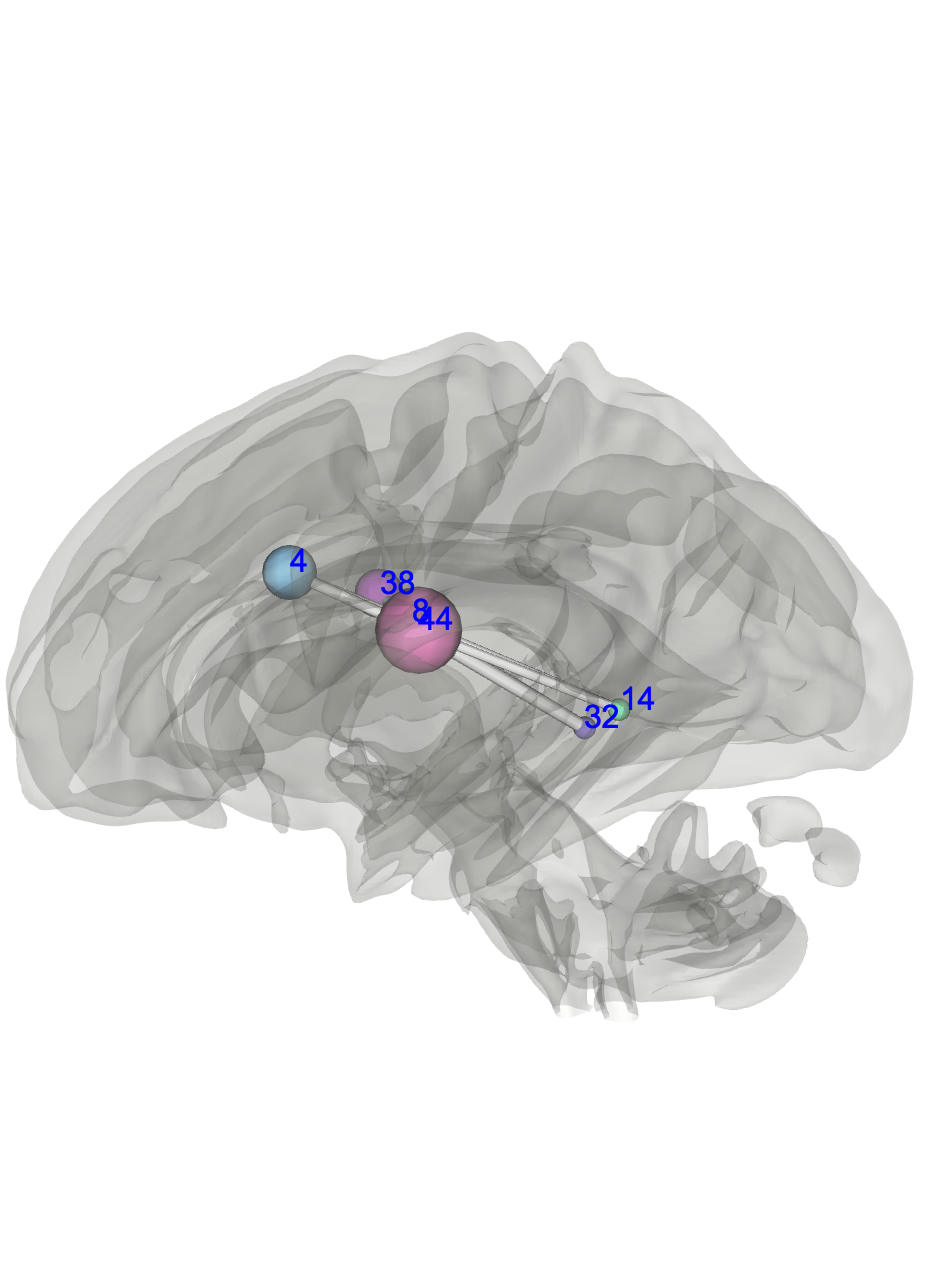} &
		   \includegraphics[width=0.225\linewidth, height=4cm]{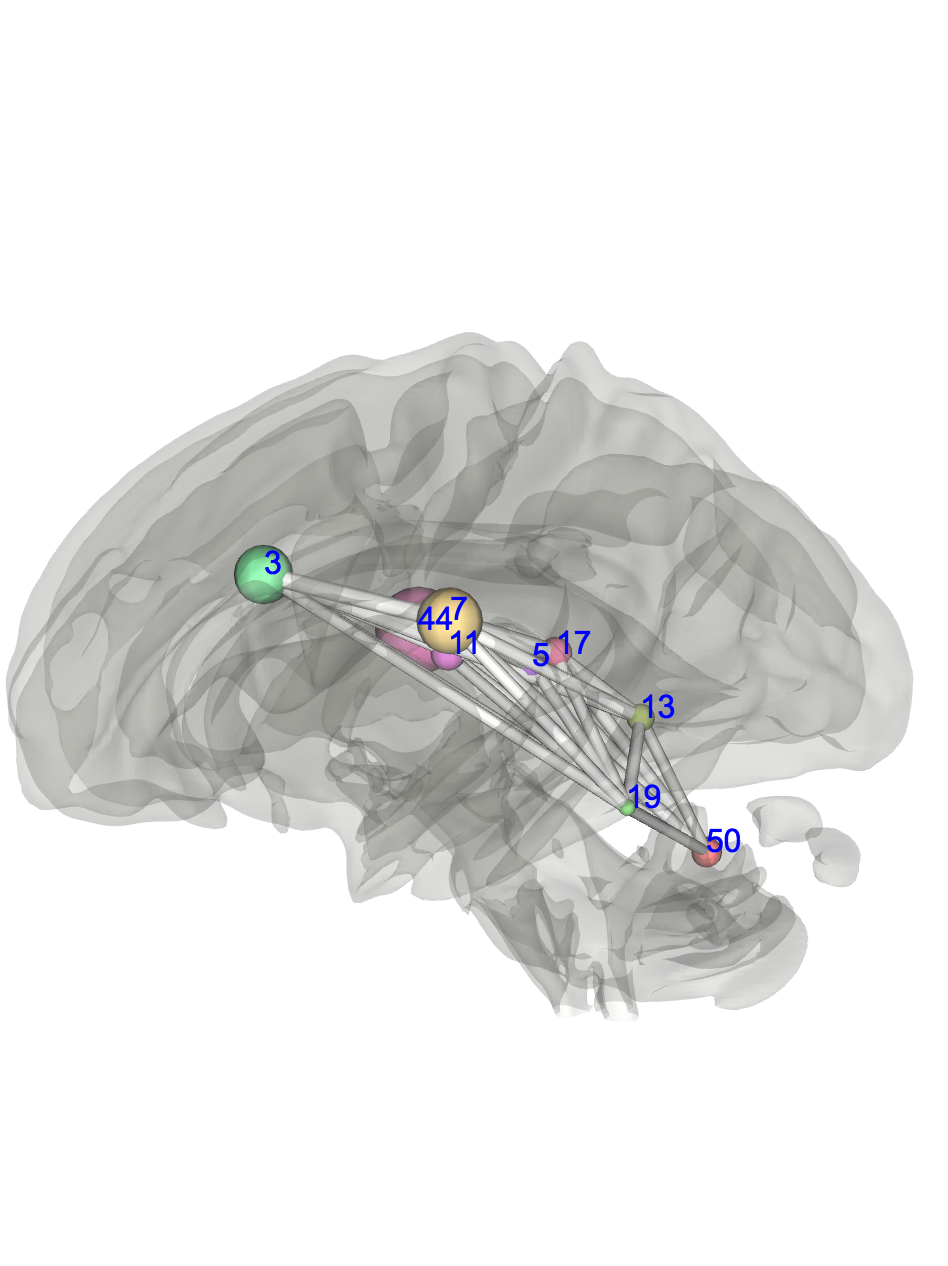} & 
		  \includegraphics[width=0.225\linewidth,height=4cm]{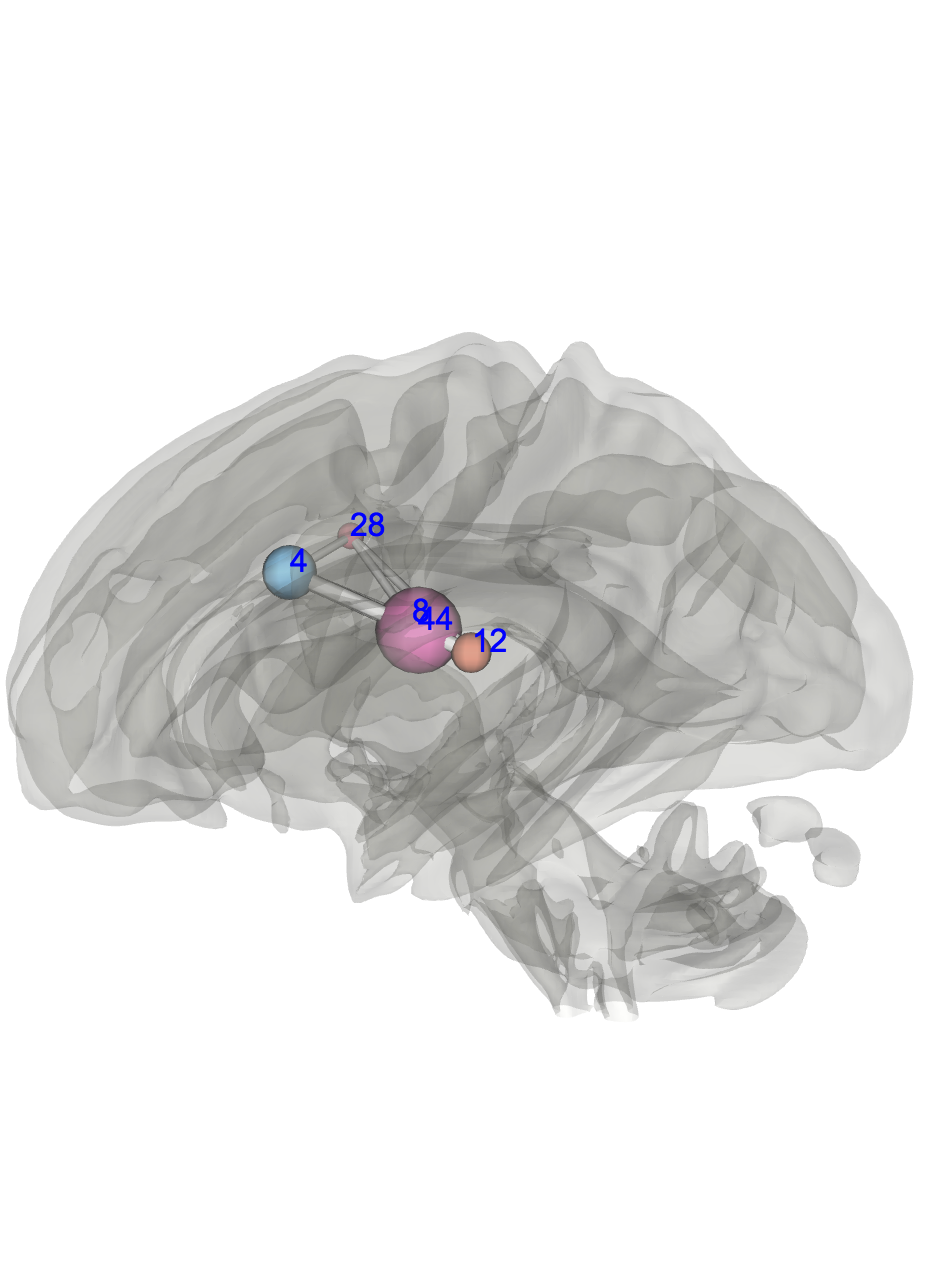} \\
		  {\small (a) $K$=50} & {\small (b) $K$=100} & {\small (c) $K$=200} & {\small (d) $K$=400}
	\end{tabular}
\caption{Visualization for trait ProcSpeed: each column represents the visualization under different $K$ ($K$=50; $K$=100; $K$=200; $K$=400) in PPA; First row represents the connectivity matrix between any two ROI regions in HCP842 tractography atlas. Second row shows anatomy of connections in an axial view. Third row shows anatomy of connections in a sagittal view.} 
\label{fig:psd} 
\end{figure*}

In this Appendix we present (1) the description of seven traits, (2) Table~\ref{table1} and Table~\ref{table2} for region names with their corresponding region numbers in HCP842 tractography atlas, 
(3) Figures \ref{compare1} and  \ref{compare2} for comparing various versions of PPA using different fiber tracking algorithms and regularization strategies, respectively,
(4) Figures \ref{compare3} and  \ref{compare4} for the predictive performance of various versions of APA using different atlases and metrics in calculating the connectivity matrix, respectively,
(5) Figure~\ref{mse_2trackings} for the comparison of 5-fold cross validation MSE of trait prediction against the number of active fibers based on PPA with the other two fiber tracking options EuDX and SFM for three traits: PicVocab, ReadEng and ListSort, and (6) Figure~\ref{fig:ps}--\ref{fig:psd} for visualization for the other three traits: PicSeq, CardSort, ProcSpeed.

The brief description of each trait is listed below. 
\begin{itemize}
	\item {\bf Receptive vocabulary score (PicVocab).} This score comes from a picture vocabulary test, in which respondents are presented with an audio recording of a word and four photographic images on the computer screen and are asked to select the picture that most closely matches the meaning of the word.
	\item {\bf Oral reading score (ReadEng).} This is from an oral reading recognition test. In this test, respondents were scored on their ability in reading and pronouncing letters and words accurately.
	\item {\bf List sorting score (ListSort).} This test assesses working memory and requires each participant to sequence different visually- and orally- presented stimuli: pictures of different foods and animals with both sound clips and written names. Concretely, participants are asked to either order items by size or report items in size order.
	\item {\bf Flanker score (Flanker).} This test measures both a subject's attention and inhibitory control. In this test, participants are asked to focus on a given stimulus while inhibiting attention to stimuli flanking it. Sometimes the middle stimulus is pointing in the same direction as the flankers (congruent) and sometimes in the opposite direction (incongruent).
	\item{\bf Picture sequence memory score (PicSeq).} This test is a measure developed for the assessment of episodic memory. It involves recalling increasingly lengthy series of illustrated objects and activities that are presented in a particular order on the computer screen. The participants are asked to recall the sequence of pictures that is demonstrated over two learning trials, where sequence length varies from 6-18 pictures, depending on age. 
	\item{\bf Card sort score (CardSort).} This test is a measure of cognitive flexibility. Two target pictures are presented that vary along two dimensions (e.g., shape and color). Participants are asked to match a series of bivalent test pictures to the target pictures, first according to one dimension and then, after a number of trials, according to the other dimension. 
	\item{\bf Processing speed score (ProcSpeed).} This test measures speed of processing by asking participants to discern whether two side-by-side pictures are the same or not. Participants' raw score is the number of items correct in a 90-second period. The items are designed to be simple to most purely measure processing speed. 
\end{itemize}

\newpage
 \begin{table}[H]\centering
\footnotesize
 \ra{1.3}
 \begin{tabular}{|cc||cc|}
\hline
\rowcolor{black!50}              
    No.       & Region Name & No. & Region Name       \\ \hline
 \rowcolor{black!20}1  & Acoustic\_Radiation\_L& 2 & Acoustic\_Radiation\_R \\
 3 & Cortico\_Striatal\_Pathway\_L & 4 & Cortico\_Striatal\_Pathway\_R \\
 \rowcolor{black!20} 5 & Cortico\_Spinal\_Tract\_L& 6 & Cortico\_Spinal\_Tract\_R \\
 7& Corticothalamic\_Pathway\_L & 8 & Corticothalamic\_Pathway\_R \\
 \rowcolor{black!20} 9& Fornix\_L& 10 & Fornix\_R \\
11 & Frontopontine\_Tract\_L & 12 & Frontopontine\_Tract\_R \\
 \rowcolor{black!20} 13 & Occipitopontine\_Tract\_L & 14 & Occipitopontine\_Tract\_R \\
  15& Optic\_Radiation\_L & 16 & Optic\_Radiation\_R \\
 \rowcolor{black!20} 17 & Parietopontine\_Tract\_L& 18& Parietopontine\_Tract\_R \\
  19 & Temporopontine\_Tract\_L & 20 & Temporopontine\_Tract\_R \\
   \rowcolor{black!20}21  & Arcuate\_Fasciculus\_L& 22 & Arcuate\_Fasciculus\_R \\
 23 & Cingulum\_L & 24 & Cingulum\_R \\
 \rowcolor{black!20} 25 & Extreme\_Capsule\_L& 26 & Extreme\_Capsule\_R \\
 27& Frontal\_Aslant\_Tract\_L& 28 & Frontal\_Aslant\_Tract\_R \\
 \rowcolor{black!20} 29& Inferior\_Fronto\_Occipital\_Fasciculus\_L& 30 & Inferior\_Fronto\_Occipital\_Fasciculus\_R \\
31 & Inferior\_Longitudinal\_Fasciculus\_L & 32 & Inferior\_Longitudinal\_Fasciculus\_R \\
 \rowcolor{black!20} 33 & Middle\_Longitudinal\_Fasciculus\_L & 34 & Middle\_Longitudinal\_Fasciculus\_R \\
  35& Superior\_Longitudinal\_Fasciculus\_L & 36 & Superior\_Longitudinal\_Fasciculus\_R \\
 \rowcolor{black!20} 37 & U\_Fiber\_L& 38& U\_Fiber\_R \\
  39 & Uncinate\_Fasciculus\_L & 40 & Uncinate\_Fasciculus\_R \\
 \bottomrule
 \end{tabular}
 \caption{80 regions' names in HCP842 tractography atlas}
 \label{table1}
 \end{table}

  \begin{table}[H]\centering
\footnotesize
 \ra{1.3}
 \begin{tabular}{|cc||cc|}
\hline
\rowcolor{black!50}              
    No.       & Region Name & No. & Region Name       \\ \hline
 \rowcolor{black!20}41  & Vertical\_Occipital\_Fasciculus\_L& 42 & Vertical\_Occipital\_Fasciculus\_R \\
 43 & Anterior\_Commissure & 44 & Corpus\_Callosum \\
 \rowcolor{black!20} 45 & Posterior\_Commissure& 46 & Cerebellum\_L \\
 47& Cerebellum\_R & 48 & Inferior\_Cerebellar\_Peduncle\_L \\
 \rowcolor{black!20} 49& Inferior\_Cerebellar\_Peduncle\_R& 50 & Middle\_Cerebellar\_Peduncle\\
51 & Superior\_Cerebellar\_Peduncle & 52 & Vermis \\
 \rowcolor{black!20} 53 & Central\_Tegmental\_Tract\_L & 54 & Central\_Tegmental\_Tract\_R \\
  55& Dorsal\_Longitudinal\_Fasciculus\_L& 56 & Dorsal\_Longitudinal\_Fasciculus\_R \\
 \rowcolor{black!20} 57 & Lateral\_Lemniscus\_L& 58& Lateral\_Lemniscus\_R \\
  59 & Medial\_Lemniscus\_L & 60 & Medial\_Lemniscus\_R \\
   \rowcolor{black!20}61  & Medial\_Longitudinal\_Fasciculus\_L& 62 & Medial\_Longitudinal\_Fasciculus\_R \\
 63 & Rubrospinal\_Tract\_L & 64 & Rubrospinal\_Tract\_R \\
 \rowcolor{black!20} 65 & Spinothalamic\_Tract\_L& 66 & Spinothalamic\_Tract\_R \\
 67& CNII\_L& 68 & CNII\_R \\
 \rowcolor{black!20} 69& CNIII\_L& 70 & CNIII\_R \\
71 & CNIV\_L& 72 & CNIV\_R \\
 \rowcolor{black!20} 73 & CNV\_L & 74 & CNV\_R \\
  75& CNVII\_L & 76 & CNVII\_R \\
 \rowcolor{black!20} 77 & CNVIII\_L& 78& CNVIII\_R \\
  79 & CNX\_L & 80 & CNX\_R \\
 \bottomrule
 \end{tabular}
 \caption{cont. 80 regions' names in HCP842 tractography atlas}
 \label{table2}
 \end{table}

\newpage

\newpage

\newpage

\end{document}